\documentclass{article}
\usepackage{epubtk}
\usepackage{graphicx}
\usepackage{amsmath}
\usepackage{booktabs}
\usepackage{color}

\newcommand{\msolar}{\,M_{\odot}}

\newcommand{\mptr}{{\cal M}}
\newcommand{\fracparen}[2]{\left(\frac{#1}{#2}\right)}
\newcommand{\qddd}{\mbox{$\stackrel{\ldots}{\mbox{\it Q}}$}}

\newcommand{\oderiv}[2]{\frac{d #1}{d #2}}
\newcommand{\oderivn}[3]{\frac{d^{#3}\!#1}{d #2^{#3}}}

\newcommand{\oderivnf}[3]{{d^{#3}\!#1}\!/{d #2^{#3}}}

\newcommand{\E}[1]{\times 10^{#1}}
\newcommand{\Eqref}[1]{Equation~(\ref{#1})}
\newcommand{\Eqnref}[1]{Equation~(\ref{#1})}
\newcommand{\Figref}[1]{Figure~\ref{#1}}

\newcommand{\Secref}[1]{Section~\ref{#1}}
\newcommand{\hide}[1]{}\newcommand{\units}{\rm\;}
\newcommand{\exhat}{\hat{e}_x}
\newcommand{\eyhat}{\hat{e}_y}

\newcommand{\htensor}{{\mathbf h}}
\newcommand{\dtensor}{{\mathbf d}}
\newcommand{\eptensor}{{\mathbf e}_+}
\newcommand{\extensor}{{\mathbf e}_\times}

\def\gsim{\mathrel{
\rlap{\raise 0.511ex \hbox{$>$}}{\lower 0.511ex
\hbox{$\sim$}}}}

\def\lsim{\mathrel{
\rlap{\raise 0.511ex \hbox{$<$}}{\lower 0.511ex
\hbox{$\sim$}}}}

\def\pbar{\overline{p}}

\newcommand{\textsub}[1]{{}_{\text{#1}}}
\newcommand{\textsup}[1]{{}^{\text{#1}}}
\newcommand{\rhogw}{\rho\textsub{gw}}
\newcommand{\Omegagw}{\Omega\textsub{gw}}

\newcommand{\ahat}{\hat{a}}

\newcommand{\Nhat}{\hat{N}}
\newcommand{\exRhat}{\hat{e}_x^R}
\newcommand{\eyRhat}{\hat{e}_y^R}
\newcommand{\alphahat}{\hat{\mathbf\alpha}}
\newcommand{\betahat}{\hat{\mathbf\beta}}
\newcommand{\epsptensor}{{\mathbf\epsilon}_+}
\newcommand{\epsxtensor}{{\mathbf\epsilon}_\times}

\begin{document}

\title{Physics, Astrophysics and Cosmology \\with Gravitational Waves}

\author{%
\epubtkAuthorData{B.S.\ Sathyaprakash}
		 {School of Physics and Astronomy, Cardiff University,\\
		   Cardiff, U.K.}
		 {B.Sathyaprakash@astro.cf.ac.uk}
		 {}
\\
\and \\
\epubtkAuthorData{Bernard F.\ Schutz}
		 {School of Physics and Astronomy, Cardiff University,\\
		   Cardiff, U.K.\\
		 and\\
		 Max Planck Institute for Gravitational Physics\\
		 (Albert Einstein Institute)\\
		 Potsdam-Golm, Germany}
		 {Bernard.Schutz@aei.mpg.de}
		 {}
}

\date{}
\maketitle

\begin{abstract}
Gravitational wave detectors are already operating at interesting
sensitivity levels, and they have an upgrade path that should result
in secure detections by 2014. We review the physics of gravitational
waves, how they interact with detectors (bars and interferometers),
and how these detectors operate. We study the most likely sources of
gravitational waves and review the data analysis methods that are used
to extract their signals from detector noise. Then we consider the
consequences of gravitational wave detections and observations for
physics, astrophysics, and cosmology.
\end{abstract}

\epubtkKeywords{Gravitational waves, Gravitational wave sources,
  Gravitational wave detectors, Data analysis}

%==================================================================
%FS for arxiv
\newpage
\tableofcontents

%==================================================================
\newpage

\section{A New Window onto the Universe}

The last six decades have witnessed a great revolution in astronomy,
driven by improvements in observing capabilities across the
electromagnetic spectrum: very large optical telescopes, radio
antennas and arrays, a host of satellites to explore the infrared,
X-ray, and gamma-ray parts of the spectrum, and the development of key
new technologies (CCDs, adaptive optics). Each new window of
observation has brought new surprises that have dramatically changed
our understanding of the universe. These serendipitous discoveries
have included:

\begin{itemize}

\item the relic cosmic microwave background radiation (Penzias and
  Wilson~\cite{PENREF}), which has become our primary tool for
  exploring the Big Bang;

\item the fact that quasi-stellar objects are at cosmological
  distances (Maarten Schmidt~\cite{SCHMIDTREF}), which has developed
  into the understanding that they are powered by supermassive black
  holes;

\item pulsars (Hewish and Bell~\cite{BELLREF}), which opened up the
  study of neutron stars and illuminated one endpoint for stellar
  evolution;

\item X-ray binary systems (Giacconi and
  collaborators~\cite{GIACREF}), which now enable us to make detailed
  studies of black holes and neutron stars;

\item gamma-ray bursts coming from immense distances (Klebesadel et
  al.~\cite{GAMMAREF}), which are not fully explained even today;

\item the fact that the expansion of the universe is accelerating (two
  teams~\cite{SNTEAMS1, SNTEAMS2}), which has led to the hunt for the
  nature of dark energy.

\end{itemize}

None of these discoveries was anticipated by the observing team, and in many
cases the instruments were built to observe completely different phenomena.

Within a few years another new window on the universe will open up, with the
first direct detection of gravitational waves. There is keen interest in
observing gravitational waves directly, in order to test Einstein's theory of
general relativity and to observe some of the most exotic objects in nature,
particularly black holes. But, in addition, the potential of gravitational wave
observations to produce more surprises is very high.

The gravitational wave spectrum is completely distinct from, and complementary
to, the electromagnetic spectrum. The primary emitters of electromagnetic
radiation are charged elementary particles, mainly electrons; because of overall
charge neutrality, electromagnetic radiation is typically emitted in small
regions, with short wavelengths, and conveys direct information about the
physical conditions of small portions of the astronomical sources. By contrast,
gravitational waves are emitted by the cumulative mass and momentum of entire
systems, so they have long wavelengths and convey direct information about
large-scale regions. Electromagnetic waves couple strongly to charges and so are
easy to detect but are also easily scattered or absorbed by material between us
and the source; gravitational waves couple extremely weakly to matter, making
them very hard to detect but also allowing them to travel to us substantially
unaffected by intervening matter, even from the earliest moments of the Big
Bang.

These contrasts, and the history of serendipitous discovery in astronomy, all
suggest that electromagnetic observations may be poor predictors of the
phenomena that gravitational wave detectors will eventually discover. Given that
96\% of the mass-energy of the universe carries no charge, gravitational waves
provide us with our first opportunity to observe directly a major part of the
universe. It might turn out to be as complex and interesting as the charged
minor component, the part that we call ``normal'' matter.

Several long-baseline interferometric gravitational-wave detectors planned 
over a decade ago [Laser Interferometer Gravitational-Wave Observatory
(LIGO)~\cite{bss:ligo}, GEO~\cite{Luck:1997hv}, 
VIRGO~\cite{bss:virgo} and TAMA~\cite{1997gwd..conf..183T}]
have begun initial operations~\cite{Abbott:2003vs, Luck:2006ug, Acernese:2006bj} 
with unprecedented sensitivity levels and wide bandwidths at acoustic frequencies
(10~Hz\,--\,10~kHz)~\cite{HOUGHREF}. These large
interferometers are superseding a world-wide network of narrow-band resonant bar
antennas that operated for several decades at frequencies near
1~kHz. Before 2020 the space-based LISA~\cite{bss:lisa} gravitational wave
detector may begin observations in the low-frequency band from 0.1~mHz to
0.1~Hz. This suite of detectors can be expected to open up the gravitational
wave window for astronomical exploration, and at the same time perform stringent
tests of general relativity in its strong-field dynamic sector.

Gravitational wave antennas are essentially omni-directional, with linearly
polarized quadrupolar antenna patterns that typically have a response better
than 50\% of its average over 75\% of the sky~\cite{HOUGHREF}. Their nearly all-sky
sensitivity is an important difference from pointed astronomical antennas and
telescopes. Gravitational wave antennas operate as a network, with the aim of
taking data continuously. Ground-based interferometers can at present
(2008) survey a
volume of order $10^4\;\text{Mpc}^3$ for inspiraling compact star binaries
-- among the most promising sources for these detectors -- and plan to enhance
their range more than tenfold with two major upgrades (to \emph{enhanced} and
then \emph{advanced} detectors) during the period 2009\,--\,2014. For the advanced
detectors, there is great confidence that the resulting thousandfold volume
increase will produce regular detections. It is this second phase of operation 
that will be more interesting from the astrophysical point of view,
bringing us physical and astrophysical insights into populations of
neutron star and black hole binaries, supernovae and formation of
compact objects, populations of isolated compact objects in our galaxy, 
and potentially even completely unexpected systems. Following
that, LISA's ability to survey the entire universe for black hole
coalescences at milliHertz frequencies will extend gravitational wave
astronomy into the cosmological arena.

However, the present initial phase of observation, or observations
after the first enhancements, may very well produce the first detections.
Potential sources include coalescences of binaries consisting of black holes
at a distance of 100\,--\,200~Mpc and spinning neutron stars in our galaxy
with ellipticities greater than about $10^{-6}$. Observations even at
this initial level may, of course, also reveal new sources not
observable in any other way. These initial detections, though
not expected to be frequent, would be important from the fundamental
physics point of view and could enable us to directly test general relativity
in the strongly nonlinear regime.

Gravitational wave detectors register gravitational waves coherently
by following the phase of the wave and not just measuring its
intensity. Since the phase is determined by large-scale motions of
matter inside the sources, much of the astrophysical information is
extracted from the phase. This leads to different kinds of data
analysis methods than one normally encounters in astronomy, based on
matched filtering and searches over large parameter spaces of
potential signals. This style of data analysis requires the input of
pre-calculated template signals, which means that gravitational wave
detection depends more strongly than most other branches of astronomy
on theoretical input. The better the input, the greater the range of
the detectors.

The fact that detectors are omni-directional and detect coherently the
phase of the incoming wave makes them in many ways more like
microphones for sound than like conventional telescopes. The analogy
with sound can be helpful, since microphones can be used to monitor
environments for disturbances in any location, and since when we
listen to sounds our brains do a form of matched filtering to allow us
to interpret the sounds we want to understand against a background of
noise. In a very real sense, gravitational wave detectors will be
listening to the sounds of a restless universe. The gravitational wave
``window'' will actually be a listening post, a monitor for the most
dramatic events that occur in the universe.

%% It turns out that useful information can be
%% extracted either when the number of events of a kind are large though
%% the signal-to-noise ratio of each event is very weak
%% or when the signal strengths are large though the number of events is
%% very small. For instance, a single binary inspiral event is enough to
%% carry out many tests of general relativity and to rule out theories
%% of gravitation while a large population of binary MACHO events, though
%% each event in itself produces no more than a snr of 5, will be
%% necessary to map the halo of the galaxy assuming that the halo is made
%% up of primordial black holes.

\subsection{Birth of gravitational astronomy}

Gravity is the dominant interaction in most astronomical systems. The big
surprise of the last three decades of the 20th century was that
\emph{relativistic gravitation} is relevant in so many of these
systems. Strong gravitational fields are Nature's most efficient
converters of mass into energy. Examples where strong-field
relativistic gravity is important include the following:

\begin{itemize}

\item neutron stars, the residue of supernova explosions, represent up to
  0.1\% (by number) of the entire stellar population of any galaxy;

\item stellar-mass black holes power many binary X-ray sources and
  tend to concentrate near the centers of globular clusters;

\item massive black holes in the range $10^6\mbox{\,--\,}10^9\msolar$ seem
  almost ubiquitous in galaxies that have central bulges, and power
  active galaxies, quasars, and giant radio jets;

\item and, of course, the Big Bang is the only naked singularity we
  expect to be able to see.

\end{itemize}

Most of these systems are either dynamical or were formed in catastrophic
events; many are or were, therefore, strong sources of gravitational radiation. As
the 21st century opens, we are on the threshold of using this radiation to gain
a new perspective on the observable universe.

The theory of gravitational radiation already makes an important contribution to
the understanding of a number of astronomical systems, such as neutron
star binaries, cataclysmic variables, young neutron stars, low-mass X-ray binaries, and
even the anisotropy of the microwave background radiation. As the understanding
of relativistic phenomena improves, it can be expected that gravitational
radiation will play a crucial role as a theoretical tool in modeling
relativistic astrophysical systems.

\subsection{What this review is about}

The first three-quarters of the 20th century were required to place the
mathematical theory of gravitational radiation on a sound footing. Many
of the most fundamental constructs in general relativity, such as null
infinity and the theory of conserved quantities, were developed at
least in part to help solve the technical problems of gravitational radiation.
We will not cover this history here, for which there are excellent
reviews~\cite{MTW, DAMOUR1987}. There are still many open questions, since it is
impossible to construct exact solutions for most interesting situations.
For example, we still lack a full understanding
of the two-body problem, and we will review the theoretical work on this
problem below.  But the fundamentals of the theory of gravitational
radiation are no longer in doubt. Indeed, the observation of the orbital
decay in the binary pulsar PSR~B1913+16 \cite{WeisbergTaylor2005} has
lent irrefutable support to the correctness of the theoretical
foundations aimed at computing gravitational wave emission, in
particular to the energy and angular momentum carried away by the
radiation.

It is, therefore, to be expected that the evolution of astrophysical systems
under the influence of strong tidal gravitational fields will be associated with
the emission of gravitational waves. Consequently, these systems are of interest
both to a physicist, whose aim is to understand fundamental interactions in
nature, their inter-relationships and theories describing them, and to an
astrophysicist, who wants to dig deeper into the environs of dense or
nonlinearly gravitating systems in solving the mysteries associated with
relativistic phenomena listed in Sections~\ref{sec:gwphysics},
\ref{sec:gwastronomy} and \ref{sec:gwcosmology}. Indeed, some of the
gravitational wave antennas that are being built are capable of
observing systems to cosmological distances, and even to the edge of
the universe. The new window, therefore, is also of interest to
cosmologists.

This is a \emph{living review} of the prospects that lie ahead for gravitational
antennas to test the predictions of general relativity as a fundamental theory,
for using relativistic gravitation as a means to understand highly energetic
sources, for interpreting gravitational waves to uncover the
(electromagnetically) dark universe, and ultimately for employing networks of
gravitational wave detectors to observe the first fraction of a second of the
evolution of the universe.

We begin in Section~\ref{sec:gwobservables} with a brief review of the physical
nature of gravitational waves, giving a heuristic derivation of the formulas
involved in the calculation of the gravitational wave observables such as the
amplitude, frequency and luminosity of gravitational waves.  This is followed in
Section~\ref{sec:gwsources} by a discussion of the astronomical sources of
gravitational waves, their expected event rates, amplitudes, waveforms and
spectra. In Section~\ref{sec:gwdetectors} we then give a detailed description of
the existing and upcoming gravitational wave antennas and their sensitivity.
Included in Section~\ref{sec:gwdetectors} are bar and interferometric antennas covering both
ground and space-based experiments. Section~\ref{sec:gwdetectors} also compares the sensitivity
of the antennas with the strengths of astronomical sources and expected
signal-to-noise ratios (SNRs). We then turn in Section~\ref{sec:gwdataanalysis} to data
analysis, which is a central component of gravitational wave astronomy,
focusing on those aspects of analysis that are crucial in gleaning physical,
astrophysical and cosmological information from gravity wave observations.

Sections~\ref{sec:gwastronomy}\,--\,\ref{sec:conclusions} treat in some detail how gravitational wave
observations will aid in a better understanding of nonlinear gravity and test
some of its fundamental predictions.  In Section~\ref{sec:gwphysics} we review
the physics implications of gravitational wave observations, including new tests
of general relativity that can be performed via gravitational wave observations,
how these observations may help in formulating and gaining insight into the
two-body problem in general relativity, and how gravitational wave observations
may help to probe the structure of the universe and the nature of dark
energy. In Section~\ref{sec:gwastronomy} we look at the astronomical information
returned by gravitational wave observations, and how these observations will
affect our understanding of black holes, neutron stars, supernovae, and other
relativistic phenomena. Section~\ref{sec:gwcosmology} is devoted to the
cosmological implications of gravitational wave observations, including placing
constraints on inflation, early phase transitions associated with spontaneous
symmetry breaking, and the large-scale structure of the universe.

This review is by no means exhaustive. We plan to expand it to include other key
topics in gravitational wave astronomy with subsequent revisions.

Unless otherwise specified we shall use a system of units
in which $c=G=1$, which means 
$1\,M_\odot\simeq 5 \times 10^{-6}\mathrm{\ s} \simeq 1.5\mathrm{\ km}$,
$1\mathrm{\ Mpc} \simeq 10^{14}\mathrm{\ s}$. 
We shall assume a universe with cold dark-matter density of
$\Omega_M = 0.3$, dark energy of $\Omega_\Lambda = 0.7$, and a
Hubble constant of $H_0 = 70\mathrm{\ km\, s^{-1}\,Mpc^{-1}}$.

%==================================================================
\newpage

\section{Gravitational Wave Observables}
\label {sec:gwobservables}

To benefit from gravitational wave observations we must first understand
what are the attributes of gravitational waves that we can observe.
This section is devoted to a short discussion of the nature of gravitational
radiation.

\subsection{Gravitational field vs gravitational waves}

Gravitational waves are propagating oscillations of the gravitational field,
just as light and radio waves are propagating oscillations of the
electromagnetic field. Whereas light and radio waves are emitted by accelerated
electrically-charged particles, gravitational waves are emitted by accelerated
masses. However, since there is only one sign of mass, gravitational waves never
exist on their own: they are never more than a small part of the overall
external gravitational field of the emitter. One may wonder, therefore, how it
is possible to infer the presence of an astronomical body by the gravitational
waves that it emits, when it is clearly not possible to sense its much larger
stationary (essentially Newtonian) gravitational potential.  There are, in fact,
two reasons:

\begin{itemize}

\item In general relativity, the effects of both the stationary field
  and gravitational radiation are described by the tidal forces they
  produce on free test masses. In other words, single geodesics alone
  cannot detect gravity or gravitational radiation; we need at least a
  pair of geodesics.  While the stationary tidal force due to the
  Newtonian potential $\phi$ of a self-gravitating source at a
  distance $r$ falls off as $\nabla\nabla\phi\sim r^{-3}$, the tidal
  force due to the gravitational wave amplitude $h$ that it emits at
  wavelength $\lambda$ decreases as $\nabla\nabla h\sim
  r^{-1}\lambda^{-2}$. Therefore, the stationary coulomb gravitational
  potential is the dominant tidal force close to the gravitating body
  (in the near zone, where $r\leq\lambda$). However, in the far zone
  ($r\gg\lambda$) the tidal effect of the waves is much stronger.

\item The stationary part of the tidal field is a DC effect, and
  simply adds to the stationary tidal forces of all other objects in
  the universe. It is not possible to discriminate one source from
  another. Gravitational waves carry time-dependent tidal forces, and
  so they can be discriminated from the stationary field if one knows
  what kind of time dependence to look for. Interferometers are ideal
  detectors in this respect because they sense only changes in the
  position of an interference fringe, which makes them insensitive to
  the DC part of the tidal field.

\end{itemize}

Because gravitational waves couple so weakly to our detectors, those
astronomical sources that we can detect must be extremely luminous in
gravitational radiation. Even at the distance of the Virgo cluster of galaxies,
a detectable source could be as luminous as the full Moon, if only for a
millisecond! Indeed, while radio astronomers deal with flux levels of Jy, mJy
and even $\mu$Jy, in the case of gravitational wave sources we encounter fluxes
that are typically $10^{20}$~Jy or larger. Gravitational wave astronomy
therefore is biased toward looking for highly energetic, even catastrophic,
events.

Extracting useful physical, astrophysical and cosmological information
from gravitational wave observations is made possible by measuring a
number of gravitational wave attributes that are related to the
properties of the source. In the rest of this section we discuss those
attributes of gravitational radiation that can be measured via
gravitational wave observations. In the process we will review the
basic formulas used in computing the gravitational wave amplitude and
luminosity of a source. These will then be used in
Section~\ref{sec:gwsources} to make an order-of-magnitude estimate of
the strength of astronomical sources of gravitational waves.

%% \subsection{Nature of gravitational radiation and its interaction
%% with matter}
\subsection{Gravitational wave polarizations}

Because of the equivalence principle, single isolated particles cannot be used
to measure gravitational waves: they fall freely in any gravitational field and
experience no effects from the passage of the wave. Instead, one must look for
inhomogeneities in the gravitational field, which are the tidal forces carried
by the waves, and which can be measured only by comparing the positions or
interactions of two or more particles.

In general relativity, gravitational radiation is represented by a second rank,
symmetric trace-free tensor. In a general coordinate system, and in an arbitrary
gauge (coordinate choice), this tensor has ten independent components. However,
as in the electromagnetic case, gravitational radiation has only two independent
states of polarization in Einstein's theory: the plus polarization and the
cross polarization (the names being derived from the shape of the equivalent
force fields that they produce). In contrast to electromagnetic waves, the angle
between the two polarization states is $\pi/4$ rather than $\pi/2$. This is
illustrated in Figure~\ref{fig:polarization}, where the response of a ring of free
particles in the $(x,y)$ plane to plus-polarized and cross-polarized
gravitational waves traveling in the $z$-direction is shown.  The effect of the
waves is to cause a tidal deformation of the circular ring into an elliptical
ring with the same area. This tidal deformation caused by passing gravitational
waves is the basic principle behind the construction of gravitational wave
antennas.

\epubtkImage{plus-cross.png}{%
  \begin{figure}[htbp]
    \centerline{
      \includegraphics[width=2in]{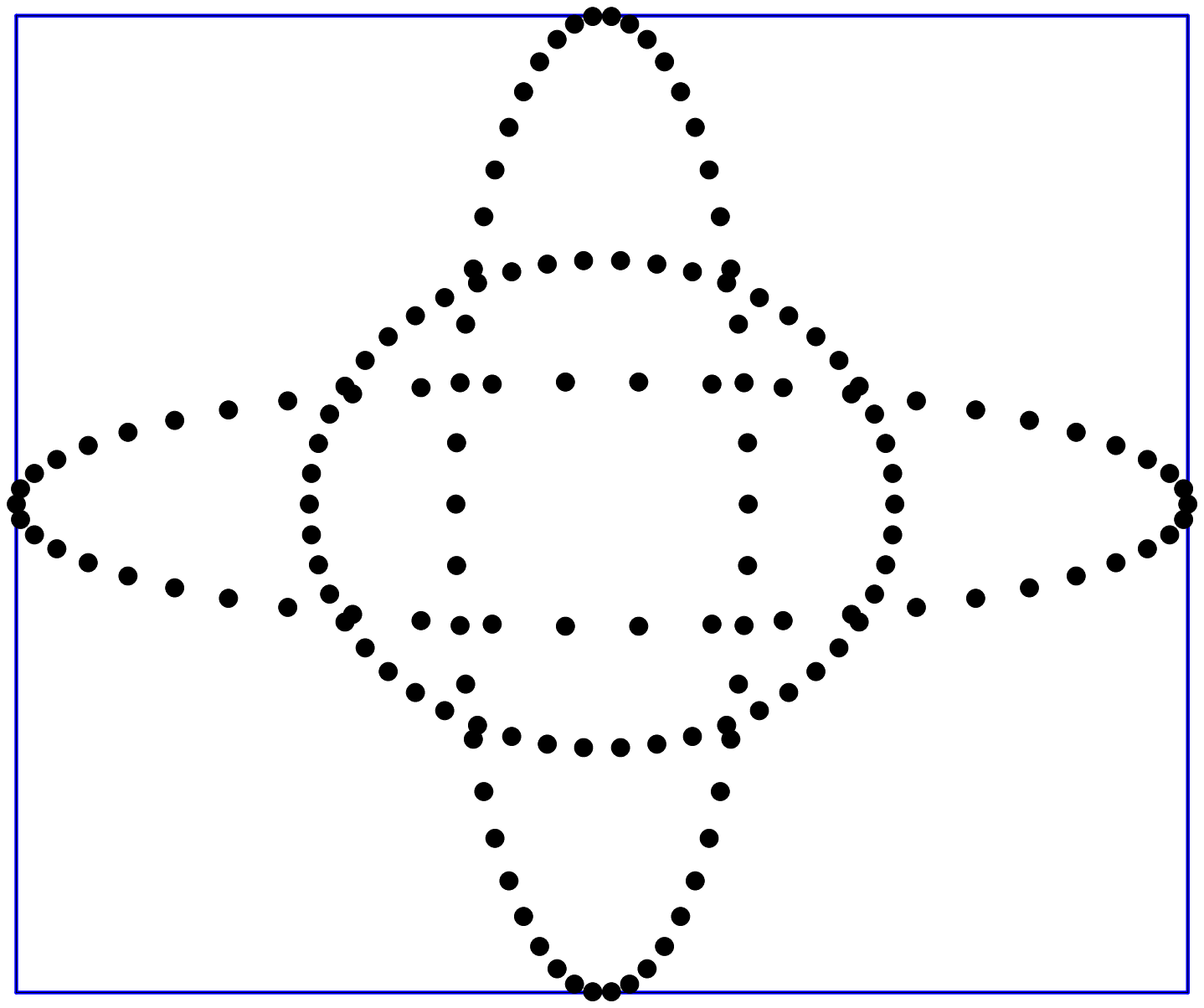}
      \includegraphics[width=2in]{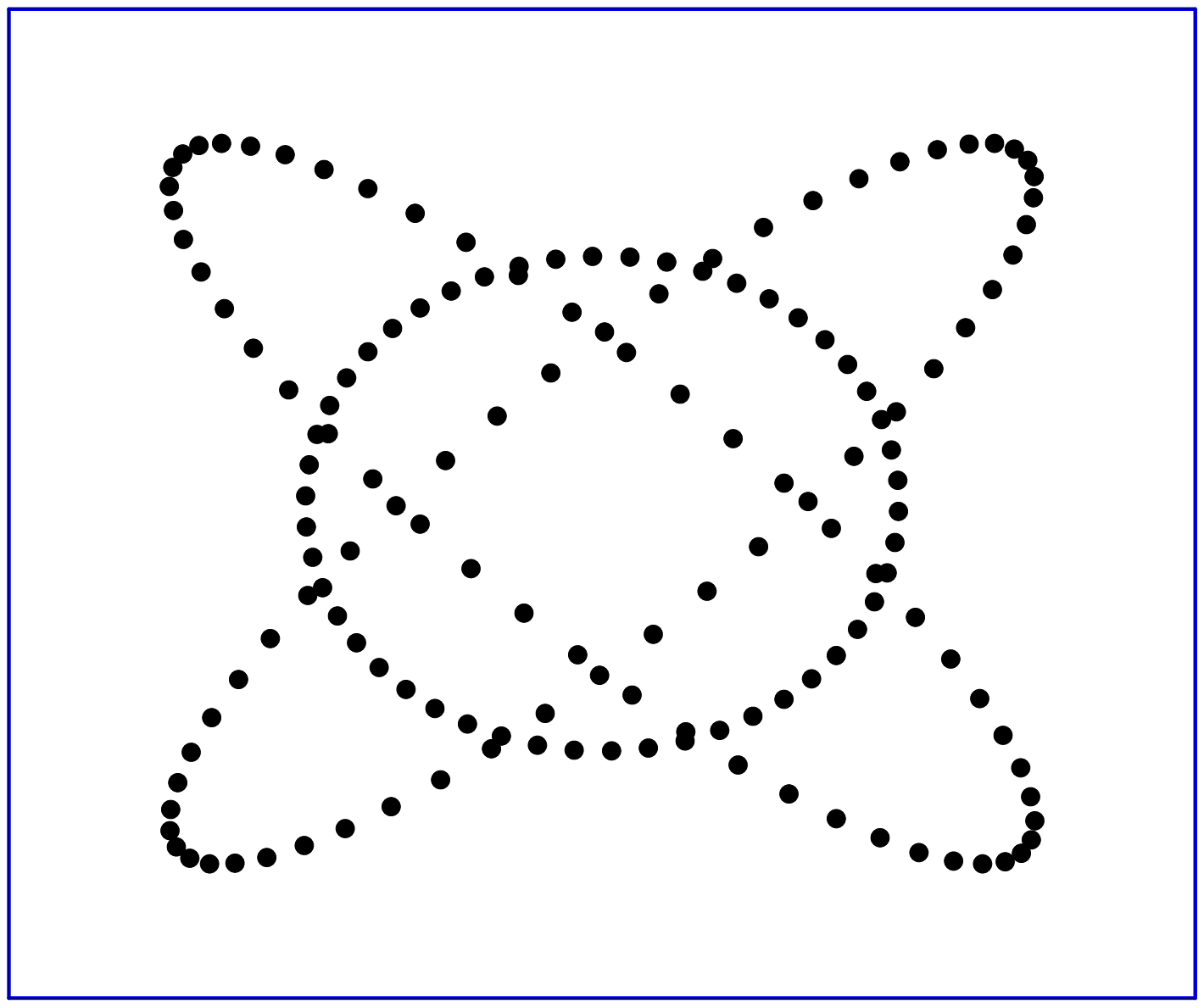}
    }
    \caption{In Einstein's theory, gravitational waves have two
      independent polarizations. The effect on proper separations of
      particles in a circular ring in the $(x,y)$-plane due to a
      plus-polarized wave traveling in the $z$-direction is shown in
      (a) and due to a cross-polarized wave is shown in (b). The ring
      continuously gets deformed into one of the ellipses and back
      during the first half of a gravitational wave period and gets
      deformed into the other ellipse and back during the next half.}
    \label{fig:polarization}
\end{figure}}

The two independent polarizations of gravitational waves are denoted $h_+$ and
$h_\times$. These are the two primary time-dependent observables of a
gravitational wave. The polarization of gravitational waves from a source, such
as a binary system, depends on the orientation of the dynamics inside the source
relative to the observer. Therefore, measuring the polarization provides
information about, for example, the orientation of the binary system.

\subsection{Direction to a source}

Gravitational wave antennas are linearly-polarized quadrupolar detectors and do
not have good directional sensitivity. As a result we cannot deduce the
direction to a source using a single antenna. One normally needs simultaneous
observation using three or more detectors so that the source can be triangulated
in the sky by measuring the time differences in signal arrival times at various
detectors in a network. Ground-based detectors have typical separation baselines
of $L\sim 3\times 10^{6}\mathrm{\ m}$, so that at a wavelength of $\lambda = 3\times
10^{5}\mathrm{\ m} = 1\mathrm{\ ms}$ (a frequency of 1~kHz) the network has a resolution of
$\delta\theta = \lambda / L = 0.1\mathrm{\ rad}$. If the amplitude SNR is
high, then one can localize the source by a factor of $1/\text{SNR}$
better than this.

For long-lived sources, however, a single antenna synthesizes many antennas by
observing the source at different points along its orbit around the
sun. The baseline for such observations is 2~AU, so that, for a source
emitting radiation at 1~kHz, the resolution is as good as $\Delta
\theta = 10^{-6}\mathrm{\ rad}$, which is smaller than an arcsecond.

For space-based detectors orbiting the sun, like LISA, the baseline is again
2~AU, but the observing frequency is some five or six orders of magnitude lower,
so the basic resolution is only of order 1 radian. However, as we shall see
later, some of the sources that a space-based detector will observe have huge
amplitude SNRs in the range of $\text{SNR}\sim 10^3\mbox{\,--\,}10^4$,
which improves the resolution to arcminute accuracies in the best cases.

\subsection{Amplitude of gravitational waves -- the quadrupole approximation}
\label{sec:amplitudes}

%The amplitude of gravitational waves falls off inversely with distance so
%that a measurement of the amplitude would help infer the luminosity-distance
%to the source, provided, of course, we know its intrinsic luminosity.
%As we shall discuss below this would indeed be possible for compact binary
%mergers.
%Sathya: this seems out of place here! Bernard

The Einstein equations are too difficult to solve analytically in the generic
case of a strongly gravitating source to compute the luminosity and amplitude of
gravitational waves from an astronomical source. We will discuss numerical
solutions later; the most powerful available analytic approach is called the
\emph{post-Newtonian} approximation scheme. This approximation is suited to
gravitationally-bound systems, which constitute the majority of expected
sources. In this scheme~\cite{BLANCHETREF, FUTAMASEREF}, solutions are
expanded in the small parameter $(v/c)^2$, where $v$ is the typical
dynamical speed inside the system. Because of the virial theorem, the
dimensionless Newtonian gravitational potential $\phi/c^2$ is of the
same order, so that the expansion scheme links orders in the expanded
metric with those in the expanded source terms. The lowest-order
post-Newtonian approximation for the emitted radiation is the
\emph{quadrupole formula}, and it depends only on the density ($\rho$)
and velocity fields of the Newtonian system. If we define the spatial
tensor $Q_{jk}$, the second moment of the mass distribution, by the
equation
\begin{equation}\label{eqn:ibar}
Q_{jk}=\int\rho x_jx_k\, \mathrm{d}^3x,
\end{equation}
then the amplitude of the emitted gravitational wave is, at lowest order, the
three-tensor
\begin{equation}\label{eqn:hjk}
h_{jk} = \frac{2}{r}\oderivn{Q_{jk}}{t}{2}.
% \label{eq:gwamplitude}
\end{equation}
This is to be interpreted as a linearized gravitational wave in the distant
almost-flat geometry far from the source, in a coordinate system (gauge) called
the Lorentz gauge.

\subsubsection{Wave amplitudes and polarization in TT-gauge}

A useful specialization of the Lorentz gauge is the TT-gauge, which
is a comoving coordinate system: free particles remain at constant
coordinate locations, even as their proper separations change. To get the
TT-amplitude of a wave traveling outwards from its source, project the tensor in
\Eqref{eqn:hjk} perpendicular to its direction of travel and remove the
trace of the projected tensor. The result of doing this to a symmetric 
tensor is to produce, in the transverse plane, a two-dimensional matrix 
with only two independent elements:
\begin{equation}\label{eqn:ttmatrix}
h_{ab}=\left(\begin{array}{cc}
h_+ & h_\times \\
h_\times & -h_+
\end{array}\right).
\end{equation}
This is the definition of the wave amplitudes $h_+$ and $h_\times$ that are 
illustrated in Figure~\ref{fig:polarization}. These amplitudes are 
referred to as the coordinates chosen for that plane. If the coordinate unit basis vectors 
in this plane are $\exhat$ and $\eyhat$, then we can define the basis 
tensors 
\begin{eqnarray}\label{eqn:TTbasistensors}
\eptensor &=& \exhat\otimes\exhat - \eyhat\otimes\eyhat,\\
\extensor &=& \exhat\otimes\eyhat + \eyhat\otimes\exhat.
\end{eqnarray}
In terms of these, the TT-gravitational wave tensor can be written as 
\begin{equation}\label{eqn:TTtensor}
\htensor = h_+\eptensor + h_\times\extensor.
\end{equation}

If the coordinates in the transverse plane are rotated 
by an angle $\psi$, then one obtains new amplitudes $h_+'$ and $h_{\times}'$ given 
by
\begin{eqnarray}\label{eqn:rotatedTT}
h_+' &= &\cos2\psi\;h_+ + \sin2\psi\;h_\times,\\
h_{\times}' &= &-\sin2\psi\;h_+ + \cos2\psi\;h_\times.
\end{eqnarray}
This shows the quadrupolar nature of the polarizations, and is consistent with our 
remark in association with Figure~\ref{fig:polarization} that a rotation of $\pi/4$ 
changes one polarization into the other. 

It should be clear from the TT projection operation
that the emitted radiation is not isotropic: it will be stronger in some
directions than in others\epubtkFootnote{In the case of an inspiraling binary, the
root mean square of the two polarization amplitudes in a direction orthogonal to the orbital plane will
be a factor $2\sqrt{2}$ larger than in the plane.}.  It should also be clear from
this that spherically-symmetric motions do not emit any gravitational radiation:
when the trace is removed, nothing remains.

\subsubsection{Simple estimates}

A typical
component of $\oderivnf{Q_{jk}}{t}{2}$ will (from \Eqref{eqn:ibar}) have
magnitude $(Mv^2)_{\text{nonsph}}$, where $(Mv^2)_{\text{nonsph}}$ is twice 
the nonspherical part of the kinetic energy
inside the source. So a bound on any component of
\Eqref{eqn:hjk} is
\begin{equation}\label{eqn:nonsph}
h\lsim \frac{2(Mv^2)_{\text{nonsph}}}{r}.
\end{equation}
It is interesting to observe that the ratio $\epsilon$ of the wave
amplitude to the Newtonian potential $\phi_{\mathrm{ext}}$ of its source at
the observer's distance $r$ is simply bounded by
\[h/\phi_{\mathrm{ext}} < 2v^2_{\text{nonsph}},\]
and this bound is attained if the entire mass of the source is 
involved in the nonspherical motions, so that $(Mv^2)_{\text{nonsph}} \sim Mv^2_{\text{nonsph}}$.
By the virial theorem for self-gravitating bodies
\begin{equation}\label{eqn:epsilonandphi}
v^2_{\text{nonsph}} \leq  \phi_{\mathrm{int}},
\end{equation}
where $\phi_{\mathrm{int}}$ is the maximum value of the Newtonian gravitational
potential \emph{inside} the system.  This provides a convenient bound in practice~\cite{BackOfEnvelope}:
\begin{equation}\label{eq:backofenvelope}
h \lsim 2\phi_{\mathrm{int}}\phi_{\mathrm{ext}}.
\end{equation}
The bound is attained if the system is highly nonspherical. An equal-mass star binary 
system is a good example of a system that attains this bound.

For a neutron star source, one has $\phi_{int} \sim 0.2$.  If the star is in the
Virgo cluster ($r\sim 18\mathrm{\ Mpc}$) and has a mass of $1.4\msolar$, and if it is 
formed in a highly-nonspherical gravitational collapse, then the upper
limit on the amplitude of the radiation from such an event is $1.5\E{-21}$. This
is a simple way to get the number that has been the goal of detector development
for decades, to make detectors that can observe waves at or below an amplitude
of about $10^{-21}$.

\subsection{Frequency of gravitational waves}

The signals for which the best waveform predictions are available have
well-defined frequencies. In some cases the frequency is dominated by an
existing motion, such as the spin of a pulsar.  But in most cases the frequency
will be related to the \emph{natural frequency} for a self-gravitating body,
defined as
\begin{equation}\label{eqn:natfreq}
\omega_0 = \sqrt{\pi G \bar{\rho}},\ \  \text{or}\ \  f_0 = \omega_0/2\pi =
\sqrt{G\bar{\rho}/4\pi},
\end{equation}
where $\bar{\rho}$ is the mean density of mass-energy in the source. This is of
the same order as the binary orbital frequency and the fundamental pulsation
frequency of the body. Even though this is a Newtonian formula, it provides a
remarkably good order-of-magnitude prediction of natural frequencies, even for
highly relativistic systems such as black holes.

The frequency of the emitted gravitational waves need not be the natural
frequency, of course, even if the mechanism is an oscillation with that
frequency. In many cases, such as binary systems, the radiation comes out at
twice the oscillation frequency. But since, at this point, we are not trying to be
more accurate than a few factors, we will ignore this distinction here. In
later sections, with specific source models, we will get the factors right.

The mean density and hence the frequency are determined by the size
$R$ and mass $M$ of the source, taking $\bar{\rho} = 3M/4\pi R^3$.
For a  neutron star of mass $1.4\msolar$ and radius 10~km, the natural
frequency is $f_0 = 1.9\mathrm{\ kHz}$. For a black hole of mass
$10\msolar$ and radius $2M = 30\mathrm{\ km}$, it is $f_0 = 1\mathrm{\
  kHz}$. And for a large black hole of mass $2.5\E6\msolar$, such as the one at
the center of our galaxy, this goes down in inverse proportion to the
mass to $f_0 = 4\mathrm{\ mHz}$. In general, the characteristic
frequency of the radiation of a compact object of mass $M$ and radius
$R$ is
\begin{equation}
f_0 = \frac{1}{4\pi} \fracparen{3M}{R^3}^{1/2} \simeq 1\,\text{kHz} \fracparen{10\,M_\odot}{M} .
\label{eq:gw frequency}
\end{equation}

\Figref{fig:freq} shows the mass-radius diagram for likely sources of
gravitational waves.  Three lines of constant natural frequency are plotted:
$f_0 = 10^4\mathrm{\ Hz}$, $f_0 = 1\mathrm{\ Hz}$, and $f_0 = 10^{-4}\mathrm{\ Hz}$.  These are interesting
frequencies from the point of view of observing techniques: gravitational waves
between 1 and $10^4$~Hz are in principle accessible to ground-based detectors, while lower
frequencies are observable only from space. Also shown is the line marking the
black-hole boundary. This has the equation $R=2M$.  There are no objects below
this line, because they would be smaller than the horizon size for their mass.
This line cuts through the ground-based frequency band in such a way as to
restrict ground-based instruments to looking at stellar-mass objects. \emph{No system with 
a mass above about $10^4\msolar$ can produce quadrupole radiation in the ground-based 
frequency band.}

\epubtkImage{dyn_full.png}{%
  \begin{figure}[htbp]
    \centerline{\includegraphics[width=4in]{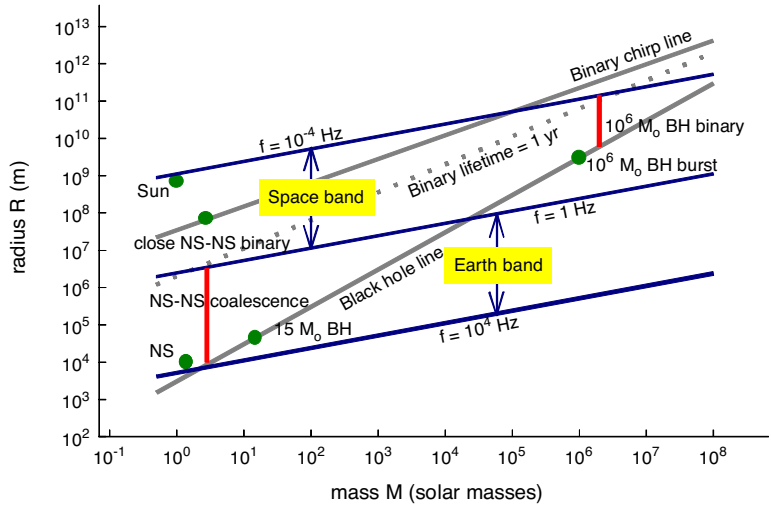}}
    \caption{Mass-radius plot for gravitational wave sources. The
    horizontal axis is the total mass of a radiating system, and the
    vertical axis is its size. Typical values from various sources for
    ground-based and space-based detectors are shown. Lines give
    order-of-magnitude constraints and relations. Characteristic
    frequencies are estimated from $f \sim (G\rho/4\pi)^{1/2}$. The
    black-hole and binary lines are described in the text.}
    \label{fig:freq}
\end{figure}}

A number of typical relativistic objects are placed in the diagram: a neutron
star, a pair of neutron stars that spiral together as they orbit, some
black holes.  Two other interesting lines are drawn. The lower (dashed) line is
the 1-year coalescence line, where the orbital shrinking timescale due to
gravitational radiation backreaction (cf.\ \Eqref{eqn:chirp}) is less than
one year. The upper (solid) line is the 1-year chirp line: if a binary lies
below this line, then its orbit will shrink enough to make its orbital frequency
increase by a measurable amount in one year.  (In a one-year observation one
can, in principle, measure changes in frequency of $1\units yr^{-1}$, or
$3\E{-8}$~Hz.)

It is clear from the Figure that \emph{any binary system that is observed from
the ground will coalesce within an observing time of one year}.  Since
pulsar binary statistics suggest that neutron-star--binary coalescences 
happen less often than once every $10^5$~years in our galaxy,
ground-based detectors must be able to register these events in a
volume of space containing at least $10^6$ galaxies in order to have a
hope of seeing occasional coalescences. That corresponds to a volume
of radius roughly 100~Mpc. For comparison, first-generation
ground-based interferometric detectors have a reach of around 20~Mpc
for such binaries, while advanced interferometers should extend that
to about 200~Mpc.

\subsection{Luminosity in gravitational waves}

The general formula for the local stress-energy of a gravitational wave field
propagating through flat spacetime, using the TT-gauge, is given by the Isaacson
expression~\cite{MTW, schutz.2009}
\begin{equation}\label{eqn:gwstressenergy}
T_{\alpha\beta} = \frac{1}{32\pi}\left<h\textsup{TT}_{jk,\alpha}
h^{{\rm TT}\,jk}_{,\beta}\right>,
\end{equation}
% FS check eq
%
where the angle brackets denote averages over regions of the size of a
wavelength and times of the length of a period of the wave. The energy flux of a
wave in the $x^i$ direction is the $T^{0i}$ component. 

The gravitational wave luminosity in the quadrupole approximation is obtained
by integrating the energy flux from \Eqref{eqn:gwstressenergy} over a
distant sphere. When one correctly takes into account the projection factors
mentioned after \Eqref{eqn:hjk}, one obtains~\cite{MTW}
\begin{equation}\label{eqn:luminosity}
L_{gw} = \frac{1}{5}\left(\sum_{j,k}\qddd_{jk}\qddd_{jk} -
\frac{1}{3}\qddd^2\right),
\end{equation}
where $Q$ is the trace of the matrix $Q_{jk}$. This equation can be used to
estimate the backreaction effect on a system that emits gravitational
radiation.

Notice that the expression for $L_{gw}$ is dimensionless when $c=G=1$.
It can be converted to
normal luminosity units by multiplying by the scale factor
\begin{equation}\label{luminosityscale}
L_0=c^5/G = 3.6\E{52}\text{ W}.
\end{equation}
This is an enormous luminosity. By comparison, the luminosity of
the sun is only $3.8\E{26}$~W, and that of a typical galaxy would be $10^{37}$~W. 
All the galaxies in the visible universe emit, in visible light, on the order of $10^{49}$~W. 
We will see that gravitational wave systems always emit at a fraction of $L_0$, but that 
the gravitational wave luminosity can come close to $L_0$ and can greatly exceed 
typical electromagnetic luminosities. Close binary systems normally 
radiate much more energy in gravitational waves than in light. Black hole mergers 
can, during their peak few cycles, compete in luminosity with the steady luminosity 
of the entire universe!

Combining Equations~(\ref{eqn:hjk}) and (\ref{eqn:luminosity}) one can
derive a simple expression for the apparent luminosity of radiation
${\cal F}$, at great distances from the source, in terms of the
gravitational wave amplitude~\cite{schutz.2009}:
\begin{equation}\label{eq:flux}
{\cal F} \sim \frac {| \dot h |^2}{16 \pi}.
\end{equation}
The above relation can be used to make an order-of-magnitude estimate of the
gravitational wave amplitude from a knowledge of the rate at which energy is
emitted by a source in the form of gravitational waves. If a source at a
distance $r$ radiates away energy $E$ in a time $T$, predominantly at a
frequency $f$, then writing $\dot h= 2\pi f h$ and noting that ${\cal
F}\sim E/(4\pi r^2T)$, the amplitude of gravitational waves is
\begin{equation} \label {eq:amplitude}
h \sim \frac{1}{\pi f r}\sqrt{ \frac{E}{T} }.
\end{equation}
When the time development of a signal is known, one can filter the detector
output through a copy of the expected signal (see \Secref{sec:gwdataanalysis}
on \emph{matched filtering}). This leads to an enhancement in the SNR, as
compared to its narrow-band value, by roughly the square root of the number of
cycles the signal spends in the detector band. It is useful, therefore, to
define an \emph{effective amplitude} of a signal, which is a better measure of
its detectability than its raw amplitude:
\begin{equation}\label{effective.ampl.def}
h_{\mathrm{eff}}\equiv \sqrt {n} h.
\end{equation}
Now, a signal lasting for a time $T$ around a frequency $f$ would produce
$n\simeq fT$ cycles. Using this we can eliminate $T$ from
Equation~(\ref{eq:amplitude}) and get the effective amplitude of the signal in terms
of the energy, the emitted frequency and the distance to the source:
\begin{equation} \label {eq:effective.amplitude}
h_{\mathrm{eff}} \sim \frac{1}{\pi r}\sqrt{\frac{E}{f}}.
\end{equation}
Notice that this depends on the energy only through the total
\emph{fluence}, or time-integrated flux $E/4\pi r^2$ of the wave. As
in many other branches of astronomy, the detectability of a source is
ultimately a function of its apparent luminosity and the observing
time. However, one should not ignore the dependence on frequency in
this formula. Two sources with the same fluence are not equally easy
to detect if they are at different frequencies: higher frequency
signals have smaller amplitudes.

%==================================================================
\newpage

\section{Sources of Gravitational Waves}
\label {sec:gwsources}

\subsection{Man-made sources}

One source can unfortunately be ruled out as undetectable: man-made
gravitational radiation.  Imagine creating a wave generator with the following
extreme properties.  It consists of two masses of $10^3$~kg each (a small car)
at opposite ends of a beam 10~m long.  At its center the beam pivots about an
axis.  This centrifuge rotates 10 times per second. All the velocity is
nonspherical, so $v^2_{\text{nonsph}}$ in \Eqref{eqn:nonsph} is about $10^5\units
m^2\;s^{-2}$.  The frequency of the waves will actually be 20~Hz, since the mass
distribution of the system is periodic in time with a period of half the
rotation period.  The wavelength of the waves will, therefore, be $1.5\E7$~m,
similar to the diameter of the earth.  In order to detect gravitational waves,
not near-zone Newtonian gravity, the detector must be at least one wavelength
from the source, say diametrically opposite the centrifuge on the Earth. Then
the amplitude $h$ can be deduced from \Eqref{eqn:nonsph}: $h \sim 5\E{-43}$.
This is far too small to contemplate detecting! The story changes, fortunately,
when we consider astrophysical sources of gravitational waves, where nature
arranges for masses that are $10^{27}$ times larger than our centrifuge to move
at speeds close to the speed of light!

Until observations of gravitational waves are successfully made, one
can only make intelligent guesses about most of the sources that will
be seen.  There are many that \emph{could} be strong enough to be seen
by the early detectors: star binaries, supernova explosions, neutron
stars, the early universe. In this section, we make rough luminosity
estimates using the quadrupole formula and other approximations, which
are usually accurate to within factors of order two, and, very
importantly, they show how key observables scale with the properties
of the systems. Where appropriate we also make use of predictions from
the much more accurate modelling that is available for some sources,
such as binary systems and black hole mergers. The detectability
depends, of course, not only on the intrinsic luminosity of the
source, but on how far away it is. Often the biggest uncertainties in
making predictions are the spatial density and event rate of any
particular class of sources. This is not surprising, since our
information at present comes from electromagnetic observations, and as
our earlier remarks about the differences between the mechanisms of
emission of gravitational and electromagnetic radiation make clear,
electromagnetic observations may not strongly constrain the source
population.

\subsection{Gravitational wave bursts from gravitational collapse}

Neutron stars and black holes are formed from the gravitational collapse 
of a highly evolved star or the core collapse of an accreting white
dwarf.  In either case, if the collapse is nonspherical, perhaps induced by
strong rotation, then gravitational waves could carry away some of the binding
energy and angular momentum depending on the geometry of the collapse. Collapse
events are thought to produce supernovae of various types, and increasingly there is 
evidence that they also produce most of the observed gamma-ray bursts~\cite{Hjorth2003} 
in \emph{hypernovae} and \emph{collapsars}~\cite{Woosley1993, MacFadyenWoosley1999}.
Supernovae of Type II are believed to occur at a rate of between 0.1 and 0.01 per year in a
milky-way equivalent galaxy (MWEG); thus, within the Virgo supercluster, we might
expect an event rate of about 30 per year. Hypernova events are considerably 
rarer and might only contribute observable gravitational-wave events in 
current and near-future detectors if they involve so much rotation that 
strong non-axisymmetric instabilities are triggered. 

Simulating gravitational collapse is a very active area of numerical 
astrophysics, and most simulations also predict the 
energy and spectral characteristics of the emitted gravitational waves~\cite{Living:Freyer}. 
However, it is still beyond the capabilities of computers to simulate a 
gravitational collapse event with all the physics that might be necessary 
to give reliable predictions: three-dimensional hydrodynamics, neutrino 
transport, realistic nuclear physics, magnetic fields, rotation. In fact, 
it is still by no means clear why Type II supernovae explode at all: simulations 
typically have great difficulty reversing the inflow and producing an explosion
with the observed light-curves and energetics. It may be that the answer lies 
in some of the physics that has to be oversimplified in order to be used in 
current simulations, or in some neutrino physics that we do not yet know, or in 
some unexplored hydrodynamic mechanism~\cite{OttPRL2006}.
In a typical supernova, simulations suggest that gravitational waves might extract 
between about $10^{-7}$ and $10^{-5}$ of the total available
mass-energy~\cite{Muller, Dimmelmeier2002, Dimmelmeier2007}, and the
waves could come off in a burst whose frequency might lie in the range
of $\sim$~200\,--\,1000~Hz.

We can use \Eqref{eq:amplitude} to make a rough estimate of the
amplitude, if the emitted energy and timescale are known.  
Using representative values for a supernova in our galaxy, lying at 10~kpc, 
emitting the energy equivalent of $10^{-7}\,M_\odot$ at a frequency of 1~kHz, 
and lasting for 1~ms, the received amplitude would be
\begin{equation}
h \sim 6 \times 10^{-21}
\left ( \frac{E}{10^{-7}\,M_\odot} \right )^{1/2}
\left ( \frac{1\mathrm{\ ms}}{T} \right )^{1/2}
\left ( \frac{1\mathrm{\ kHz}}{f} \right )
\left ( \frac{10\mathrm{\ kpc}}{r} \right ).
\label {eq:amplitudeB}
\end{equation}
The upper bound in \Eqref{eq:backofenvelope} would give the same amplitude 
for a source 60 times further away, which reflects the fact that simulations 
find it difficult to put significant energy into gravitational waves.  
This amplitude is large enough for current ground-based detectors to observe with a
reasonably high confidence, but of course the event rate within 10~kpc is
expected to be far too small to make an early detection likely.

\subsection{Gravitational wave pulsars}

Some likely gravitational wave sources behave like the centrifuge example 
we used in the first paragraph of this section,
only on a grander scale.  Suppose a  neutron star of
radius $R$ and mass $M$ spins with a frequency $f$ and has an irregularity, a
deformation of its otherwise axially symmetric shape. We idealize
this as a ``bump'' of mass $m$ on its surface, although of course it
will really be a distribution of mass leading to an
asymmetrical quadrupole tensor. The moment of inertia of the bump 
will be $mR^2$, and it is conventional to parameterize the bump in
terms of the fractional asymmetry it creates in the moment of
inertia tensor itself. If we idealize the star as having uniform 
density, then the spherical moment of inertia is $2\,MR^2/5$, and so 
the bump has fractional asymmetry
\begin{equation}\label{eqn:bumpepsilon}
\epsilon = \frac{5}{2}\frac{m}{M}, \qquad m = 0.4\epsilon M.
\end{equation}
The bump will emit gravitational radiation at frequency
$2f$ because the star spins about its net center of mass, so it
effectively has mass excesses on both sides of the star. The
nonspherical velocity will be just $v_{\text{nonsph}} = 2\pi Rf$.
The radiation amplitude will be, from \Eqref{eqn:nonsph},
\begin{equation}\label{eqn:bump}
h_{bump}\sim (4/5)(2\pi Rf)^2\epsilon M/r,
\end{equation}
and the luminosity, from \Eqref{eqn:luminosity} (assuming that
roughly four comparable components of $Q_{jk}$ contribute to the sum),
\[L_{\text{bump}}\sim (16/125)(2\pi f)^6\epsilon^2M^2R^4.\]
The radiated energy would presumably come from the rotational
energy of the star $Mv^2/5$. This would lead to a spindown of the star
on a timescale
\begin{equation}\label{eqn:spindown}
t_{\text{spindown}} \sim \frac{1}{5} Mv^2/L_{\text{bump}}
\sim \frac{25}{32\pi}\epsilon^{-2}f^{-1}\fracparen{M}{R}^{-1}v^{-3}.
\end{equation}
It is believed that neutron star crusts are not strong enough to
support fractional asymmetries larger than about
$\epsilon\sim10^{-6}$~\cite{CBU2000}, and realistic asymmetries 
may be much smaller. 

From these considerations one can estimate the
likelihood that the observed spindown timescales of pulsars are
due to gravitational radiation. In most cases, it seems that
gravitational wave losses could account for a substantial amount 
of the spindown: the required asymmetries are much smaller 
than $10^{-4}$, often smaller than $10^{-7}$. But an interesting 
exception is the Crab pulsar, PSR~J0534+2200, whose young age 
and consequently short spindown time (measured to be  
$8.0\E{10}$~s, about 2500~yr) would require an exceptionally large 
asymmetry. If we take the neutron star's 
radius to be 10~km, so that $M/R\sim0.21$ and the 
speed of any irregularity is $v/c\sim6.2\E{-3}$, then 
\Eqref{eqn:spindown} would require an asymmetry of  $\epsilon\sim1.4\E{-3}$.
Of course, we have made a lot of approximations to get here, only 
keeping our estimates of amplitudes and energies correct to within factors 
of two, but a more careful calculation reduces this only by a factor of 
two to $\epsilon\sim7\E{-4}$ \cite{Abbott:2008fx}. What makes this interesting 
is the fact that 
an asymmetry this large would produce radiation detectable by 
first-generation interferometers. Conversely, an upper limit 
from first-generation interferometers would provide  direct observational 
limits on the asymmetry and on the fraction of energy lost by the 
Crab pulsar to gravitational waves.

From \Eqref{eqn:bump} the Crab pulsar would, if its spindown is dominated
by gravitational wave losses, produce an amplitude at the Earth of 
$h\sim1.5\E{-24}$, if its distance is 2~kpc. Is this detectable when 
present instruments are only capable of seeing millisecond bursts of radiation 
at levels of $10^{-21}$? The answer is yes, if the observation time 
is long enough. Indeed, the latest LIGO observations have not detected any 
gravitational waves from the Crab pulsar, which has been used to set an
upper limit on the asymmetry in its mass distribution~\cite{Abbott:2008fx}. 
The limit depends on the model assumed for the pulsar. If one assumes that
gravitational waves are produced at exactly twice the pulsar 
spin frequency and uses the inferred values of the pulsar orientation 
and polarization angle, then for a canonical value of 
the moment-of-inertia $I=10^{38}\mathrm{\ kg}\mathrm{\ m^2}$, one gets an 
upper limit on the ellipticity of $\epsilon \le 1.8 \times 10^{-4}$, 
assuming the pulsar is at 2~kpc. This is a factor of 4.2 
below the spindown limit~\cite{Abbott:2008fx}. If, however, one
assumes that gravitational waves are emitted at a frequency
close, but not exactly equal, to twice the spin frequency and
one uses a uniform prior for the orientation and polarization angle,
then one gets $\epsilon \le 9 \times 10^{-4}$, which is 0.8
of the limit derived from the spin-down rate.

Indeed, even signals weaker than the amplitude determined by the
Crab spindown rate will be observable by present
detectors, and these may be coming from a larger
variety of neutron stars, in particular
low-mass X-ray binary systems (LMXBs).  The neutron stars in
them are accreting mass and angular momentum, so they should
be spinning up.  Observations suggest
that most neutron stars are spinning at speeds between
about 300 and 600~Hz, far below their maximum, which is greater
than 1000~Hz. The absence of faster stars suggests that something
stops them from spinning up beyond this range.
Bildsten suggested~\cite{Bildsten:1998ey} that the limiting mechanism
may be the re-radiation of the accreted angular momentum in gravitational
waves, possibly due to a quadrupole moment created by asymmetrical heating 
induced by the accreted matter.  Another possible
mechanism~\cite{Melatos:2003} is that a ``bump'' of the kind we have
treated is formed by accreting matter channeled onto the surface by
the star's magnetic field. It is also possible that accretion drives
an instability in the star that leads to steady
emission~\cite{Reisenegger:2003cq, Nayyar:2005th}. In either case, the
stars could turn out to be long-lived sources of gravitational
waves. This idea, which is a variant of one proposed long ago by
Wagoner~\cite{Wagoner:1984pv}, is still speculative, but the numbers
make a plausible case. We discuss it in more detail in
\Secref{sec:LMXB}.

\subsection{Radiation from a binary star system}
\label{sec:rr}

\subsubsection{Radiation from a binary system and its backreaction}

A binary star system can also be treated as a ``centrifuge''. Two stars of
the same mass $m$ in a circular orbit of radius $R$ have all their mass in 
nonspherical motion, so that 
\[(Mv^2)_{\text{nonsph}} = M(\Omega R)^2 = \frac{M^2}{R},\] 
where $\Omega$ is the orbital angular velocity.  The gravitational wave amplitude
can then be written 
%\red{It looks like $M$ is being used to represent
%both the component and total mass in this section. Correct any mistakes
%and adopt one notation.}
%
\begin{equation}\label{eqn:binary}
h_{\text{binary}} \sim 2\frac{M}{r}\frac{M}{R}.
\end{equation}
Since the internal radius $R$ of the orbit is not an observable, it is 
sometimes convenient to replace $R$ by the orbital angular frequency $\Omega$ 
using the above orbit equation, giving
\begin{equation}\label{eqn:binary2}
h_{\text{binary}} \sim \frac{2}{r} M^{5/3}\Omega^{2/3}.
\end{equation}

The gravitational wave luminosity of such a system is, by a
calculation analogous to that for bumps on neutron stars (assuming
that four components of $Q_{ij}$ to be significant),
\begin{equation}\label{eqn:binarylum}
L_{\text{binary}} \sim \frac{4}{5}\fracparen{M}{R}^5,
\end{equation}
in units given by the fundamental luminosity $L_0$ 
in \Eqref{luminosityscale}.
This shows that self-gravitating systems always
emit at a fraction of $L_0$, since $M/R$ is always smaller than 1, 
but it can approach $L_0$ for highly-relativistic systems where
$M/R\sim 1$.

The radiation of energy by the orbital motion causes the
orbit to shrink. The shrinking will make any observed
gravitational waves increase in frequency
with time.  This is called a \emph{chirp}. The
timescale\epubtkFootnote{In Sections~\ref{sec:matched filtering} we
  will use parameters called \emph{chirp times}, instead of the masses,
  to characterize a binary. The timescale defined here is closely
  related to the chirp times.} for this in a binary system with equal
masses is
\begin{equation}\label{eqn:chirp}
t_{\text{chirp}} = \frac{Mv^2}{2}/L_{\text{binary}} \sim \frac{5M}{8}\fracparen{M}{R}^{-4}.
\end{equation}
As the binary evolves, the frequency and amplitude of the wave
grow and this drives the binary to evolve even more rapidly. The signal's frequency,
however, will not increase indefinitely; the slow inspiral phase 
ends either when the stars begin to interact and merge or (if they are very 
compact) when the distance between the stars is roughly at the last stable 
orbit (LSO) $R=6M$, which corresponds to a gravitational wave frequency of 
\begin{equation}\label{eqn:LSO}
f_{\text{LSO}} \sim 220 \fracparen{20\,M_\odot}{M}\text{ Hz}, 
\end{equation}
where we have normalized this to
a binary with $M=20\,M_\odot$. This is the \emph{last stable orbit} (LSO) frequency.

A compact-object binary that coalesces after passing through the 
last stable orbit is a powerful source of gravitational waves, with 
a luminosity that approaches the limiting luminosity $L_0$. This is 
called a \emph{coalescing binary} in gravitational wave searches. Since 
a typical search might last on the order of one year, a coalescing binary 
can be defined as a system that has a chirp time smaller than one year. In 
\Figref{fig:freq} the coalescence line is shown as a straight line with 
slope 3/4 (set $t_{\text{chirp}}$ to a constant in \Eqref{eqn:chirp}). 
Binary systems below this line have a chirp time smaller than one year. It is 
evident from the figure that all
binary systems observable by ground-based detectors will coalesce in less 
than a year.

As mentioned for gravitational wave pulsars, the raw amplitude of the 
radiation from a long-lived system is not by itself
a good guide to its detectability, if the waveform can be predicted. 
Data analysis techniques like matched filtering are able
to eliminate most of the detector noise and allow the
recognition of weaker signals. The improvement in amplitude
sensitivity is roughly proportional to the square root
of the number of cycles of the waveform that one observes.
For neutron stars that are observed from a frequency of
10~Hz until they coalesce, there could be on the order of
$10^4$ cycles, meaning that the sensitivity of a
second-generation interferometric detector would effectively be 100 times
better than its broadband (prefiltering) sensitivity.  Such detectors could
see typical coalescences at $\sim$~200~Mpc.  The event rate for second-generation 
detectors is estimated at around 40 events per year, with 
rather large error bars~\cite{Burgay:2003jj, Kalogera:2003tn, Living:Lorimer}.

\subsubsection{Chirping binaries as standard sirens}
\label{sec:sirens}

When we consider real binaries we must do the calculation for systems that 
have unequal masses. Still assuming for the moment that the binary orbit 
is circular, the quadrupole amplitude turns out to be 
\begin{equation}\label{eqn:unequalbinary}
h_{\text{binary}} \sim \frac{1}{r} \mptr^{5/3}\Omega^{2/3},
\end{equation}
where we define the \emph{chirp mass} $\mptr$ as
\begin{equation}\label{eqn:chirpmassdef}
\mptr = \mu^{3/5}M^{2/5} = \nu^{3/5} M,\quad \nu=\frac{\mu}{M},
\end{equation}
with $\mu$ the reduced mass, $M$ the total mass and $\nu$
the symmetric mass ratio. We have left out of \Eqref{eqn:unequalbinary}
a factor of order one that depends on the angle from which the binary system
is viewed. The two polarization amplitudes can be found in 
\Eqref{eqn:binaryradiation}.

Remarkably, the other observable, namely the shrinking of the orbit as 
measured by the rate of change of the orbital frequency $P_b$ also 
depends on the masses just through $\mptr$~\cite{Peters:1963ux}:
\begin{equation}\label{eqn:circularPM}
\dot {P_b} = -\frac{192\pi}{5} \left ( \frac{2\pi \cal M}{P_b} \right )^{5/3}.
\end{equation}
In this case, the chirp time is
\begin{equation}\label{eqn:chirptimegeneral}
t_{\text{chirp}} = \frac{5M}{96}\frac{1}{\nu}\left(\frac{M}{R}\right)^{-4}.
\end{equation}
This is just the equal-mass chirp time of \Eqref{eqn:chirp} 
scaled inversely with the symmetric mass ratio $\nu = m_1m_2/M^2$. From 
this equation it is clear that systems with large mass ratios between 
the components can spend a long time in highly relativistic orbits, whereas
equal-mass binaries can be expected to merge after only a few orbits in 
the highly relativistic regime.

If one observes $P_b$ and $\dot {P_b}$, one can infer $\mptr$
from \Eqref{eqn:circularPM}. Then, from 
the observed amplitude in \Eqref{eqn:unequalbinary}, the only remaining
unknown is the distance $r$
to the source. Gravitational wave observations of orbits
that shrink because of gravitational energy losses can
therefore directly determine the distance to the source~\cite{SCHUTZ1986}.
By analogy with the ``standard candles'' of electromagnetic astronomy, these 
systems are now being called ``standard sirens''.  Although our calculation 
here assumed an equal-mass circular system, the conclusion is robust: 
any binary, even with ellipticity and extreme mass ratio, encodes its 
distance in its gravitational wave signal.

This is another way in which gravitational wave observations
are complementary to electromagnetic ones, providing information
that is hard to obtain electromagnetically.  One consequence
is the possibility that observations of coalescing compact
object binaries could allow one to measure the Hubble
constant~\cite{SCHUTZ1986} or other cosmological parameters.
This will be particularly interesting for the LISA project,
whose observations of black hole binaries could  contribute
an independent measurement of the acceleration of the 
universe~\cite{HH, BBHSirens, Arun:2007hu}.

Because chirping systems are so interesting we have also drawn, 
in \Figref{fig:freq}, a line where the chirp time can be measured
in one year. 
This means that the change in frequency due to the chirp must 
be larger than the frequency resolution $1\text{ yr}^{-1}$. A 
little algebra shows that the condition for the chirp to be 
resolved in an observation time $T$ in a binary with period $P_b$ is 
\begin{equation}\label{eqn:chirpresolve}
P_b t_{\text{chirp}} = T^2.
\end{equation}
Since $P_b\propto R^{3/2}M^{-1/2}$, this condition leads to a line
of slope 7/11 in the logarithmic plot in \Figref{fig:freq}. The 
line drawn there corresponds to a resolution time $T$ of one year. 
All binaries below this line will chirp in a short enough time 
for their distances to be measured.

\subsubsection{Binary pulsar tests of gravitational radiation theory}
\label{sec:binpsr}

The most famous example of the effects
of gravitational radiation on an orbiting system is the Hulse--Taylor
Binary Pulsar, PSR~B1913+16.  In this system, two neutron stars orbit in a
close eccentric orbit.  The pulsar provides a regular clock that allows
one to deduce, from post-Newtonian effects, all the relevant orbital
parameters and the masses of the stars.  
The key to the importance of this binary system is that all
of the important parameters of the system can be measured
before one takes account of the orbital shrinking due to 
gravitational radiation reaction.  This
is because a number of post-Newtonian effects on
the arrival time of pulses at the Earth, such as the
precession of the position of the periastron and the
time-dependent gravitational redshift of the pulsar period
as it approaches and recedes from its companion, can be 
measured accurately, and they fully determine the masses,
the semi-major axis and the eccentricity of their
orbit~\cite{Living:Will, Living:Stairs}.

\Eqref{eqn:chirp} for the chirp time predicts that this system would 
change its orbital period $P_b=7.75\mathrm{\ hrs}$ on the timescale (taking $M=1.4\msolar$ and 
$R=10^{6}\mathrm{\ km}$)
\[t_{\text{chirp}}= P_b/\dot{P_b} \sim 1.9\E{18}\text{ s}.\]
From this one can infer that $\dot{P_b}\sim 1.5\E{-14}$. But this 
has to be corrected for our oversimplification of the orbit as 
circular: an eccentric orbit evolves much faster because, during the 
phase of closest approach, the velocities are much higher, and the 
emitted luminosity is a very strong function of the velocity. Using 
equations first computed by Peters and Mathews~\cite{Peters:1963ux}, for the
actual eccentricity of 0.62, one finds (see 
\Eqref{fig:period-decay} below) $\dot P_{\mathrm{T}}=-(2.40242\pm 0.00002)
\times 10^{-12}$. Observations~\cite{Living:Will, WeisbergTaylor2005} currently give
$\dot P_{\mathrm{O}}= -(2.4184\pm 0.0009) \times 10^{-12}$. There 
is a significant discrepancy between these, but it can be removed by 
realizing that the binary system is accelerating toward the center of 
our galaxy, which produces a small period change. Taking this into 
account gives a corrected prediction of $-(2.4056\pm 0.0051)\times 10^{-12}$, 
and this agrees with the observation within the uncertainties~\cite{Living:Will, TaylorWeisberg1989}.
This is the most sensitive test that we have of the correctness of
Einstein's equations with respect to gravitational radiation, and it leaves
little room for doubt in the validity of the quadrupole formula for other
systems that may generate detectable radiation.

A number of other binary systems are now known in which such 
tests are possible~\cite{Living:Stairs}. The most important of the other systems is 
the ``double pulsar'' in which both neutron stars are seen as 
pulsars~\cite{Lyne:doublepulsar}. This system will soon overtake 
the Hulse--Taylor binary as the most accurate test of gravitational 
radiation.

\subsubsection{White-dwarf binaries}

Binary systems at lower frequencies are much more abundant
than coalescing binaries, and they have much longer lifetimes.
LISA will look for close white-dwarf binaries in our galaxy, and will probably see thousands of them.  White dwarfs
are not as compact as black holes or neutron stars. Although their
masses can be similar to that of a neutron star their sizes 
are much larger, typically 3,000~km in
radius. As a result, white-dwarf binaries never reach the
last stable orbit, which would occur at roughly 1.5~kHz for these
masses.  We will discuss the implications of multi-messenger astronomy
for white-dwarf binaries in Section~\ref{sec:multimessanger}.

The maximum amplitude of the radiation from a white-dwarf binary will
be several orders of magnitude smaller than that of a neutron
star or black hole binary at the same distance but close to coalescence.
However, a binary system with a short period is long lived, so the
effective amplitude (after matched filtering) improves as the
square root of the observing time. Besides that, these sources are
nearer: there are many thousands of such systems in our galaxy
radiating in the LISA frequency window above about 1~mHz, and LISA
should be able to see all of them. Below 1~mHz there are even more
sources, so many that LISA will not resolve them individually, but
will see them blended together in a stochastic background of
radiation, as shown in \Figref{fig:noise-curves}.

\subsubsection{Supermassive black hole binaries}

Observations indicate that the center of every galaxy probably hosts a 
black hole whose mass is in the range of
$10^6\mbox{\,--\,}10^9\,M_\odot$~\cite{ReesMBH}, with the black holes
mass correlating well with the mass of the galactic bulge. A black
hole whose mass is in the above range is called a \emph{supermassive
  black hole} (SMBH). There is now abundant observational evidence
that galaxies often collide and merge, and there are good reasons to believe that 
when that happens, friction between the SMBHs and the stars and gas of the 
irregular merged galaxy will lead the SMBHs to spiral into a common nucleus and 
(on a timescale of some $10^8$~yr) even get close enough to be
driven into complete orbital decay by gravitational radiation 
reaction. In many systems this should lead to a black hole merger within a
Hubble time~\cite{KomossaEtAl}.  For a binary with two 
nonspinning $M=10^6\,M_\odot$ black holes, the frequency
of emitted gravitational waves at the last stable orbit is, from \Eqref{eqn:LSO}, 
$f_{\text{LSO}} = 4\mathrm{\ mHz}$;
during and after the merger the frequency rises from 4~mHz to the quasi-normal-mode 
frequency of 24~mHz (if the spin of the final black hole is negligible). (Naturally, 
all these frequencies simply scale inversely with the mass for other mass ranges.)
This is in the frequency range of LISA, and observing these mergers is one 
of the central purposes of the mission.

SMBH mergers are so spectacularly strong that they will be visible in LISA's 
data stream even before applying any matched filter,
although good models of the inspiral and particularly the 
merger radiation will be needed to extract source parameters. Because the 
masses of such black holes are so large, LISA can see essentially any 
merger that happens in its frequency band anywhere in the universe, even out 
to extremely high redshifts. It can thereby 
address astrophysical questions about the origin, growth and population of
SMBHs.  The recent discovery of an SMBH binary~\cite{KomossaEtAl}
and the association of X-shaped radio lobes with the merger of SMBH binaries~\cite{Merritt:2002hc} has further raised the optimism concerning
SMBH merger rates, as has the suggestion that an SMBH has been observed to have been expelled from the center of its
galaxy, an event that could only have happened as a result of a merger
between two SMBHs~\cite{Komossa:2008qd}. The rate at
which galaxies merge is about 1~yr$^{-1}$ out to a red-shift of
$z=5$~\cite{Haehnelt}, and LISA's detection rate for SMBH mergers
might be roughly the same.

Modelling of the merger of two black holes requires numerical relativity, and 
the accuracy and reliability of numerical simulations is now becoming good enough 
that they will soon become an integral part of gravitational wave searches.

\subsubsection{Extreme and intermediate mass-ratio inspiral sources}

The SMBH environment of our own galaxy is known to contain
a large number of compact objects and white dwarfs. Near the central SMBH  
there is a disproportionately large number of stellar-mass
black holes, which have sunk there through random gravitational 
encounters with the normal stellar population (dynamical friction). Three body
interaction will occasionally drive one of these compact objects into a capture orbit of the central
SMBH.  The compact object will sometimes be captured~\cite{ReesMBH, bss:mbh.mbh.coalescence2, Sigurdsson:1997vc}
into a highly eccentric trajectory
($e > 0.99$) with the periastron close to the last stable orbit of the SMBH.
Since the mass of the captured object will be about $1\mbox{\,--\,}100\,M_\odot$, while
the SMBH will have a far greater mass, we essentially have a ``test mass'' falling
in the geometry of a Kerr black hole. By \Eqref{eqn:chirptimegeneral} we would 
expect that the small body would spend many orbits in the relativistic regime near 
the horizon of the large black hole: a $10\msolar$ black hole falling into a $10^6\msolar$ black hole 
would require on the order of $10^5$ orbits.  The emitted gravitational radiation~\cite{Ryan:1995wh, Glampedakis:2002ya, Glampedakis:2002cb, BarackCutler2003, Gair:2005ih, Babak:2006uv} would consist of a very long wave train that carries 
information about the nearly geodesic trajectory  of the test body, thereby
providing a very clean probe to survey the spacetime geometry of the central
object (which could be a Kerr black hole or some other compact object) and
test whether or not this geometry is as predicted by general
relativity~\cite{Ryan:1997hg, Hughes:2000ssa, Glampedakis:2005cf, Glampedakis:2005hs, Barack:2006pq}.

This kind of event happens occasionally to every SMBH that lives in the 
center of a galaxy. Indeed, since the SNR from
matched filtering builds up in proportion to the square root of the
observation time $t_{\mathrm{chirp}} \propto \nu^{-1} = (\mu/M)^{-1}$ 
[cf.~\Eqref{eqn:chirptimegeneral}]
and the inherent amplitude of the radiation is linear in $\nu$ 
[cf.~\Eqref{eqn:unequalbinary}], the SNR varies with the symmetric mass ratio
 as $\sqrt{\nu}$ and typical SNR will be about ten to 
 a thousand times smaller than an SMBH binary at the same distance.
LISA will, therefore, be able to see such sources only to much
smaller distances, say between 1 to 10~Gpc depending on the mass ratio.
For events at such distances LISA's SNR after matched filtering 
could be in the range 10\,--\,100, but matched filtering will be very 
difficult because of the complexity of the orbit, especially of its 
evolution due to radiation effects.  However, 
this volume of space contains a large number of galaxies, and the event 
rate is expected to be several tens to hundreds per year~\cite{BarackCutler2003}.
There will be a background from more distant sources that might in the end 
set the limit on how much sensitivity LISA has to these events.

Due to relativistic frame dragging, for each passage of the apastron the
test body could execute several nearly circular orbits at its periastron.
Therefore, long periods of low-frequency, small-amplitude
radiation will be followed by several cycles of high-frequency, large-amplitude
radiation~\cite{Ryan:1995wh, Glampedakis:2002ya, Glampedakis:2002cb, BarackCutler2003, Gair:2005ih, Babak:2006uv}.  
The apastron slowly shrinks, while the periastron remains more or less at the same
location, until the final plunge of the compact object before merger. Moreover,
if the central black hole has a large spin then spin-orbit coupling leads to precession
of the orbital plane thereby changing the polarization of the wave as seen by LISA.

Thus, there is a lot of structure in the waveforms owing to a number
of different physical effects: contribution from higher-order
multipoles due to an eccentric orbit, precession of the orbital plane,
precession of the periastron, etc., and gravitational radiation
backreaction plays a pivotal role in the dynamics of these systems. If
one looks at the time-frequency map of such a signal one notices that
the signal power is greatly \emph{smeared} across the
map~\cite{BSS:Moriond}, as compared to that of a sharp chirp from a
nonspinning black-hole binary. For this reason, this \emph{spin
  modulated chirp} is sometimes referred to as a
\emph{smirch}~\cite{BSSAndBFS}. More commonly, such sources are called
\emph{extreme mass ratio inspirals} (EMRIs) and represent systems
whose mass ratio is in the range of $\sim
10^{-3}\mbox{\,--\,}10^{-6}$.  Inspirals of systems with their mass
ratio in the range $\sim 10^{-2}\mbox{\,--\,}10^{-3}$ are termed
intermediate mass ratio inspirals or IMRIs. These latter systems
correspond to the inspiral of intermediate mass black holes of mass
$\sim 10^3\mbox{\,--\,}10^4\,M_\odot$ and might constitute a prominent
source in LISA provided the central SMBH grew in mass as a result of a
number of mergers of small black holes~\cite{AmaroSeoane:2006pz,
  AmaroSeoane:2006py, AmaroSeoane:2007aw}.

While black hole perturbation theory with a careful treatment
of radiation reaction is necessary for the description of EMRIs, IMRIs may be
amenable to a description using a hybrid scheme of post-Newtonian approximations
and perturbation theory. This is an area that requires more study.

\subsection{Quasi-normal modes of a black hole}

In 1970, Vishveshwara~\cite{bss:vishu} discussed a \emph{gedanken} experiment,
similar in philosophy to Rutherford's (real) experiment with the atom. In 
Vishveshwara's experiment, he scattered gravitational radiation off a black hole to explore
its properties. With the aid of such a gedanken experiment,
he demonstrated for the first time that gravitational waves
scattered off a black hole will have a characteristic waveform, when the
incident wave has frequencies beyond a certain value, depending on the
size of the black hole. It was soon realized that
perturbed black holes have \emph{quasi-normal modes} (QNMs) of vibration and
in the process emit gravitational radiation,  whose amplitude, frequency and
damping time are characteristic of its mass and angular
momentum~\cite{bss:press, Living:KokkotasSchmidt}. We will discuss
in~\Secref{sec:black hole spectroscopy} how observations of QNMs 
could be used in testing strong field predictions of general relativity.

We can easily estimate the amplitude of gravitational waves emitted
when a black hole forms at a distance $r$ from Earth as a result of
the coalescence of compact objects
in a binary. The effective amplitude is given
by~\Eqref{eq:effective.amplitude}, which involves the energy
$E$ put into gravitational waves and the frequency $f$ at which the waves
come off. By dimensional arguments $E$ is proportional to the total mass
$M$ of the resulting black hole. The efficiency at which the energy
is converted into radiation depends on the symmetric mass ratio $\nu$
of the merging objects. One does not know the fraction of the
total mass emitted nor the exact dependence on $\nu$. Flanagan
and Hughes~\cite{Flanagan1998} argue that
$E \sim 0.03 (4\nu)^2 M$. The frequency $f$ is inversely proportional to
$M$; indeed, for Schwarzschild black holes $f=(2\pi M)^{-1}$. Thus, the
formula for the effective amplitude takes the form
\begin{equation}
h_{\mathrm{eff}} \sim \frac {4 \alpha \nu M}{\pi r},
\end{equation}
where $\alpha$ is a number that depends on the
(dimensionless) angular momentum $a$ of the black hole and has a value between
0.7 (for $a=0$, Schwarzschild black hole) and 0.4 (for $a=1$, maximally
spinning Kerr black hole). For stellar mass black holes at a distance
of 200~Mpc the amplitude is:
\begin{equation}
h_{\mathrm{eff}} \simeq 10^{-21}
\left (\frac{\nu}{0.25} \right )
\left (\frac{M}{20\,M_\odot} \right )
\left (\frac{r}{200\mathrm{\ Mpc}} \right )^{-1}.
\end{equation}
For SMBHs, even at cosmological distances, the
amplitude of quasinormal mode signals is pretty large:
\begin{equation}
h_{\mathrm{eff}} 
\simeq 3 \times 10^{-17}
\left (\frac{\nu}{0.25} \right )
\left (\frac{M}{2 \times 10^6\,M_\odot} \right )
\left (\frac{r}{6.5\mathrm{\ Gpc}} \right )^{-1}.
\end{equation}
In the first case we have a pair of $10\,M_\odot$ black holes
inspiraling and merging to form a single black hole. In this case
the waves come off at a frequency of around 
500~Hz [cf. \Eqref{eq:gw frequency}]. The initial
ground-based network of detectors might be able to pick these
waves up by matched filtering,  especially when an inspiral
event precedes the ringdown signal. A $100\,M_\odot$ black hole
plunging into a $10^{6}\,M_\odot$  black hole at a distance of 6.5~Gpc 
($z\simeq 1$) gives out radiation at a frequency of about 15 mHz.
Note that in the latter case the radiation is redshifted from 
30~mHz to 15~mHz. 
Such an event produces an amplitude just large enough to be detected
by LISA. At the same distance, a pair of $10^6\,M_\odot$ SMBHs spiral in and merge to produce a fantastic amplitude of 
$3 \times 10^{-17}$, way above the LISA background noise. In this 
case, the signals would be given off at about 7.5~mHz and will be loud 
and clear to LISA.  It will not only be possible to detect these 
events, but also to accurately measure the masses and spins of the objects
before and after merger and thereby test the black hole no-hair theorem
and confirm whether the result of the merger is indeed a black hole
or some other exotic object (e.g., a boson star or a naked singularity).

\subsection{Stochastic background}
\label{sec:stochastic bg introduction}

In addition to radiation from discrete sources, the universe should have 
a random gravitational wave field that results from a superposition of 
countless discrete systems and also from fundamental processes, such as 
the Big Bang. Observing any of these backgrounds would bring useful 
information, but the ultimate goal of detector development is the observation
of the background radiation from the Big Bang. It is expected to be very weak, 
but it will come to us unhindered from as early as $10^{-30}$~s, and it could 
illuminate the nature of the laws of physics at energies far higher than we 
can hope to reach in the laboratory. 

It is usual to characterize the intensity of a random field of gravitational 
waves by its energy density as a function of frequency.   Since the energy density 
 of a plane wave is the same as its flux (when $c=1$), we have from \Eqref{eq:flux}
\[\rho_{\text{gw}} = \frac{\pi}{4}f^2h^2. \]
But the wave field in this case is a random variable, so we must replace $h^2$ 
by a statistical mean square amplitude per unit frequency 
(Fourier transform power per unit frequency) called $S_{\mathrm{gw}}(f)$, 
so that the energy density \emph{per unit frequency} 
is proportional to $f^2 S_{\mathrm{gw}}(f)$.  It is then conventional to talk about the 
energy density per unit logarithm of the frequency, which means multiplying by $f$.    
The result, after being careful about averaging over all directions of the 
waves and all independent polarization components, is~\cite{Allen1997, Thorne1987}
\[\oderiv{\rho_{\text{gw}}}{\ln f} = 4\pi^2 f^3 S_{\mathrm{gw}}(f).\]
Finally, what is of most interest is the energy density as a fraction of the 
closure or critical cosmological density, given by the Hubble constant 
$H_0$ as $\rho_c = 3H_0^2/8\pi$.  The resulting ratio is called $\Omega_{\text{gw}}(f)$:
\[\Omega_{\text{gw}}(f) = \frac{10\pi^2}{3H_0^2}f^3S_{\mathrm{gw}}(f).\]

The only tight constraint on  
$\Omega_{\text{gw}}$ from non--gravitational-wave astronomy is that it 
must be smaller than $10^{-5}$, in order not to disturb the agreement 
between the standard Big Bang model of nucleosynthesis (of helium and
other light elements) and observation. If the universe contains
this much gravitational radiation today, then at the time of
nucleosynthesis the (blue-shifted) energy density of this radiation
would have been comparable to that of the photons and the three
neutrino species. Although the radiation would not have participated
in the nuclear reactions, its extra energy density would have required
that the expansion rate of the universe at that time be
significantly faster, in order to evolve into the universe we see
today. In turn, this faster expansion would have provided less time
for the nuclear reactions to ``freeze out'', altering the abundances
from the values that are observed today~\cite{Pagel:2000,
  Steigman:2007xt}. First-generation interferometers should be able to
set direct limits on the cosmological background at around this
level. Radiation in the lower-frequency LISA band, from galactic and
extra-galactic binaries, is expected to be much smaller than this
bound.

Random radiation is indistinguishable from instrumental noise in a single 
detector, at least for short observing times. If the random field is produced by
an anisotropically-distributed set of astrophysical sources (the binaries in our 
galaxy, for example) then over a year, as the detector changes its orientation, the 
noise from this background should rise and fall in a systematic way, allowing 
it to be identified. But this is a rather crude way of detecting the radiation, and 
a better way is to perform a cross-correlation between two detectors, if available. 

In cross-correlation, which amounts to multiplying the outputs and integrating, the 
random signal in one detector essentially acts as a template for the signal in the 
other detector. If they match, then there will be a stronger-than-expected correlation. 
Notice that they can only match well if the wavelength of the gravitational waves is longer
than the separation between the detectors: otherwise time delays for waves reaching one 
detector before the other degrade the match. The outcome is not like standard 
matched filtering, however, since the ``filter'' of the first detector has as much 
noise superimposed on its template as the other detector. As a result, the 
amplitude SNR of the correlated field grows only with observing time $T$ as $T^{1/4}$, 
rather than the square root growth that characterizes matched filtering~\cite{Thorne1987}.

%==================================================================
\newpage

\section{Gravitational Wave Detectors and Their Sensitivity}
\label {sec:gwdetectors}

Detectors of gravitational waves generally divide into two classes:
\emph{beam detectors} and \emph{resonant mass detectors}. In beam
detectors, gravitational waves interact with a beam of electromagnetic
radiation, which is monitored in some way to register the passage of
the wave. In resonant mass detectors, the gravitational wave transfers
energy to a massive body, from which the resultant oscillations are observed.

Both classes include a variety of systems. The principal beam
detectors are the large ground-based laser interferometers currently
operating in several locations around the globe, such as the LIGO
system in the USA. The ESA--NASA LISA mission aims to put a laser
interferometer into space to detect milliHertz gravitational
waves. But beam detectors do not need to involve interferometry: the
radio beams transponded to interplanetary spacecraft can carry the
signature of a passing gravitational wave, and this method has been
used to search for low-frequency gravitational waves. And radio
astronomers have for many years monitored the radio beams of distant
pulsars for evidence of gravitational waves; new radio instrumentation
is turning this into a powerful and promising method of looking for
stochastic backgrounds and individual sources. And at ultra-low
frequencies, gravitational waves in the early universe may have left
their imprint on the polarization of the cosmic microwave background.

Resonant mass detectors were the first kind of detector built in the
laboratory to detect gravitational waves: Joseph
Weber~\cite{Weber1967} built two cylindrical aluminum bar detectors
and attempted to find correlated disturbances that might have been
caused by a passing impulsive gravitational wave. His claimed
detections led to the construction of many other bar detectors of
comparable or better sensitivity, which never verified his
claims. Some of those detectors were not developed further, but others
had their sensitivities improved by making them cryogenic, and today there
are two ultra-cryogenic detectors in operation (see Section~\ref{sec:barprinciples}).

%future revision: In addition there have been attempts to observe gravitational waves through 
%the resonant oscillations of the Earth, Moon, and sun. 

In the following, we will examine the principal detection methods that
hold promise today and in the near future.

\subsection{Principles of the operation of resonant mass detectors}
\label{sec:barprinciples}

A typical ``bar'' detector consists of a cylinder of aluminum with a
length $\ell \sim 3\mathrm{\ m}$,   a very narrow resonant frequency
between $f\sim 500\mathrm{\ Hz}$ and 1.5~kHz, and a mass $M\sim
1000\mathrm{\ kg}$. A short gravitational wave burst with $h\sim
10^{-21}$ will make the bar vibrate with an amplitude 
\begin{equation}
\delta \ell_{gw} \sim h\ell \sim 10^{-21}\mathrm{\ m}. 
\end{equation}
To measure this, one must fight against three main sources of noise.

\begin{enumerate}

\item {\bf Thermal noise.} The original Weber bar operated at room temperature, 
but the most advanced detectors today, Nautilus~\cite{Astone:2002ej} and 
Auriga~\cite{AURIGA}, are ultra-cryogenic, operating at $T = 100\mathrm{\ mK}$.  At this temperature 
the root mean square (rms) amplitude of vibration is 
\begin{equation}
\langle \delta \ell^2 \rangle^{1/2}_{th} = 
\left(\frac{kT}{4 \pi^2 M f^2} \right)^{1/2} \sim 6\times 10^{-18}\mathrm{\ m}. 
\end{equation}
This is far larger than the gravitational wave amplitude expected from
astrophysical sources.  But if the material has a high $Q$ (say,
$10^6$) in its fundamental mode, then that changes its thermal
amplitude of vibration in a random walk with very small steps, taking
a time $Q/f\sim 1000\mathrm{\ s}$ to change by the full amount.
However, a gravitational wave burst will cause a change in 1~ms. In
1~ms, thermal noise will have random-walked to an expected amplitude
change  $(1000\mathrm{\ s}/1\mathrm{\ ms})^{1/2} = Q^{1/2}$ times
smaller, or (for these numbers)
\begin{equation}
\langle \delta \ell^2 \rangle^{1/2}_{th:\;1\mathrm{\ ms}} = 
\left(\frac{kT}{4 \pi^2 M f^2Q} \right)^{1/2} \sim 6\times 10^{-21}\mathrm{\ m}. 
\end{equation}
So ultra-cryogenic bars can approach the goal of 
detection near $h=10^{-20}$ despite thermal noise.

\item {\bf Sensor noise.} A transducer converts the bar's mechanical 
energy into electrical energy, and 
an amplifier  increases the electrical 
signal to record it.  If sensing of the vibration could 
be done perfectly, then the detector would be broadband: 
both thermal impulses and gravitational wave forces are mechanical forces, 
and the ratio of their induced vibrations would be the same at all 
frequencies for a given applied impulsive force.  

But sensing is not 
perfect: amplifiers introduce noise, and this makes small 
amplitudes harder to measure.  The amplitudes of vibration are largest in
the resonance band near $f$, so amplifier noise limits the detector sensitivity
to gravitational wave frequencies near $f$.  But if the noise is small, then 
the measurement bandwidth about $f$ can be much larger 
than the resonant bandwidth $f/Q$.  Typical measurement
bandwidths are 10~Hz, about $10^4$ times larger than the 
resonant bandwidths, and   100~Hz is not out of the question~\cite{Baggio2005}.  

\item {\bf Quantum noise.} The zero-point vibrations of a bar with a frequency 
of 1~kHz are 
\begin{equation}
\langle \delta \ell^2 \rangle^{1/2}_{quant} = 
\left(\frac{\hbar}{2 \pi M f} \right)^{1/2} \sim 4 \times 10^{-21}\mathrm{\ m}. 
\end{equation}
This is comparable to the thermal 
limit over 1~ms.  So, as detectors improve their thermal limits, 
they  run into the quantum limit, which must be breached before a signal 
at $10^{-21}$ can be seen with such a detector.  

It is  not impossible to do better than the quantum limit. The uncertainty principle 
only sets the limit above if a measurement tries to determine the excitation 
energy of the bar, or equivalently the phonon number.  But one is not interested 
in the phonon number, except in so far as it allows one to determine the 
original gravitational wave amplitude. 
It is possible to define other observables that also respond to the 
gravitational wave and can be measured more accurately by {\bf
  squeezing} their uncertainty at the expense of greater errors in
their conjugate observable~\cite{Caves1980}. It is not yet clear
whether squeezing will be viable for bar detectors, although squeezing
is now an established technique in quantum optics and will soon be
implemented in interferometric detectors (see below).

\end{enumerate}

Reliable gravitational wave detection, whether with bars or with other detectors, 
requires coincidence observations, in which two or more detectors confirm each other's 
findings. The principal bar detector projects around the world formed the International
Gravitational Event Collaboration (IGEC)~\cite{IGEC} to arrange for long-duration 
coordinated observations and joint data analysis. A report in 2003 of an analysis of 
a long period of coincident observing over three years found no evidence of significant events~\cite{IGEC2003}. The ALLEGRO bar~\cite{ALLEGRO} at Louisiana 
State University  made joint data-taking 
runs with the nearby LIGO interferometer, setting an upper limit on the 
stochastic gravitational-wave background at around 900~Hz of $h_{100}^2\Omega_{gw}(f) \le 0.53$~\cite{ALLEGRO_LIGO2007}. 
More recently, because funding for many of the bar detector projects has
 become more restricted,  only two groups continue to operate bars at present (end of 
2008): the Rome~\cite{ROG} and Auriga~\cite{AURIGA} groups. The latest 
observational results from IGEC may be found in~\cite{Astone2007}.

It is clear from the above discussion that bars have great difficulty achieving 
the sensitivity goal of $10^{-21}$.  This limitation was apparent even in the 1970s, and that 
motivated a number of groups to explore the intrinsically 
wide-band technique of laser interferometry, leading to the projects described in 
Section~\ref{sec:present} below. However,  the excellent sensitivity 
of resonant detectors within their narrow bandwidths 
makes them suitable for specialized, high-frequency searches, including cross-correlation 
searches for stochastic backgrounds~\cite{ComptonSchutz1997}. Therefore, 
novel and imaginative designs for resonant-mass 
detectors continue to be proposed. For example, it is possible  to 
construct large spheres of a similar 
size (1 to 3~m diameter) to existing cylinders. This increases the 
mass of the detector and also improves its direction-sensing. One can  in 
principle push to below $10^{-21}$ with spheres~\cite{COCCIA1995e}. A spherical 
prototype called MiniGRAIL\cite{GRAIL} has been operated in 
the Netherlands\cite{MiniGRAIL2007}. A similar prototype called the Schenberg
detector\cite{Schenberg} is being built in Brazil~\cite{Schenberg2005}. 
Nested cylinders or spheres, or masses designed to sense multiple modes of vibration
may also provide a clever way to improve on bar sensitivities~\cite{Dual2006}. 

While these ideas have interesting potential, funding for them is at present (2008) 
very restricted, and the two remaining bar detectors are likely to be shut down in the 
near future, when the interferometers begin operating at sensitivities clearly better than 
$10^{-21}$.

\subsection{Principles of the operation of beam detectors}
\label{sec:beamprinciples}

Interferometers use laser light to measure changes in the difference 
between the lengths of two perpendicular (or nearly perpendicular) arms. 
Typically, the arm lengths respond differently to 
a given gravitational wave, so an interferometer is a natural 
instrument to measure gravitational waves.  But other detectors also 
use electromagnetic radiation, for example, ranging to spacecraft 
in the solar system and even pulsar timing.

The basic equation we need is for the effect of a plane linear gravitational wave 
on a beam of light. Suppose the angle between the direction of the beam 
and the direction of the plane wave is $\theta$. We imagine a very simple experiment in which 
the light beam originates at a clock, whose proper time is called $t$, and is 
received by a clock, whose proper time is $t_f$. The beam and gravitational-wave travel directions determine a plane, and we denote the polarization component
of the gravitational wave that acts in this plane by $h_+(t)$, as measured at 
the location of the originating clock.  The proper distance between the 
clocks, in the absence of the wave, is $L$. If the originating clock puts 
timestamps onto the light beam, then the receiving clock can measure the rate 
of arrival of the timestamps. If there is no gravitational wave, and if the 
clocks are ideal, then the rate will be constant, which can be normalized to unity.
The effect of the gravitational wave is to change the arrival rate as a function 
of the emission rate by 
\begin{equation}\label{eqn:onewaylight}
\oderiv{t_f}{t} = 1 + \frac{1}{2}(1+\cos\theta)\left\{h_+[t+(1-\cos\theta)L]-h_+(t)\right\}.
\end{equation}
This is very simple: the beam of light leaves the emitter at the time when the 
gravitational wave of phase $t$ passes the emitter, and it reaches the receiver
at the time when the gravitational wave of phase $t+(1-\cos\theta)L$ is passing 
the receiver. So in the plane wave case, only the amplitudes of the wave at the 
emitting and receiving events affect the time delay. 

In order to use such an arrangement to detect gravitational waves, one needs 
two very stable clocks. The best clocks today are stable to a few parts 
in $10^{16}$~\cite{Living:Armstrong}, which implies that the minimum 
amplitude of gravitational waves that could be detected by such 
a two-clock experiment is $h\sim10^{-15}$. However, this equation is also
fundamental to the detection of gravitational waves by pulsar timing, in which 
the originating `clock' is a pulsar. By correlating many pulsar signals,
one can beat down the single-pulsar noise. This is described below 
in \Secref{sec:pulsartiming}.

An arrangement that uses only one clock is one that sends a beam out to 
a receiver, which then reflects or retransmits (transponds) the beam back 
to the sender. The sender has the clock, which measures variations in the 
round-trip time. This method has been used with interplanetary spacecraft, 
which has the advantage that the only clock is on the ground, which can be 
made more stable than one carried in a spacecraft (see \Secref{sec:ranging}). 
For the same arrangement as above, the return time $t_\text{return}$ varies 
at the rate
\begin{eqnarray}
\oderiv{t_\text{return}}{t}&=&1 + \frac{1}{2}\left\{(1-\cos\theta)\, h_+(t+2L) 
- (1+\cos\theta)\, h_+(t) 
\right . \nonumber \\
& +& \left . 2\, \cos\theta\, h_+[t+L(1-\cos\theta)]\right\}. \label{eqn:threeterm}
\end{eqnarray}
This is known as the \emph{three-term relation}, the third term being the wave 
strength at the time the beam returns back to the sender. 

But the sensitivity of such a one-path system as a gravitational wave detector 
is still limited by the stability of the clock. For that reason, interferometers
have become the most sensitive beam detectors: effectively one arm of the 
interferometer becomes the `clock', or at least the time standard, that variations
in the other arm are compared to. Of course, if both arms are affected by a 
gravitational wave in the same way, then the interferometer will not see the 
wave. But this happens only in very special geometries. For most wave 
arrival directions and polarizations, the arms are affected differently, and 
a simple interferometer measures the difference between the round-trip travel time 
variations in the two arms. For the triangular space array LISA, the measured 
signal is somewhat more complex (see \Secref{sec:LISA} below), but it still preserves the principle
that the time reference for one arm is a combination of the others.

\subsubsection{The response of a ground-based interferometer}
\label{sec:beam factors}

Ground-based interferometers are the most sensitive detectors operating 
today, and are likely to make the first direct detections~\cite{HOUGHREF}.
The largest detectors operating today are the LIGO 
detectors~\cite{RAAB1995}, two of which have arm lengths of 
4~km. This is much 
smaller than the wavelength of the gravitational wave, so the interaction 
of one arm with a gravitational wave can be well approximated 
by the small-$L$ approximation to \Eqref{eqn:threeterm}, namely
\begin{equation}\label{eqn:shortarm}
\oderiv{t_\text{return}}{t} = 1 + \sin^2\theta L\dot{h}_+(t).
\end{equation}
(See~\cite{Grishchuk2004} for first corrections to the short-arm approximation.)
To analyze the full detector, where the second arm will normally point out of 
the plane we have been working in up till now, it is helpful to go over to a 
tensorial expression, independent of special coordinate orientations. The gravitational 
wave will act in the plane transverse to the propagation direction; let us call
that direction $\Nhat$ and let us set up radiation basis vectors $\exRhat$ and $\eyRhat$ 
in the transverse plane, such that $\exRhat$ lies in the plane formed by the wave propagation direction and the arm of our gravitational wave sensor, which lies along the $x$-axis of the detector plane, whose unit vector is $\exhat$. (For a picture 
of this geometry, see the left-hand panel of \Figref{fig:polConvention}, where for the moment
we are ignoring the $y$-arm of the detector shown there.)

With these definitions, the wave amplitude $h_+$ is the one that has $\exRhat$ and 
$\eyRhat$ as the axes of its ellipse. The full wave amplitude is described, 
as in \Eqnref{eqn:TTtensor}, by the wave tensor 
\begin{equation}\label{eqn:wavetensor}
\htensor(t) = h_+(t) \eptensor + h_\times(t) \extensor,
\end{equation}
where $\eptensor$ and $\extensor$ are the polarization tensors associated with these 
basis vectors (compare \Eqnref{eqn:TTbasistensors}):
\begin{equation}\label{eqn:poltensor}
\eptensor = (\exRhat \otimes \exRhat - \eyRhat \otimes \eyRhat), \quad 
\extensor = (\exRhat \otimes \eyRhat + \eyRhat \otimes \exRhat).
\end{equation}
The unique way of expressing \Eqnref{eqn:shortarm} in terms of $\htensor$ is
\begin{equation}\label{eqn:invariantshortarm}
\left(\oderiv{t_\text{return}}{t}\right)_{\mathrm{x-arm}} = 1 + L \exhat \cdot \dot{\htensor} \cdot \exhat.
\end{equation}
This does not depend on any special orientation of the arm relative to the wave 
direction, and does not depend on the basis we chose in the transverse plane, so 
we can use it as well for the second arm of the interferometer, no matter what its 
orientation. Let us assume it lies along the unit vector by $\eyhat$. (We do not, in fact, have to assume that the two arms are perpendicular to each other, but it simplifies the diagram a little.) The return-time derivative along the second arm is then given by
\[\left(\oderiv{t_\text{return}}{t}\right)_{\mathrm{x-arm}} = 1 + L \eyhat \cdot \dot{\htensor} \cdot \eyhat\].
The interferometer responds to the difference between these times, 
\[\left(\oderiv{\delta t_\text{return}}{t}\right)\ = \left(\oderiv{t_\text{return}}{t}\right)_{\mathrm{x-arm}} - \left(\oderiv{t_\text{return}}{t}\right)_{\mathrm{y-arm}} = L \exhat \cdot \dot{\htensor} \cdot \exhat - L \eyhat \cdot \dot{\htensor} \cdot \eyhat\].
By analogy with the wave tensor, we define the \emph{detector tensor} $\dtensor$ by~\cite{dt88}
\begin{equation}\label{eqn:interferometerdetectortensor}
\dtensor = L (\exhat \otimes \exhat - \eyhat \otimes \eyhat).
\end{equation}
(If the arms are not perpendicular this expression would still give
the correct tensor if the unit vectors lie along the actual arm
directions.) Then we can express the differential return time rate in
the simple invariant form
\begin{equation}
\label{interferometerrate}
\left(\oderiv{\delta t_\text{return}}{t}\right)\ = \dtensor:\dot{\htensor},
\end{equation}
where the notation ${\mathbf d}:{\mathbf h}\equiv d_{lm} h^{lm}$
denotes the Euclidean scalar product of the tensors ${\mathbf d}$ and
${\mathbf h}$. Equation~(\ref{interferometerrate}) can be integrated
over time to give the instantaneous path-length (or time-delay, or
phase) difference between the arms, as measured by the central
observer's proper time clock:
\begin{equation}\label{interferometerphase}
\delta t_\text{return}(t) = \dtensor:\htensor.
\end{equation}
This is a robust and compact expression for the response of any interferometer to any wave 
in the long-wavelength (short-arm) limit. Its dependence on the wave direction is 
called its antenna pattern.

It is conventional to re-express this measurable in terms of the stretching of the arms of the 
interferometer. Within our approximation that the arms are shorter than a wavelength, 
this makes sense: it is possible to define a local inertial coordinate system that 
covers the entire interferometer, and within this coordinate patch (where light 
travels at speed 1) time differences measure proper length differences. The 
differential return time is twice the differential length change of the arms: 
\begin{equation}\label{eqn:interferometerstretch}
\delta L(t) = \frac{1}{2}\,\dtensor:\htensor.
\end{equation}

For a bar detector of length $L$ lying along the director $\ahat$, the detector 
tensor is 
\begin{equation}\label{eqn:bardetectortensor}
\dtensor = L \ahat\otimes\ahat,
\end{equation}
although one must be careful that the change in proper length of a bar is not simply 
given by \Eqnref{eqn:interferometerstretch}, because of the restoring forces in 
the bar.

When dealing with observations by more than one detector, it is not convenient 
to tie the alignment of the basis vectors in the sky plane with 
those in the detector frame, as we have done in the left-hand panel of 
\Figref{fig:polConvention}, since the detectors will have different orientations. 
Instead it will usually be more convenient to choose polarization tensors in the 
sky plane according to some universal reference, e.g., using a convenient 
astronomical reference frame. The right-hand panel of \Figref{fig:polConvention} 
shows the general situation, where the basis vectors $\alphahat$ and $\betahat$ are 
rotated by an angle $\psi$ from the basis used in the left-hand panel. The 
polarization tensors on this new basis,
\begin{equation}\label{eqn:epspoltensor}
\epsptensor = (\alphahat \otimes \alphahat - \betahat \otimes \betahat), \quad 
\epsxtensor = (\alphahat \otimes \betahat + \betahat \otimes \alphahat),
\end{equation}
are found by the following transformation from the previous ones:
\begin{eqnarray}\label{eqn:poltensortransform}
\epsptensor & = & 
\eptensor \cos 2\psi + \extensor \sin 2\psi,\nonumber \\
\epsxtensor & = &
-\eptensor \sin 2\psi + \extensor \cos 2\psi.
\end{eqnarray}

Then one can write \Eqnref{eqn:interferometerstretch} as
\begin{equation}\label{eqn:interferometerantenna}
\frac{\delta L(t)}{L} = F_+(\theta,\,\phi,\,\psi) h_+(t) + 
F_\times(\theta,\,\phi,\,\psi) h_\times(t), 
\end{equation}
where $F_+$ and $F_\times$ are the \emph{antenna pattern} functions for the 
two polarizations defined on the sky-plane basis by
\begin{equation}\label{eqn:antennapatternsDef}
F_+ \equiv \dtensor:\eptensor,\quad  F_\times \equiv \dtensor:\extensor.
\end{equation}
Using the geometry in the right-hand panel of \Figref{fig:polConvention},
one can show that 
\begin{eqnarray}\label{eqn:interferometerantennapatterns}
F_+ & = & \frac{1}{2}\left (1+\cos^2\theta \right )\cos 2\phi \cos 2\psi - 
\cos\theta\sin 2\phi \sin 2\psi, \nonumber \\
F_\times & = & \frac{1}{2}\left (1+\cos^2\theta \right )\cos 2\phi \sin 2\psi + 
\cos\theta\sin 2\phi \cos 2\psi.
\end{eqnarray}

\epubtkImage{Detector_sky_2arms-Detector_sky_full.png}{%
  \begin{figure}[htbp]
    \centerline{
      \includegraphics[width=2.6in]{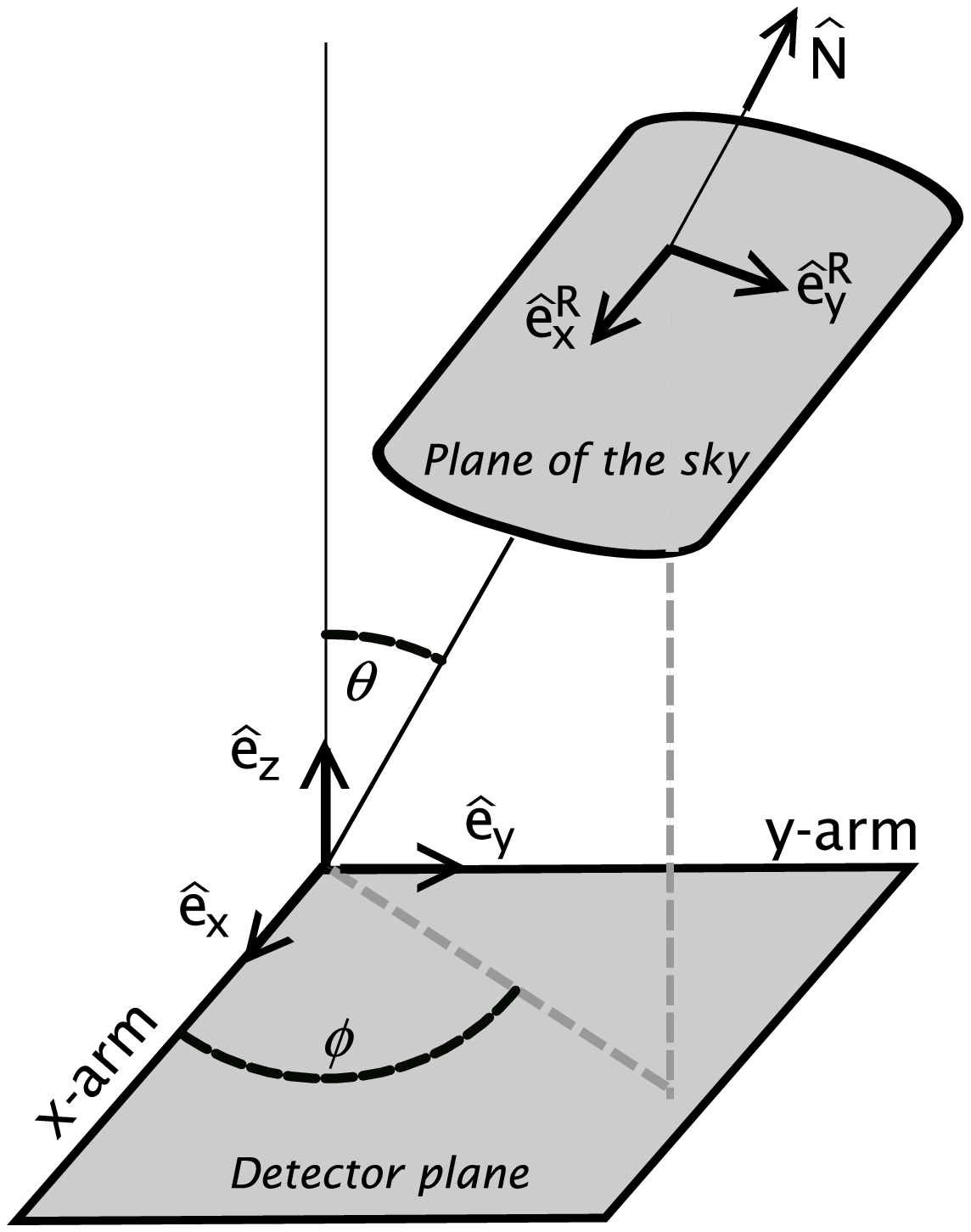}
      \includegraphics[width=2.6in]{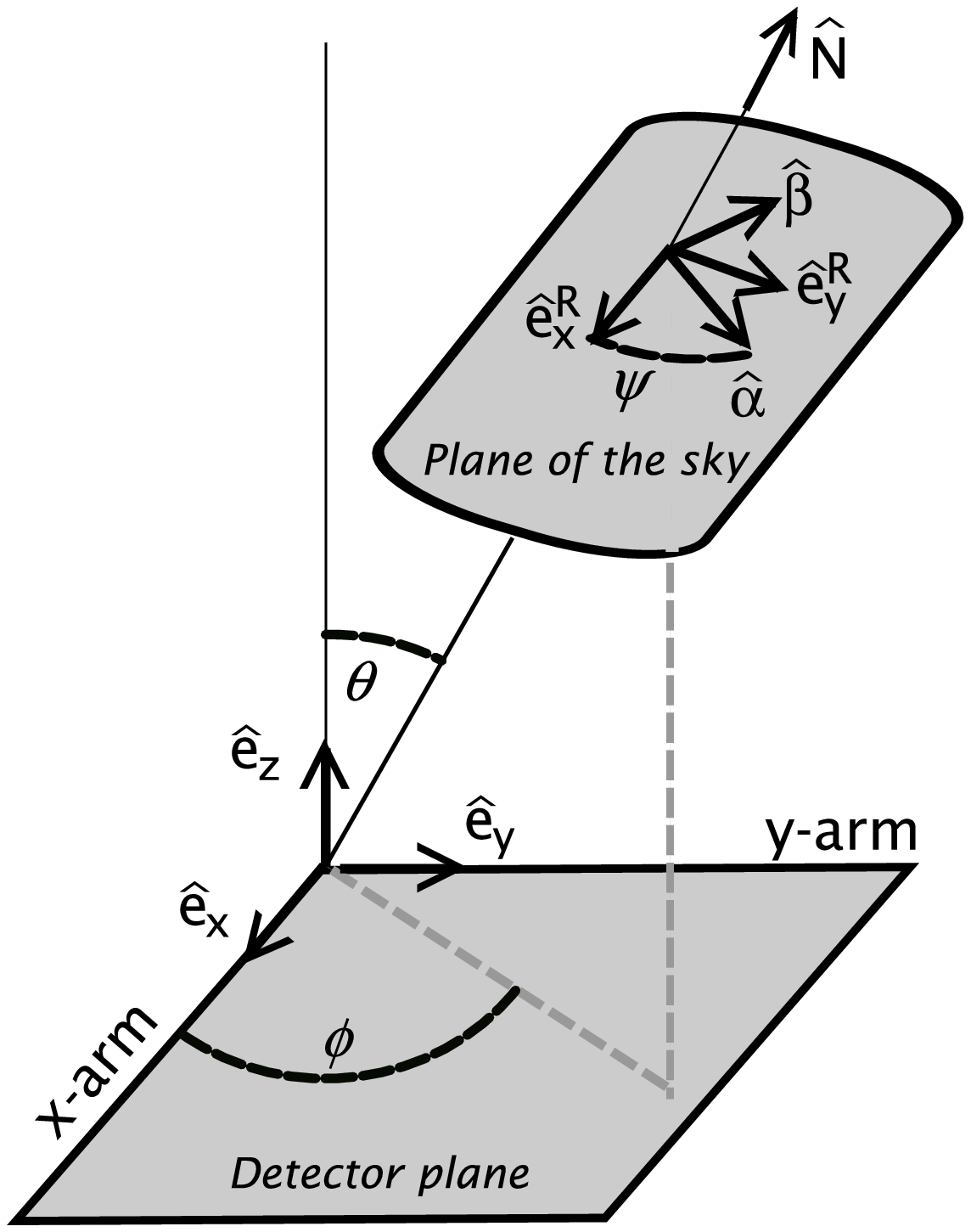}
    }
    \caption{The relative orientation of the sky and detector frames
    (left panel) and the effect of a rotation by the angle $\psi$ in
    the sky frame (left panel).}
    \label{fig:polConvention}
\end{figure}}

These are the antenna-pattern response functions of the interferometer to the 
two polarizations of the wave as defined in the sky plane~\cite{Thorne1987}. If one wants 
the antenna pattern referred to the detector's own axes, then one just sets 
$\psi=0$. If the arms of the interferometer are not perpendicular to each other, 
then one defines the detector-plane coordinates $x$ and $y$ in such a way that the 
bisector of the angle between the arms lies along the bisector of the angle between 
the coordinate axes~\cite{schutz.tinto}. Note that the maximum value of either $F_+$ or 
$F_\times$ is 1. 
%% If the wave comes in with overall amplitude $h$ and 
%% polarization angle $\psi$ relative to the polarization 
%% convention of the detector, so that the wave has $+$ polarization when $\psi=0$ and 
%% $\times$ orientation when $\psi=\pi/4$, then we can write 
%% \[h_+ = h\cos2\psi, \quad h_\times = h\sin2\psi,\]
%% where both $h$ and $\psi$ may be functions of time. Now 
%% the detector response in \Eqnref{eqn:interferometerantenna} 
%% becomes~\cite{schutz.tinto}
%% \begin{equation}\label{eqn:polarizedresponse}
%% \delta L = hL\left( F_+\cos2\psi + F_\times\sin2\psi\right).
%% \end{equation}
%% \newcommand{\Fcomplex}{{\cal F}}
%% \newcommand{\hcomplex}{{\cal H}}
%% Note that this can rather elegantly be written in complex notation by defining 
%% a complex antenna pattern function $\Fcomplex$ 
%% and a complex wave amplitude function $\hcomplex$:
%% \begin{equation}\label{eqn:complexpattern}
%% \Fcomplex = F_+ + i F_\times, \quad \hcomplex = he^{2\pi i \psi}, 
%% \end{equation}
%% in terms of which \Eqnref{eqn:polarizedresponse} becomes the more 
%% compact~\cite{dt88}
%% \begin{equation}\label{eqn:complexresponse}
%% \delta L = L\Re(\Fcomplex\hcomplex^*).
%% \end{equation}

The corresponding antenna-pattern functions of a bar detector whose longitudinal axis is aligned 
along the $z$ direction, are 
\begin{equation}\label{eqn:barantennapatterns}
F_+ = \sin^2\theta \cos 2 \psi,\quad F_\times = \sin^2\theta \sin 2\psi.
\end{equation}

%% It is interesting to average the response over the sky and over 
%% polarizations of the incoming wave: the mean square
%% response of an interferometer with perpendicular arms to 
%% a randomly linearly polarized 
%% gravitational wave of amplitude $h$ is~\cite{schutz.tinto}
%% \[\left<\left(\frac{\delta L/L}{h}\right)^2 \right> = <F_+^2\cos^22\psi> + 
%% <F_\times^2\sin^22\psi> = 1/5.\]
%% 

%\subsection{Antenna Pattern and the Response of a Detector}
%\label{sec:beam factors}
Any one detector cannot directly measure both independent polarizations 
of a gravitational wave at the same time, but responds rather to a 
linear combination of the two that depends on the geometry of the detector
and source direction.  
%Our 
%expressions for the interferometer response in Section~\ref{sec:ground}, as 
%well as analogous expressions for other detectors (see below), may elegantly be 
%put into the form of a complex response $R$ of the 
%detector~\cite{schutz.tinto,dt88}:
%\begin{equation}
%R = h_+ F_+ + i h_\times F_\times,\ \ F_+ = \Re(F),\ \ F_\times=\Im(F),
%\end{equation}
%where $F_+$ and $F_\times$ are the real and imaginary parts of a complex
%antenna pattern function, which is a function of the direction 
%$(\theta,\phi)$ to the source and the polarization angle $\psi$
%of the wave, in a coordinate system `attached' to the detector. 
If the wave lasts only a short time, then the responses of three
widely-separated detectors, together with two independent differences
in arrival times among them, are, in principle, sufficient to fully
reconstruct the source location and gravitational wave polarization. A
long-lived wave will change its location in the antenna pattern as the
detector moves, and it will also be frequency modulated by the motion
of the detector; these effects are in principle sufficient to
determine the location of the source and the polarization of the wave.

\epubtkImage{fracArea.png}{%
  \begin{figure}[htbp]
    \centerline{\includegraphics[width=2.5in]{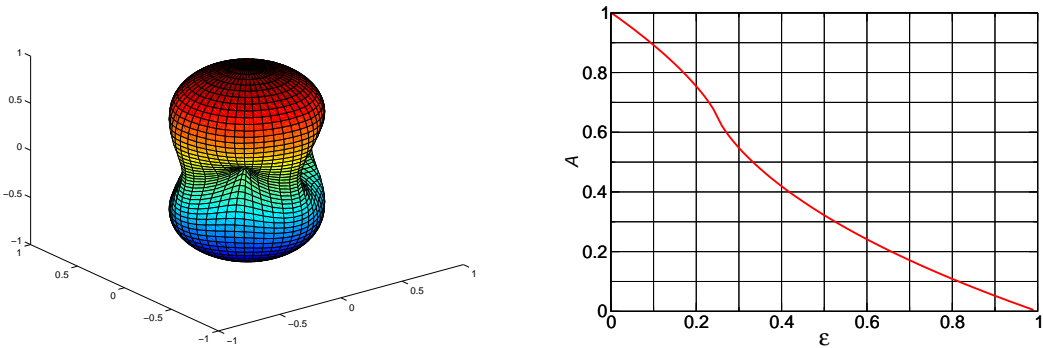}\hspace{1cm}
      \includegraphics[width=2.5in]{fracArea}}
    \caption{The antenna pattern of an interferometric detector (left
      panel) with the arms in the $x$-$y$ plane and oriented along the
      two axes. The response $F$ for waves coming from a certain
      direction is proportional to the distance to the point on the
      antenna pattern in that direction. Also shown is the fractional
      area in the sky over which the response exceeds a fraction
      $\epsilon$ of the maximum (right panel).}
    \label{fig:response}
\end{figure}}

Since the polarization angle of an incoming gravitational wave would generally 
be expected to be unrelated to its direction of arrival, depending rather on 
the internal orientations in the source, it is useful to characterize the 
directional sensitivity of a detector by averaging over the polarization angle 
$\psi$. If the wave has a given amplitude $h$ and is linearly polarized, then, 
if we are interested in a single detector's response, it is always possible to 
align the polarization angle $\psi$ in the sky plane with that of the wave, 
so that the wave has pure $+$-polarization. Then the rms response function
of the detector is 
\begin{equation}\label{eqn:rmsresponse}
\overline{F} =  \left(\int F_+^2\mathrm{\ d}\psi \right)^{1/2}.
\end{equation}
The function $\overline{F}$ is often simply called the \emph{antenna pattern}. 
For a resonant bar, the antenna pattern is 
\begin{equation}
\overline{F} = \sin^2\theta,
\end{equation}
and for an interferometer, it is given by
\begin{equation}
\overline{F}^2 = \frac{1}{4} \left ( 1 + \cos^2\theta \right)^2
\cos^2 2\phi + \cos^2\theta \sin^2 2\phi.
\end{equation}
The antenna pattern of an interferometric detector is plotted in the left panel of 
\Figref{fig:response}, which clearly shows the quadrupolar 
nature of the detector.  Note that the response of an interferometer is
the best for waves coming from a direction orthogonal to the plane 
containing the detector, and it is zero for waves in the plane of an interferometer's 
arms (i.e., $\theta=\pi/2$) that arrive from a direction bisecting the two arms 
(i.e., $\phi=\pi/4$)  or from directions differing from this by a multiple of 
$\pi/2$.
%% \marginpar{Sathya, I have changed some numbers here. Not sure 
%% what your original text meant. Think that one has to define 
%% the averaging carefully. Should we do more -- I have done 
%% linearly pol waves, but maybe we should do the case of binaries
%% where we average over the inclination of the orbit???}
What is the response of an antenna to a linearly-polarized source at a
random location in the sky? This is given by the rms value of
$\overline{F}$ over the sky,
\begin{equation}
\label{eq:F Average}
\left [ \frac{1}{4\pi} \int \overline{F}^2 \sin\theta\, \mathrm{d}\theta\, 
\mathrm{d}\phi \right ]^{1/2},
\end{equation}
which is smaller than the maximum response by a factor of 
$2/\sqrt{15}$ (52\%) for a bar detector and by $\sqrt{2/5}$ (63\%) for an 
interferometer.  

The polarization amplitudes of the radiation from an inspiraling 
binary, a rotating neutron star, or a ringing black hole, take a
simple form as follows:
\begin{equation*}
h_+ = \frac{h_0}{2} \left ( 1 + \cos^2 \iota \right ) \cos \Phi(t),\quad 
h_\times= h_0 \cos\iota \sin\Phi(t) ,
\end{equation*}
where $h_0$ is an overall (possibly time-dependent) amplitude, $\Phi(t)$ is
the signal's phase and $\iota$ is the angle made by the characteristic
direction in the source (e.g., the orbital or the spin angular momentum)
with the line of sight. In this case, the response takes a particularly
simple form:
\begin{equation}
h(t) = F_+ h_+ + F_\times h_\times = A h_0 \cos (\Phi(t) - \Phi_0),
\end{equation}
where 
\begin{equation*}
A = \left ( A_+^2 + A_\times^2 \right )^{1/2},\quad 
\tan\Phi_0 = \frac{A_\times}{A_+},\quad
A_+ = \frac{1}{2} F_+ (1 + \cos \iota^2), \quad A_\times = F_\times
\cos\iota .
\end{equation*}
Note that $A$, just as $F$, takes values in the range [0,\,1]. 
In this case the average response has to be worked out by considering
all possible sky locations, polarizations (which drops out of the
calculation) and source orientations. More precisely, the rms response is 
\begin{equation}
\overline {A} = \frac{1}{8\pi^2} \int_0^\pi \sin\iota\, \mathrm{d}\iota 
\int_0^\pi \sin\theta\, \mathrm{d}\theta \int_0^{2\pi}\, \mathrm{d}\varphi \left (A_+^2 + A_\times^2
\right).
\label{eq:abar}
\end{equation}
For an interferometer the above integral gives 2/5. Thus, the rms
response is still 40\% of the peak response.

The right-hand panel of \Figref{fig:response} shows the percentage area of
the sky over which the antenna pattern of an interferometric  
detector is larger than a certain fraction $\epsilon$ of the peak value.  
%\marginpar{Can't be right: figure would show that the response is 
%never better than the rms, which is a contradiction. Is the figure 
%the fraction better than epsilon of the peak response??
%And is it on the rms response or the squared response??}
The response is better than the rms value over 40\% of the sky, 
implying that gravitational wave detectors are fairly omni-directional. 
In comparison, the sky coverage of most conventional telescopes 
(radio, infrared, optical, etc.) is a tiny fraction of the area
of the sky.

\subsection{Practical issues of ground-based interferometers}
\label{sec:ground}

A detector with an arm length of 4~km responds to a gravitational wave 
with an amplitude of $10^{-21}$ with
\[\delta l_{gw}\sim hl \sim 4\E{-18}\mathrm{\ m}. \]
Light takes only about $10^{-5}$~s to go up and down one arm, much less than the 
typical period of gravitational waves of interest. Therefore, it is beneficial 
to arrange for the light to remain in an arm longer than this, say for 100 
round trips.  This increases its effective path length by 100 and hence 
the shift in the position of a given phase of the light beam will be of 
order $10^{-16}$~m. Most interferometers keep the light in the arms for 
this length of time by setting up optical cavities in the arms with 
low-transmissivity mirrors; these are called Fabry--P\'erot cavities.

The main sources of noise against which a measurement must compete
are:

\begin{enumerate}

\item {\bf Ground vibration.}  External mechanical vibrations must be 
screened out. These are a problem for bar detectors, too, but are more 
serious for interferometers, not least because interferometers 
bounce light back and forth between the mirrors, and so each reflection 
introduces further vibrational noise.  Suspension/isolation systems are 
based on pendulums. A 
pendulum is a good mechanical filter for frequencies above its natural 
frequency.  By hanging the mirrors on pendulums of perhaps 0.5~m length, 
one achieves filtering above a few Hertz.  Since the spectrum of ground noise 
falls at higher frequencies, this provides suitable isolation. But these 
systems can be very sophisticated; the GEO600~\cite{DANZMANN1995} detector has a three-stage 
pendulum and other vibration isolation components~\cite{Plissi1998}.
The most ambitious multi-stage isolation system has been developed for the Virgo 
detector~\cite{GIAZOTTO1995}.

\item {\bf Thermal noise.} Vibrations of the mirrors and of the 
suspending pendulum can mask gravitational waves. As with vibrational noise, 
this is increased by the bouncing of the light between the mirrors.
Opposite to bars, interferometers perform measurements  
at frequencies \emph{far} from the resonant frequency, where the amplitude of 
thermal vibration is largest. Thus, the pendulum suspensions have thermal noise at a few Hertz, 
so measurements will be made above 40~Hz in the first detectors. 
Internal vibrations of the mirrors have natural frequencies of several 
kHz, which sets an effective upper limit to the observing band. 
By ensuring that both kinds of oscillations have very high $Q$, 
one can confine most of the vibration energy to a small bandwidth around the resonant 
frequency, so that at the measurement frequencies the vibration amplitudes 
are extremely small. This allows interferometers to operate at 
room temperature.  But mechanical $Q$s of $10^7$ or higher are required, and 
this is technically demanding.

Thermal effects produce other disturbances besides vibration. Some of the 
mirrors in interferometers are partly transmissive, as is the beam splitter. 
A small amount of light power is absorbed during transmission, which raises 
the temperature of the mirror and changes its index of refraction. The 
resulting ``thermal lensing'' can ruin the optical properties of the system, 
and random fluctuations in lensing caused by time-dependent thermal fluctuations 
(thermo-refractive noise) can appear at measurement frequencies. These effects
can limit the amount of laser power that can be used in the 
detector. Other problems can arise from heating effects in the multiple-layer
coatings that are applied to the reflective surfaces of mirrors. 
 
\item {\bf Shot noise.}  The photons that are used to do interferometry 
are quantized, and so they arrive at random and make random fluctuations 
in the light intensity that can look like a gravitational wave signal. 
The more photons one uses, the smoother the interference signal will be.
As a random process, the error improves with the square root of the number 
$N$ of photons.  Using infrared light with a wavelength 
$\lambda\sim 1\;\mu\mathrm{m}$, one can expect to measure to an accuracy of
\[\delta l_{shot} \sim \lambda/(2\pi\sqrt{N}).\]
To measure at a frequency $f$, one has to make at least $2f$ measurements 
per second, so one can accumulate photons for a time $1/2f$.  
With light power $P$, one gets $N=P/(hc/\lambda)/(2f)$ photons.  In order that $\delta 
l_{shot}$ should be below $10^{-16}$~m, one needs high light power, 
far beyond the output of any continuous laser.

Light-recycling techniques overcome this problem, by using light efficiently.
An interferometer actually has two places where light leaves.  One is where 
the interference is measured, the difference port. The other is the sum 
of the two return beams on the beam splitter, which goes back towards the input laser. 
Normally one makes sure that no light hits the interference sensor, so that 
only when a gravitational wave passes does a signal register there. This 
means that all the light normally returns toward the laser input, apart from small 
losses at the mirrors.  Since mirrors are of good quality, only one part 
in $10^3$ or less of the light is lost during a 1~ms storage time. By 
placing a power-recycling 
mirror in front of the laser, one can reflect this wasted light back in, 
allowing power to build up in the arms until the laser merely resupplies
the mirror losses~\cite{DREVER1983b}. This can dramatically reduce
the power requirement for the laser.  The first interferometers work 
with laser powers of 5\,--\,10~W.  Upgrades will use ten or more times
this input power.

\item {\bf Quantum effects.} Shot noise is a quantum noise, and like 
all quantum noises there is a corresponding conjugate noise. As laser 
power is increased to reduce shot noise, the position sensing accuracy 
improves, and one eventually comes up against the Heisenberg uncertainty 
principle: the momentum transferred to the mirror by the measurement leads
to a disturbance that can mask a gravitational wave. To reduce this 
backaction pressure fluctuation, scientists are experimenting with a 
variety of interferometer configurations that modify the quantum state 
of the light, by ``squeezing'' the Heisenberg uncertainty ellipse, 
in order to reduce the effect of this uncertainty on the variable being 
measured, at the expense of its (unmeasured) conjugate. The key point 
here is that we are using a quantum field (light) to measure an effectively
classical quantity (gravitational wave amplitude), so we do not need to 
know everything about our quantum system: we just need to reduce the 
uncertainty in that part of the quantum field that responds to the 
gravitational wave at the readout of our interferometer. The best 
results on squeezing so far~\cite{SchnabelSqueezing} 
have been obtained during preparations for 
the GEO-HF upgrade of the GEO600 detector~\cite{Willke2006}.
These techniques may be needed for the second-generation 
advanced detectors and will certainly be needed for advances beyond
that.

\item {\bf Gravity gradient noise.} One noise that cannot be screened 
out is that due to changes in the local Newtonian gravitational field on 
the timescale of the measurements. A gravitational wave detector will 
respond to tidal forces from local sources just as well as 
to gravitational waves. Environmental noise comes not only from man-made 
sources, but even more importantly from natural ones: seismic waves are 
accompanied by changes in the gravitational field, and changes in air 
pressure are accompanied by changes in air density.  The spectrum 
falls steeply with increasing frequency, so for first-generation 
interferometers this will not be a problem, but it may limit the 
performance of more advanced detectors. And it is the primary 
reason that detecting gravitational waves in the low-frequency band 
below 1~Hz must be done in space.

\end{enumerate}

\subsubsection{Interferometers around the globe}
\label{sec:present}

The two largest interferometer projects are LIGO~\cite{RAAB1995} 
and VIRGO~\cite{GIAZOTTO1995}.  LIGO has built three detectors at two sites. 
At Hanford, Washington, there is a 4~km and a 2~km detector in the same 
vacuum system. At Livingston, Louisiana, there is a single 4~km detector, 
oriented to be as nearly parallel to the Hanford detector as possible.  
After a series of ``engineering'' runs, which helped to debug
the instruments,  interspersed with several ``science runs'', which
helped to test and debug the data acquisition system and various analysis
pipelines, LIGO reached its design sensitivity goal in the final months
of 2005. In November 2005, LIGO began a two-year data-taking run, called S5,
which acquired a year's worth of triple coincidence
data among the three LIGO detectors. S5 ended on 30 September 2007. Although 
interferometers are pretty stable detectors,
environmental disturbances and instrumental malfunctions can cause
them to lose lock during which the data quality will be either poor
or ill defined. The typical duty cycle at one of the LIGO sites is about 
80\%, and hence about two years of operation was required to
accumulate a year's worth of triple coincident data. Up to date 
information on LIGO can be found on the project's website~\cite{LIGOLab}. A 
recent review of LIGO's status is~\cite{Raab2006}.

VIRGO finished commissioning its single 3-km detector at Cascina, near Pisa,
in early 2007 and began taking 
data in coincidence with LIGO in May 2007, thus joining for the last 
part of S5. VIRGO is a collaboration among research laboratories in 
Italy and France, and its umbrella organization EGO looks after 
the operation of the site and planning for the future. There are 
websites for both VIRGO~\cite{VIRGOLab} and EGO~\cite{EGO}. A recent 
review of VIRGO's status is~\cite{Acernese2007}.

A smaller 600-m detector, GEO600, has been operational near Hanover, 
Germany, since 2001 \cite{DANZMANN1995}. It is a collaboration among 
research groups principally in Germany and Britain. Although smaller, 
GEO600 has developed and installed \emph{second-generation} 
technology (primarily in its suspensions, 
mirror materials and interferometer configuration) that help it achieve 
a higher sensitivity.  GEO600 technology is being 
transferred to LIGO and VIRGO as part of their planned upgrades, described below. 
Full information about GEO can be found on its website~\cite{GEO600HomePage}.  A 
recent review of GEO600's status is~\cite{Willke2007}.

LIGO and GEO have worked together under the umbrella of the LIGO Scientific 
Collaboration (LSC) since the beginning of science data runs in 2001. The LSC 
contains dozens of groups from universities around the world, which contribute 
to data analysis and technology development. The two detector groups pool their 
data and analyze it jointly. The LSC has a website containing detailed information, 
and providing access to the publications and open-source software of the 
collaboration~\cite{LSC}.

VIRGO has signed an agreement with the LSC to pool data and analyze it 
jointly, thereby creating a single 
worldwide network of long-baseline gravitational wave detectors. VIRGO is not, 
however, a member of the LSC.

The LSC has already published many papers on the analysis of data
acquired during its science runs, and many more can be expected. The results from these
science runs, which will be discussed later, are already becoming astrophysically
interesting. The LSC maintains a public repository of its papers and 
contributions to conference proceedings~\cite{LSCPapers}.

For instance, although the search for continuous waves from known
pulsars has not found any definitive candidates, it has been possible
to set stringent upper limits $\epsilon \le \mathrm{few} \times 10^{-6}$
on the magnitude of the ellipticity  of some of these systems~\cite{Abbott:2007ce}. 
In particular, in the case of the Crab pulsar, gravitational wave observations
have begun to improve~\cite{Abbott:2008fx} the upper limit on the strength of 
radiation obtained by radio observations of the spin-down rate. 

A yet smaller detector in Japan, TAMA300~\cite{TSUBONO1995}, with 300~m arms, 
was the first large-scale interferometer to achieve continuous operation, 
at a sensitivity of about $10^{-19}\mbox{\,--\,}10^{-20}$.  TAMA is seen as 
a development prototype, and its sensitivity will be 
confined to higher frequencies (above $\sim$~500~Hz). An ambitious follow-on 
detector called the Large-scale Cryogenic Gravitational-Wave Telescope
(LCGT) is being planned in Japan, and, as its name suggests, it will 
be the first to use cooled mirrors to reduce the effects of thermal 
noise. TAMA~\cite{TAMAProject} and the LCGT~\cite{LCGTProject} have 
websites where one can get more information. A recent review of 
TAMA's status is~\cite{Tatsumi2007}.

There are plans for a detector in Australia, and a small interferometer 
is operating in Western Australia~\cite{McClelland}. The Australian 
Interferometric Gravitational Observatory (AIGO)~\cite{AIGO} is a proposal of the 
Australian Consortium for Interferometric Gravitational Astronomy (ACIGA)~\cite{ACIGA}.
The ACIGA collaboration is a member
of the LSC and assists in mirror and interferometry development, but 
it is not yet clear whether and when
a larger detector might be funded.  From the point of view of 
extracting information from observations, it is very desirable to have 
large-scale detectors in Japan and Australia, because of their very long
baselines to the USA and Europe. But the future funding of both LCGT 
and AIGO is not secure as of this writing (2008).  

The initial sensitivity levels achieved by LIGO, VIRGO, and GEO are just 
the starting point. Detailed plans exist for upgrades for all three 
projects. In October 2007, both LIGO and VIRGO began upgrading to
\emph{enhanced} detectors, which should improve on LIGO's S5 sensitivity by 
a factor of roughly two. These should come online in 2009. After a further 
observing run, called S6, the detectors will again shut down for a much 
more ambitious upgrade to \emph{advanced} detectors, to operate 
around 2014. This will 
provide a further factor of five in sensitivity, and hence in range. Altogether 
the two upgrades will extend the volume of space that can be surveyed for 
gravitational waves by a factor of 1000, and this will make regular detections a 
virtual certainty. Advanced LIGO has a website giving the plans for the 
upgrade in the context of development from the initial sensitivity~\cite{AdvancedLIGO}.

GEO600 will remain in science mode during the upgrade to enhanced detectors, 
just in case a nearby supernova or equally spectacular event should occur when 
the larger detectors are down. But, when the enhanced detectors begin operating, 
GEO will upgrade to GEO-HF~\cite{Willke2006}, a modification designed to improve its sensitivity 
in the high-frequency region above 1~kHz, where its short arm length does not 
prevent it being competitive with the larger instruments. GEO is also a 
partner in the Advanced LIGO project, contributing high-power lasers and high-Q 
suspensions for controlling thermal noise.

Beyond that, scientists are now studying the technologies that may be 
needed for a further large step in sensitivity to third-generation 
detectors. This may involve cooling mirrors, using ultra-massive 
substrates of special materials, using purely nontransmissive optics, 
and even circumventing the quantum limit in interferometers, as has 
been studied for bars. The goal of third-generation detectors would 
be to be limited just by gravity-gradient noise and quantum effects. 
A design study for a concept called the ``Einstein Telescope'' started
in Europe in 2008.

\subsubsection{Very-high--frequency detectors}

The gravitational wave spectrum above the detection band of conventional interferometers, 
say above 10~kHz, may not be empty, and stochastic gravitational waves from the Big Bang
may be present up to megaHertz frequencies and beyond. It is exceedingly difficult 
to build sensitive detectors at these high frequencies, but two projects have made
prototypes: a microwave-based detector that senses the change in polarization 
as the electromagnetic waves follow a waveguide circuit as a gravitational wave passes
by~\cite{Cruise2006}, and a more conventional light-based interferometer~\cite{Akutsu2008}.

\subsection{Detection from space}

Space offers two important ingredients for beam detectors: long 
arms and a free vacuum. In this section, we describe the three ways
that space has been and will be used for gravitational wave detection: ranging 
to spacecraft (\Secref{sec:ranging}), pulsar timing (\Secref{sec:pulsartiming}), 
and direct detection using space-based 
interferometers (\Secref{sec:LISA}).

\subsubsection{Ranging to spacecraft}
\label{sec:ranging}

Both NASA and ESA perform 
experiments in which they monitor the return time of communication 
signals with interplanetary spacecraft for the characteristic effect 
of gravitational waves.  For missions to Jupiter and Saturn, for 
example, the return times are of order $2\mbox{\,--\,}4\E3$~s.  Any gravitational 
wave event shorter than this will, by \Eqnref{eqn:threeterm}, 
appear three times in the time delay: once when the wave passes the 
Earth-based transmitter, once when it passes the spacecraft, and 
once when it passes the Earth-based receiver. Searches use a form 
of data analysis using pattern matching.  Using two transmission 
frequencies and very stable atomic clocks, it is possible to 
achieve sensitivities for $h$ of order $10^{-13}$, and 
even $10^{-15}$ may soon be reached~\cite{Living:Armstrong}.

\subsubsection{Pulsar timing}
\label{sec:pulsartiming}

Many pulsars, particularly the 
old millisecond pulsars, are extraordinarily regular clocks when 
averaged over timescales of a few years, with 
random timing irregularities too small for the best atomic 
clocks to measure.  If one assumes that they emit pulses perfectly 
regularly, then one can use observations of timing irregularities 
of single pulsars to set upper limits on the background
gravitational-wave field.  Here, the one-way formula
\Eqnref{eqn:onewaylight} is appropriate. The transit time of a signal
to the Earth from the pulsar may be thousands of years, so we cannot
look for correlations between the two terms in a given signal.
Instead, the delay is a combination of the effects of waves at the
pulsar when the signal was emitted and waves at the Earth when it is
received.  If one observes a single pulsar, then because not enough is
known about the intrinsic irregularity in pulse emission, the most one
can do is to set upper limits on a background of gravitational
radiation at very low frequencies~\cite{Living:Lorimer,
  Living:Stairs}.

If one simultaneously observes two or more pulsars, then the
Earth-based part of the delay is correlated, and this offers, in
addition, a means of actually detecting strong gravitational waves
with periods of several years that pass the Earth (in order to achieve
the long-period stability of pulse arrival times). Observations are
currently underway at a number of observatories. The most stringent
limits to date are from the Parkes Pulsar Timing
Array~\cite{Jenet:2006sv}, which sets an upper limit on a stochastic
background of $\Omega_{\text{gw}} \le 2\times10^{-8}$. Two further
collaborations for timing have been formed: the European Pulsar Timing
Array (EPTA)~\cite{EPTA} and NanoGrav~\cite{nanograv}. Astrophysical
backgrounds in this frequency band are likely (see \Secref{sec:agwb}),
so these arrays have a good chance of making early detections. Future
timing experiments will be even more powerful, using new phased arrays
of radio telescopes that can observe many pulsars simultaneously, such
as the Low Frequency Array (LOFAR)~\cite{LOFAR} and the Square Kilometer
Array~\cite{SKA}.

Pulsar timing can also be used to search for individual events, not 
just a stochastic signal. The first example of an upper limit from 
such a search was the exclusion of a black-hole--binary model for
3C66B~\cite{Jenet2004}.

\subsubsection{Space interferometry}
\label{sec:LISA}

Gravity-gradient noise on the Earth is much larger than the 
amplitude of any expected waves from astronomical sources at 
frequencies below about 1~Hz, but this noise falls off rapidly 
as one moves away from the Earth.  A detector in space would 
not notice the Earth's noisy environment. The Laser Interferometer
Space Antenna (LISA) project, 
currently being developed in collaboration by ESA and NASA 
with a view toward launching in 2018, would open up the frequency 
window between 0.1~mHz and 0.1~Hz for the first time~\cite{HOUGH1995, LISA2003}. 
There are several websites that provide full information about this 
project~\cite{LISAHannover, LISAESA, LISANASA}.

We will see below that there are many exciting sources expected 
in this wave band, for example the coalescences of giant black holes 
in the centers of galaxies. LISA would see such events with 
extraordinary sensitivity, recording typical SNRs of 1000 or more for events at redshift one.  

A space-based interferometer can have arm lengths much greater 
than a wavelength. LISA, for example, would have arms 
$5\times10^6$~km long, and that would be longer than half a 
wavelength for any gravitational waves above 30~mHz. In 
this regime, the response of each arm will follow the 
three-term formula we encountered earlier. The short-arm 
approximation we used for ground-based interferometers 
works for LISA only at the lowest frequencies in its observing
band.

LISA will consist of three free-flying spacecraft, 
arranged in an array that orbits the 
sun at 1~AU, about 20 degrees behind the Earth in its orbit. They 
form an approximately equilateral triangle in a plane tilted at 
$60^\circ$ to the ecliptic, and their simple Newtonian elliptical orbits
around the sun preserve this arrangement, with the array rotating 
backwards once per year as the spacecraft orbit the sun. 
By passing light along each of the arms, one can construct 
three different Michelson-type interferometers, one for each 
vertex. With this array one can measure the polarization 
of a gravitational wave directly.  The spacecraft are too far 
apart to use simple mirrors to reflect light back along 
an arm: the reflected light would be too weak. Instead, 
LISA will have optical transponders: light from one spacecraft's 
on-board laser will be received at another, which will then 
send back light from its own laser locked exactly to the 
phase of the incoming signal. 

The main environmental disturbances to LISA are forces from 
the sun: solar radiation pressure and 
pressure from the solar wind.  To minimize these, 
LISA incorporates drag-free technology.  Interferometry is 
referenced to an internal proof mass that falls freely, unattached to the spacecraft. The job of the spacecraft is 
to shield this mass from external disturbances. It does 
this by sensing the position of the mass and firing its 
own jets to keep itself (the spacecraft) stationary relative 
to the proof mass. To do this, it needs thrusters of very 
small thrust that have accurate control.  The key technologies 
that have enabled the LISA mission are the availability of 
such thrusters, accelerometers needed to sense disturbances 
to the spacecraft, and lasers capable of continuously 
emitting 1~W of infrared light for many years. ESA is planning 
to launch a satellite called LISA Pathfinder to test all of these 
technologies in 2010~\cite{Pathfinder}.

LISA is not the only proposal for an interferometer in space for
gravitational wave detection. The DECIGO proposal is a more ambitious
design, positioned at a higher frequency to fill the gap between LISA
and ground-based detectors~\cite{DECIGO}. Even more ambitious, in the
same frequency band, is the Big Bang Observer, a NASA concept study to
examine what technology would be needed to reach the ultimate
sensitivity of detecting a gravitational wave background from
inflation at these frequencies~\cite{BBO}.

\subsection{Characterizing the sensitivity of a gravitational wave
  antenna}
\label{sec:sens}

The performance of a gravitational wave detector is characterized by the 
\emph{power spectral density} (henceforth denoted PSD) of its noise
background.  One can construct the
noise PSD as follows; a gravitational wave detector outputs a dimensionless data train, 
say $x(t)$, which in the case of an interferometer is the relative 
strain in the two arms, scaled to represent the value of $h$ that would 
produce that strain if the wave is optimally oriented with respect to the 
detector. In the absence of any gravitational wave signal, the detector 
output is just an instance of noise $n(t)$, that is, $x(t)=n(t)$. 
The noise auto-correlation function $\kappa$ is defined as 
\begin{equation}
\kappa \equiv \overline {n(t_1) n(t_2)},
\label{eq:noise correlation}
\end{equation}
where an overline indicates the average over an ensemble of noise realizations.
In general, $\kappa$ depends both on $t_1$ and $t_2$. However, if the detector
output is a stationary noise process, i.e., its performance is, statistically
speaking, independent of time, then $\kappa$ depends only on 
$\tau\equiv |t_1 - t_2|$. 

The assumption of stationarity is not strictly valid in the case of real gravitational-wave
detectors; however, if their performance doesn't vary greatly over 
time scales much larger than typical observation time scales, stationarity
could be used as a working rule.  While this may be good enough in the case
of binary inspiral and coalescence searches, it is a matter of concern for
the observation of continuous and stochastic gravitational waves. In this review, for 
simplicity, we shall assume that the detector noise is
stationary. In this case the \emph{one-sided} noise PSD, defined only at
positive frequencies, is the Fourier transform of the noise 
auto-correlation function:
\begin{eqnarray}
S_h(f) & \equiv & \frac{1}{2} \int_{-\infty}^{\infty} \kappa(\tau) 
e^{2\pi i f \tau}\, \mathrm{d}\tau,\ \ f\ge 0, \nonumber\\
       & \equiv & 0, \ \ f<0,
\label{eq:psd1}
\end{eqnarray}
where a factor of 1/2 is included by convention. By using the Fourier
transform of $n(t)$, that is 
$\tilde n(f) \equiv \int_{-\infty}^{\infty} n(t) e^{2\pi i ft}\, \mathrm{d}t$,
in \Eqnref{eq:noise correlation} and substituting the resulting
expression in \Eqnref{eq:psd1}, it is easy to shown that 
for a stationary noise process
background
\begin{equation}
\overline {\tilde n(f)\tilde n^*(f')} = \frac {1}{2} S_h(f) \delta (f-f'),
\label{eq:psdinfourier}
\end{equation}
where $\tilde n^*(f)$ denotes the complex 
conjugate of $\tilde n(f)$.
The above equation justifies the name PSD given to
$S_h(f)$. 

It is obvious that $S_h(f)$ has dimensions of time but it is conventional
to use the dimensions of Hz$^{-1}$, since it is a quantity defined in the
frequency domain. The square root of $S_h(f)$ is 
the noise amplitude, $\sqrt{S_h(f)}$, and has dimensions of Hz$^{-1/2}$. 
Both noise PSD and noise amplitude measure the noise 
in a linear frequency bin. It is often useful to define
the power per logarithmic bin $h_n^2(f) \equiv fS_h(f)$, where 
$h_n(f)$ is called the \emph{effective gravitational-wave noise}, and
it is a dimensionless quantity. In gravitational-wave--interferometer literature one also comes 
across \emph{gravitational-wave displacement noise} or \emph{gravitational-wave strain noise} defined 
as $h_\ell (f) \equiv \ell h_n(f)$, and the corresponding noise spectrum
$S_\ell(f) \equiv \ell^2 S_h(f)$, where $\ell$ is the arm length of the interferometer. 
The displacement noise gives the smallest strain $\delta \ell/\ell$ in the arms
of an interferometer that can be measured at a given frequency.

\epubtkImage{lisa-ground.png}{%
  \begin{figure}[htbp]
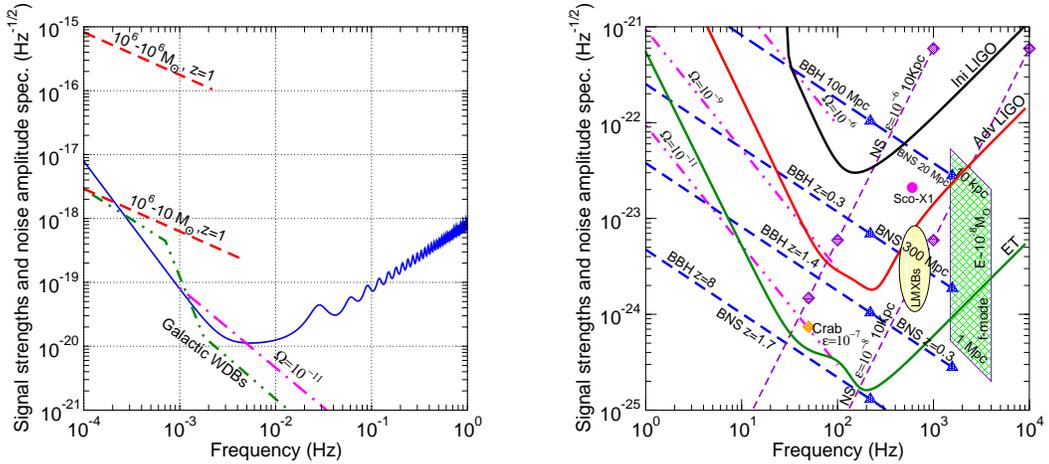

    \centerline{
      \includegraphics[width=2.5in]{lisa}\hspace{1cm}
      \includegraphics[width=2.5in]{ground}
    }
    \caption{The right panel plots the noise amplitude spectrum,
      $\sqrt{fS_h(f)}$, in three generations of ground-based
      interferometers. For the sake of clarity, we have only plotted
      initial and advanced LIGO and a possible third generation
      detector sensitivities. VIRGO has similar sensitivity to LIGO at
      the initial and advanced stages, and may surpass it at lower
      frequencies. Also shown are the expected amplitude spectrum of
      various narrow and broad-band astrophysical sources.  The left
      panel is the same as the right except for the LISA detector. 
      %% The curve labelled ``LISA Noise (total)'' 
      %% is the effective noise assuming that confusion from binaries 
      %% cannot be removed; see Equation~\ref{LISAfullnoise}. 
      The SMBH sources are assumed to lie at a redshift of $z=1$, but
      LISA can detect these sources with a good SNR practically
      anywhere in the universe. The curve labelled ``Galactic WDBs''
      is the confusion background from the unresolvable Galactic
      population of white dwarf binaries.}
    \label{fig:noise-curves}
\end{figure}}

\subsubsection{Noise power spectral density in interferometers}
\label {sec:noise psd}

As mentioned earlier,  the performance of a gravitational wave
detector is characterized by the one-sided noise PSD. The noise PSD
plays an important role in signal analysis.  In this review we will
only discuss the PSDs of interferometric gravitational-wave
detectors. 

The sensitivity of ground based detectors is limited at
frequencies less than a Hertz by the time-varying local gravitational 
field caused by a variety of different noise sources, e.g., low frequency 
seismic vibrations, density variation in the atmosphere due to winds, etc.
Thus, for data analysis purposes, the noise PSD is assumed to be essentially
infinite below a certain lower cutoff $f_s$. Above this cutoff,  i.e., 
for $f\ge f_s$, Table~\ref{table:psd} lists the noise PSD $S_h(f)$ for
various interferometric detectors and some of these are plotted in 
\Figref{fig:noise-curves}.

For LISA, Table~\ref{table:psd} gives the internal instrumental noise
only, taken from~\cite{FinnThorne}. It is based on the noise budget
obtained in the LISA Pre-Phase A Study~\cite{PPA}. However, in the frequency range
$10^{-4}\mbox{\,--\,}10^{-2}$~Hz, LISA will be affected by source confusion
from astrophysical backgrounds produced by several populations of
galactic binary systems,  such as closed white-dwarf binaries,
binaries consisting of Cataclysmic Variables, etc. At frequencies
below about 1~mHz, there are too many binaries for LISA to resolve in,
say, a 10-year mission, so that they form a Gaussian noise. Above this
frequency range, there will still be many resolvable binaries which
can, in principle, be removed from the data.

\begin{table}
  \caption[Noise power spectral densities $S_h(f)$ of various
  interferometers in operation and under construction: GEO600, Initial
  LIGO (ILIGO), TAMA, VIRGO, Advanced LIGO (ALIGO), Einstein Telescope
  (ET) and LISA (instrumental noise only). For each detector the noise
  PSD is given in terms of a dimensionless frequency $x=f/f_0$ and
  rises steeply above a lower cutoff $f_s$.]{Noise power spectral
  densities $S_h(f)$ of various interferometers in operation and under
  construction: GEO600, Initial LIGO (ILIGO), TAMA, VIRGO, Advanced
  LIGO (ALIGO), Einstein Telescope (ET) and LISA (instrumental noise
  only). For each detector the noise PSD is given in terms of a
  dimensionless frequency $x=f/f_0$ and rises steeply above a lower
  cutoff $f_s$. The parameters in the ET design sensitivity curve are
  $\alpha=-4.1$, $\beta=-0.69$, $a_0=186$, $b_0=233$, $b_1 = 31$, $b_2
  = -65$, $ b_3 = 52$, $ b_4 = -42$, $b_5 = 10$, $ b_6 = 12$, $c_1 =
  14$, $ c_2 = -37$, $ c_3 = 19$, $c_4 = 27$. (See also
  Figure~{\ref{fig:noise-curves}}.)}
  \label{table:psd}
  \vskip 4mm

  \centering
  \begin{tabular}{ccccc}
    \toprule
    && & \\[-0.1cm]
    Detector & $f_s$/Hz & $f_0$/Hz &  $S_0$/Hz$^{-1}$ & $S_h(x)/S_0$ \\
    && & \\[-0.1cm]
    \midrule
    & & & \\
    GEO     & 40  &  150     & $1.0 \times 10^{-46}$ & $(3.4 x)^{-30} + 34 x^{-1} + 
    \frac{20(1-x^2+ 0.5 x^4)}{(1+ 0.5 x^2)} $\\[0.2cm]
    ILIGO   & 40  &  150     & $9.0 \times 10^{-46}$ & $(4.49 x)^{-56} + 0.16 x^{-4.52} + 0.52 + 0.32 x^2 $ \\[0.2cm] 
    TAMA    & 75  &  400     & $7.5 \times 10^{-46}$ & $x^{-5} + 13x^{-1}+9(1+x^2)$ \\[0.2cm]
    VIRGO   & 20  &  500     & $3.2 \times 10^{-46}$ & $(7.8 x)^{-5} + 2 x^{-1} + 0.63 + x^2 $\\[0.2cm]
    ALIGO   & 20  &  215     & $1.0 \times 10^{-49}$ & $x^{-4.14} - 5x^{-2} + \frac{ 111 (1 - x^2 + 
      0.5 x^4)} {1 + 0.5 x^2}$\\[0.2cm]
    ET      & 10  &  200     & $1.5 \times 10^{-52}$ & $ x^{\alpha} + a_0 x^{\beta} + 
    \frac{b_0( 1 + b_1 x + b_2 x^2 + b_3 x^3 + b_4 x^4 + b_5 x^5 + b_6 x^6 )}
         {1 + c_1 x + c_2 x^2 + c_3 x^3 + c_4 x^4} $\\[0.2cm]
	 LISA    & $10^{-5}$ &  $10^{-3}$ & $9.2 \times 10^{-44}$ & $
    (x/10)^{-4} + 173 + x^{2} $\\[0.2cm]
   \bottomrule
  \end{tabular}
\end{table}

Nelemans et al.~\cite{Nelemans2001} estimate that the effective noise
power contributed by binaries in the galaxy is
\begin{equation}\label{eqn:cwdbpower}
S_h^{\text{gal}} = 2.1\times10^{-38}\left(\frac{f}{f_s}\right)^{7/3}\quad\text{Hz}^{-1}, \qquad f_s = 10^{-3}\;\text{Hz},
\end{equation}
normalized to the same $f_s$ as we use for LISA in
Table~\ref{table:psd}. This power is a mean frequency average based on
projections of the population LISA will find, but, of course, above
about 1~mHz, LISA will resolve many binaries and identify most of the
members of this population. Barack and Cutler~\cite{BarackCutler2003}
have provided a prescription for including this effect when adding in
the confusion noise. They make the conservative assumption that
individual binaries contaminate the instrumental noise
$S_h^{\text{instr}}$ (see Table~\ref{table:psd}) in such a way that,
effectively, one or a few frequency resolution bins need to be cut out
and ignored when detecting other signals, including, of course, other
binary signals. This would have approximately the same effect as if
the overall instrumental noise at that frequency were raised by an
amount obtained simply by dividing the noise by the fraction $\eta$ of
bins free of contamination. Of course, when this fraction reaches zero
(below 1~mHz), this approximation is not valid, and instead one should
just add the full binary confusion noise in \Eqnref{eqn:cwdbpower} to
the instrumental noise. A smooth way of merging these two regimes is
to set
\begin{equation}\label{LISAfullnoise}
S_h^{\text{full}} = \min\left(\frac{1}{\eta}S_h^{\text{instr}},\quad S_h^{\text{instr}} + S_h^{\text{gal}}\right),
\end{equation}
where $S_h^{\text{instr}}$ is from Table~\ref{table:psd} and
$S_h^{\text{gal}}$ is from \Eqnref{eqn:cwdbpower}. This prescription
uses the contaminated instrumental noise, when it is below the total
noise power from the binaries, but then uses the total binary
confusion power when the prescription for allowing for contamination
breaks down.

The fraction $\eta$ of uncontaminated frequency bins as a function of
frequency remains to be specified. Let $dN/df$ be the number of
binaries in the galaxy per unit frequency. Since the size of the
frequency bin for an observation that lasts a time $T_{\text{obs}}$ is
$1/T_{\text{obs}}$, the expected number of binaries per frequency bin
is 
\[\Delta N(f) = \frac{1}{T_{\text{obs}}}\frac{dN(f)}{df}.\]
Barack and Cutler multiply this by a ``fudge factor'' $\kappa>1$ to
allow for the fact that any binary may contaminate several bins, so
that $\kappa\Delta N(f)$ is the expected number of contaminated bins
per binary. If this is small, then it will equal the fractional
contamination at frequency $f$. In that case, the fraction of
uncontaminated bins is just $1-\kappa\Delta N(f)$. However, if the
expected contamination per bin approaches or exceeds one, then we have
to allow for the fact that the binaries are really randomly
distributed in frequency, so that the expected fraction not
contaminated comes from the Poisson distribution, 
\begin{equation}\label{eqn:uncontaminated}
\eta = \exp(-\kappa\Delta N).
\end{equation}
Inserting this into \Eqnref{LISAfullnoise} gives a reasonable
approximation to the effective instrumental noise if binaries cannot
be removed in a clean way from the data stream when looking for other
signals.

Because LISA will observe binaries for several years, the accuracy
with which it will know the frequency, say, of a binary, will be much
better than the frequency resolution of LISA during the observation of
a transient source, such as many of the IMBH events considered by
Barack and Cutler. Therefore, there is a good chance that, in the
global LISA data analysis, the effective noise can be reduced below
the one-year noise levels that are normally used in projecting the
sensitivity of LISA and the science it can do.

\subsubsection{Sensitivity of interferometers in units of energy flux}

In radio astronomy one talks about the sensitivity of a telescope
in terms of the limiting detectable energy flux from an astronomical
source. We can do the same here too. Given the gravitational wave amplitude
$h$ we can use \Eqnref{eq:flux} to compute the flux of gravitational
waves. One can translate the noise power spectrum $S_h(f)$, given 
in units of Hz$^{-1}$ at frequency $f$, to Jy (Jansky), with the conversion 
factor $4c^3f^2/(\pi G)$. In \Figref{fig:psd-jy}, the left panel shows the noise
power spectrum in astronomical units of Jy and the right panel depicts
the noise spectrum in units of Hz$^{-1}$ together with lines of constant
flux.

\epubtkImage{NoiseCurves-Jy-NoiseCurves-Sh.png}{%
  \begin{figure}[htbp]
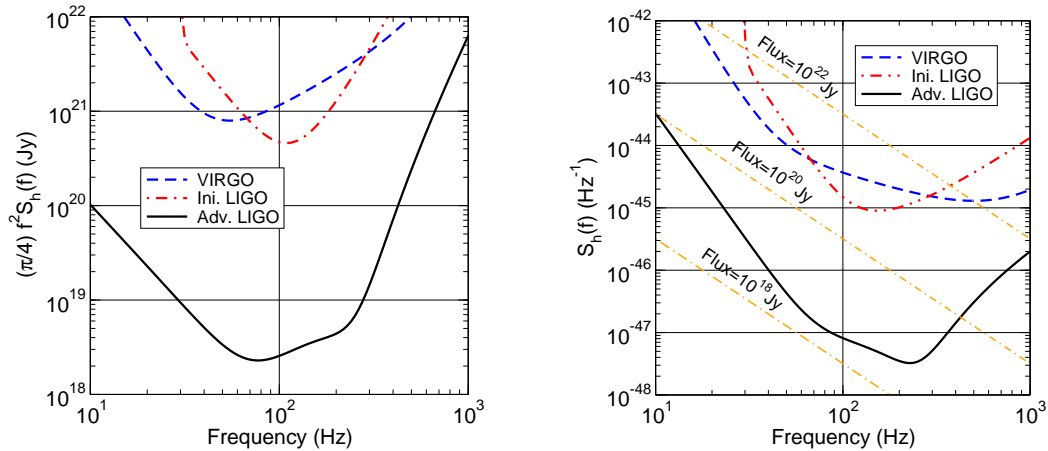

    \centerline{
    \includegraphics[width=2.5in]{NoiseCurves-Jy}\hspace{1cm}
    \includegraphics[width=2.5in]{NoiseCurves-Sh}}
    \caption{The sensitivity of interferometers in terms of the
      limiting energy flux they can detect, Jy/Hz, (left panel) and in
      terms of the gravitational wave amplitude with lines of constant
      flux levels (right panel).}
    \label {fig:psd-jy}
\end{figure}}

What is striking in \Figref{fig:psd-jy} is the magnitude of flux. While
modern radio interferometers are sensitive to flux levels of
milli and micro-Jy the gravitational wave interferometers need
their sources to be 24\,--\,27 orders of magnitude 
brighter. Turning this argument around, the gravitational wave
sources we expect to observe are not really weak, but rather
extremely bright sources. The difficulty in detecting them is
due to the fact that gravitation is the weakest of all 
known interactions.

\subsection{Source amplitudes vs sensitivity}

How does one compare the gravitational wave amplitude of astronomical
sources with the instrumental sensitivity and assess what sort of
sources will be observable against noise? Comparisons are almost
always made in the frequency domain, since stationary noise is most
conveniently characterized by its PSD.

The simplest signal to characterize is a long-lasting periodic signal
with a given fixed frequency $f_0$. In an observation time T, all the
signal power $|\tilde{h}(f_0)|^2$ is concentrated in a single
frequency bin of width $1/T$. The noise against which it competes is
just the noise power in the same bin, $S_h(f_0)/T$. The power
SNR is then $T|\tilde{h}(f_0)|^2/S_h(f_0)$, and
the amplitude SNR is $\sqrt{T}|\tilde{h}(f_0)|/|S_h(f_0)|^{1/2}$. This
improves with observation time as the square root of the time. The
reason for this is that the noise is stationary, but longer and longer
observation times permit the signal to compete only with noise in
smaller and smaller frequency windows.

Of course, no expected gravitational-wave signal would have a single fixed frequency
in the detector frame, because the detector is attached to the Earth,
whose motion produces frequency modulations. But the principle of this
SNR increase with time can still be maintained if one has a signal
model that allows one to exclude more and more noise from competing
with the signal over longer and longer periods of time. This happens
with \emph{matched filtering}, which we return to in
Section~\ref{sec:gwdataanalysis}.

Short-lived signals have wider bandwidths, and long observation times
are not relevant. To characterize their SNR, it is useful to define
the dimensionless noise power per logarithmic bandwidth, $fS_h(f)$,
which we earlier called $h_n^2(f)$. The signal Fourier amplitude
$\tilde h(f) \equiv \int_{-\infty}^\infty\, \mathrm{d}t~h(t) e^{2\pi i
  ft}$ has dimensions of Hz$^{-1}$ and so the Fourier amplitude per
logarithmic frequency, which is called the \emph{characteristic signal
  amplitude} $h_c = f |\tilde h(f)|$, is dimensionless. This quantity
should be compared with $h_n(f)$ to obtain a rough estimate of the SNR
of the signal: SNR $\sim h_c/h_n$.

\subsection{Network detection}

Gravitational wave detectors are almost omni-directional. As discussed in
Section~\ref{sec:beam factors}, both interferometers and bars have good
sensitivity over a large area of the sky. In this regard, gravitational wave
antennas are unlike conventional astronomical telescopes, e.g., optical, 
radio, or infrared bands, which observe only a very small fraction 
of the sky at any given time. The good news is that gravitational wave
interferometers will have good sky coverage and therefore only a small
number (around six) are enough to survey the sky. The bad news, however,
is that gravitational wave observations will not automatically provide 
the location of the source in the sky. It will either be necessary
to observe the same source in several non--co-located detectors 
and triangulate the position of the source using the information 
from the delay in the arrival times of the signals to different
detectors, or observe for a long time and use the location-dependent Doppler
modulation caused by the motion of the detector relative to the source
to infer the source's position in the sky. The latter is a well-known technique in radio astronomy of synthesizing a long-baseline 
observation to gain resolution, and only possible for sources, such
as rotating neutron stars or stochastic backgrounds, that
last for a long enough duration.

A network of detectors is, therefore, essential for source reconstruction.
Network observation is not only powerful in identifying a source
in the sky, but independent observation of the same source in several
detectors adds to the detection confidence, especially since the
noise background in the first generation of interferometers is not
well understood and is plagued by nonstationarity and non-Gaussianity.

\subsubsection{Coherent vs coincidence analysis}
\label{sec:CoherentVsCoincidence}

The availability of a network of detectors offers two different methods
by which the data can be combined. One can either first bring
the data sets together, combine them in a certain way, and then
apply the appropriate filter to the network data and integrate the
signal coherently, coherent detection~\cite{Pai:2000zt, Bose:1999bp, Finn:2000hj, Arnaud:2003zq}, or first analyze
the data from each detector separately by applying the relevant
filters and then look for coincidences in the multi-dimensional space
of intrinsic (masses of the component stars, their spins, $\ldots$) 
and extrinsic (arrival times, a constant phase, source location, $\ldots$) 
parameters, coincidence detection~\cite{Jaranowski:1994xd, Jaranowski:1996hs, Finn:2000hj, Arnaud:2001my, Tagoshi:2007ni,
Abbott:2003pj, Abbott:2005pe, Abbott:2005pf, Abbott:2005kq}.

A recent comparison of coherent 
analysis vis-a-vis coincidence analysis under the assumption that 
the background noise is Gaussian and stationary has concluded that 
coherent analysis, as one might expect, is far better than coincidence 
analysis~\cite{CoherentvsCoincident}. These authors also explore,
to a limited extent, the effect of nonstationary noise and reach
essentially the same conclusion.

At the outset, coherent analysis sounds like a good idea, since in a 
network of $N_D$ similar detectors the visibility of a signal improves by 
a factor of $\sqrt{N_D}$ over that of a single detector. One can take
advantage of this enhancement in SNR to either lower the false alarm
rate by increasing the detection threshold, while maintaining the same
detection efficiency, or improve detection efficiency at a given false
alarm rate.

However, there are two reasons that current data-analysis 
pipelines prefer coincidence analysis over coherent analysis. 
Firstly, since the detector 
noise is neither Gaussian nor stationary, coincidence analysis can 
potentially reduce the background rate far greater than one might 
think otherwise. Secondly, coherent analysis is computationally far 
more expensive than coincidence analysis and it is presently not 
practicable to employ coherent analysis.

Coincidence analysis is indeed a very powerful method to veto
out spurious events.  One can associate with each event in a given 
detector an ellipsoid, whose location and orientation depends on where
in the parameter space and when the event was found, and the SNR can be
used to fix the size of the ellipsoid~\cite{Robinson:2008un}. One is associating
with each event a `sphere' of influence in the multi-dimensional
space of masses, spins, arrival times, etc., and there is a stringent
demand that the spheres associated with events from different detectors 
should overlap each other in order to claim a detection. Since random
triggers from a network of detectors are less likely to be consistent
with one another, this method serves as a very powerful veto.

It is probably not possible to infer beforehand which method might
be more effective in detecting a source, as this might depend on the
nature of the detector noise, on how the detection statistic is
constructed, etc. An optimal approach might
be a suitable combination of both of these methods. For instance,
a coherent follow-up of a coincidence analysis (as is currently done
by searches for compact binaries within the LSC)
or to use coincidence criteria on candidate events from a coherent search.

Coherent addition of data improves the visibility of the signal, but 
`coherent subtraction' of the data in a detector network should
lead to data products that are devoid of gravitational wave signals. 
This leads us naturally to the introduction of the null stream veto.

\subsubsection{Null stream veto}
\label{sec:Null Stream}

Data from a network of detectors, when suitably shifted in time
and combined linearly with coefficients that depend on the source 
location, will yield a time series that, in the ideal case, will be
entirely be devoid of the gravitational signal. Such a combination
is called a \emph{null stream}. For instance, for a set of three
misaligned detectors, each measuring a data stream $x_k(t)$, $k=1,2,3$,
the combination $x(t) = A_{23}(\theta, \varphi) h_1(t+\tau_1) +  
A_{31}(\theta, \varphi) h_2(t+\tau_2) +  A_{12}(\theta, \varphi) 
h_3(t+\tau_3)$, where $A_{ij}$ are functions of the
responses of the antennas $i$ and $j$, and $\tau_k$'s, $k=1,2,3$, are 
time delays that depend on the source location and the location of
the antenna, is a null stream. If $x_k(t)$, $k=1,2,3$, contain a 
gravitational wave signal from
an astronomical source, then $x(t)$ will not contain the signature
of this source. In contrast, if $x(t)$ and $x_k(t)$ both contain
the signature of a gravitational wave event, then that is an 
indication that one of the detectors has a glitch. 

The existence and usefulness of a null stream was first 
pointed out by G\"ursel and Tinto~\cite{GurselTinto1989}.  
Wen and Schutz~\cite{WenSchutz2005} proposed implementing 
it in LSC data analysis as a veto, 
and this has been taken up now by several search groups.

\subsubsection{Detection of stochastic signals by cross-correlation}
\label{sec:crosscorrelation}

Stochastic background sources and their detection is discussed in more 
detail in \Secref{sec:gwcosmology}. Here we will briefly  mention the 
problem in the context of detector networks.  As mentioned in 
\Secref{sec:stochastic bg introduction}, the universe might be 
filled with stochastic gravitational waves that were either generated 
in the primeval universe or by a population of background sources. 
For point sources, although each source in a population 
might not be individually detectable, they could collectively produce a confusion 
background via a random superposition of the waves from that population. 
Since the waves are random in nature, it is not possible to use the
techniques described in Sections~\ref{sec:CoherentVsCoincidence},
\ref{sec:Null Stream} and \ref{sec:matched filtering} to
detect a stochastic background. However, we might use the noisy 
stochastic signal in one of the detectors as a ``matched-filter'' for the data
in another detector~\cite{Thorne1987, Flanagan1993, Allen1997, Allen:1997ad}. 
In other words, it should be possible to detect
a stochastic background by cross-correlating the data from a pair of 
detectors; the common gravitational-wave background will grow in 
time more rapidly than the random backgrounds in the two instruments,
thereby facilitating the detection of the background. 

If two instruments with identical spectral noise density $S_h$ are 
cross-correlated over a bandwidth $\Delta f$ for a total time $T$, 
the spectral noise density of the output is reduced by a factor of 
$(T\Delta f)^{1/2}$. Since the noise amplitude is proportional to the 
square root of $S_h$, the amplitude of a signal that can be detected 
by cross-correlation improves only with the fourth root of the observing 
time. This should be compared with the square root improvement that 
matched filtering gives.

The \emph{cross-correlation technique}  works well when the two
detectors are situated close to one another. When separated, only
those waves whose wavelength is larger than or comparable to the
distance between the two detectors, or which arrive from a direction
perpendicular to the separation between the detectors, can contribute
coherently to the cross-correlation statistic. Since the instrumental
noise builds up rapidly at lower frequencies, detectors that are
farther apart are less useful in cross-correlation. However, very
near-by detectors (as in the case of two LIGO detectors within the
same vacuum tube in Hanford) will suffer from common background noise
from the control system and the environment, making it rather difficult
to ascertain if any excess noise is due to a stochastic background of
gravitational waves.

%% I will write another paragraph giving the relevant equations for
%% for  the cross-correlation statistic and discuss and show the overlap
%% reduction function for one or two paris of interferometers. 

\subsection{False alarms, detection threshold and coincident observation}
\label{sec:false alarms}

Gravitational-wave event rates in initial interferometers is 
expected to be rather low: about a few per year. Therefore, one has to set a high threshold, so that the noise-generated false alarms mimicking an event are 
negligible. 

For a detector output sampled at 1 kHz and processed through 
a large number of filters, say $10^3$, one has  
$\sim 3 \times 10^{13}$ instances of noise in a year. If the filtered noise is Gaussian,
then the probability $P(x)$ of observing an amplitude in the range of $x$ to $x+dx$ is
\begin{equation}
P(x)\ dx = \frac{1}{\sqrt{2\pi}\sigma} \exp\left( \frac{-x^2}{2\sigma^2} \right )\ dx,
\end{equation}
where $\sigma$ is the standard deviation.  The above probability-distribution
function implies that the probability that the noise amplitude is greater than
a given threshold $\eta$ is
\begin{equation}
P(x | x \ge \eta) = \int_\eta^\infty P(x)\, \mathrm{d}x = \frac{1}{\sqrt{2\pi} \sigma} \int_\eta^\infty 
\exp\left( \frac{-x^2}{2\sigma^2} \right )\, \mathrm{d}x.
\end{equation}
Demanding that no more than one noise-generated false alarm occur in 
a year's observation means that $P(x | x \ge \eta) = 1/(3 \times 10^{13})$.
Solving this equation for $\eta$, one finds that $\eta \simeq 7.5\sigma$ in
order that false alarms are negligible in a year's observation.
Therefore, a source is detectable  only if its amplitude is 
significantly larger than the effective noise amplitude, i.e., 
$f\tilde h(f) \gg h_n(f)$. 
  
The reason for accepting only such high-sigma events is that 
the event rate of a transient source, i.e., a source lasting for a few seconds to
minutes, such as a binary inspiral, could be as low as a few per year, and
the noise generated false alarms, at low SNRs $\sim 3$-4, over a period of a 
year, tend to be quite large. Setting higher thresholds for detection helps in
removing spurious, noise generated events.
However, signal enhancement techniques 
(cf. Section~\ref{sec:gwdataanalysis}) make it possible to detect a signal of
relatively low amplitude, provided there are a large number of
wave cycles and the shape of the wave is known accurately.

Real detector noise is neither Gaussian nor stationary and therefore the filtered 
noise cannot be expected to obey these properties either. One of the most 
challenging problems is how to remove or veto the false alarm generated by a
non-Gaussian and/or nonstationary background. There has been some effort to address
the issue of non-Gaussianity~\cite{Creighton:1999qw} and nonstationarity~\cite{Mohanty:1999qn}; 
more work is needed in this direction. However, it is expected that
the availability of a network of gravitational wave detectors alleviates the
problem to some extent. This is because a high amplitude gravitational wave event will 
be coincidentally observed in several detectors, although not necessarily
with the same SNR,  while false alarms are, in general, not coincident, as
they are normally produced by independent sources located close to the detectors. 

We have seen that coincident observations help to reduce the false alarm rate 
significantly. The rate can be further reduced, and possibly even nullified,
by subjecting coincident events to further consistency 
checks in a detector network consisting of four or more detectors. 
As discussed in Section~\ref{sec:gwobservables}, each gravitational wave event
is characterized by five kinematic (or extrinsic) observables: location
of the source with respect to the detector $(D, \theta, \varphi)$ and the
two polarizations $(h_+, h_\times)$. Each detector in a network measures
a single number, say the amplitude of the wave. In addition, in a network of 
$N$ detectors, there are $N-1$ independent time delays in the arrival times
of the wave at various detector locations, giving a total of $2N-1$ observables.
Thus, the minimum number of detectors needed to reconstruct the wave and
its source is $N=3$. More than three detectors in a network will have 
redundant information that will be consistent with the quantities inferred
from any three detectors, provided the event is a true coincident event
and not a chance coincidence, and most likely a true gravitational wave event.
In a detector network consisting of $N (\ge 4)$ detectors, one can perform
$2N-6$ consistency checks. Such consistency checks further reduce the number
of false alarms. 

When the shape of a signal is known, matched filtering is the optimal 
strategy to pull out a signal buried in Gaussian, stationary 
noise (see Section~\ref{sec:matched filtering}). The presence of high-amplitude 
transients in the data can render the background nonstationary and 
non-Gaussian, therefore matched filtering is not necessarily an optimal 
strategy. However, the knowledge of a signal's shape, especially when it has a broad bandwidth, can be used beyond matched filtering to construct 
a $\chi^2$ veto~\cite{Allen:2004gu} to distinguish between triggers caused 
by a true signal from those caused by high-amplitude transients or 
other artifacts.  One specific implementation of the $\chi^2$
veto compares the expected signal spectrum with the real spectrum to quantify
the confidence with which a trigger can be accepted to be caused by a true
gravitational wave signal and has been the most powerful method for 
greatly reducing the false alarm rate. We shall discuss the $\chi^2$ 
veto in more detail in Section~\ref{sec:matched filtering}.

%==================================================================
\newpage

\section{Data Analysis}
\label{sec:gwdataanalysis}

Observing gravitational waves requires a data analysis strategy, which
is in many ways different from conventional astronomical data analysis.
There are several reasons why this is so:

\begin{itemize}

\item Gravitational wave antennas are essentially omni-directional, with
their response better than 50\% of the root mean square over 75\% of the sky
(see \Figref{fig:response}, right panel, recalling that the rms
response is 2/5 of the peak).
Hence, data analysis systems will have to carry out all-sky searches
for sources. 

\item Interferometers are typically broadband covering three to four
orders of magnitude in frequency. While this is obviously to our
advantage, as it helps to track sources whose frequency 
may change rapidly, it calls for searches to be carried out over 
a wide range of frequencies. 

\item In Einstein's theory, gravitational 
radiation has two independent states of polarization. 
Measuring polarization is of fundamental importance (as there are other 
theories of gravity in which the number of polarization states 
is more than two and in some theories even dipolar and scalar 
waves exist~\cite{bss:will.93}) and has astrophysical  
implications too (for example, gravitational-wave--polarization measurement is
one way to resolve the mass-inclination 
degeneracy of binary systems observed electromagnetically, as
discussed in \Secref{sec:inclination}).
Polarization measurements would be possible with a network of 
detectors, which means analysis algorithms that work with data 
from multiple antennas will have to be developed. This should
also benefit coincidence analysis and the efficiency of event recognition. 

\item Unlike typical detection techniques for electromagnetic 
radiation from astronomical sources, most astrophysical gravitational waves
are detected coherently, by following the phase of the radiation, 
rather than just the energy. That is, the SNR 
is built up by coherent superposition of many wave cycles emitted 
by a source. The phase evolution contains more information than the 
amplitude does and the signal structure is a rich source of the 
underlying physics.  
Nevertheless, tracking a signal's phase means searches will have to be made not
only for specific sources but over a huge region of the parameter
space for each source, placing severe demands both on the theoretical
understanding of the emitted waveforms as well as on the data analysis
hardware.

\item Finally, gravitational wave detection is computationally intensive. Gravitational wave antennas
acquire data continuously for many years at the rate of several
megabytes per second. About 1\% of
this data is signal data; the rest is housekeeping data that 
monitors the operation of the detectors. The large parameter space 
mentioned above requires that the signal data be filtered many times 
for different searches, and this puts big demands on computing 
hardware and algorithms.

\end{itemize}

Data analysis for broadband detectors has been strongly developed since 
the mid 1980s~\cite{Thorne1987, SCHUTZ1991a, SCHUTZ_GWDAW}. 
The field has a regular series of annual 
Gravitational Wave Data Analysis Workshops; the published proceedings
are a good place to find current thinking and challenges. Early 
coincidence experiments with interferometers~\cite{Nicholson1996} and 
bars~\cite{ICEG2000} provided the first opportunities to apply 
these techniques. Although the theory is now fairly well 
understood~\cite{Living:JaranowskiKrolak}, strategies for 
implementing data analysis depend on available computer resources, 
data volumes, astrophysical knowledge, and source modeling, and so are under
constant revision.

We will begin with a discussion of the matched filtering algorithm
and next use it to estimate the SNRs for binary
coalescences in various 
detectors.  After that, we will develop the theory of matched filtering
further to work out
the computational costs to carry out online searches, that is
to search at the same rate as the data is acquired. In the
final section, we will use the formalism developed in earlier
sections to discuss parameter estimation. The foundations of signal 
analysis lie in the statistics of making ``best estimates'' of 
whether a signal is present in noisy data or not. See the Living Review 
by Jaranowski and Kr{\'o}lak~\cite{Living:JaranowskiKrolak} for a 
discussion of this in the gravitational wave context.

\subsection{Matched filtering and optimal signal-to-noise ratio}
\label{sec:matched filtering}

Matched filtering is a data analysis
technique that efficiently searches for a signal of known shape 
buried in noisy data~\cite{cwh68}.
The technique consists in correlating the 
output of a detector with a waveform, variously known as a
template or a filter. Given a signal $h(t)$ buried in noise
$n(t)$, the task is to find an `optimal' template $q(t)$ that 
would produce, on the average, the best possible
SNR. In this review, we shall
treat the problem of matched
filtering as an operational exercise. However, this intuitive
picture has a solid basis in the theory of hypothesis testing.
The interested reader may consult any standard text
book on signal analysis, for example Helstrom~\cite{cwh68},
for details. 

Let us first fix our notation. We shall use $x(t)$ to denote
the detector output, which is assumed to consist of a background
noise $n(t)$ and a useful gravitational wave signal $h(t)$.
The Fourier transform of a quantity $x(t)$ will be denoted $\tilde x(f)$ 
and is defined as
\begin{equation}
\tilde x(f) = \int_{-\infty}^\infty x(t) e^{2\pi i f t}\, \mathrm{d}t.
\end{equation}

\subsubsection{Optimal filter}

The detector output $x(t)$ is just a realization of noise $n(t)$, 
i.e., $x(t)=n(t)$, when no signal is present. In the presence of a 
signal $h(t)$ with an \emph{arrival time} $t_a$, $x(t)$ takes the form,
\begin{equation}
x(t) = h(t-t_a) + n(t).
\end{equation}
The correlation $c$ of a template $q(t)$ with the detector output is 
defined as
\begin{equation}
c(\tau) \equiv \int_{-\infty}^{\infty} x(t) q(t+\tau)\mathrm{\ d}t.
\label {eq:correlation}
\end{equation}
In the above equation, $\tau$ is called the \emph{lag}; it denotes the
duration by which the filter function lags behind the detector output.
The purpose of the above correlation integral is to concentrate 
all the signal energy at one place. The following analysis reveals how
this is achieved; we shall work out the \emph{optimal}
filter $q(t)$ that maximizes the correlation
$c(\tau)$ when a signal $h(t)$ is present in the detector output.
To do this let us first write the correlation integral in the Fourier
domain by substituting for 
$x(t)$ and $q(t)$, in the above integral, 
their Fourier transforms $\tilde x(f)$ and $\tilde q(f)$, i.e.,
$x(t) \equiv \int_{-\infty}^\infty \tilde x(f) 
\exp \left (-2 \pi i ft \right ) \, \mathrm{d}f$ and $q(t) \equiv 
\int_{-\infty}^\infty \tilde q(t) \exp \left (-2 \pi i ft \right )\, \mathrm{d}f$, 
respectively.  After some straightforward algebra, one obtains
\begin{equation}
c(\tau) = \int_{-\infty}^{\infty} \tilde x(f) \tilde q^*(f)
e^{-2\pi i f \tau }\, \mathrm{d}f,
\end{equation}
where $\tilde q^*(f)$ denotes the complex conjugate of $\tilde q(f)$.

Since $n$ is a random process, $c$ is also a random process. Moreover, 
correlation is a linear operation and hence 
$c$ obeys the same probability distribution function as $n$. In particular,
if $n$ is described by a Gaussian random process with zero mean, then 
$c$ is also described by a Gaussian distribution function, although 
its mean and variance will, in general, differ from those of $n$. 
The mean value of $c$, denoted by
$S\equiv \overline c$, is, clearly, the correlation of the template $q$ 
with the signal $h$, since the mean value of $n$ is zero:
\begin{equation}
S \equiv \overline{c}(\tau) = \int_{-\infty}^{\infty} \tilde h(f) 
         \tilde q^*(f) e^{-2\pi i f (\tau-t_a)}\, \mathrm{d}f. 
\label{eq:signal}
\end{equation}
The variance of $c$, denoted $N^2 \equiv \overline {(c- \overline c)^2}$, 
turns out to be,
\begin{equation}
N^2 = \overline{ (c-\overline c)^2} = \int_{-\infty}^{\infty} S_h(f) \left | \tilde q(f) \right |^2\, \mathrm{d}f.
\label{eq:noise}
\end{equation}
Now the SNR $\rho$ is defined by 
$\rho^2 \equiv S^2/N^2$. 

The form of integrals in Equations~(\ref{eq:signal}) and~(\ref{eq:noise}) leads naturally to the definition 
of the scalar product of waveforms. Given two functions, 
$a(t)$ and $b(t)$, we define their scalar product 
$\left <a, b\right >$ to be~\cite{Finn:1992wt, Finn:1992xs, Chernoff:1993th, Cutler:1994ys}
\begin{equation}
\left < a, b \right > \equiv 2 \int_{0}^\infty \frac{\mathrm{d}f}{S_h(f)} 
\left [\tilde a(f) \tilde b^*(f) + \tilde a^*(f) \tilde b(f)\right].
\label {eq:scalar product}
\end{equation}
Note that $S_h(f) \ge 0$ [cf.\ Equation~(\ref{eq:psdinfourier})],
consequently, the scalar product is real and positive definite.

Noting that the Fourier transform of a real function $h(t)$ obeys
$\tilde h(-f) = \tilde h^*(f)$, we can write down the SNR 
in terms of the above scalar product:
\begin{equation}
\rho^2 = \frac {\left <he^{2\pi i f (\tau-t_a)}, S_h q\right > }
               {\sqrt {\left <S_h q, S_h q\right >}}.
\label{eq:snr1}
\end{equation}
From this it is clear that the template $q$ that obtains the
maximum value of $\rho$ is simply
\begin{equation}
\tilde {q}(f) = \gamma \frac {\tilde {h}(f)e^{i2\pi f(\tau-t_a)}} {S_h(f)},
\label{eq:optimal}
\end{equation}
where $\gamma$ is an arbitrary constant.
From the above expression for an optimal filter we note two important
things. First, the SNR is maximized when the lag parameter $\tau$ is
equal to the time of arrival of the signal $t_a$. Second, 
the optimal filter is not just a copy of the signal, but rather it is
weighted down by the noise PSD. We will see below
why this should be so.

\subsubsection{Optimal signal-to-noise ratio}

We can now work out the optimal SNR by substituting Equation~(\ref{eq:optimal}) for the
optimal template in Equation~(\ref{eq:snr1}),
\begin{equation}
\rho_{\mathrm{opt}} = \left < h, h \right >^{1/2} = 2\left [
  \int_{0}^\infty \, \mathrm{d}f\frac{ \left |\tilde h(f)\right|^2 }{S_h(f)} \right]^{1/2}.
\label{eq:snr}
\end{equation}
%
%FS 
\begin{sloppypar}
We note that the optimal SNR is not just the total energy of the signal
(which would be $2\int_{0}^\infty \, \mathrm{d}f |\tilde h(f) |^2$),
but rather the integrated signal power weighted down by the noise
PSD. This is in accordance with what we would guess intuitively:
the contribution to the SNR from a frequency
bin where the noise PSD is high is smaller than from a bin
where the noise PSD is low. Thus, an optimal filter automatically 
takes into account the nature of the noise PSD.
\end{sloppypar}

The expression for the optimal SNR Equation~(\ref{eq:snr}) suggests
how one may compare signal strengths with the noise performance of a
detector.  Note that one cannot directly compare $\tilde h(f)$ with
$S_h(f)$, as they have different physical dimensions. In gravitational
wave literature one writes the optimal SNR in one of the following
equivalent ways
\begin{equation}
\rho_{\mathrm{opt}} = 
2\left [ \int_{0}^\infty
\frac {\mathrm{d}f}{f} \frac {\left |\sqrt{f} \tilde h(f)\right|^2}{S_h(f)}  \right]^{1/2} =
2\left [ \int_{0}^\infty
\frac {\mathrm{d}f}{f} \frac {\left |f \tilde h(f)\right|^2}{f S_h(f)}  \right]^{1/2},
\label{eq:snr2}
\end{equation}
which facilitates the comparison of signal strengths
with noise performance. One can compare the dimensionless quantities,
$f|\tilde h(f)|$ and $\sqrt{f S_h(f)}$, or dimensionful quantities,
$\sqrt{f} |\tilde h(f)|$ and $\sqrt{S_h(f)}$.

Signals of interest to us are characterized by several (a priori unknown)
parameters, such as the masses of component stars in a binary, their 
intrinsic spins, etc., and an optimal filter must agree with 
both the signal shape and its parameters.
A filter whose parameters are slightly mismatched
with that of a signal can greatly degrade the SNR.
For example, even a mismatch of one cycle in $10^4$ cycles
can degrade the SNR by a factor two. 

When the parameters of a filter and its shape are precisely 
matched with that of a signal, what is the improvement brought 
about, as opposed to the case when no knowledge of the signal 
is available?  
Matched filtering helps in enhancing the SNR in proportion to the 
square root of the number of signal cycles in the detector band,
as opposed to the case in which the signal shape is unknown and
all that can be done is to Fourier transform the detector output
and compare the signal energy in a frequency bin to noise energy
in that bin.  We shall see below that, in initial interferometers,
matched filtering leads to an enhancement of order 30\,--\,100 for 
compact binary inspiral signals.

\subsubsection{Practical applications of matched filtering}

Matched filtering is currently being applied to mainly two sources: detection of 
(1)~chirping signals from compact binaries consisting of black 
holes and/or neutron stars and (2)~continuous waves from rapidly-spinning
neutron stars. 

\paragraph{Coalescing binaries.} In the case of chirping binaries, post-Newtonian 
theory (a perturbative approximation to Einstein's equations in which the relevant
quantities are expanded as a power-series in $1/c$, where $c$ is the speed
of light) has been used to model the dynamics of these systems to a very high order
in $v/c$, where $v$ is the relative speeds of the objects in the binary (see also
\Secref{sec:two body problem}, in which binaries are discussed in more detail).
This is an approximation that can be effectively used to match filter the signal
from binaries whose component bodies are of equal, or nearly equal, masses and
the system is still ``far'' from coalescence. In reality, one takes the
waveform to be valid until the last stable circular orbit (LSCO). In the case of
binaries consisting of two neutron stars, or a neutron star and a black hole,
tidal effects might affect the evolution significantly before reaching the 
LSCO. However, this is likely to occur at frequencies well-above the sensitivity
band of the current ground-based detectors, so that for all practical purposes
post-Newtonian waveforms are a good approximation to low-mass ($M<10\,M_\odot$)
binaries.

As elucidated in \Secref{sec:numerical relativity},
progress in analytical and numerical relativity has made it possible to have
a set of waveforms for the merger phase of compact binaries too. 
The computational cost in matched filtering the merger phase, however, will not be 
high, as there will only be on the order of a few 100 cycles in this phase. But it is important
to have the correct waveforms to enhance signal visibility and, more importantly,
to enable strong-field tests of general relativity.

In the general case of black-hole--binary inspiral the search space is characterized
by 17 different parameters. These are the two masses of the bodies, their
spins, eccentricity of the orbit, its orientation at some fiducial time, 
the position of the binary in the sky and its distance from the Earth, the 
epoch of coalescence and phase of the signal at that epoch, and the 
polarization angle. However, not all these parameters are important in a 
matched filter search. Only those parameters that change the shape of the
signal, such as the masses, orbital eccentricity and spins, or cause a 
modulation in the signal due to the motion of the detector relative to 
the source, such as the direction to the source, are to be searched for 
and others, such as the epoch of coalescence and the phase at that epoch,
are simply reference points in the signal that can be determined without
any significant burden on computational power. 

For binaries consisting of nonspinning objects that are either observed 
for a short enough period that the detector motion can be neglected,
or last for only a short time in the sensitive part of a detector's sensitivity
band, there are essentially two search parameters -- the component masses
of the binary. It turns out that the signal manifold in this case is nearly
flat, but the masses are curvilinear coordinates and are not good 
parameters for choosing templates.  Chirp times, which are nonlinear functions
of the masses, are very close to being Cartesian coordinates and 
template spacing is more or less uniform in terms of these
parameters.  Chirp times are post-Newtonian contributions at different orders to
the duration of a signal starting from a time when
the instantaneous gravitational-wave frequency has a fiducial value 
$f_{\mathrm{L}}$ to a time when the gravitational wave 
frequency formally diverges and system coalesces. 
For instance, the chirp times $\tau_0$ and $\tau_3$ 
at Newtonian and 1.5~PN orders, respectively, are
\begin{equation}
\tau_0 = \frac{5}{256\,\pi\,\nu\,f_{\mathrm{L}}}
\left (\pi M f_{\mathrm{L}}\right )^{-5/3}, \quad
\tau_3 = \frac{1}{8\,\nu\,f_{\mathrm{L}}}
\left (\pi M f_{\mathrm{L}}\right )^{-2/3},  
\end{equation}
where $M$ is the total mass and $\nu=m_1m_2/M^2$ is the symmetric mass 
ratio. The above relations can be inverted to obtain $M$ and $\nu$ in
terms of the chirp times:
\begin{equation}
M = \frac{5}{32\,\pi^2\,f_{\mathrm{L}}} \frac{\tau_3}{\tau_0},\quad
\nu = \frac{1}{8\,\pi\,f_{\mathrm{L}}\,\tau_3} \left ( \frac{32\,\pi\,\tau_0}
{5\,\tau_3} \right )^{2/3}.
\end{equation}

There is a significant amount of literature on the computational
requirements to search for compact
binaries~\cite{Sathyaprakash:1991mt, Dhurandhar:1992mw, owen97,
  Owen:1998dk}. The estimates for initial detectors are not alarming
and it is possible to search for these systems online. Searches for
these systems by the LSC (see, for example, \cite{Abbott:2005kq})
employs a hexagonal lattice of templates~\cite{Cokelaer:2007kx} in the
two-dimensional space of chirp times. For the best LIGO detectors we
need several thousand templates to search for component masses in the
range $[m_{\mathrm{low}},m_{\mathrm{high}}] =
[1,100]\,M_\odot$~\cite{Owen:1998dk}.  Decreasing the lower-end of the
mass range leads to an increase in the number of templates that goes
roughly as $m^{-8/3}_{\mathrm{low}}$ and most current
searches~\cite{Abbott:2003pj, Abbott:2005pe} only begin at
$m_{\mathrm{low}}=1\,M_\odot$, with the exception of one that looked
for black hole binaries of primordial origin~\cite{Abbott:2005pf}, in
which the lower end of the search was $0.2\,M_\odot$.

Inclusion of spins is only important when one or both of the
components is \emph{rapidly} spinning~\cite{Apostolatos:1995pj, BCV03b}. 
Spins effects are unimportant for neutron star binaries, for which the 
dimensionless spin parameter $q$, that is the ratio of its spin magnitude to 
the square of its mass, is tiny: $q=J_{\mathrm{NS}}/M_{\mathrm{NS}}^2\ll 1$. 
For ground-based detectors, even after including spins, 
the computational costs, while high, are not formidable and
it should be possible to carry out the search on large computational clusters
in real time~\cite{BCV03b}. Recently, the LSC has successfully carried out
such a search~\cite{Abbott:2007ai}.

\paragraph{Searching for Continuous Wave Signals.}
In the case of continuous waves (CWs), the signal shape is pretty trivial:
a sinusoidal oscillation with small corrections to take account of the
slow spin-down of the neutron star/pulsar to account for the loss of
angular momentum to gravitational waves and other radiation/particles.
However, what leads to an enormous computational cost here is the Doppler
modulation of the signal caused by Earth's rotation, the motion of the 
Earth around the solar system barycenter and the moon. The number of
independent patches that we have to observe so as not to lose appreciable
amounts of SNR can be worked out in the following manner.
The baseline of a gravitational wave detector for CW sources is
essentially $L = 2 \times 1 AU \simeq 3 \times 10^{11}\mathrm{\ m}$. For a source that
emits gravitational waves at 100~Hz, the wavelength of the radiation
is $\lambda=3\times 10^{6}\mathrm{\ m}$, and the angular resolution
$\Delta \theta$ of the antenna at an SNR of 1 is $\Delta\theta \simeq \lambda/L = 10^{-5}$, or a
solid angle of $\Delta \Omega \simeq (\Delta \theta)^2 = 10^{-10}$. In other
words, the number of patches one should search for is $N_{\mathrm{patches}}
\sim 4\pi/\Delta\Omega \simeq 10^{11}$. Moreover, for an observation that lasts
for about a year $(T\simeq 3\times 10^{7}\mathrm{\ s})$ the frequency resolution is
$\Delta f = 1/T \simeq  3\times 10^{-8}$. Searching over a frequency band of
300~Hz, around the best sensitivity of the detector, gives the number of 
frequency bins to be about $10^{10}$. Thus, it is necessary to search
over roughly $10^{11}$ patches in the sky for each of the $10^{10}$ frequency
bins. This is a formidable task and one can only perform a matched
filter search over a short period (days/weeks) of the data or over a
restricted region in the sky, or just perform targeted searches for known
objects such as pulsars, the galactic center, etc.~\cite{Brady1998}.

The severe computational burden faced in the case of CW searches has
led to the development of specialized searches that look for signals
from known pulsars~\cite{Abbott:2004ig, Abbott:2007ce, Abbott:2008fx} 
using an efficient search algorithm that makes
use of the known parameters~\cite{Christensen:2004jm, Dupuis:2005xv} 
and hierarchical algorithms that add power incoherently with the
minimum possible loss in signal visibility~\cite{Krishnan:2004sv,
  Abbott:2005pu, Sintes:2006uc, Abbott:2007tda}.  The most ambitious
project in this regard is the Einstein@Home
project~\cite{EinsteinAtHome}. The goal here is to carry out coherent
searches for CW signals using wasted CPUs on idle computers at homes,
offices and university departments around the world.  The project has
been successful in attracting a large number of subscriptions and
provides the largest computational infrastructure to the LSC for the
specific search of CW signals and the first scientific results from
such are now being published by the LSC~\cite{Abbott:2008uq}.

\paragraph{$\chi^2$ veto.}
Towards the end of Section~\ref{sec:false alarms} we discuss a powerful
way of rejecting triggers, whose root cause is not gravitational wave signals
but false alarms due to instrumental and environmental artifacts. In this
section we will further quantify the $\chi^2$ veto \cite{Allen:2004gu} by 
using the scalar product introduced in the context of matched filtering. 

The main problem with real data is that it can be glitchy in the form
of high amplitude transients that might look like damped sinusoids. An
inspiral signal and a template employed to detect it are both
broadband signals. Therefore, the matched-filter SNR for such signals
has contributions from a wide range of frequencies. However, the
statistic of matched filtering, namely the SNR, is an integral over
frequency and it is not sensitive to contributions from different
frequency regions. Imagine dividing the frequency range of integration
into a finite number of bins $f_k\le f < f_{k+1}$, $k=1,\ldots, p$,
such that their union spans the entire frequency band, $f_1=0$ and
$f_{p+1}=\infty$, and further that the contribution to the SNR from
each frequency bin is the same, namely,
\begin{equation}
4 \int_{f_k}^{f_{k+1}} \frac{| \tilde h(f) |^2}{S_h(f)} \mathrm{d}f = \frac{4}{p}
\int_{0}^{\infty} \frac{| \tilde h(f) |^2}{S_h(f)} \mathrm{d}f.
\end{equation}
Now, define the contribution to the matched filtering statistic coming 
from the $k$-th bin by \cite{Allen:2004gu}
\begin{equation}
z_k \equiv \left < q , x \right >_k \equiv 2 \int_{f_k}^{f_{k+1}} \left [
\tilde q^*(f) \tilde x(f) + \tilde q(f) \tilde x^*(f) \right ]\,\frac{\mathrm{d}f}{S_h(f)},
\end{equation}
where, as before,  $\tilde x(f)$ and $\tilde q(f)$ are the Fourier transforms
of the detector output and the template, respectively. Note that
the sum $z=\sum_k z_k$ gives the full matched filtering statistic \cite{Allen:2004gu}:
\begin{equation}
z = \left < q , x \right > \equiv 2 \int_{0}^{\infty} \left [
\tilde q^*(f) \tilde x(f) + \tilde q(f) \tilde x^*(f) \right ]\,\frac{\mathrm{d}f}{S_h(f)}.
\end{equation}

Having chosen the bins and quantities $z_k$ as above, one can construct a 
statistic based on the measured SNR in each bin as 
compared to the expected value, namely
\begin{equation}
\chi^2 = p \sum_{k=1}^p \left (z_k - \frac{z}{p} \right )^2.
\end{equation}
When the background noise is stationary and Gaussian, the quantity
$\chi^2$ obeys the well-known chi-square distribution with $p-1$
degrees of freedom. Therefore, the statistical properties of the
$\chi^2$ statistic are known. Imagine two triggers with identical
SNRs, but one caused by a true signal and the other
caused by a glitch that has power only in a small frequency range.
It is easy to see that the two triggers will have very different
$\chi^2$ values; in the first case the statistic will be far 
smaller than in the second case. This statistic has served as a
very powerful veto in the search for signals from coalescing
compact binaries and it has been instrumental in cleaning up the
data (see, e.g., \cite{Abbott:2003pj,Abbott:2005pe}).

\subsection{Suboptimal filtering methods}

It is not always possible to compute the shape of the signal from a source.
For instance, there is no computation, numerical or analytical, that reliably gives 
us the highly relativistic and nonlinear dynamics of gravitational collapse,
the supernova that follows it and the emitted gravitational signal. The biggest 
problem here is the unknown physical state of the pre-supernova star and the 
complex physics that is involved in the collapse and explosion.  Thus, matched
filtering cannot be used to detect signals from supernovae.

Even when the waveform is known, the great variety
in the shape of the emitted signals might render matched filtering ineffective.
In binaries, in which one of the component masses is much smaller than the other,
the smaller body will evolve on a highly precessing and in some cases eccentric 
orbit, due to strong spin-orbit coupling. Moreover, the radiation backreaction 
effects, which in the case of equal mass binaries are computed in
an approximate way by averaging over an orbital time scale, should be computed
much more accurately. The resulting motion of the small body in the Kerr spacetime
of the larger body is extremely complicated, leading to a waveform that is rather 
complex and matched filtering would not be a practical approach.

Suboptimal methods can be used in such cases and they have a twofold advantage: 
they are less sensitive to the shape of the signal and are computationally 
significantly cheaper than matched filtering. Of course, the price is a loss
in the SNR. The best suboptimal methods are sensitive to
signal amplitudes a factor of two to three larger than that required by matched filtering
and a factor of 10 to 30 in volume.

Most suboptimal techniques are one form of time-frequency transform or the
other. They determine the presence or absence of a signal by comparing the 
power over a small volume in the time-frequency plane in a given segment of 
data to the average power in the same volume over a large segment of data.
The time-frequency transform $q(\tau, f)$ of data $x(t)$ using a window $w(t)$
is defined as
\begin{equation}
q(\tau, f) = \int_{-\infty}^{\infty} w(t-\tau) x(t)  e^{2\pi i f t} \, \mathrm{d}t.
\end{equation}
The window function $w(t-\tau)$ is centered at $t=\tau$, and one obtains
a time-frequency map by moving the window from one end of a data segment 
to the other. The window is not unique and the effectiveness of a window
depends on the signal one is looking for. Once the time-frequency map
is constructed, one can look for excess power (compared to average)
in different regions~\cite{Anderson:2000cs}, or look for certain patterns. 

The method followed depends
on the signal one is looking for. For instance, when looking for unknown
signals, all that can be done is to look for a departure from averaged
behavior in different regions of the map~\cite{Anderson:2000cs}.
However, when some knowledge of the spectral and temporal content of
the signal is known, it is possible to tune the algorithm to
improve efficiency.  The wavelet-based \emph{waveburst} algorithm is
one such example~\cite{Klimenko:2004qh} that has been applied to
search for unstructured bursts in LIGO data~\cite{Abbott:2007wu}.

One can employ strategies that improve 
detection efficiency over a simple search for excess power. For example, 
chirping signals will leave a characteristic track in the time-frequency 
plane, with increasing frequency and power as a function of time. Time
frequency map of a chirp signal buried in noisy data is shown in 
Figure~\ref{fig:tf-map}. An algorithm that optimizes the search for
specific shapes in the time-frequency plane is discussed in~\cite{Heng:2004tv}. These and other methods have been applied to
understand how to analyze LISA data~\cite{Gair:2006nj, Wen:2005xn}. 

\epubtkImage{BH-0_0-10-BH-0_9-10.png}{%
  \begin{figure}[htbp]
    \centerline{
    \includegraphics[width=2.5in]{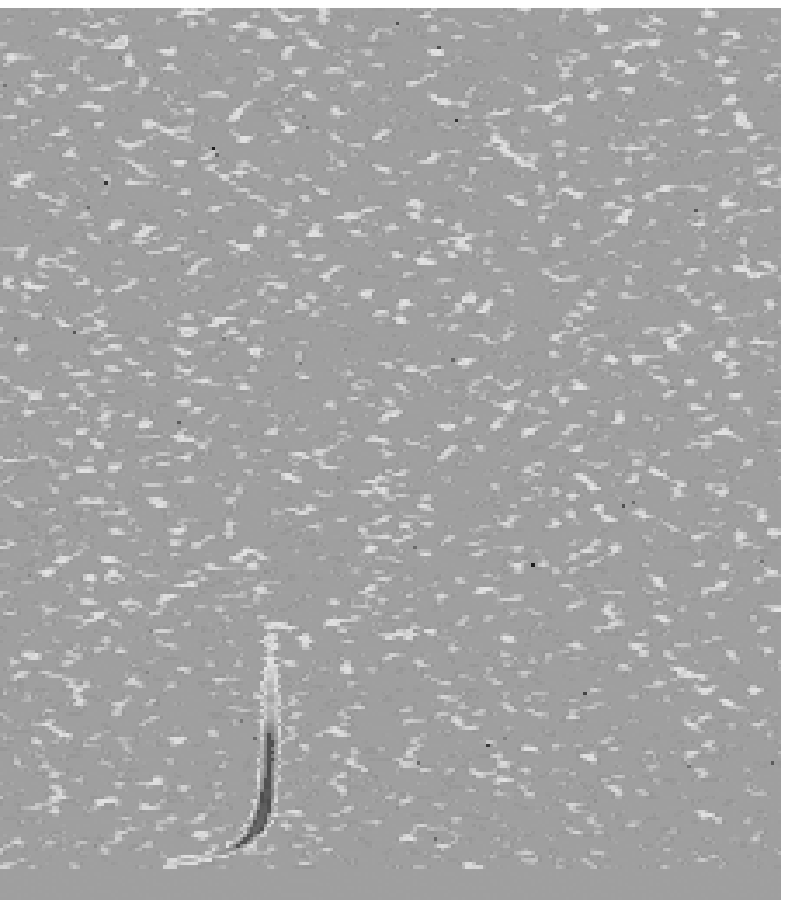}\hspace{1cm}
    \includegraphics[width=2.5in]{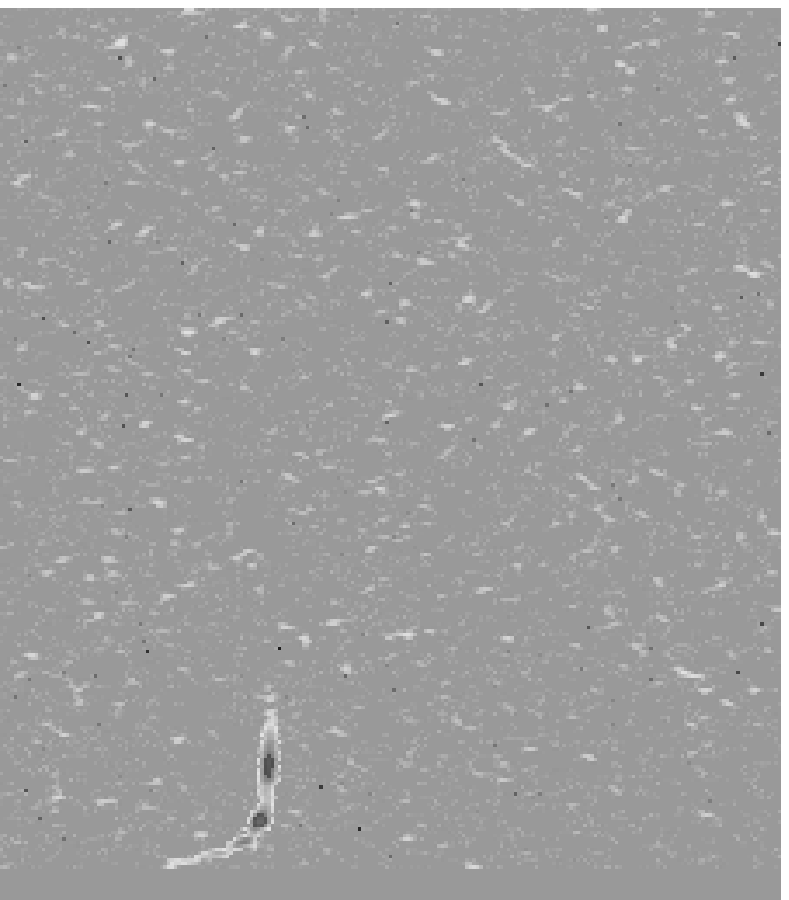}}
    \caption{Time-frequency maps showing the track left by the inspiral
    of a small black hole falling into an SMBH as
    expected in LISA data. The left panel is for a central black hole
    without spin and the right panel is for a central black hole whose
    dimensionless spin parameter is $q=0.9$.}
    \label{fig:tf-map}
\end{figure}}

More recently, there has been a lot of progress in extending burst search
algorithms for a network of detectors~\cite{Chatterji:2006nh, Klimenko:2008fu}, as well
as exploring new Bayesian-based methods to search for unknown transients~\cite{Searle:2007uv}.

\subsection{Measurement of parameters and source reconstruction}

We have so far focused on the problem of detection and have not discussed
parameter estimation in any concrete way. In principle, parameter estimation
should not be considered to be distinct from detection. From a practical 
point of view, however, certain methods might be computationally 
efficient in detecting a signal but not necessarily the best for parameter 
estimation, while the best parameter estimation methods might
not be computationally efficient.  Thus, the problem of parameter 
estimation is often treated separately from detection.  

The first thing to realize is that we can never
be absolutely certain that a signal is present in a data train~\cite{Finn:1992wt, Finn:1992xs}; we can
only give confidence levels about its presence, which could be close
to 100\% for high values of the SNR. The next thing to realize is that,
whatever the SNR may be, we cannot be absolutely certain about the
true parameters of the signal: at best we can make an estimate and these
estimates are given in a certain range. The width of the range
depends on the confidence level required, being larger for higher 
confidence levels~\cite{Finn:1992wt}.

\emph{Maximum likelihood estimates} have long been used to measure
the parameters of a known signal buried in noisy data. The method
consists in maximizing the \emph{likelihood ratio} -- the ratio of the
probability that a given signal is present in the data to the
probability that the signal is absent~\cite{cwh68, Finn:1992wt}.
Maximum likelihood estimates are not always 
minimum uncertainty estimates, as has been particularly demonstrated in
the case of binary inspiral signals by Balasubramanian, et al.~\cite{BSD1, BSD2}.  However, until recently, this is the method that 
has been very widely followed in the gravitational wave literature.
But what is important to note is that maximum likelihood estimates
are unbiased when the SNR is large\epubtkFootnote{How large the SNR should
be to presume that there is no bias in the estimation of parameters
depends on the number of parameter-space dimensions and strictly
speaking the statement is true only in the limit as 
$\mathrm{SNR}\rightarrow \infty$.}, and the mean of the distribution 
of measured values of the parameters will be centered around the
true parameter values. This is an important quality that will
be useful in our discussion below.

\emph{Bayesian estimates}, which take into account any prior 
knowledge that may be available about the distribution of the source
parameters, often give much better estimates and do not rely on 
the availability of an ensemble of detector outputs~\cite{Sivia:1996, Nicholson:1997qh}.  However, they are computationally a lot
more expensive than maximum likelihood estimates.

In any one measurement, any estimated parameters, however efficient, 
robust and accurate, are unlikely to be the actual parameters of the signal, 
since, at any finite SNR, noise alters the input signal. In the geometric 
language, the signal vector is being altered
by the noise vector and our matched filtering aims at computing the
projection of this altered vector onto the signal space. 
The true parameters are expected to lie within an ellipsoid of 
$p$ dimensions at a certain confidence level -- the volume of 
the ellipsoid increasing with the confidence level at a given
SNR but decreasing with the SNR at a given confidence level.

\subsubsection{Ambiguity function}
\label{sec:estimation}

The ambiguity function, well known in the statistical theory of
signal detection~\cite{cwh68}, is a very powerful tool in signal analysis:
it helps one to assess the number of templates required to span the parameter 
space of the signal~\cite{Sathyaprakash:1991mt}, to 
make estimates of variances and covariances involved
in the measurement of various parameters, to compute biases introduced in
using a family of templates whose shape is not the same as that of a family of
signals intended to be detected, etc. We will
see below how the ambiguity function can be used to compute the required number
of templates. Towards the end of this section we will use the ambiguity
function for the estimation of parameters.

The ambiguity function is defined (see \Eqnref{defcorr} below) as the scalar product of
two normalized waveforms maximized over the initial phase of the
waveform, in other words, the absolute value of the 
scalar product\epubtkFootnote{Working with analytic signals 
$h(t)=a(t) e^{\phi(t)+ i\phi_0}$, where $a(t)$ and $\phi(t)$
are the time-varying amplitude and phase of the signal, respectively,
we see that the initial phase $\phi_0$ of the signal simply factors
out as a constant phase in the Fourier domain and we can maximize
over this initial phase by simply taking the absolute value of the
scalar product of a template with a signal.}.
A waveform $e$ is said to be \emph{normalized} if 
$\left < e, e \right >^{1/2}  = 1$, where the inner product is inversely 
weighted by the PSD, as in \Eqnref{eq:scalar product}. Among other things, normalized 
waveforms help in defining signal strengths: a signal is said to 
be of strength $h_0$ if $h=h_0 e$. Note that the optimal SNR 
for such a signal of strength $h_0$ is, $\left < h, h \right >^{1/2} = h_0$. 

Let $e(t; { \alpha})$, where 
$ {\alpha} =\{\alpha^i | i=0,\ldots,p\}$ is the 
parameter vector comprised of $p+1$ parameters, denote a 
normalized waveform. It is conventional to choose the parameter
$\alpha^0$ to be the lag $\tau$, which simply corresponds to a
coordinate time when an event occurs and is therefore called 
an \emph{extrinsic} parameter, while the rest of the $p$ parameters
are called the \emph{intrinsic} parameters and characterize the
gravitational wave source.  

Given two normalized waveforms $e(t; {\alpha})$
and $e(t; {\beta})$, whose parameter vectors are not necessarily
the same, the ambiguity ${\cal A}$ is defined as
\begin{equation}
{\cal A} ({ \alpha}, { \beta}) \equiv 
\left | \left <e({ \alpha}), e({ \beta})\right > \right |.
\label {defcorr}
\end{equation}
Since the waveforms are normalized,
${\cal A} ({\alpha}, {\alpha}) = 1$ and
${\cal A} ({\alpha}, {\beta}) < 1$, 
if ${\alpha} \ne {\beta}$.
Here, ${\alpha}$ can be thought of
as the parameters of a (normalized) template while ${\beta}$ those of
a signal. With the template parameters ${\alpha}$ fixed,
the ambiguity function is a function of $p$ signal parameters $\beta^i$,
giving the SNR obtained by the template for different signals. 
The region in the signal parameter space for which a 
template obtains SNRs larger than a chosen value 
(called the \emph{minimal match}~\cite{owen97}) is the \emph{span}
of that template. Template families should be chosen so that altogether
they span the entire signal parameter space of interest with
the least overlap of
one other's spans. One can equally well interpret the ambiguity function as
the SNR obtained for a given signal by filters of different parameter values.

It is clear that the ambiguity function is a local maximum at the
``correct'' set of parameters, $\beta = \alpha$. Search methods that
vary $\beta$ to find the best fit to the parameter values make use of
this property in one way or another. But the ambiguity function will
usually have secondary maxima as a function of $\beta$ with fixed
$\alpha$. If these secondaries are only slightly smaller than the
primary maximum, then noise can lead to confusion: it can, at random,
sometimes elevate a secondary and suppress a primary. These can lead
to false measurements of the parameters. Search methods need to be
designed carefully to avoid this as much as possible. One way would be
to fit the known properties of the ambiguity function to an ensemble
of maxima. This would effectively average over the noise on individual
peaks and point more reliably to the correct one.

It is important to
note that in the definition of the ambiguity function there is
no need for the functional forms of the template and signal to be
the same; the definition holds true for any signal-template
pair of waveforms. Moreover, the number of template parameters need 
not be identical (and usually aren't) to the number of parameters 
characterizing the signal. For instance, a binary can be characterized
by a large number of parameters, such as the masses, spins, eccentricity 
of the orbit, etc., while we may take as a model waveform 
the one involving only the masses. 
In the context of inspiral waves, $e(t;{\beta})$ is the exact
general relativistic waveform emitted by a binary, whose form we do not
know, while the template family is a post-Newtonian, or some other,
approximation to it, that will be used to detect the true waveform. 
Another example would be signals emitted by spinning neutron stars, 
isolated or in binaries, whose 
time evolution is unknown, either because we cannot anticipate all the
physical effects that affect their spin, or because the parameter
space is so large that we cannot possibly take into account all of them
in a realistic search.  

Of course, in such cases we cannot compute the ambiguity function,
since one of the arguments to the ambiguity function is unknown. These
are, indeed, issues where substantial work is called for. What are all
the physical effects to be considered so as not to miss out a waveform
from our search? How to make a choice of templates when the functional
form of templates is different from those of signals? For this review
it suffices to assume that the signal and template waveforms are of
identical shape and the number of parameters in the two cases is the
same.

\subsubsection{Metric on the space of waveforms}
\label{sec:metric}

The computational cost of a search and the estimation of
parameters of a signal afford a lucid geometrical picture developed by
Balasubramanian et al.~\cite{BSD2} and Owen~\cite{owen97}.
Much of the discussion below is borrowed from their work.

Let $x_k$, $k=1,2,\ldots,N$, denote the discretely sampled output
of a detector. The set of all possible detector outputs satisfy 
the usual axioms of a vector space. Therefore, $x_k$ can be thought 
of as an $N$-dimensional vector. It is more convenient to work 
in the continuum limit, in which case we have infinite dimensional
vectors and the corresponding vector space.  However, all the
results are applicable to the realistic case in which
detector outputs are treated as finite dimensional vectors.
 
Amongst all vectors, of particular interest are those corresponding
to gravitational waves from a given astronomical source. While every signal
can be thought of as a vector in the infinite-dimensional vector space of
the detector outputs, the set of all such signal vectors do not, by themselves, 
form a vector space. However, the set of all normed signal vectors 
(i.e., signal vectors of unit norm) form a manifold, the parameters of 
the signal serving as a coordinate system~\cite{BSD1, BSD2, owen97, Owen:1998dk}. 
Thus, each class of an astronomical source 
forms an $n$-dimensional manifold ${\cal S}_n$, where $n$ is the number of 
independent parameters characterizing the source. For instance, the 
set of all signals from a binary on a quasi-circular orbit inclined
to the line of sight at an angle $\iota$, consisting of nonspinning 
black holes of masses $m_1$, and $m_2$, located a distance $D$ from 
the Earth\epubtkFootnote{Even though we deal with normed signals 
(which amounts to fixing $D$), astrophysical gravitational wave 
signals are characterized by this additional parameter.} 
initially in the direction $(\theta,\varphi)$ and expected 
to merge at a time $t_C$ with the phase of the signal at merger 
$\varphi_C$, forms a nine-dimensional manifold with coordinates 
$\{D,\,\theta,\, \varphi,\, m_1,\, m_2,\, t_C,\, \varphi_C,\,
\iota,\,\psi\}$, where $\psi$ is the polarization angle of the signal.
In the general case of a signal characterized by $n$ parameters we 
shall denote the parameters by $p^{\alpha}$, where $\alpha=1,\ldots,n$.

The manifold ${\cal S}_n$ can be endowed with a metric
$g_{\alpha\beta}$ that is induced by the scalar product defined in
\Eqref{eq:scalar product}. The components of the metric in a
coordinate system $p^\alpha$ are defined by\epubtkFootnote{We have
  followed the definition of the metric as is conventional in
  parameter estimation theory (see, e.g., \cite{Finn:1992wt,
    Finn:1992xs, Chernoff:1993th, BSD2}), which differs from that used
  in template placement algorithms (see, e.g., \cite{owen97}) by a
  factor of two. This difference will impact the relationship between
  the metric and the match, as will be apparent in what follows.}
\begin{equation}
g_{\alpha\beta} \equiv 
\left < \partial_\alpha \hat h,\, \partial_\beta \hat h \right >,
\ \ \ \ \partial_\alpha \hat h \equiv \frac{\partial \hat h}{\partial p^\alpha}.
\label{eq:metricDef}
\end{equation}
The metric can then be used on the signal manifold as a measure of the proper 
distance $\mathrm{d}\ell$ between nearby signals with coordinates $p^\alpha$ and 
$p^\alpha+\mathrm{d}p^\alpha$, that is signals $\hat h(p^\alpha)$ and 
$\hat h(p^\alpha+\mathrm{d}p^\alpha)$, 
\begin{equation}
\label{eq:properDistance}
\mathrm{d}\ell^2 = g_{\alpha\beta} \mathrm{d}p^\alpha \mathrm{d}p^\beta.
\end{equation}

Now, by Taylor expanding $\hat h(p^\alpha+\mathrm{d}p^\alpha)$ around $p^\alpha$,
and keeping only terms to second order in $\mathrm{d}p^\alpha$,
it is straightforward to see that the overlap $\cal O$ of two infinitesimally 
close signals can be computed using the metric:
\begin{eqnarray}
  \label{eq:overlap}
  {\cal O}(\mathrm{d}p^\alpha;\, p^\alpha) & \equiv & \left <\hat
  h(p^\alpha),\, \hat h(p^\alpha+\mathrm{d}p^\alpha) \right >\nonumber
  \\
  & = & 1 - \tfrac{1}{2} g_{\alpha\beta} \mathrm{d}p^\alpha \mathrm{d}p^\beta.
\end{eqnarray}

The metric on the signal manifold is nothing but the well-known
Fisher information matrix usually denoted $\Gamma_{\alpha\beta}$,
(see, e.g.,~\cite{cwh68,PBCV04})
but scaled down by the square of the SNR, i.e.,
$g_{\alpha\beta} = \rho^{-2}\Gamma_{\alpha\beta}$.
The information matrix is itself the inverse of the covariance 
matrix $C_{\alpha\beta}$ and is a very useful quantity in signal analysis.

\subsubsection{Covariance matrix}

Having defined the metric, we next consider the application of the
geometric formalism in the estimation of statistical errors involved
in the measurement of the parameters. We closely follow the notation
of Finn and Chernoff~\cite{Finn:1992wt, Finn:1992xs, Chernoff:1993th}.

Let us suppose a signal of known shape with parameters $p^\alpha$ is
buried in background noise that is Gaussian and stationary. Since the
signal shape is known, one can use matched filtering to dig the signal
out of the noise. The measured parameters $\pbar^\alpha$ will, in
general, differ from the true parameters of the
signal\epubtkFootnote{In what follows we shall use an over-line to
  distinguish the measured parameters from the true parameters
  $p^\alpha$.}. Geometrically speaking, the noise vector displaces the
signal vector and the process of matched filtering projects the (noise
+ signal) vector back on to the signal manifold. Thus, any nonzero
noise will make it impossible to measure the true parameters of the
signal. The best one can hope for is a proper statistical estimation
of the influence of noise.

The posterior probability density function ${\cal P}$ of 
the parameters $\pbar^\alpha$ is given by a multivariate Gaussian 
distribution\epubtkFootnote{A Bayesian interpretation of 
${\cal P}(\Delta p^\alpha)$ is the probability of having the 
true signal parameters lie somewhere inside the ellipsoidal 
volume centered at the Maximum Likelihood point $\pbar^\alpha$.}:
\begin{equation}
\label{eq:multivariatePDF}
{\cal	P}(\Delta p^\alpha)\, \mathrm{d}^n\Delta p = 
\frac{\mathrm{d}^n\Delta p}{(2\pi)^{n/2}\sqrt{C}} 
\exp \left [-\frac{1}{2}C^{-1}_{\alpha\beta}\, \Delta p^\alpha \Delta \, p^\beta \right ],
\end{equation}
where $n$ is the number of parameters, $\Delta p^\alpha = p^\alpha -
\pbar^\alpha$, and $C_{\alpha\beta}$ is the covariance matrix, $C$
being its determinant. Noting that $C^{-1}_{\alpha\beta}=\rho^2
g_{\alpha\beta}$, we can rewrite the above distribution as
\begin{equation}
\label{eq:multivariatePDFinMetric}
{\cal	P}(\Delta p^\alpha)\, \mathrm{d}^n\Delta p = 
\frac{\rho^n\, \sqrt{g}\, \mathrm{d}^n\Delta p}{(2\pi)^{n/2}} 
\exp \left [-\frac{\rho^2}{2}\, g_{\alpha\beta}\, 
\Delta p^\alpha \Delta \, p^\beta \right ],
\end{equation}
where we have used the fact that $C=1/(\rho^{2n}\, g)$, $g$ being the
determinant of the metric $g_{\alpha\beta}$.  Note that if we define
new parameters $p'^\alpha = \rho p^\alpha$, then we have exactly the
\emph{same} distribution function for all SNRs, except that the
deviations $\Delta p^\alpha$ are scaled by $\rho$.

Let us first specialize to one dimension to illustrate the region of
the parameter space with which one should associate an event at a
given confidence level. In one dimension the distribution of the
deviation from the mean of the measured value of the parameter $p$ is
given by
\begin{equation}
\label{eq:prob}
{\cal P}(\Delta p) \mathrm{d}\Delta p 
= \frac{\mathrm{d}\Delta p}{\sqrt{2\pi}\sigma} \exp\left (-\frac{\Delta p^2}{2\sigma^2}\right )
= \frac{\rho\,\sqrt{g_{pp}} \mathrm{d}\Delta p}{\sqrt{2\pi} } 
\exp\left (-\frac{\rho^2}{2}{g_{pp}\Delta p^2} \right ),
\end{equation}
where, analogous to the $n$-dimensional case, we have used $\sigma^2=1/(\rho^2g_{pp})$.
Now, at a given SNR, what is the volume $V_P$ in the parameter space, 
such that the probability of finding the measured parameters $\pbar$ inside this 
volume is $P?$ This volume is defined by
\begin{equation}
P = \int_{\Delta p \in V_P}{\cal P}(\Delta p) \mathrm{d}\Delta p.
\label{eq:volumeInsideP}
\end{equation}
Although $V_P$ is not unique, it is customary to choose it to be centered around $\Delta p=0$:
\begin{equation}
P = \int_{(\Delta p/\sigma )^2 \le r^2(P)} \frac{\mathrm{d}\Delta p}{\sqrt{2\pi} \sigma} 
\exp\left (-\frac{\Delta p^2}{2\sigma^2} \right ) 
= \int_{\rho^2 g_{pp}\Delta p^2 \le r^2(P)}\frac{\rho\, \sqrt{g_{pp}} \mathrm{d}\Delta p}{\sqrt{2\pi}} 
\exp\left (-\frac{\rho^2\,g_{pp}\Delta p^2}{2} \right ),
\end{equation}
where, given $P$, the above equation can be used to solve for $r(P)$ and it
determines the range of integration: $-r\sigma\le \Delta p \le r\sigma$. 
For instance, the volumes $V_P$ corresponding to $P\simeq 0.683, 0.954, 0.997,\ldots$, are the 
familiar intervals $[-\sigma,\, \sigma]$, $[-2\sigma,\, 2\sigma]$, $[-3\sigma,\, 3\sigma]$, $\ldots$,
and the corresponding values of $r$ are 1, 2, 3.
Since $\sigma=1/\sqrt{\rho^2 g_{pp}}$, we see that in terms of $g_{pp}$ the above intervals
translate to 
\begin{equation}
\frac{1}{\rho} \left[-\frac{1}{\sqrt{g_{pp}}},\, \frac{1}{\sqrt{g_{pp}}}\right ],\,\,
\frac{1}{\rho} \left[-\frac{2}{\sqrt{g_{pp}}},\, \frac{2}{\sqrt{g_{pp}}}\right ],\,\,
\frac{1}{\rho} \left[-\frac{3}{\sqrt{g_{pp}}},\, \frac{3}{\sqrt{g_{pp}}}\right ],\ldots.
\end{equation}
Thus, for a given probability $P$, the volume $V_P$ shrinks as $1/\rho$. The maximum
distance $d_{\mathrm{max}}$ within which we can expect to find ``triggers'' 
at a given $P$ depends inversely on the SNR $\rho$:
$d{\ell} = \sqrt{g_{pp}\Delta p^2} = r/\rho$.  Therefore, for $P\simeq 0.954$, 
$r=2$ and at an SNR of $5$ the maximum distance is 0.4, which
corresponds to a match of $\epsilon=1- \tfrac{1}{2}d{\ell}^2
= 0.92$. In other words, in one dimension 95\% of the time
we expect our triggers to come from templates that have an overlap  
greater than or equal to 0.92 with the buried signal when the SNR is five.
This interpretation in terms of the match is a good approximation as long as
$d{\ell} \ll 1$, which will be true for large SNR events. However, for
weaker signals and/or greater values of $P$ we can't interpret the results
in terms of the match, although Equation~(\ref{eq:volumeInsideP}) can be
used to determine $r(P)$. As an example, at $P\simeq 0.997$, $r=3$ and at an
SNR of $\rho=4$, the maximum distance is $d\ell=0.75$
and the match is $\epsilon=23/32\simeq 0.72$, which is significantly smaller than one
and the quadratic approximation is not good enough to compute the match.

These results generalize to $n$ dimensions. In $n$-dimensions the volume $V_P$ 
is defined by
\begin{equation}
\label{eq:VolumeDefiningEqn}
P = \int_{\Delta p^\alpha \in V_P} {\cal P}(\Delta p^\alpha)\, 
\mathrm{d}^n\Delta p. 
\end{equation}
Again, $V_P$ is not unique but it is customary to center the volume around the point 
$\Delta p^\alpha=0$: 
\begin{equation}
P = \int_{\rho^2 g_{\alpha\beta}\, \Delta p^\alpha \Delta \, p^\beta \le r^2(P,n)} 
\frac{\rho^{n}\, \sqrt{g}\, \mathrm{d}^{n}\Delta p}{(2\pi)^{n/2}} 
\exp \left [-\frac{\rho^2}{2}\, g_{\alpha\beta}\, 
\Delta p^\alpha \Delta \, p^\beta \right ].
\label{eq:Vp}
\end{equation}
Given $P$ and the parameter space dimension $n$, one can iteratively
solve the above equation for $r(P,n)$. The volume $V_P$ is the surface
defined by the equation
\begin{equation}
g_{\alpha\beta} \Delta p^\alpha \Delta p^\beta = 
\left( \frac{r}{\rho} \right)^2.
\label{eq:scaledVolume}
\end{equation}
This is the equation of an $n$-dimensional ellipsoid  
whose size is defined by $r/\rho$. For a given $r$ (which
determines the confidence level), the size of the ellipsoid
is inversely proportional to the SNR, the volume decreasing
as $\rho^n$. However, the size is not small enough for all 
combinations of $P$ and $\rho$ to interpret the distance 
from the center of the ellipsoid to its surface, in terms of 
the overlap or match of the signals at the two
locations, except when the distance is close to zero. This is because the
expression for the match in terms of the metric is based on the quadratic
approximation, which breaks down when the matches are small. However, the 
region defined by Equation~(\ref{eq:scaledVolume})
always corresponds to the probability $P$ and there is no approximation
here (except that the detector noise is Gaussian). 

When the SNR $\rho$ is large and $1-P$ is not close to zero, the triggers 
are found from the signal with matches greater than or equal to
$1-\tfrac{r^2(P,n)}{2\rho^2}$. Table~\ref{table:one} lists the value of 
$r$ for several values of $P$ in one, two and three-dimensions and 
the minimum match $\epsilon_{\mathrm{MM}}$ for SNRs 5, 10 and 20.

\begin{table}
  \caption{The value of the (squared) distance $d{\ell}^2=r^2/\rho^2$
  for several values of $P$ and the corresponding smallest match that
  can be expected between templates and the signal at different values
  of the SNR.}
  \label{table:one}
  \vskip 4mm

  \centering
  \begin{tabular}{c|cccccccc}
    \toprule
    & \multicolumn{2}{c}{$P=0.683$}& \vline
    & \multicolumn{2}{c}{$P=0.954$} & \vline
    & \multicolumn{2}{c}{$P=0.997$} \\
    \midrule
    $\rho$ & $d\ell^2$ & $\epsilon_{\mathrm{MM}}$ & \vline & $d\ell^2$ & $\epsilon_{\mathrm{MM}}$ & \vline &
    $d\ell^2$ & $\epsilon_{\mathrm{MM}}$ \\
    \midrule
    \multicolumn{8}{c}{$n=1$} \\
    \midrule
    5      & 0.04      & 0.9899  &\vline & 0.16      & 0.9592  &\vline & 0.36      & 0.9055  \\
    10     & 0.01      & 0.9975  &\vline & 0.04      & 0.9899  &\vline & 0.09      & 0.9772  \\
    20     & 0.0025    & 0.9994  &\vline & 0.01      & 0.9975  &\vline & 0.0225    & 0.9944  \\
    \midrule
    \multicolumn{8}{c}{$n=2$} \\
    \midrule
    5      & 0.092      & 0.9767  &\vline & 0.2470      & 0.9362  &\vline & 0.4800      & 0.8718 \\
    10     & 0.023      & 0.9942  &\vline & 0.0618      & 0.9844  &\vline & 0.1200      & 0.9695 \\
    20     & 0.00575    & 0.9986  &\vline & 0.0154      & 0.9961  &\vline & 0.0300      & 0.9925 \\
    \midrule
    \multicolumn{8}{c}{$n=3$} \\
    \midrule
    5      & 0.1412     & 0.9641  &\vline & 0.32      & 0.9165  &\vline & 0.568      & 0.8462 \\
    10     & 0.0353     & 0.9911  &\vline & 0.08      & 0.9798  &\vline & 0.142      & 0.9638 \\
    20     & 0.00883    & 0.9978  &\vline & 0.02      & 0.9950  &\vline & 0.0355     & 0.9911 \\
    \bottomrule  
  \end{tabular}
\end{table}

Table~\ref{table:one} should be interpreted in light of the fact that
triggers come from an analysis pipeline in which the templates are laid
out with a certain minimal match and one cannot, therefore,
expect the triggers from different detectors to be matched better than the
minimal match. 

From Table~\ref{table:one}, we see that, when the SNR is large (say 
greater than about 10), the dependence of the match $\epsilon_{\mathrm{MM}}$ on 
$n$ is very weak; in other words, irrespective of the number of dimensions, 
we expect the match between the trigger and the true signal (and for our 
purposes the match between triggers from different instruments) to be 
pretty close to 1, and mostly larger than a minimal match of about 0.95 
that is typically used in a search. Even when the SNR is in the region of 
5, for low $P$ again there is a weak dependence 
of $\epsilon_{\mathrm{MM}}$ on the number of parameters. For large $P$ and low
SNR, however, the dependence of $\epsilon_{\mathrm{MM}}$ on the number of 
dimensions becomes important.  At an SNR of $5$ and $P\simeq 0.997$, 
$\epsilon_{\mathrm{MM}}=0.91, 0.87, 0.85$ for $n=1, 2, 3$ dimensions, respectively.

Bounds on the estimation computed using the covariance matrix are called
Cram\'er--Rao bounds. Cram\'er--Rao bounds
are based on local analysis and do not take into consideration
the effect of distant points in the parameter space on the errors computed
at a given point, such as the secondary maxima in the likelihood.
Though the Cram\'er--Rao bounds are in disagreement with
maximum likelihood estimates, global analysis, taking the effect of distant
points on the estimation of parameters, does indeed give results in
agreement with maximum likelihood estimation as shown by
Balasubramanian and Dhurandhar~\cite{Balasubramanian:1997qz}.

\subsubsection{Bayesian inference}

A good example of an efficient detection algorithm that is not a reliable
estimator is the time-frequency transform of a chirp. For signals that are
loud enough, a time-frequency transform of the data would be a very
effective way of detecting the signal, but the transform contains hardly
any information about the masses, spins and other information about the
source. This is because the time-frequency transform of a chirp is a
mapping from the multi-dimensional (17 in the most general case) space 
of chirps to just the two-dimensional space of time and frequency.
Even matched filtering, which would use templates that are defined
on the full parameter space of the signal, would not give the parameters 
at the expected accuracy. This is because the templates are defined
only at a certain minimal match and might not resolve the signal well 
enough, especially for signals that have a high SNR. 

In recent times Bayesian inference techniques have been applied with
success in many areas in astronomy and cosmology. These techniques
are probably the most sensible way of estimating the parameters, and the
associated errors, but cannot be used to efficiently search for signals.
Bayesian inference is among the simplest of statistical measures to
state, but is not easy to compute and is often subject to controversies.
Here we shall only discuss the basic tenets of the method and refer
the reader for details to an excellent treatise on the subject (see,
e.g., Sivia~\cite{Sivia:1996}).

To understand the chief ideas behind Bayesian inference, let us begin 
with some basic concepts in probability theory. Given two hypothesis
or statements $A$ and $B$ about an observation, let $P(A,B)$ denote 
the joint probability of $A$ and $B$ being true.  For the sake of clarity,
let $A$ denote a statement about the universe and $B$ some observation
that has been made. Now, the joint
probability can be expressed in terms of the individual probability densities
$P(A)$ and $P(B)$ and conditional probability densities $P(A|B)$
and $P(B|A)$ as follows:
\begin{equation}
P(A,B) = P(A)P(B|A) \qquad \mathrm{or} \qquad  P(A,B) = P(B)P(A|B).
\label{eq:Bayes0}
\end{equation}
The first of these equations says \emph{the joint probability of $A$ and $B$
both being true is the probability that $A$ is true times the probability
that $B$ is true given that $A$ is true} and similarly the second.  
We can use the above equations to arrive at Bayes theorem:
\begin{equation}
P(B)P(A|B) = P(A)P(B|A) \qquad \mathrm{or} \qquad  
P(A|B) = \frac{P(A)P(B|A)}{P(B)}. 
\label{eq:Bayes1}
\end{equation}
The left-hand side of the above equation can be interpreted as a statement
about $A$ (the universe) given $B$ (the data). This is the \emph{posterior}
probability density. The right-hand side contains $P(B|A)$, which is the
probability that $B$ is obtained given that $A$ is true and is called the 
\emph{likelihood}, $P(A)$, which is the probability of $A$, called the
\emph{prior} probability of $A$, and $P(B)$ (the prior of $B$),
which is simply a normalizing constant often ignored in Bayesian analysis.

For instance, if $A$ denotes the statement \emph{it is going to rain} and 
$B$ \emph{the amount of humidity in the air} then the above equation gives
us the posterior probability that it rains when the air contains a certain 
amount of humidity. Clearly, the posterior depends on what is the likelihood of
the air having a certain humidity when it rains and the prior probability 
of rain on a given day. If the prior is very small (as it would be in a 
desert, for example) then you would need a rather large likelihood for 
the posterior to be large. Even when the prior is not so small, 
say a 50\% chance of rain on any given day (as it would be if you are 
in Wales), the likelihood has to be large for posterior probability to say  
something about the relationship between the level of humidity and the 
chance of rain.

As another example, and more relevant to the subject of this
review, let $s$ be the statement \emph{the data contains a chirp} (signal), 
$n$ the statement \emph{the data contains an instrumental transient}, (noise),
and let $t$ be a test that is performed to infer which of the
two statements above are true. Let us suppose $t$ is a very good test,
in that it discriminates between $s$ and $n$ very well, and say the
detection probability is as high as $P(t|s)=0.95$ with a low
false alarm rate $P(t|n)=0.05$ (note that $P(t|s)$ and $P(t|n)$ need not
necessarily add up to 1). Also, the expected event rate of a 
chirp during our observation is low, $P(s)=10^{-5}$, but the chance 
of an instrumental transient is relatively large, $P(n)=0.01$. 
We are interested in knowing what the posterior probability of 
the data containing a chirp is, given that the test has been passed. 
By Bayes theorem this is
\begin{equation}
P(s|t) = \frac{P(t|s)P(s)}{P(t)}= \frac{P(t|s)P(s)} {P(t|s)P(s)+P(t|n)P(n)},
\label{eq:Bayes2}
\end{equation}
where $P(t)$ (the probability of the test being positive) is taken to
result from either the chirp or the instrumental transient.
Substituting for various quantities in the above equation we find
\begin{equation}
P(s|t) = \frac{0.95 \times 10^{-5}} {0.95 \times 10^{-5} + 0.05 \times 0.01}
\simeq 0.02.
\label{eq:BayesExample}
\end{equation}
There is only a 2\% chance that the data really contains a chirp when the
test was taken. On the contrary, for the same data we find that the
chance of an instrumental transient for a positive test result is
$P(n|t)\sim 98\%$. Thus, though there is a high (low) probability for
the test to be positive in the presence of a signal (noise) when the
test is indeed positive, we cannot necessarily conclude that a signal 
is present.
This is not surprising since the prior probability of the signal being
present is very low. The situation can be remedied by designing a test
that gives a far lower probability for the test to give a positive
result in the case of an instrumental transient (i.e., a very low false alarm
rate).

Thus, Bayesian inference neatly folds the prior knowledge about 
sources in the estimation process. One might worry
that the outcome of a measurement process would be seriously biased by
our preconception of the prior. To understand this better, let us rewrite
Equation~(\ref{eq:Bayes2}) as follows:
\begin{equation}
P(s|t) = \frac{1} {1 + P(t|n)P(n)/P(t|s)P(s)} 
= \frac{1} {1 + L_{\mathrm{NS}}p_{\mathrm{SN}}}, 
\label{eq:Bayes3}
\end{equation}
where $L_{\mathrm{NS}} = P(t|n)/P(t|s)$ is the ratio of the two likelihoods
and $p_{\mathrm{SN}} = P(s)/P(n)$ is the ratio of the priors.
The latter is not in the hands of a data analyst; it is determined 
by the nature of the source being searched for and the property 
of the instrument.  The only way an analyst can make the posterior probability large
is by choosing a test that gives a small value for the ratio of the
two likelihoods. When $L_{\mathrm{NS}} \ll p_{\mathrm{SN}}$ (i.e., the likelihood of
the test being positive when the signal is present is far larger, depending
on the priors, than when the transient is present) the posterior will be
close to unity. 

The above example tells us why we have to work with 
unusually-small false-alarm probability in the case of gravitational wave
searches. For instance, to search for binary coalescences in ground-based 
detectors we use a (power) 
SNR threshold of about 30 to 50. This is because the 
expected event rate is about 0.04 per year. 

Computing the posterior involves multi-dimensional integrals and these are 
highly expensive computationally, when the number of parameters involved
is large.  This is why it is often not possible to apply 
Bayesian techniques to continuously streaming data; it would be sensible
to reserve the application of Bayesian inference only for candidate events
that are selected from inexpensive analysis techniques. 
Thus, although Bayesian analysis is not used in current detection pipelines, 
there has been a lot of effort in evaluating its ability to search
for~\cite{Christensen:2004jm, Stroeer:2006ye, Cornish:2006dt, Cornish:2007if} and measure the 
parameters of~\cite{Christensen:1998gf, Cornish:2006ry, Veitch:2008wd} a signal and in
follow-up analysis~\cite{Veitch:2008ur}.

%==================================================================
\newpage

\section{Physics with Gravitational Waves}
\label{sec:gwphysics}

%% \blue{Discuss the covariance between the amplitude
%% and angular measurements: especially important for
%% LISA.}

Classical general relativity has passed all possible experimental
and observational tests so far. The theory is elegant, self-consistent
and mathematically complete  (i.e., its equations are, in principle,
solvable). However,  theorists are uncomfortable with general
relativity because it has so far eluded all efforts of quantization, making it
a unique modern theory, whose quantum mechanical analogue is unknown.
Although general relativity arises as a by-product
in certain string theories, the physical relevance of such theories is
unclear. Therefore, it has been proposed that general relativity is 
a low-energy limit of a more general theory, which in itself is amenable
to both quantization and unification. There are also other theoretical
motivations to look for modifications of general relativity 
or new theories of gravity. While there are some alternative
candidates (including the Brans--Dicke theory), none has predictions
that contradict general relativistic predictions in linear and mildly
nonlinear gravitational fields. More precisely, the extra parameters of
these other theories of gravity are constrained by the present experimental
and astronomical observations, however, they are expected to significantly
deviate from general relativistic predictions under conditions of 
strong gravitational fields.

Gravitational wave observations provide a unique opportunity to test
strongly nonlinear and highly relativistic gravity and hence provide 
an unprecedented testbed for confronting different theories of gravity.
Every nonlinear gravitational effect in general relativity will have
a counterpart in alternative theories and therefore a measurement of 
such an effect would provide an opportunity to compare the performance
of general relativity with its competitors. Indeed, a single measurement
of the full polarization of an incident gravitational wave can potentially
rule out general relativity.  This is a field that  would benefit from 
an in-depth study. What we are lacking is a systematic study of higher-order 
post-Newtonian effects in alternative theories of gravity. For instance,
we do not know how tails of gravitational waves or tails of tails would
appear in any theory other than general relativity. 

In what follows we present strong field tests of general 
relativity afforded by future gravitational wave observations. 
We will begin with observations of single black holes followed 
by black hole binaries (more generally, coalescing binaries of 
compact objects).

%% In this Section we will review the major predictions that have 
%% beem made about likely gravitational wave sources, and we will 
%% examine the potential science that can be extracted by careful 
%% interpretation of the observations.  The discussions of this 
%% Section are put into the context of the sensitivity of the 
%% detectors by Figure~\ref{fig:noise-curves}.
%% \epubtkImage{}{%
%% \centerline{\includegraphics[width=4.5in]{gw_int_all}}
%% \caption{Sensitivity to sources of gravitational wave interferometers in
%% space- and Earth-based detectors across
%% 10 decades of frequency.  The solid curves show the current and
%% planned sensitivities of 3 representative interferometers, initial
%% LIGO, advanced LIGO and EGO.  The vertical axis is the strain 
%% spectral density.  The strength of the signals 
%% expected from coalescing systems are also shown as spectral densities, in 
%% such as way that the area between a signal curve and an instrumental 
%% noise curve indicates the signal-to-noise ratio that could be 
%% achieved with perfect matched filtering.}
%% \label{fig:sens}
%% \end{figure}

\subsection{Speed of gravitational waves}

Association of a gravitational wave event with an electromagnetic
event, such as the observation of a gamma or X-ray burst coincidentally
with a gravitational wave event, would help to
deduce the speed of gravitational waves to a phenomenal accuracy.
The best candidate sources for the simultaneous observations of both are
the well-known extra-galactic gamma-ray bursts (GRBs).  Depending 
on the model that produces the GRB, the delay between the emission of
a GRB and gravitational waves might be either a
fraction of a second (as in GRBs generated by internal shocks in
a fireball~\cite{Rees:1994nw}) or 100's of seconds (as in 
GRBs generated when the fireball is incident on an external
medium~\cite{MeszarosAndRees93}). It is unlikely that high-redshift
gamma-ray observations will be visible in the gravitational wave band,
since the amplitude of gravitational waves might be rather low.
However, advanced detectors might see occasional low-redshift events,
especially if the GRB is caused by black-hole--neutron-star mergers.
Third generation detectors would be sensitive to such events
up to $z=2$. A single unambiguous association can verify the
speed of gravitational waves relative to light to a fantastic precision.

For instance, even a day's delay in the arrival times of 
gravitational and electromagnetic radiation from a source at a 
distance of one giga light year (distance to a low-redshift
GRB detectable by advanced detectors) would determine 
the relative speeds to better than one part in $10^{11}$  
$(1\mathrm{\ day}/10^{9}\mathrm{\ yr} \sim 3 \times
10^{-12})$. Coincident detection of GRBs and gravitational waves would
require good timing accuracy to determine the direction of the source
so that astronomical observations of associated gamma rays (and
afterglows in other spectral bands) can be made. Consequently,
gravitational wave antennas around the globe will have to make a
coincident detection of the event.

If the speed of gravitational waves is less than that of light, then this 
could indicate that the graviton has an effective nonzero mass. 

This would have other observable effects, in particular dispersion;
different frequencies should move at different
speeds. Will~\cite{bss:will.98} pointed out that LISA's observations
of coalescences of SMBHs at high redshifts will place extremely tight
constraints on dispersion, and may, therefore, indirectly set the best
available limits on the speed of gravitational waves. This and other
bounds on the graviton mass are discussed in
Section~\ref{sec:gravitonmass}.

\subsection{Polarization of gravitational waves}

As noted in Section~\ref{sec:gwobservables}, in Einstein's theory
gravitational waves have two independent polarizations, usually 
denoted $h_+$ and $h_\times$~\cite{MTW}.  A general wave will be a 
linear combination of both. Rotating sources typically emit both polarizations 
with a phase delay between them, leading to elliptical polarization 
patterns.  Depending on the nature of the source such polarizations can 
be detected either with a single detector (in the case of continuous 
wave sources) or with a network of detectors (in the case of burst sources).

While Einstein's general relativity predicts
only two independent polarizations, there are other theories of gravitation
in which there are additional states of polarization.  For instance, in
Fierze--Jordan--Brans--Dicke theory~\cite{Living:Will} there are 
four polarization degrees
of freedom more than in Einstein's theory. Therefore, an 
unambiguous determination of the polarization of the waves will be 
of fundamental importance. 

In the case of a burst source, to determine two polarization states, source
direction and amplitude requires three detectors, observing other polarizations
would require the use of more than three detectors (see, for example, 
Will~\cite{Living:Will}). The scalar polarization mode of Brans--Dicke, for 
example, expands a transverse ring of test particles without changing its shape. 
This is the \emph{breathing mode}, or \emph{monopole} polarization. 
If such a wave is incident from above on an interferometer, it will not register 
at all. But if it comes in along one of the arms, then, since it acts transversely, 
it will affect only the other arm and leave a signal. If the wave is seen with 
enough detectors, then it is possible to determine that it has scalar polarization.
Note that a measurement such as this 
can make a qualitative change in physics: \emph{a single
measurement could put general relativity in jeopardy}.

Polarization measurements have an important application in astronomy.
The polarization of the waves contains orientation information. 
For example, a binary system emits purely circular polarization 
along the angular momentum axis, but purely linear polarization in its equatorial 
plane. By measuring the polarization of waves from a binary (or 
from a spinning neutron star) one can determine the orientation and 
inclination of its spin axis.  This is a piece of information that is 
usually very hard to extract from optical observations.
We will return to this discussion in Section~\ref{sec:inclination}.

\subsection{Gravitational radiation reaction}

In 1974, Hulse and Taylor discovered the first double neutron star 
binary PSR~B1913+16, a system in which the emission of gravitational 
radiation has an observable effect~\cite{HulseNobel, TaylorNobel}.  
General relativity predicts
that the loss of energy and angular momentum due to the emission 
of gravitational waves should cause the period of the system to
decrease and, by carefully monitoring the orbital period of the binary,
that it would be possible to measure the rate at which the period changes.
The rate at which the period decays can be computed
using the quadrupole formula for the luminosity of the emitted
radiation combined with the energy-balance equation; namely that
the energy carried away by the waves comes at the expense of the
binding energy of the system. 

For a binary consisting of 
stars of masses $m_1$ and $m_2$, in an orbit of eccentricity 
$e$ and period $P_b$, the period decay is given by the generalization 
of \Eqnref{eqn:circularPM}~\cite{Peters:1963ux}:
\begin{equation}
\dot {P_b} = -\frac{192\pi}{5} \left ( \frac{2\pi \cal M}{P_b} \right )^{5/3}
\left ( 1 + \frac{73}{24}e^2 + \frac{37}{96}e^4 \right ) \left ( 1 -e^2 \right )^{-7/2},
\label{fig:period-decay}
\end{equation}
where we recall that 
\begin{equation}\label{eqn:chirpmass}
{\cal M} = \left(m_1 m_2\right)^{3/5} \left (m_1 + m_2 \right )^{-1/5} = \mu^{3/5}M^{2/5}
\end{equation}
is the \emph{chirpmass} of the binary that we defined in 
\Eqnref{eqn:chirpmassdef}. (In the third expression here, $\mu$ is the 
reduced mass of the binary and $M$ its total mass.) Since the masses of the 
binary and the eccentricity of the orbit can be measured by other
means, one can use these parameters in the above equation to infer
the rate at which the period is predicted to decrease according to
general relativity. For the Hulse--Taylor binary the relevant values are:
$m_1 = 1.4414\,M_\odot, m_2 = 1.3867\,M_\odot, e = 0.6171338, P_b = 2.790698 \times 10^{4}\mathrm{\ s}$. 
The predicted value $\dot {P_b}^{\mathrm{GR}} = -(2.40242 \pm 0.00002) \times 10^{-12}$,
while the observed period decay (after subtracting the apparent decay due to the
acceleration of the pulsar in the gravitational field of our galaxy, as 
described in Section~\ref{sec:binpsr}) 
is $\dot{P_b}^{\mathrm{Obs}} =-(2.4056 \pm 0.0051) \times 10^{-12}$ and the
two are in agreement to better than a tenth of a percent~\cite{Living:Will}.

Observation of the decay of the orbital period in PSR~B1913+16 is an 
unambiguous direct observation of the effect of gravitational radiation backreaction 
on the dynamics of the system.  PSR B1913+16 was the first system in which the 
effect of gravitational radiation reaction force was measured.
In 2004, a new binary pulsar PSR~J0737-3039 was discovered~\cite{Burgay:2003jj, DoublePulsar}. 
J0737 is in a tighter orbit than PSR~B1913+16; with an orbital period of only
2.4~hrs, the orbit is shrinking by about 7~mm each day in good agreement
with the general relativistic prediction. Several other systems are also 
known~\cite{Living:Lorimer}. In Sections~\ref{sec:two body problem},
\ref{sec:numerical relativity} and \ref{sec:PNA} we will discuss in
some detail the dynamics of relativistic binaries and the radiation
reaction as predicted by post-Newtonian theory and numerical
relativity simulations.

\subsection{Black hole spectroscopy}
\label{sec:black hole spectroscopy}

An important question relating to the structure of a black hole is its
stability. Studies that began in the 1970s~\cite{ReggeWheeler1957,
  bss:vishu, vishu:1970, Zerilli:1970, bss:press, Teukolsky:1972my,
  Teukolsky:1973ha, Press:1973zz} showed that a black hole is stable
under external perturbation. A formalism was developed to study how a
black hole responds to generic external perturbations, which has come
to be known as \emph{black hole perturbation
  theory}~\cite{ChandraBook}. What we now know is that a distorted
Kerr black hole relaxes to its axisymmetric state by partially
emitting the energy in the distortion as gravitational radiation. The
radiation consists of a superposition of QNMs, whose frequency and
damping time depend \emph{uniquely} on the mass $M$ and spin angular
momentum $J$ of the parent black hole and not on the nature of the
external perturbation. The amplitudes and damping times of different
modes, however, are determined by the details of the perturbation and
are not easy to calculate, except in some simple cases.

The uniqueness of the QNMs is related to the ``no-hair'' theorem of
general relativity according to which a black hole is completely
specified by its mass and spin\epubtkFootnote{A black hole can, in principle,
carry an electric charge in addition to mass and spin angular momentum.
However, astrophysical black holes are believed to be electrically neutral}. 
Thus, observing QNMs would not only
confirm the source to be a black hole, but would be an unambiguous
proof of the uniqueness theorem of general relativity. 

The end state of a black hole binary will lead to the formation of
a single black hole, which is initially highly distorted. Therefore,
one can expect coalescing black holes to end their lives with the
emission of QNM radiation, often called \emph{ringdown} radiation. 
It was realized quite early on~\cite{Flanagan1998} that the 
energy emitted during the ringdown phase of a black-hole--binary 
coalescence could be pretty large. Although, the initial quantitative 
estimates~\cite{Flanagan1998} have proven to be rather high, the 
qualitative nature of the prediction has proven to be correct. 
Indeed, numerical relativity simulations show that about 1--2\% 
of a binary's total mass would be emitted in QNMs~\cite{Pretorius05}. 
The effective one-body (EOB) model~\cite{BuonD98, BuonD00}, the only
analytical treatment of the merger dynamics, gives the energy
in the ringdown radiation to be about 0.7\% of the total mass,
consistent with numerical results. Thus, it is safe to expect 
that the ringdown will be as luminous an event as the inspiral 
and the merger phases.  The fact that QNMs can be 
used to test the no-hair theorem puts a great emphasis on 
understanding their properties, especially the frequencies, 
damping times and relative amplitudes of the different modes 
that will be excited during the merger of a black hole binary 
and how accurately they can be measured.

\epubtkImage{fmodes-Qmodes.png}{%
  \begin{figure}[htbp]
    \centerline{
      \includegraphics[width=2.25in,angle=-90]{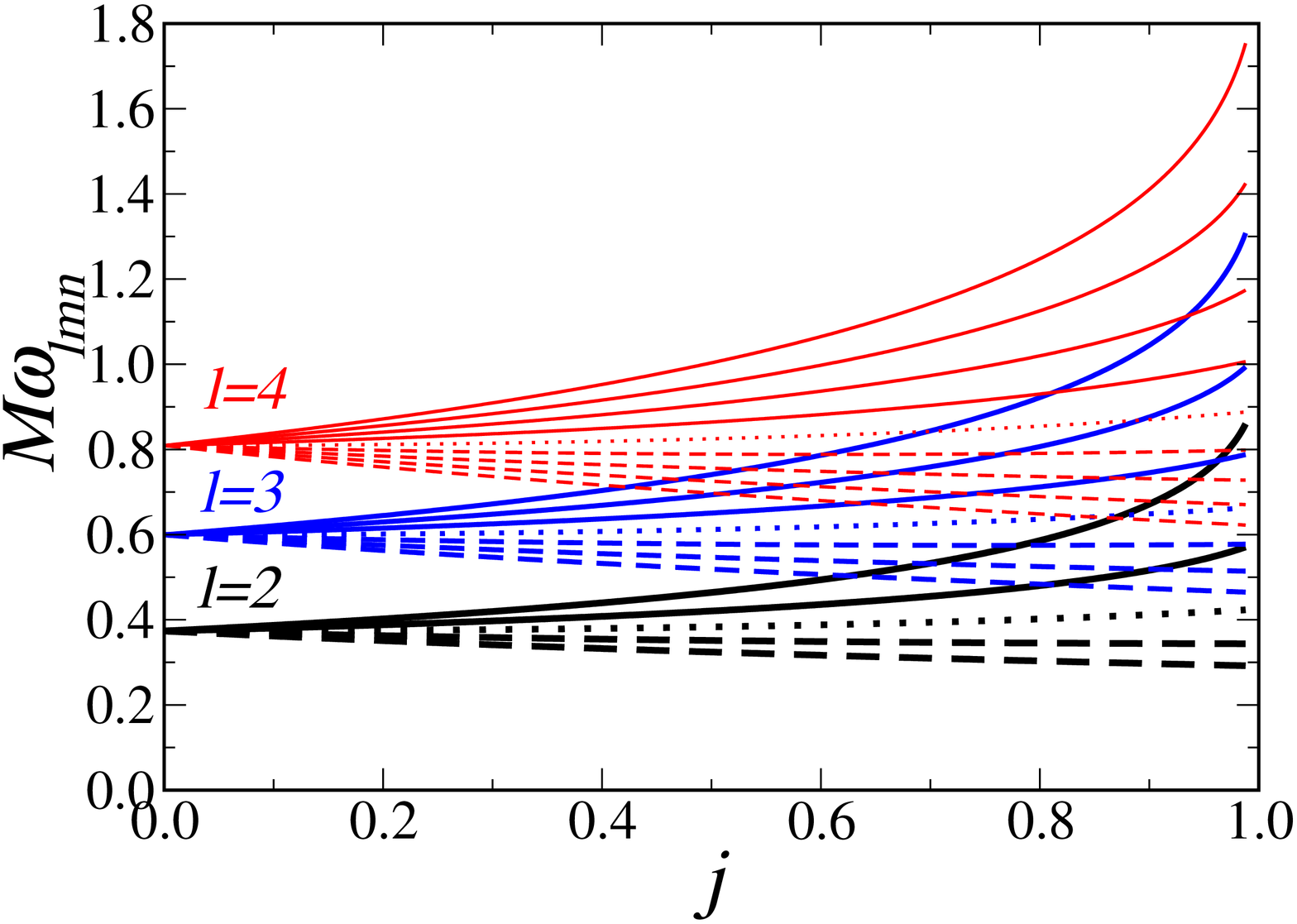}
      \includegraphics[width=2.25in,angle=-90]{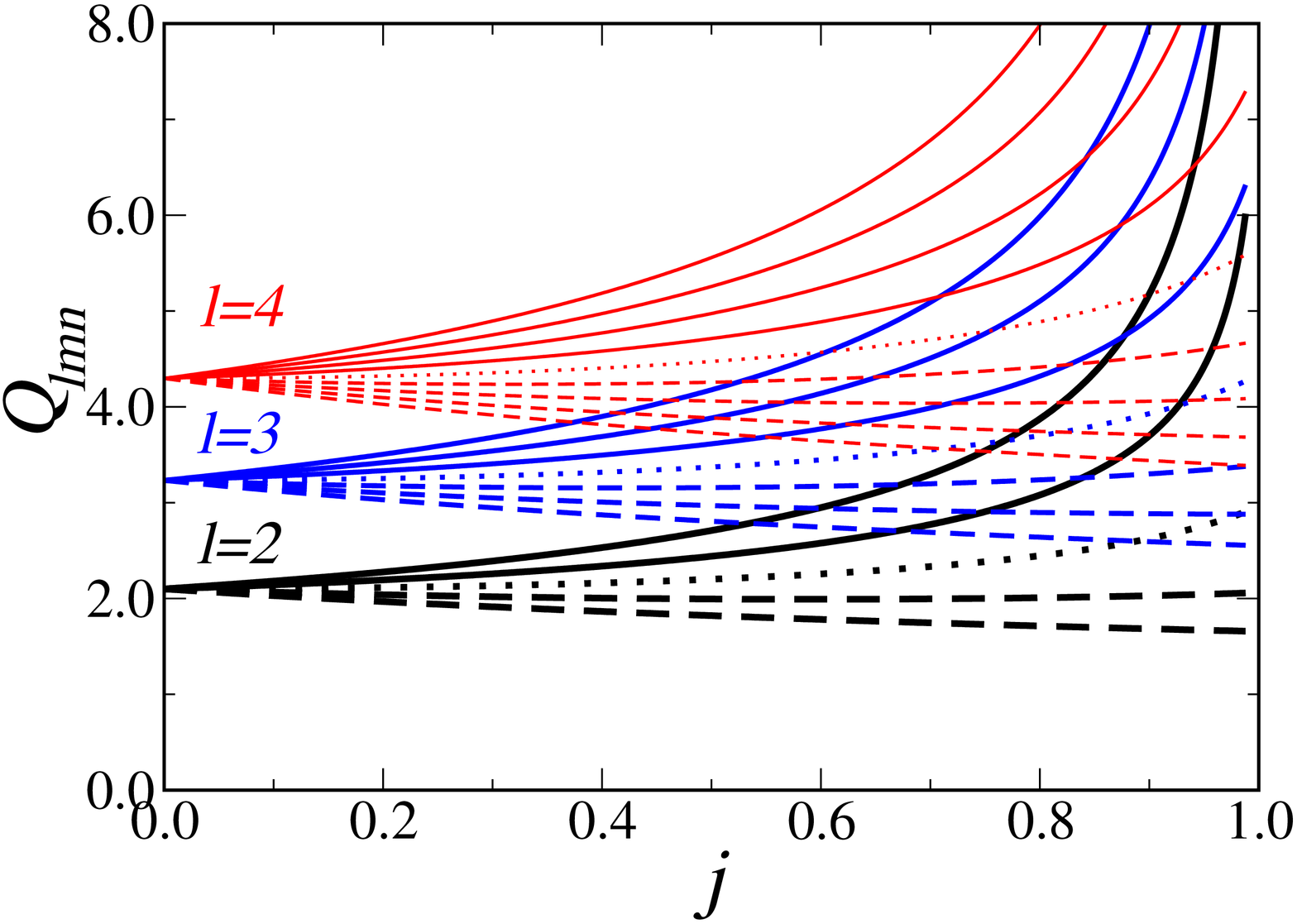}
    }
    \caption{Normal mode frequencies (left) and corresponding quality
      factors (right) of fundamental modes with $l=2,3,4$, as a
      function of the dimensionless black hole spin $j$, for different
      values of $m=l,\ldots,0,\ldots -l$ (for each $l$, different line
      styles from top to bottom correspond to decreasing values of
      $m$). Figure reprinted with permission
      from~\cite{BCW05}. \copyright\
      \href{http://link.aps.org/abstract/PRD/v73/e064030}{The American
      Physical Society}.}
    \label{fig:qnmFQ}
\end{figure}}

QNMs are characterized by a complex frequency $\omega$
that is determined by three ``quantum'' numbers, $(l,\,m,\,n)$
(see, e.g.,~\cite{BCW05}). 
Here $(l,\,m)$ are indices that are similar to those for standard 
spherical harmonics. For each pair of $(l,\,m)$ there are an 
infinitely large number of resonant modes characterized by another 
integer $n$. The time dependence of the oscillations is given by
$\exp(i\omega t)$, where $\omega$ is a complex frequency, its real part 
determining the mode frequency and the imaginary part (which
is always positive) giving the damping time: 
$\omega = \omega_{lmn}  + i / \tau_{lmn}$,  $\omega_{lmn}
=2\pi f_{lmn}$ defining the angular frequency and $\tau_{lmn}$ 
the damping time.  The ringdown wave will appear in a detector
as the linear combination $h(t)$ of the two polarizations 
$h_+$ and $h_\times$, that is $h(t)=F_+h_+ + F_\times h_\times$,
$F_+$ and $F_\times$ being the antenna pattern functions as 
defined in \Eqnref{eqn:interferometerantennapatterns}. The
polarization amplitudes for a given mode are given by 
\begin{eqnarray}
h_+ & = & \frac{A(f_{lmn}, Q_{lmn}, \epsilon_{\mathrm{rd}})}{r} (1 + \cos^2\iota)
\exp \left ( \frac{-\pi f_{lmn} t}{Q_{lmn}} \right) 
\cos \left ( 2\pi f_{lmn} t + \varphi_{lmn}\right ),
\nonumber \\
h_\times & = & \frac{A(f_{lmn}, Q_{lmn}, \epsilon_{\mathrm{rd}})}{r} 2 \cos\iota
\exp \left ( \frac{-\pi f_{lmn} t}{Q_{lmn}} \right) 
\sin \left ( 2\pi f_{lmn} t + \varphi_{lmn}\right ),
\end{eqnarray}
where $\iota$ is the angle between the black hole's spin axis and the 
observer's line of sight and $\varphi_{lmn}$ is an unknown constant phase.
The quality factor $Q_{lmn}$ of a mode is defined as $Q_{lmn} = 
\omega_{lmn}\tau_{lmn}/2$ and gives roughly the number of oscillations 
that are observable before the mode dies out.  Figure~\ref{fig:qnmFQ}~\cite{BCW05} plots frequencies and quality factors for
the first few QNMs as a function of the dimensionless spin parameter
$j=J/M^2$. The mode of a Schwarzschild black hole corresponding
to $l=2,\, m=n=0$, is given by
\begin{equation}
f_{200} = \pm 1.207 \times 10^3 \frac{10\,M_\odot}{M} \mathrm{\ Hz},
\quad
\tau_{200} = 5.537 \times 10^{-4}  \frac{M}{10\,M_\odot} \mathrm{\ s}.
\end{equation}
For stellar-mass--black-hole coalescences expected to be observed in 
ground-based detectors the ringdown signal is a transient that lasts
for a very short time. However, for space-based LISA
the signal would last several
minutes for a black hole of $M = 10^{7}\,M_\odot$. In the latter case, 
the ringdown waves could carry the energy equivalent
of $10^{5}\,M_\odot$ converted to gravitational waves -- a phenomenal
amount of energy compared even to the brightest quasars and gamma ray
bursts. Thus, LISA should be able to see QNMs from black hole 
coalescences anywhere in the universe, provided the final (redshifted) 
mass of the black hole is larger than about $10^{6}\,M_\odot$, as 
otherwise the signal lasts for far too short a time for the detector to 
accumulate the SNR.

Berti et al.~\cite{BCW05} have carried out an exhaustive study, in 
which they find that the LISA observations of SMBH binary mergers could be an excellent testbed for the no-hair theorem. 
Figure~\ref{fig:qnm-SNR-Errors} (left panel) plots the fractional energy 
$\epsilon_{\mathrm{rd}}$ that must be deposited in the ringdown mode 
so that the event is observable at a distance of 3~Gpc.  Black holes at 
3~Gpc with mass $M$ in the range of $10^{6}\mbox{\,--\,}10^{8}\,M_\odot$ would be 
observable (i.e., will have an SNR of 10 or more) even if a fraction 
$\epsilon_{\mathrm{rd}} \simeq 10^{-7}M$ of energy is in the ringdown phase.
Numerical relativity predicts that as much as 1\% of the energy 
could be emitted as QNMs, when two black holes merge, implying that 
the ringdown phase could be observed with an SNR of 100 or greater
all the way up to $z \sim 10$, provided their mass lies in the 
appropriate range\epubtkFootnote{Note that a black hole of physical 
mass $M$ at a redshift of $z$ will appear as a black hole of 
mass $M_z = (1+z) M$. This shifts the frequency of the QNM to 
the lower end of the spectrum. Assuming a frequency cutoff
of $10^{-4}$~Hz for LISA, this means that only black holes of 
intrinsic mass $M<1.2 \times 10^{8}\,M_\odot /(1+z)$ can be observed at
a redshift $z$.}. Furthermore, they find that at this redshift it
should be possible to resolve the fundamental $l=2, $ $m=2$ mode. 
Since black holes forming from primordial gas clouds at $z=10--15$
could well be the seeds of galaxy formation and large-scale structure,
LISA could indeed witness their formation through out the cosmic
history of the universe.

\epubtkImage{l2m2eps-l2m2all.png}{%
  \begin{figure}[htbp]
    \centerline{
      \includegraphics[width=2.25in,angle=-90]{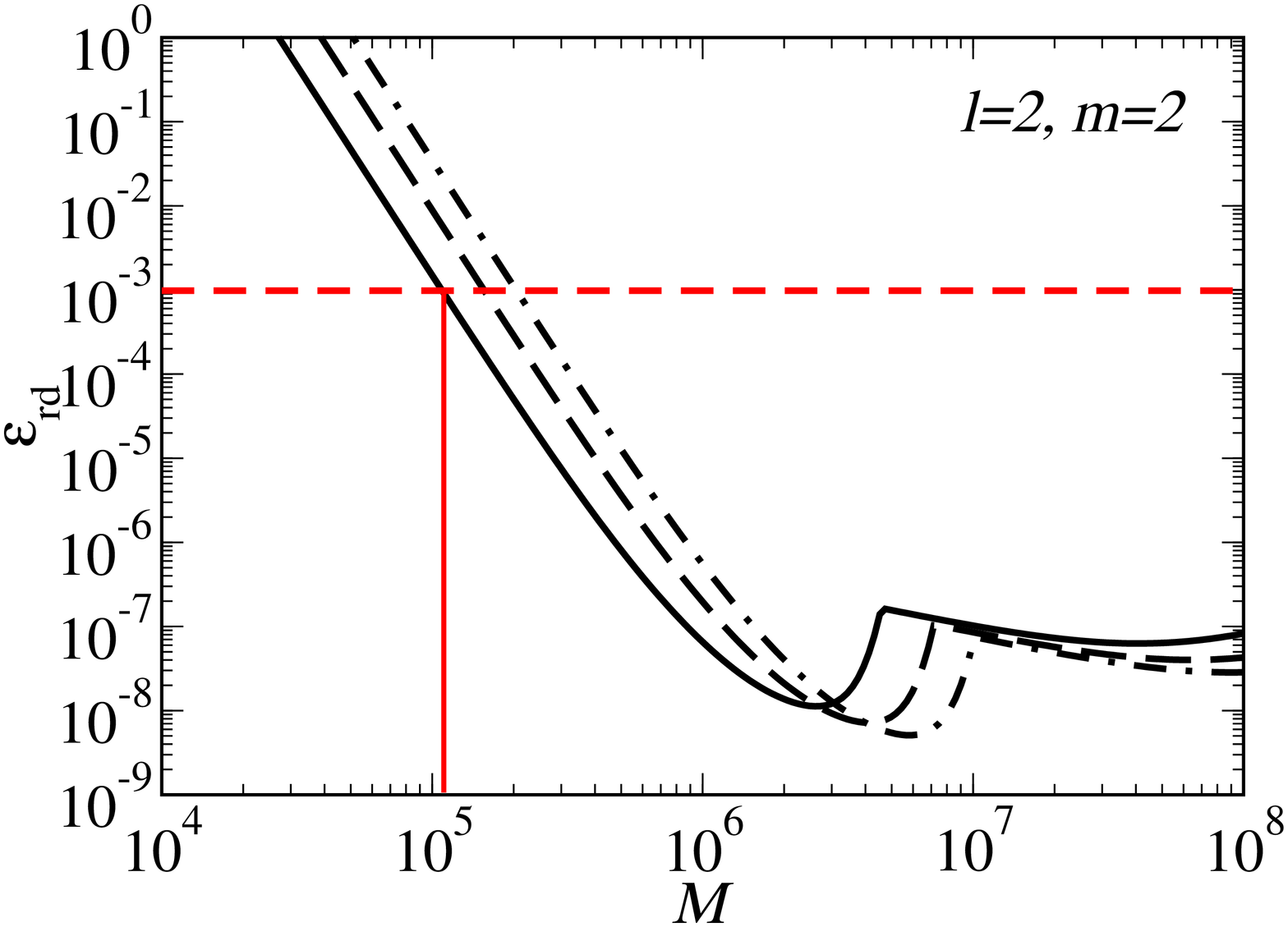}
      \includegraphics[width=2.25in,angle=-90]{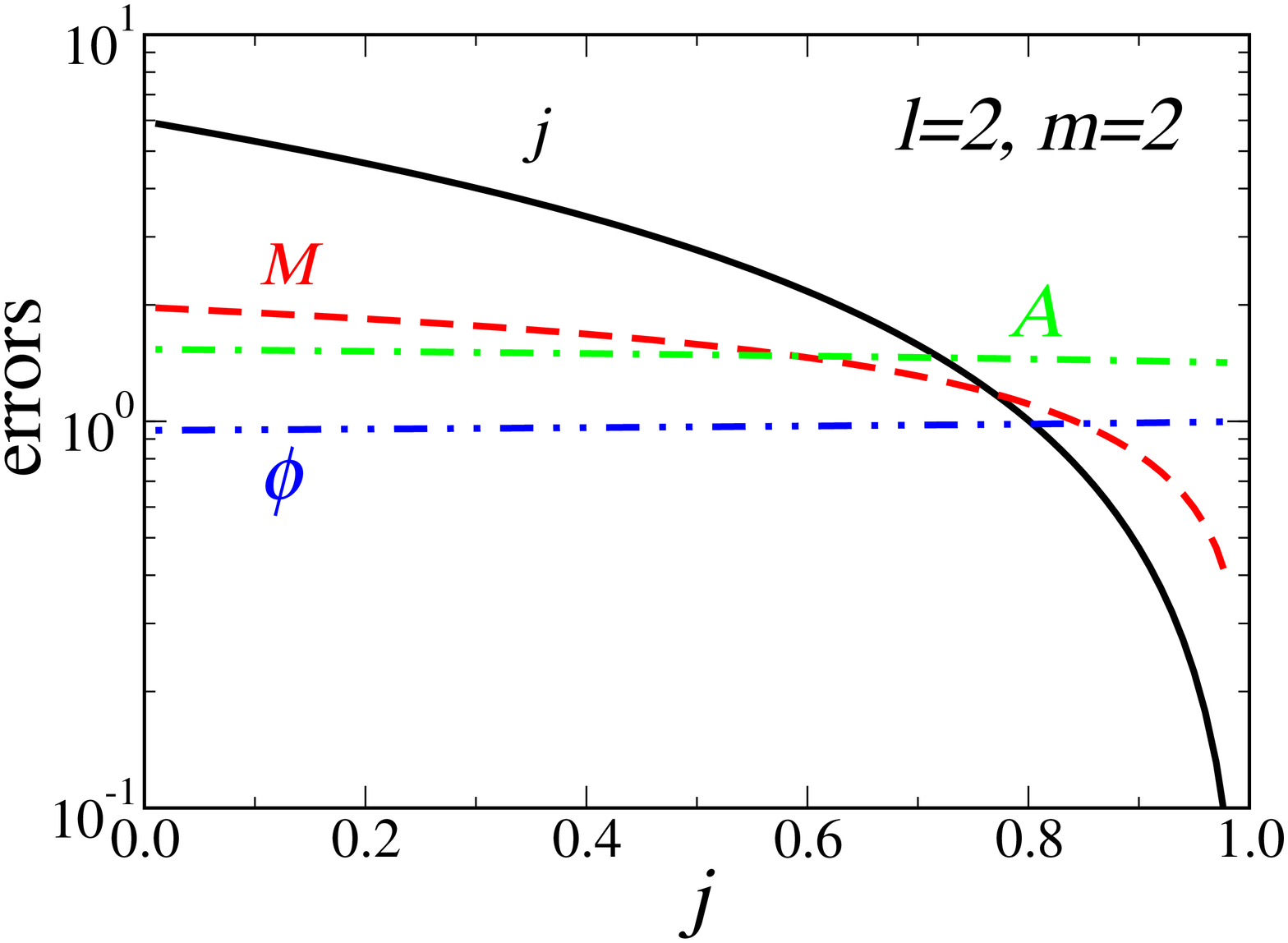}
    }
    \caption{The smallest fraction of black hole mass in ringdown
      waveforms that is needed to observe the fundamental mode at a
      distance of 3~Gpc (left) for three values of the black hole
      spin, $j=0$ (solid line) $j=0.80$ (dashed line) and $j=0.98$
      (dot-dashed line) and the error in the measurement of the
      various parameters as a function of the black hole spin for the
      same mode (right). Figure reprinted with permission
      from~\cite{BCW05}. \copyright\
      \href{http://link.aps.org/abstract/PRD/v73/e064030}{The American
      Physical Society}.}
    \label{fig:qnm-SNR-Errors}
\end{figure}}

Figure~\ref{fig:qnm-SNR-Errors} (right panel) shows SNR-normalized 
errors (i.e.,  one-sigma deviations  multiplied by the SNR) in the 
measurement of the various QNM parameters (the mass of the hole 
$M$, its spin $j$, the QNM amplitude $A$ and phase $\varphi$) for
the fundamental $l=m=2$ mode.  We see that, for expected ringdown
efficiencies of $\epsilon_{\mathrm{rd}} \simeq 10^{-2}M$ into the 
fundamental mode of an a-million--solar-mass black hole with spin
$j=0.8$ at 3~Gpc ($\rho \sim 2000$), the mass and spin of the black 
hole can measured to an accuracy of a tenth of a percent.

By observing a mode's frequency and damping time, one can deduce
the (redshifted) mass and spin of the black hole. However, this
is not enough to test the no-hair theorem. It would be necessary, although 
by no means sufficient, to observe at least one other mode (whose damping 
time and frequency can again be used to find the black hole's mass and spin)
to see if the two are consistent with each other. Berti et al~\cite{BCW05} 
find that such a measurement should be possible if the event occurs within a 
redshift of $z\sim 0.5$.

\subsection{The two-body problem in general relativity}
\label{sec:two body problem}

The largest effort in gravitational radiation theory in recent years 
has been to study the two-body problem using various approximations.  
The reason is that gravitationally bound binary systems are likely to be important 
gravitational wave sources, and until the evolution of such a 
system is thoroughly understood, it will not be possible to 
extract the maximum possible information from the observations. 

From \Figref{fig:freq}, we see that ground-based 
detectors will be sensitive to compact binaries with mass in
the range of $[1,\,10^4]\,M_\odot$ while LISA will be sensitive to the mass
range $[10^4,\,10^8]\,M_\odot$. As we have seen 
in Section~\ref{sec:gwsources}, most classes of binary sources will 
follow orbits that evolve strongly due to gravitational radiation 
reaction. In the case of ground-based detectors, they will
all merge within a year of entering the observation band.
In the case of LISA, we might observe sources (both stellar mass
binaries as well as SMBH binaries), whose frequency 
hardly changes. 

In contrast to Newtonian gravity, modeling a bound binary in general 
relativity is complicated by the existence of gravitational 
radiation and the nonlinearity of Einstein's equations. 
It must therefore be done approximately.  The three most important
approximation methods for solving gravitational wave problems are:

\begin{itemize}

\item {\bf The post-Newtonian scheme.} This is a combination of a 
low-velocity expansion ($v/c$ small) and a weak-field expansion 
($M/R$ small), in which the two small parameters are linked because 
a gravitationally-bound binary satisfies the virial relation $v^2\sim M/R$,
even in relativity. The zero-order solution is the Newtonian 
binary system. The post-Newtonian (PN) approximation has now been 
developed to a very high order in $v/c$ because the velocities 
in late-stage binaries, just before coalescence, are very high.

\item {\bf Perturbation theory.} This is an expansion in which the 
small parameter is the mass-ratio of the binary components. The 
zero-order solution is the field of the more massive component, 
and linear field corrections due to the second component determine 
the binary's orbital motion and the emitted radiation. This 
approximation is fully relativistic at all orders. It is 
being used to study the signals emitted by compact stars and 
stellar-mass black holes as they fall into SMBHs, 
an important source for LISA.

\item {\bf Numerical approaches.} With numerical relativity one can in 
principle simulate any desired relativistic system, no matter how strong the 
fields or high the velocities. It is being used to study the final stage
of the evolution of binaries, including their coalescence, after the 
PN approximation breaks down. Although it deals with 
fully relativistic and nonlinear general relativity, the method needs to 
be regarded as an approximate one, since spacetime is not 
resolved to infinite precision. The accuracy of a numerical 
simulation is normally judged by performing convergence tests, 
that is by doing the simulation at a variety of resolutions 
and showing that there are no unexpected differences between them.
\end{itemize}

We will review the physics that can be learned from models using 
each of these approximation schemes. But first we treat a subject 
that is common to all binaries that evolve due to radiation reaction, 
which is that one can estimate their distance from a gravitational 
wave observation.

\subsubsection{Binaries as standard candles: distance estimation}

Astronomers refer to systems as standard candles if their intrinsic 
luminosity is known, so that when the apparent luminosity of a 
particular system is measured, then its distance can be deduced. As 
mentioned in Section~\ref{sec:sirens}, radiating 
binaries have this property, if one can measure the 
effects of radiation reaction on their orbits~\cite{SCHUTZ1986}. 
Because of the one-dimensional nature of gravitational wave data, 
some scientists have begun calling these standard sirens~\cite{HH}. Over 
cosmological distances, the distance measured from the observation 
is the luminosity distance. We discuss in Section~\ref{sec:gwcosmology} 
below how this can be used to determine the Hubble constant and even 
the acceleration of the universe in methods independent of any cosmic 
distance ladder.

\subsubsection{Numerical approaches to the two-body problem}  
\label{sec:numerical relativity}

From the point of view of relativity, the simplest two-body problem 
is that of two black holes.  There are no matter fields and 
no point particles, just pure gravity.  Therefore, the physics
is entirely governed by Einstein's equations, which are highly
nonlinear and rather difficult to solve.  A number of teams 
have worked for over three decades towards developing accurate 
numerical solutions for the coalescence of two black holes, 
using fully three-dimensional numerical simulations.  

A breakthrough came in early 2005 with Pretorius~\cite{Pretorius05}
announcing the results from the first stable simulation ever, followed
by further breakthroughs by two other
groups~\cite{Campanelli:2005dd, Baker:2005vv} with successful
simulations. The main results from numerical simulations of
nonspinning black holes are rather simple. Indeed, just as the EOB had
predicted, and probably contrary to what many people had expected, the
final merger is just a continuation of the adiabatic inspiral, leading
on smoothly to merger and ringdown. In
Figure~\ref{fig:CompareWaveforms} we show the results from one of the
numerical simulations (right panel) and that of the EOB (left panel),
both for the same initial conditions. There is also good agreement in
the prediction of the total energy emitted by the system, being 5.0\%
($\pm$~0.4\%) (for a review see~\cite{Pretorius:2007nq}) and
3.1\%~\cite{BuonD00}, by numerical simulations and EOB, respectively,
as well as the spin of the final black hole (respectively, 0.69 and
0.8) that results from the merger.

The total energy emitted and the spin angular momentum of the 
black hole both depend on the spin angular momenta of the 
parent black holes and how they are aligned with respect to
the orbital angular momentum. In the test-mass limit, it is well known 
that the last stable orbit of a test particle in prograde orbit 
will be closer to, and that of a retrograde orbit will be farther from,
the black hole as compared to the Schwarzschild case. Thus, prograde
orbits last longer and radiate more compared to retrograde orbits.
The same is true even in the case of spinning black holes of comparable
masses; the emitted energy will be greater when the spins are aligned with 
the orbital angular momentum and least when they are anti-aligned.
For instance, for two equal mass black holes, each with its spin
angular momentum equal to 0.76, the total energy radiated in the
aligned (anti-aligned) case is 6.7\% (2.2\%) and the spin of the 
final black hole is 0.89 (0.44)~\cite{Campanelli:2007ew, Pollney:2007ss}. 
Heuristically, in the aligned case the black holes experience a
repulsive force, deferring the merger of the two bodies 
to a much later time than in the anti-aligned case, where they
experience an attractive force, accelerating the merger.

Detailed comparisons~\cite{Damour:2007yf, Pan:2007nw, Boyle:2007ft} 
show that we should be able to deploy the analytical templates from
EOB~\cite{Buonanno:2007pf, Damour:2007xr, Damour:2007vq,
  Damour:2008te} (and other approximants~\cite{Ajith:2007qp}) that
better fit the numerical data in our searches.
With the availability of merger waveforms from numerical simulations
and analytical templates, it will now be possible to search for 
compact binary coalescences with a greater sensitivity. The visibility 
of the signal improves significantly for binaries with their component 
masses in the range $[10,\,100]\,M_\odot$.  Currently, an effort
is underway to evaluate how to make use of numerical relativity 
simulations in gravitational wave searches~\cite{Ninja}, which should
help to increase the distance reach of interferometric detectors by a
factor of two and correspondingly nearly an order-of-magnitude increase
in event rate.

\epubtkImage{fig5a-fig5b.png}{%
  \begin{figure}[htbp]
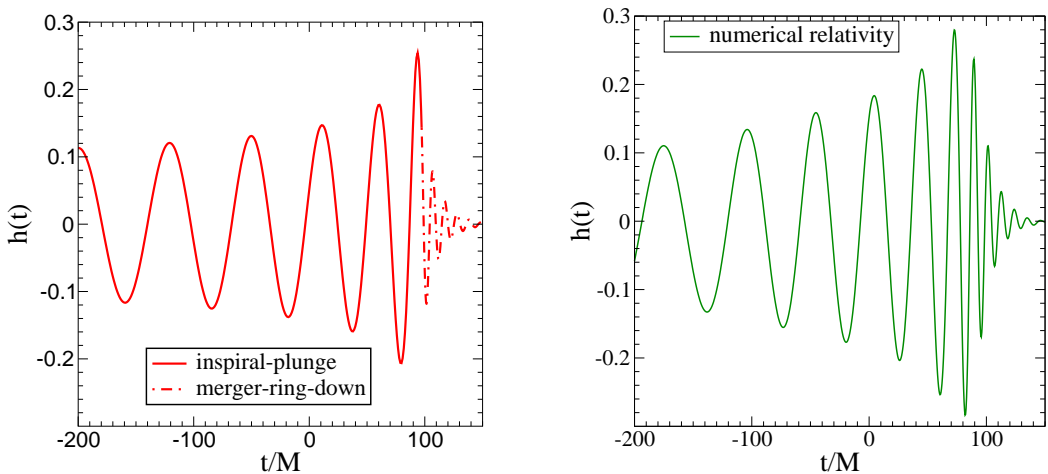

    \centerline{
      \includegraphics[width=2.5in]{fig5a}\hspace{1cm}
      \includegraphics[width=2.5in]{fig5b}
    }
    \caption{Comparison of waveforms from the analytical EOB approach
    (left) and numerical relativity simulations (right) for the same
    initial conditions. The two approaches predict very similar values
    for the total energy emitted in gravitational waves and the final
    spin of the black hole. Figure from~\cite{Buonanno:2007}.}
    \label{fig:CompareWaveforms} 
\end{figure}}

Numerical relativity simulations have now greatly matured, allowing
a variety of different studies. Some are studying the effect of 
the spin orientations of the component black holes on the linear
momentum carried away by the final black hole, fancifully called
\emph{kicks}~\cite{Herrmann:2007ac, Brugmann:2007zj, Baker:2006vn,
  Campanelli:2007ew, Gonzalez:2006md, Pollney:2007ss}; some have
focused on the dependence of the emitted waveform phase and energy on
the mass ratio; and yet others have strived to evolve the system with
high accuracy and for a greater number of cycles so as to push the
techniques of numerical relativity to the limit~\cite{Boyle:2006ne,
  Boyle:2007ft}.

Of particular interest are the numerical values of black hole kicks
that have been obtained for certain special configurations of the
component spins. Velocities as large as 4000~km~s$^{-1}$ have been
reported by several groups, but such velocities are only achieved when
both black holes have large\epubtkFootnote{By large spins we mean
  values that are close to the maximum value allowed by general
  relativity. If ${\mathbf J}$ is the magnitude of the spin angular
  momentum then general relativity requires that $|{\mathbf J}|\le
  M^2$} spins. Such velocities are in excess of escape velocities
typical of normal galaxies and are, therefore, of great astronomical
significance. These high velocities, however, are not seen for generic
geometries of the initial spin orientations; therefore, their
astronomical significance is not yet clear.

What is the physics behind kicks? Beamed emission
of radiation from a binary could result in imparting a net
linear momentum to the final black hole.  The radiation could be 
beamed either because the masses of the two black holes are
not the same (resulting in asymmetric emission in the orbital
plane) or because of the precession of the orbital plane arising
from spin-orbit and spin-spin interactions, or both. In the
case of black holes with unequal masses, the largest kick 
one can get is around 170~km~s$^{-1}$, corresponding
to a mass ratio of about 3:1. It was really with the advent of
numerical simulations that superkicks begin to be realized, but
only when black holes had large spins.
The spin-orbit configurations that produce large kicks are rather
unusual and at first sight unexpected. When the component black 
holes are both of the same mass and have equal but opposite spin 
angular momenta that lie in the orbital plane, frame dragging 
can lead to tilting and oscillation of the orbital plane, which, 
in the final phases of the evolution, could result in a rather 
large kick~\cite{Pretorius:2007nq}. SMBHs
are suspected to have large spins and, therefore, the effect of spin
on the evolution of a binary and the final spin and kick velocity
could be of astrophysical interest too. 

Curiously, a recent optical observation of a distant quasar,
SDSS J0927 12.65+294344.0, could well be the first identification
of a superkick, causing the SMBH to escape from
the parent galaxy~\cite{Komossa:2008qd}.  From a fundamental physics 
point of view, kicks offer a new way of testing frame dragging in the 
vicinity of black holes, but much work is needed in this direction. 

More recently, there has been an effort to understand 
and predict~\cite{Buonanno:2007sv, Rezzolla:2007rz, Damour:2007cb}
the spin of the final black hole, which should help in further exploring
interesting regions of the spin parameter space. In the relatively
simple case of two black holes with equal and aligned spins of magnitude
$a$, but unequal masses, with the symmetric mass ratio being $\nu=m_1m_2/(m_1+m_2)^2$, 
Rezzolla et al.~\cite{Rezzolla:2007rz} have obtained an excellent fit for the 
final spin $a_{\mathrm{fin}}$ of the black hole by enforcing basic
constraints from the test-mass limit: 
\begin{equation*}
a_{\mathrm{fin}} = a + (2\sqrt{3} + t_0 a + s_4 a^2) \nu  + 
(s_5 a + t_2) \nu^2 + t_3 \nu^3 ,
\end{equation*}
where $t_0=-2.686\pm0.065$, $t_2=-3.454\pm0.132$, $t_3=2.353\pm0.548$,
$s_4=-0.129\pm 0.012$, and $s_5=-0.384\pm0.261$.
The top and middle panels of \Figref{fig:finalspin} 
compare as functions of black hole spin and the symmetric mass ratio
the goodness of their fit (blue short-dashed line, top panels) with the
predictions of numerical simulations (circles and stars) from different
groups (AEI~\cite{Rezzolla:2007xa}, FAU--Jena~\cite{Marronetti:2007wz}, 
Jena~\cite{Berti:2007fi} and Goddard~\cite{Buonanno:2007pf}). 
Their residuals (red dotted lines, bottom panels) are less than 
a percent over the entire parameter space observed. These figures
also show the fits obtained for the equal-mass but variable-spin case 
(green long-dashed line, left panel)~\cite{Buonanno:2007sv} 
and for the nonspinning but unequal-mass case (green long-dashed line, 
middle panel)~\cite{Damour:2007cb}.

For the simple case of two equal mass black holes with aligned spins, 
the above analytical formula predicts that minimal and 
maximal final spin values of $a_{\mathrm{fin}} = 0.35\pm0.03$ and 
$a_{\mathrm{fin}} = 0.96\pm0.03$, respectively~\cite{Rezzolla:2007rz}. 
More interestingly, one can now ask what initial configurations 
of the mass ratios and spins would lead to the formation of a
Schwarzschild black hole (i.e.,
$a_{\mathrm{fin}}(a,\nu)=0$)~\cite{Hughes:2002ei}, which defines the
boundary of the region on one side of which lie systems for which the
spin of the final black hole flips relative to the initial total
angular momentum (bottom panel in Figure~\ref{fig:finalspin}).

\epubtkImage{FinalSpin-f2-FinalSpin-f3.-FinalSpin-f5.png}{%
  \begin{figure}[htbp]
    \centerline{\includegraphics[width=2.5in]{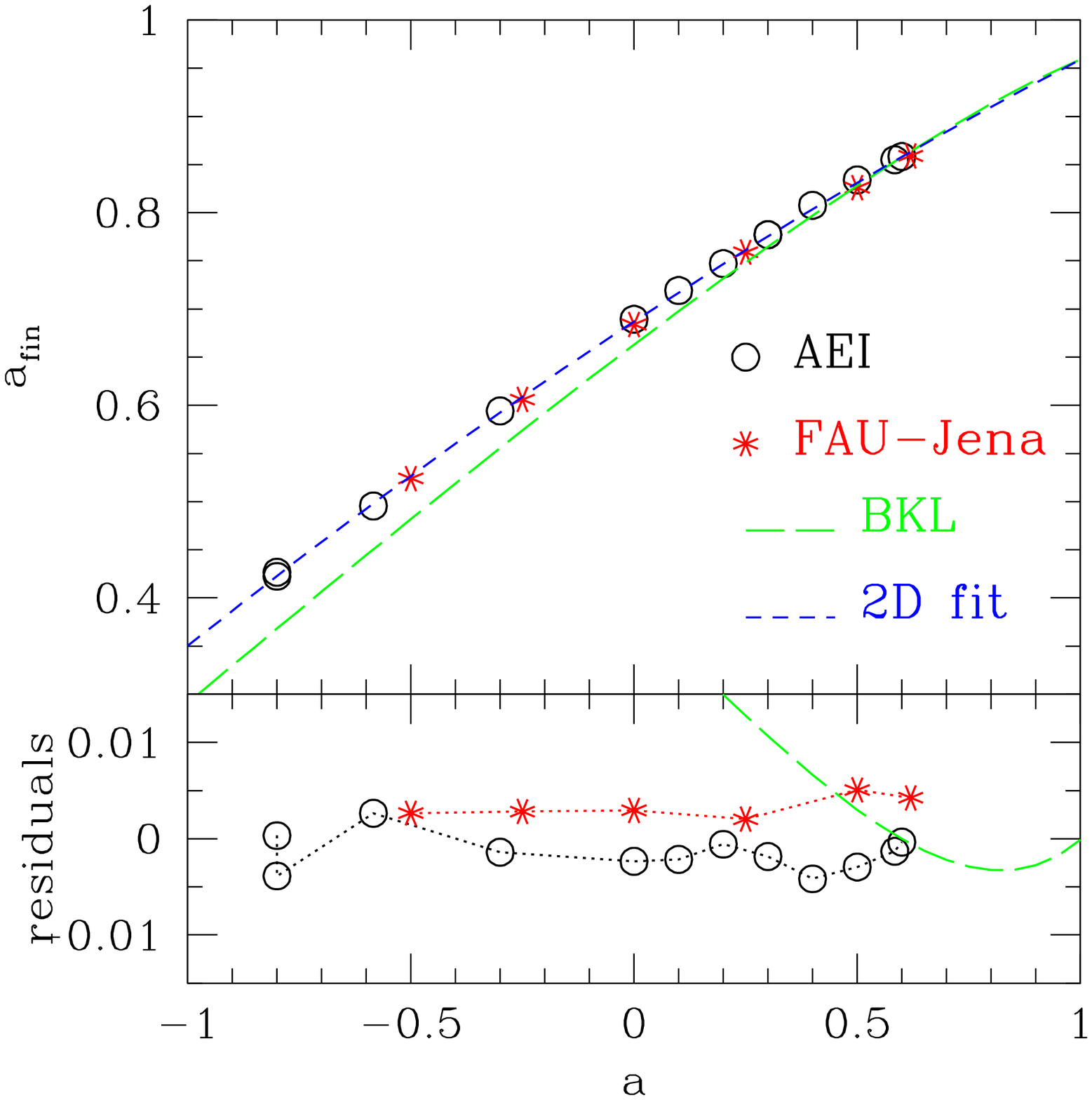}}
    \centerline{\includegraphics[width=2.5in]{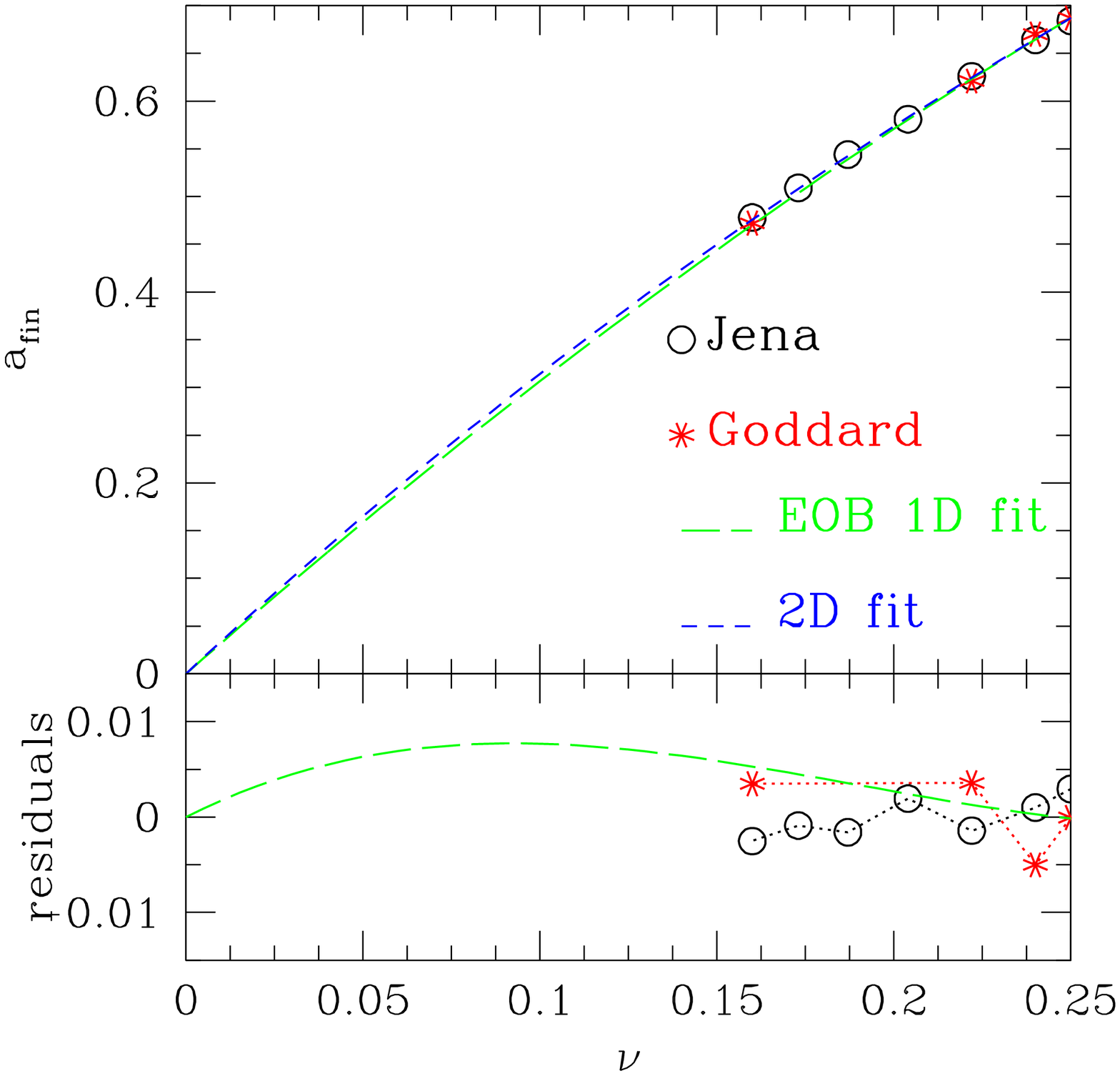}}
    \centerline{\includegraphics[width=2.5in]{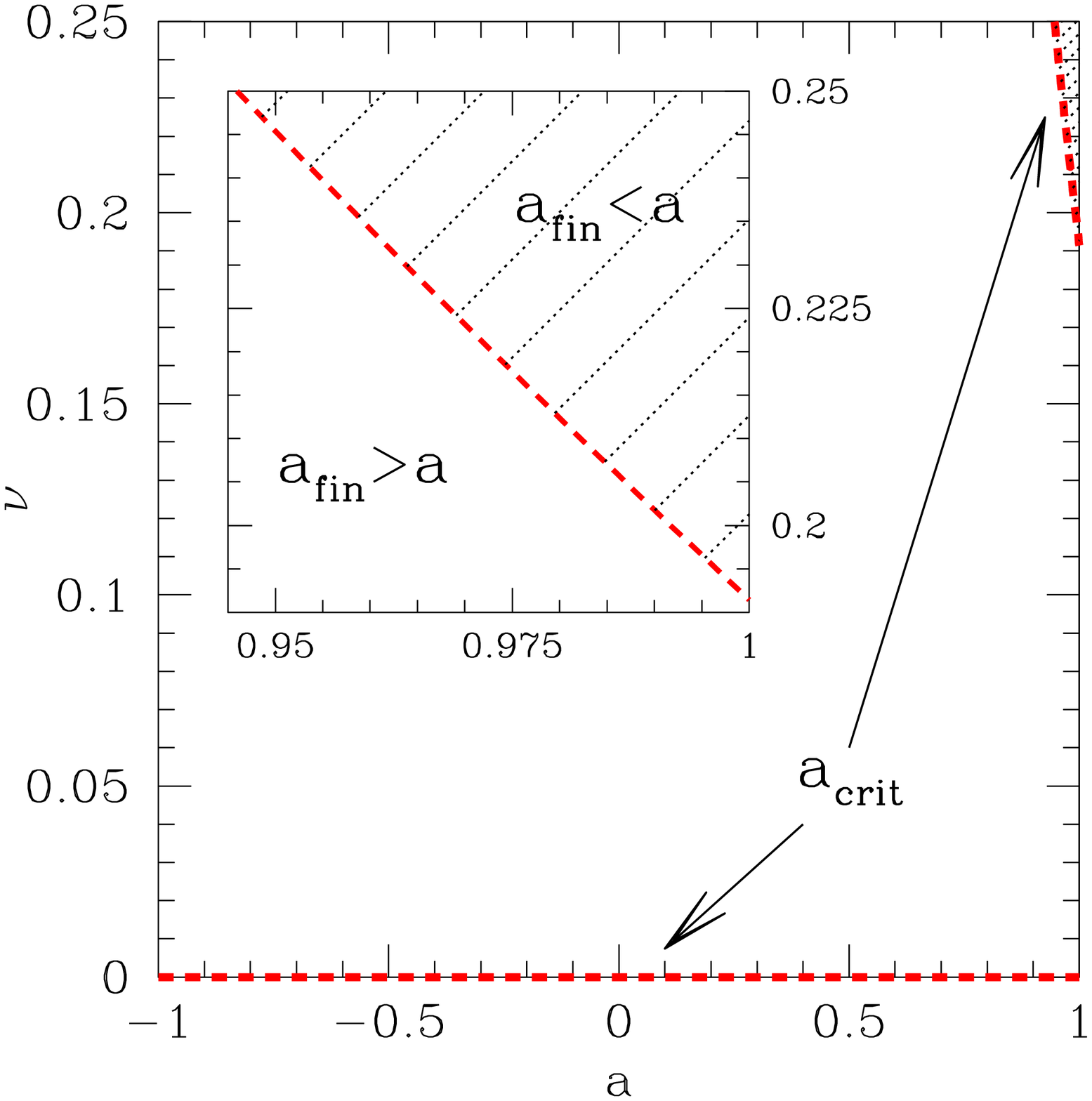}}
    \caption{The final spin of a black hole that results from the
      merger of two equal mass black holes of aligned spins (top
      panel) and nonspinning unequal mass black holes (middle
      panel). The bottom panel shows the region in the parameter space
      that results in an overall flip in the spin-orbit orientation of
      the system. Figure reprinted with permission
      from~\cite{Rezzolla:2008}. See text for details. \copyright\
      \href{http://stacks.iop.org/1538-4357/674/L29}{The American
      Astronomical Society}.}
    \label{fig:finalspin}
\end{figure}}

Finally, the evolution of binaries composed of nonspinning bodies is
characterized by a single parameter, namely the ratio of the masses of
the two black holes. The study of systems with different mass ratios
has allowed relativists to fit numerical waveforms with
phenomenological waveforms~\cite{Ajith:2007qp}. The advantage of the
latter waveforms is that one is able to more readily carry out data
analysis in any part of the parameter space without needing the
numerical data over the entire signal manifold.

Numerical relativity is still in its infancy and the parameter space 
is quite large. In the coming years more accurate simulations should 
become available, allowing the computation of waveforms with more 
cycles and less systematic errors.  However, the challenge remains 
to systematically explore the effect of different spin orientations, 
mass ratios and eccentricity. One area that has not been explored using
perturbative methods or post-Newtonian theory is that of 
intermediate--mass-ratio inspirals. These are systems with moderate 
mass ratios of order 100:1, where neither black-hole perturbation theory 
nor post-Newtonian approximation might be adequate. Yet, the prospect for
detecting such systems in ground and space-based detectors is
rather high. Numerical relativity simulations might be the only
way to set up effectual search templates for such systems.

\subsubsection{Post-Newtonian approximation to the two-body problem} 
\label{sec:PNA}

For the interpretation of observations of neutron-star--binary 
coalescences, which might be detected within five years by upgraded 
detectors that are now taking data, it is necessary to understand their 
orbital evolution to a high order in the PN
expansion.  The first effects of radiation reaction are seen 
at 2.5~PN order (i.e., at order $(v/c)^5$ beyond Newtonian gravity), 
but we probably have to have control in the equations of motion over the 
expansion at least to 3.5~PN order beyond the first radiation reaction
(i.e., to order $(v/c)^{12}$ beyond Newtonian dynamics).  There are many approaches 
to this, and we can not do justice here to the enormous effort 
that has gone into this field in recent years and refer the
reader to the Living Reviews by Blanchet~\cite{BLANCHETREF} and
by Futamase \& Itoh~\cite{FUTAMASEREF}.

Most work on this problem so far has treated a 
binary system as if it were composed of two point masses.  
This is, strictly speaking, inconsistent in general relativity, 
since the masses should form black holes of finite size. 
Blanchet, Damour, Iyer, and collaborators~\cite{Blanchet} have 
avoided this problem by a method that involves generalized 
functions.  They first expand in the nonlinearity parameter, 
and, when they have reached sufficiently high order, they obtain 
the velocity expansion of each order.  By ordering terms in the 
post-Newtonian manner they have developed step-by-step the 
approximations up to 3.5 PN order. 

A different team, led by Will, works with a different method of 
regularizing the 
point-particle singularity and compares its results with those 
of Blanchet et al. at each order~\cite{Blanchet1995}.  There 
is no guarantee that either method can be continued successfully 
to any particular order, but so far they have worked well 
and are in agreement.  Their results form the basis of the 
templates that are being designed to search for binary coalescences.  

An interesting way of extending the validity of the expansion that is
known to any order is to use Pad\'{e}
approximants~\cite{Damour1998,DIS01} (rational polynomials) of the
fundamental quantities in the theory, namely the orbital energy and
the gravitational wave luminosity. This has worked rather well in
improving the convergence of PN theory.  Buonanno and
Damour~\cite{BuonD98, BuonD00} have proposed an EOB approach to
two-body dynamics, which makes it possible to compute the orbit of the
binary and hence the phasing of the gravitational waves emitted beyond
the last stable orbit into the merger and  ringdown phases in the
evolution of the black hole binary. This analytical approach has been
remarkably successful and gained a lot of ground after the recent
success in numerical relativity (see Section~\ref{sec:numerical
  relativity}).

Other methods have been applied to this problem.  
Futamase~\cite{Futamase:1986ve} introduced a limit that combines the 
nonlinearity and velocity expansions in different ways in different 
regions of space, so that the orbiting bodies themselves have 
a regular (finite relativistic self-gravity) limit, while 
their orbital motion is treated in a Newtonian limit.  This 
should not fail at any order~\cite{FUTAMASEREF}, and has demonstrated 
its robustness by arriving at the same results as the other 
approaches, at least through 3~PN order. But it has a degree of arbitrariness 
in choosing initial data (see~\cite{SchutzStatistical}) 
that could cause problems for gravitational 
wave search templates that integrate orbits for a long period 
of time.  

Linear calculations of point particles around black 
holes are of interest in themselves 
and also for checking results of the full two-body 
calculations.  These are well-developed for certain 
situations, e.g.,~\cite{Tagoshi1996, Mino:1995fm}.  But the general 
equation of motion for such a body, taking into account all 
nongeodesic effects, has not yet been cast into a form 
suitable for practical calculations~\cite{Capon1998, Quinn1997}. This 
field is reviewed by two separate Living Reviews~\cite{Living:Poisson, Living:Sasaki}.

Matched filtering, discussed in Section~\ref{sec:matched filtering},
is a plausible method of testing the validity of different approaches
to computing the inspiral and merger waveforms from binary systems.
Though a single observation is not likely to settle the question
as to which methods are correct, a catalogue of events will help
to evaluate the accuracy of different approaches by studying the
statistics of the SNRs they measure.

\paragraph{Post-Newtonian expansions of energy and luminosity.}
Post-Newtonian calculations yield the expansion of the gravitational
binding energy $E$ and the gravitational wave luminosity $\cal F$
as a function of the post-Newtonian expansion
parameter\epubtkFootnote{In Newton's theory a two-body problem can be
  reduced to a one-body problem, in which a body of reduced mass $\mu$
  moves in an effective potential.  The parameter $v$ is the velocity
  of the reduced mass, if the orbit is circular. In the extreme mass
  ratio limit $\nu\rightarrow 0$, $v$ is the velocity of the smaller
  mass.} $v$. This is related to the frequency $f_{\mathrm{gw}}$ of the
dominant component of gravitational waves emitted by the binary system
by \[ v^3 = \pi M f_{\mathrm{gw}},\] where $M$ is the total mass of the
system. The expansions for a circular binary
are~\cite{Blanchet:2004ek, Blanchet:2005tk, BLANCHETREF}
\begin{eqnarray}\label{eqn:BinaryEnergyAndLuminosity}
E & = & -\frac{\nu M v^2}{2} \left \{ 
1 + \left ( - \frac{9+\nu}{12} \right ) v^2 + 
\left ( \frac{-81 + 57 \nu - \nu^2}{24} \right ) v^4 \right . \nonumber \\
& + & \left . \left ( -\frac {675}{64} + 
\left[ \frac{34445}{576} - \frac{205\pi^2}{96} \right ]\nu
- \frac{155}{96}\nu^2 - \frac{35}{5184}\nu^3  \right ) v^6 + {\cal O}(v^8)
\right \},\\[0.2cm]
\mathrm{and}\nonumber \\[0.2cm]
{\cal F} & = & \frac{32\nu^2 v^{10}}{5} \left \{1 
- \left ( \frac {1247}{336} + \frac{35}{12}\nu \right ) v^2 + 4\pi v^3 
+ \left ( -\frac{44711}{9072} + \frac{9271}{504}\nu 
+ \frac{65}{18}\nu^2 \right ) v^5  \right . \nonumber \\
& - & \left (\frac{8191}{672} + \frac{583}{24} \right ) \pi v^5 
+ \left [ \frac{6643739519}{69854400} + \frac{16}{3}\pi^2 - \frac{1712}{105}
\left ( \gamma + \ln(4 v) \right )  \right . \nonumber \\ 
& + & \left . \left ( -\frac{4709005}{272160} + \frac{41}{48} \pi^2 \right ) \nu 
-\frac{94403}{3024}\nu^2 - \frac{775}{324} \nu^3 \right ] v^6 \nonumber \\
& + & \left . \left ( -\frac{16285}{504} + \frac{214745}{1728} \nu + \frac{193385}{3024} 
\nu^2 \right ) \pi v^7 + {\cal O}(v^8) \right \},
\end{eqnarray}
where $\gamma=0.577\ldots$ is Euler's constant.

\paragraph{Evolution equation for the orbital phase.}
Starting from these expressions, one requires that gravitational 
radiation comes at the expense of the binding energy of the system
(see, e.g.,~\cite{DIS01}):
\begin{equation}
{\cal F} = - \frac{dE}{dt},
\end{equation}
the \emph{energy balance} equation. This can then be used
to compute the (adiabatic) evolution of the various quantities as a function
of time. For instance, the rate of change of the orbital velocity
$\omega(t) = v^3/M$ ($M$ being the total mass) can be computed using: 
\begin{equation}
\frac{d\omega(t)}{dt} = \frac{d\omega}{dv}\frac{dv}{dE}\frac{dE}{dt}
= \frac{3v^2}{M}\frac{{\cal F}(v)}{E'(v)}, \quad 
\frac{dv}{dt} = \frac{dv}{dE}\frac{dE}{dt} = \frac{-{\cal F}(v)}{E'(v)},
\end{equation}
where $E'(v) = dE/dv$. Supplemented with a differential equation for
$t$,
\begin{equation}
dt = \frac{dt}{dE}\frac{dE}{dv} = -\frac{E'(v)}{\cal F},
\end{equation}
one can solve for the evolution of the system's orbital velocity.
Similarly, the evolution of the orbital phase $\varphi(t)$ can be computed
using
\begin{equation}
\frac{d\varphi(t)}{dt} = \frac{v^3}{M}, \quad 
\frac{dv}{dt} = \frac{-{\cal F}(v)}{E'(v)}.
\label{eq:phasing1}
\end{equation}

\paragraph{Phasing formulas.}
The foregoing evolution equations for the orbital phase can be
solved in several equivalent ways~\cite{DIS01}, each correct to the 
required post-Newtonian order, but numerically different
from one another. For instance, one can retain the rational
polynomial ${\cal F}(v)/E(v)$ in Equation (\ref{eq:phasing1}) and
solve the two differential equations numerically, thereby obtaining
the time evolution of $\varphi(t)$. Alternatively, one might 
re-expand the rational function ${\cal F}(v)/E(v)$ as a polynomial
in $v$, truncate it to order $v^n$ (where $n$ is the order to which
the luminosity is given), thereby obtaining a parametric representation
of the phasing formula in terms of polynomial expressions in $v$:
\begin{equation}
\varphi(v) = \varphi_{\mathrm{ref}} + \sum_{k=0}^n \varphi_k v^k, \quad 
t(v) = t_{\mathrm{ref}} + \sum_{k=0}^n t_k v^k,
\end{equation}
where $\varphi_{\mathrm{ref}}$ and $t_{\mathrm{ref}}$ are a reference
phase and time, respectively.
The standard post-Newtonian phasing formula goes one step further
and inverts the second of the relations above to express $v$ as
a polynomial in $t$ (again truncated to appropriate order),
which is then substituted in the first of
the expressions above to obtain a phasing formula as an explicit
function of time:
\begin{eqnarray}
\varphi(t) & = &  
\frac{-1}{\nu\tau^5}\left \{ 1+\left(\frac{3715}{8064}+\frac
    {55}{96}\nu\right)\tau^2-\frac{3\pi}{4}\tau^3 
    + \left(\frac{9275495}{14450688}+\frac{284875}{258048}\nu 
    + \frac{1855}{2048}\nu^2\right)\tau^4 \right . \nonumber \\ 
& + & \left ( -\frac{38645}{172032} + \frac{65}{2048} \nu \right )\pi 
    \tau^5 \ln \tau + \left [ \frac{831032450749357}{57682522275840} 
    - \frac{53}{40} \pi^2 
    - \frac{107}{56} \left (\gamma + \ln(2\tau) \right ) \right . \nonumber \\
& + & \left . \left (-\frac{126510089885}{4161798144} 
    + \frac{2255}{2048}\pi^2 \right )\nu + \frac{154565}{1835008} \nu^2 
    - \frac{1179625}{1769472} \nu^3 \right ] \tau^6 \nonumber \\
& + & \left . \left (\frac{188516689}{173408256} + \frac{488825}{516096} \nu 
    - \frac{141769}{516096}\nu^2 \right ) \pi \tau^7 \right \},
\label{eqn:Phi}
\end{eqnarray}

\begin{eqnarray}
v^2 & = & \frac{\tau^2}{4} \left \{
1 + \left ( \frac{743}{4032} + \frac{11}{48}\nu \right ) \tau^2
-\frac{\pi}{5} \tau^3 + \left (\frac{19583}{254016} +
 \frac{24401}{193536} \nu + \frac{31}{288} \nu^2 \right )
  \tau^4 \right . \nonumber \\
  & + & \left (-\frac{11891}{53760} + \frac{109}{1920} \nu \right )
  \pi \tau^5 + \left [-\frac{10052469856691}{6008596070400} 
  +\frac{\pi^2}{6} + \frac{107}{420}(\gamma + \ln 2\tau ) \right .
   \nonumber\\ 
  & + & \left . \left ( \frac{3147553127}{780337152} - \frac{451}{3072}\pi^2
  \right ) \nu - \frac{15211}{442368}\nu^2 + \frac{25565}{331776}
  \nu^3 \right ] \tau^6 \nonumber \\
  & + & \left . \left ( - \frac{113868647}{433520640} - \frac{31821}{143360}
  \nu + \frac{294941}{3870720}\nu^2 \right ) \pi \tau^7 \right \}.
\end{eqnarray}
In the above formulas $v=\pi M f_{\mathrm{gw}}$ and $\tau=[\nu
  (t_C-t)/(5\,M)]^{-1/8}$, $t_C$ being the time at which the two stars
merge together and the gravitational wave frequency $f_{\mathrm{gw}}$
formally diverges.

\paragraph{Waveform polarizations.} The post-Newtonian formalism also gives
the two polarizations $h_+$ and $h_\times$ as multipole expansions
in powers of the parameter $v$. 
To lowest order, the two polarizations of the radiation from a binary 
with a circular orbit, located at a distance $D$, with total mass $M$
and symmetric mass ratio $\nu=m_1m_2/M^2$, are given by
\begin{equation}
h_+ = \frac{2\nu M}{D} v^2 ( 1 + \cos^2 \iota )\, \cos [2\varphi(t)],\quad
h_\times = \frac {4\nu M}{D} v^2 \cos \iota\, \sin [2\varphi(t)],
\label{eq:rpn spinless wave}
\end{equation}
where $\iota$ is the inclination of the orbital plane with the
line of sight and $v$ is the velocity parameter introduced earlier.

An interferometer will record a certain combination of the 
two polarizations given by 
\begin{equation}
h(t) = F_+ h_+ + F_\times h_\times,
\end{equation}
where the beam pattern functions $F_+$ and $F_\times$ are those discussed 
in Section~\ref{sec:beam factors}. In the case of ground-based instruments, the signal duration 
is pretty small, at most 15~min for neutron star binaries and smaller for
heavier systems. Consequently, one can assume the source direction to be unchanging
during the course of observation and the above combination produces essentially
the same functional form of the waveforms as in Equation~(\ref{eq:rpn spinless wave}). Indeed,
it is quite straightforward to show that
\begin{equation}
h(t) = 4\nu M \frac{\cal C}{D}v^2 \cos[2\varphi(t) + 2\varphi_0],
\label{eq:antenna response}
\end{equation}
where 
\begin{equation}
{\cal C} = \sqrt{A^2+B^2}, \ \ 
A=\frac{1}{2} (1+\cos^2\iota)F_+,\ \ B=\cos\iota\, F_\times, \ \ 
\tan 2\varphi_0 = \frac{B}{A}.
\end{equation}
The factor ${\cal C}$ is a function of the various angles and lies 
in the range [0,\,1] with an RMS value of 2/5 (see \Secref{sec:beam factors},
especially the discussion following \Eqref{eq:F Average}).

These waveforms form the basis for evaluating the science that can be extracted 
from future observations of neutron star and black hole binaries.
We will discuss the astrophysical and cosmological measurements 
that are made possible with such high precision waveforms in several
sections that follow (\ref{sec:harmonics} and \ref{sec:cosmography}).  
It is clear from the expressions for
the waveform polarizations that, at the lowest order, the radiation 
from a binary is predominantly emitted at twice the orbital frequency.  
However, even in the case of quasi-circular orbits the waves come 
off at other harmonics of the orbital frequency.  As we shall see 
below, these harmonics are very important for estimating the parameters
of a binary, although they do not seem to contribute much to the
SNR of the system.

\subsubsection{Measuring the parameters of an inspiraling binary}

\epubtkImage{MeasurementAccuracy.png}{%
  \begin{figure}[htbp]
    \centerline{\includegraphics[width=4in]{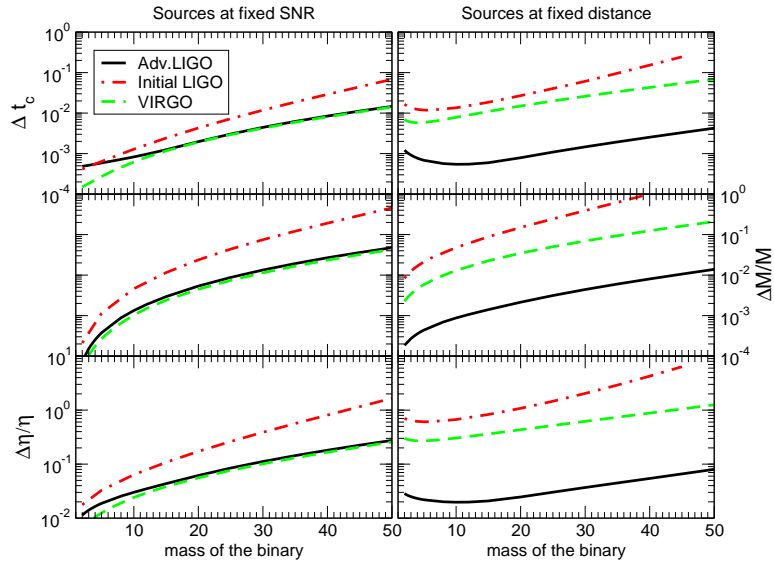}}
    \caption{One-sigma errors in the time of coalescence, chirpmass
      and symmetric mass ratio for sources with a fixed SNR (left
      panels) and at a fixed distance (right panels). The errors in
      the time of coalescence are given in ms, while in the case of
      chirpmass and symmetric mass ratio they are fractional
      errors. These plots are for nonspinning black hole binaries; the
      errors reduce greatly when dynamical evolution of spins are
      included in the computation of the covariance matrix. Slightly
      modified figure from~\cite{Arun:2005}.}
    \label{fig:covariance-matrix-spinless-case}
\end{figure}}

The issue of parameter estimation in the context of black hole
binaries has received a lot of attention~\cite{Chernoff:1993th,
  Cutler:1994ys, Flanagan:1997kp, BSD1, PoissonAndWill, BSD2,
  Arun:2006if}. Most authors have used the covariance matrix for this
purpose, although Markov Chain Monte Carlo (MCMC) techniques have also
been used occasionally~\cite{Christensen:1998gf, Rover:2006bb,
  Rover:2006ni, Cornish:2007if}, especially in the context of
LISA~\cite{Umstatter:2005jd, Cornish:2006ry, Cornish:2006dt,
  Cornish:2007jv, Crowder:2007ft}. Covariance matrix is often the
preferred method, as one can explore a large parameter space without
having to do expensive Monte Carlo simulations. However, when the
parameter space is large, covariance matrix is not a reliable method
for estimating parameter accuracies, especially at low
SNRs~\cite{Balasubramanian:1997qz, BSD1, Vallisneri:2007ev}; but at
high SNRs, as in the case of SMBH binaries in LISA, the problem might
be that our waveforms are not accurate enough to facilitate a reliable
extraction of the source parameters~\cite{Cutler:2007mi}. Although
MCMC methods can give more reliable estimates, they suffer from being
computationally extremely expensive. However, they are important in
ascertaining the validity of results based on the covariance matrix,
at least in a small subset of the parameter space, and should probably
be employed in assessing parameter accuracies of candidate
gravitational wave events.

In what follows we shall summarize the most recent work on parameter 
estimation  in ground and space-based detectors for binaries
with and without spin and the improvements brought about by including
higher harmonics.

\paragraph{Ground-based detectors -- nonspinning components.}
In Figure~\ref{fig:covariance-matrix-spinless-case} we have 
plotted the one-sigma uncertainty in the measurement of the 
time of coalescence, chirpmass and symmetric mass ratio
for initial and advanced LIGO and VIRGO~\cite{Arun:2006if}. 
The plots show errors for sources all producing a fixed SNR of 10 
(left panels) or all at a fixed distance of 300~Mpc (right panels). 
The fractional error in chirpmass, even at a modest SNR of 10, can be
as low as a few parts in ten thousand for stellar mass binaries, but the
error stays around 1\%, even for heavier systems that have only 
a few cycles in a detector's sensitivity band. Error in the mass
ratio is not as small, increasing to 100\% at the higher end of the mass
range explored. Thus, although the chirpmass can be measured to a good
accuracy, poor estimation of the mass ratio means that the individual 
masses of the binary cannot be measured very well. Note also that
the time of coalescence of the signal is determined pretty well, which
means that we would be able to measure the location of the system
in the sky quite well.

At a given SNR the accuracy is better in the case of low-mass binaries, 
since they spend a longer duration and a greater number of cycles in the 
detector band and the chirpmass can be determined better than the mass ratio,
since to first order the frequency evolution of a binary is determined
only by the chirpmass. 

\epubtkImage{Mccomparison-mucomparison.png}{%
  \begin{figure}[htbp]
    \centerline{\includegraphics[width=3in]{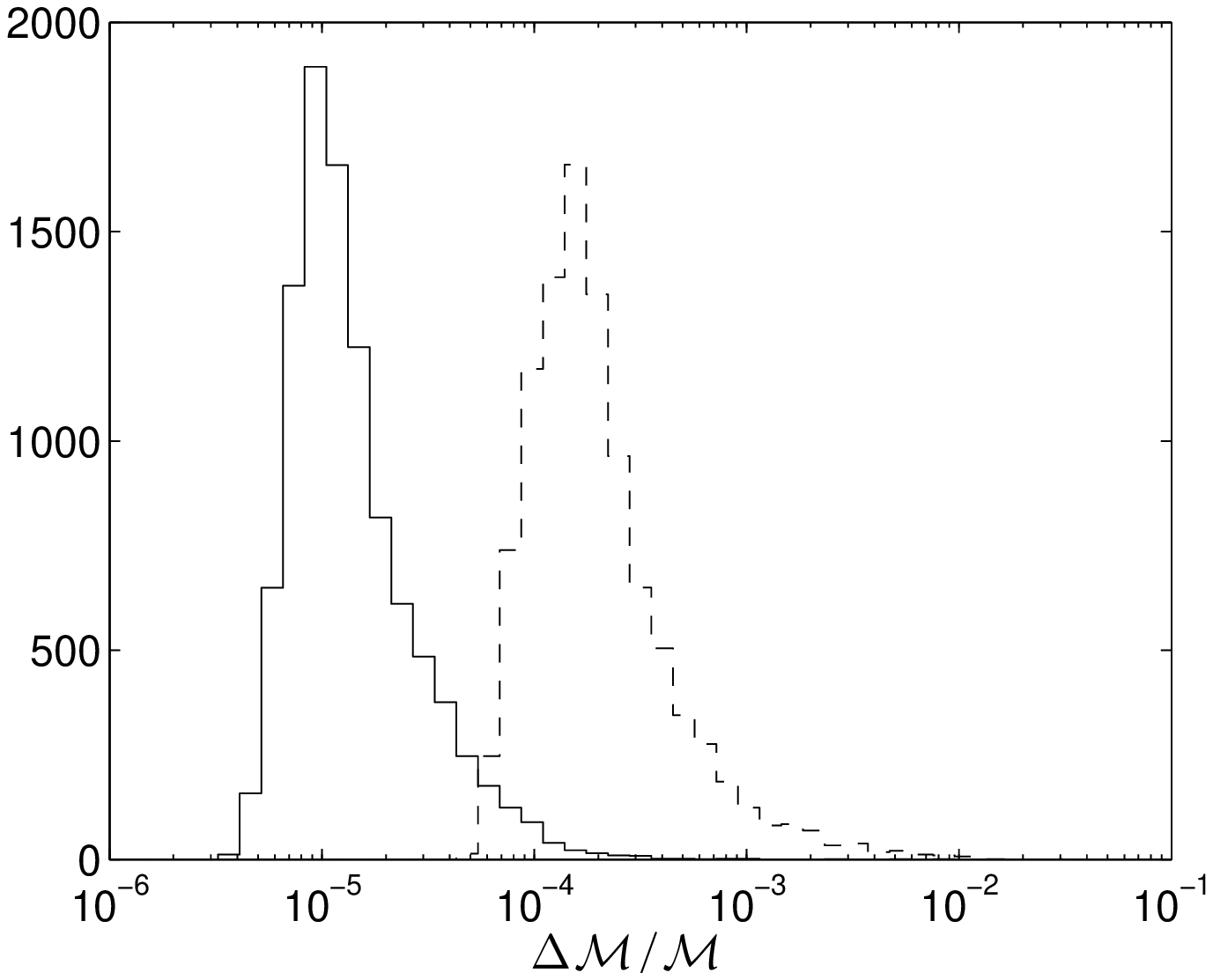}}
    \centerline{\includegraphics[width=3in]{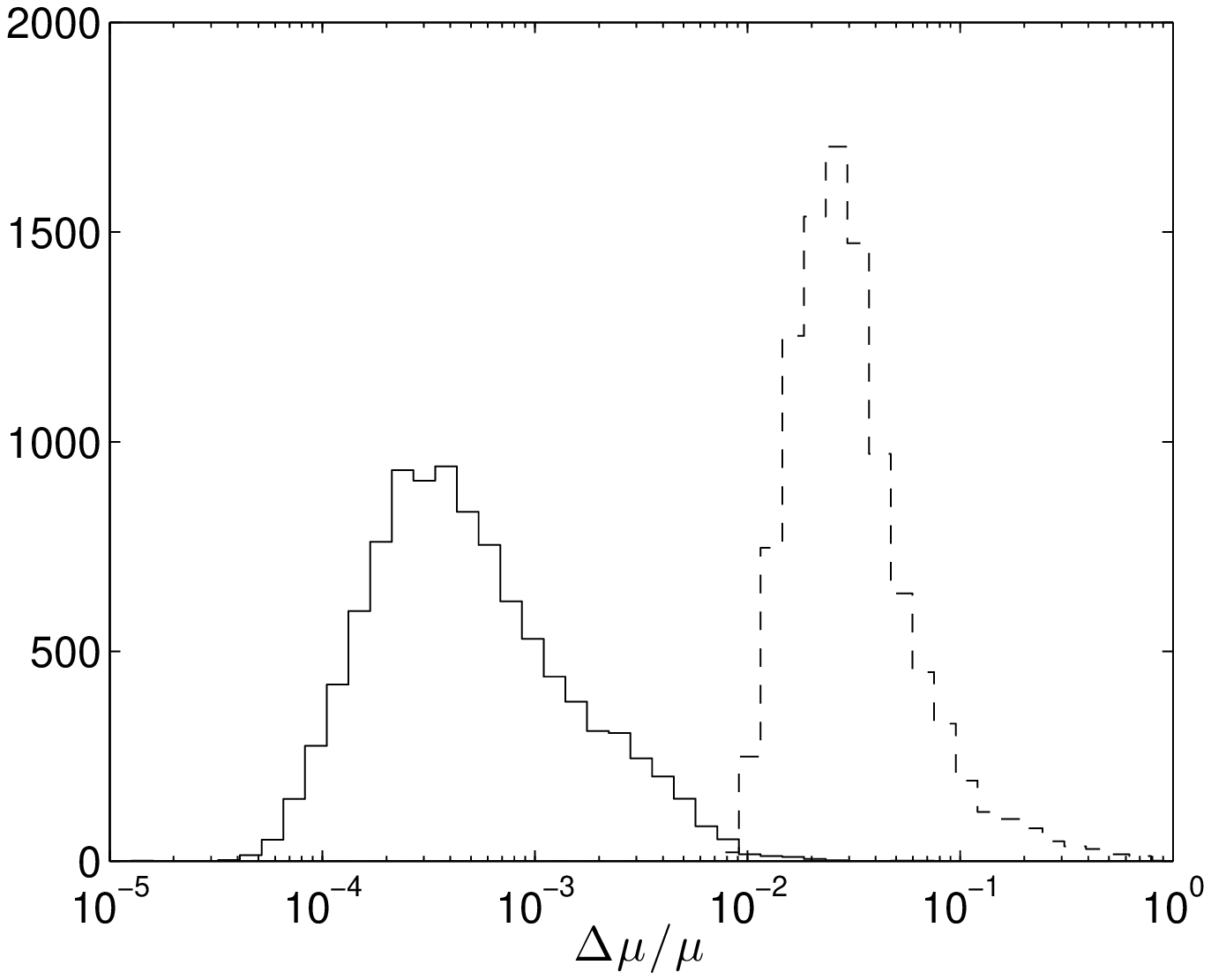}}
    \caption{Distribution of measurement accuracy for a binary merger
      consisting of two black holes of masses $m_1=10^{6}\,M_\odot$
      and $m_2=3\times 10^{5}\,M_\odot$, based on 10,000 samples of
      the system in which the sky location and orientation of the
      binary are chosen randomly. Dashed lines are for nonspinning
      systems and solid lines are for systems with spin. Figure
      reprinted with permission from~\cite{Lang:1900bz}. \copyright\
      \href{http://link.aps.org/abstract/PRD/v74/e122001}{The American
      Physical Society}.}
    \label{fig:param-accuracy-LISA}
\end{figure}}

\paragraph{Measuring the parameters of supermassive 
black hole binaries in LISA.} 
In the case of LISA, the merger of SMBHs produces 
events with extremely large SNRs, even at a redshift of $z=1$
(100s to several thousands depending on the chirpmass of the 
source). Therefore, one expects to measure the parameters
of a merger event in LISA to a phenomenal accuracy. 
Figure~\ref{fig:param-accuracy-LISA} depicts the distribution of 
the errors for a binary consisting of two SMBHs
of masses $(10^6,\,3\times 10^5)\,M_\odot$ at a redshift of $z=1$~\cite{Lang:2007ge}.
The distribution was obtained for ten thousand samples of the system
corresponding to random orientations of the binary at random
sky locations with the starting frequency greater than $3 \times
10^{-5}$~Hz and the ending frequency corresponding to the last stable
orbit.

Each plot in Figure~\ref{fig:param-accuracy-LISA} shows the results 
of computations for binaries consisting of black holes with and without 
spins. Even in the absence of spin-induced modulations in the waveform,
the parameter accuracies are pretty good.  Note that spin-induced
modulations in the waveform enable a far better estimation of parameters,
chirpmass accuracy improving by more than an order of magnitude and
reduced mass accuracy by two orders of magnitude.  It is because of 
such accurate measurements that it will be possible to use SMBH mergers to test general relativity in the strong field
regime of the theory (see below). 

Although Figure~\ref{fig:param-accuracy-LISA} corresponds to a binary
with specific masses, the trends shown are found to be true more 
generically for other systems too, the actual parameter accuracies and
improvements due to spin both depending on the specific system studied.

\epubtkImage{s45-3-res-s45-3-full.png}{%
  \begin{figure}[thbp]
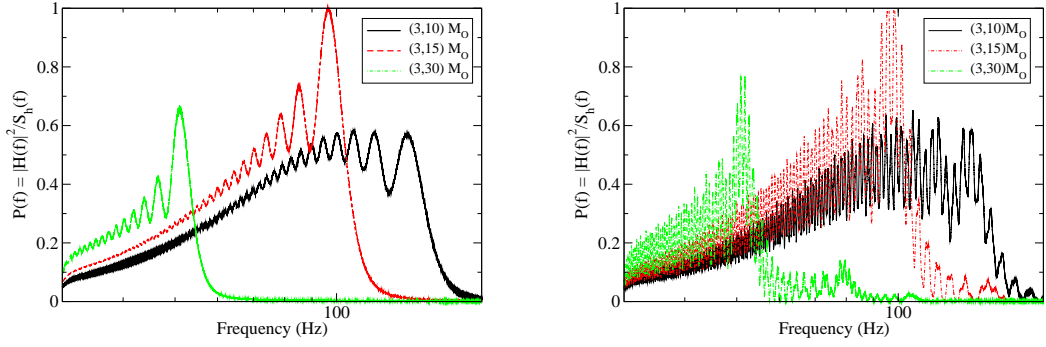

    \centerline{\includegraphics[width=2.5in]{s45-3-res}\hspace{1cm}
      \includegraphics[width=2.5in]{s45-3-ful}}
    \caption{The SNR integrand of a restricted (left panel) and full
      waveform (right panel) as seen in initial LIGO. We have shown
      three systems, in which the smaller body's mass is the same, to
      illustrate the effect of the mass ratio. In all cases the system
      is at 100~Mpc and the binary's orbit is oriented at $45^{\circ}$
      with respect to the line of sight.}
    \label{fig:harmonics}
\end{figure}}

\subsubsection{Improvement from higher harmonics}
\label{sec:harmonics}

The results discussed so far use the \emph{restricted post-Newtonian
  approximation} in which the waveform polarizations contain only
twice the orbital frequency, neglecting \emph{all} higher-order
corrections (including those to the second harmonic). The full
waveform is a post-Newtonian expansion of the two polarizations as a
power-series in $v/c$ and consists of terms that have not only the
dominant harmonic at twice the orbital frequency, but also other
harmonics of the waveform. Schematically, the full waveform can be
  written as~\cite{BLANCHETREF, VanDenBroeck:2006ar}
\begin{equation}
h(t) = \frac{4M\eta}{D_L} \sum_{k=1}^7 \sum_{n=0}^5 A_{(k,n/2)}
v^{n+2}(t) \cos \left [ k\varphi(t) + \varphi_{(k,n/2)} \right ],
\end{equation}
where $\nu=m_1m_2/M^2$ is the symmetric mass ratio, the first sum
(index $k$) is over the different harmonics of the waveform and the
second sum (index $n$) is over the different post-Newtonian
orders. Note that post-Newtonian order weighs down the importance of
higher-order amplitude corrections by an appropriate factor of the
small parameter $v$. In the restricted post-Newtonian approximation
one keeps only the lowest-order term. Since $A_{1,0}$ happens to be
zero, the dominant term corresponds to $k=2$ and $n=0$, containing
twice the orbital frequency.

The various signal harmonics, and the associated additional 
structure in the 
waveform, can potentially enhance our ability to measure the
parameters of a binary to a greater accuracy. 
The reason we can expect to do so can be
seen by looking at the spectra of gravitational waves with and
without these harmonics.  For a binary that is oriented face on
with respect to a detector only the second harmonic is seen, while
for any other orientation the radiation is emitted at all other 
harmonics, the influence
of the harmonics becoming more pronounced as the inclination
angle changes from 0 to $\pi/2$.  Figure~\ref{fig:harmonics}
compares, in the frequency band of ground-based detectors, the
spectrum of a source using the restricted post-Newtonian approximation
(left panel) to the full waveform. In both cases the source is
inclined to the line of sight at 45~degrees.

Following is a list of improvements brought about by higher harmonics. 
In the case of ground-based detectors Van Den Broeck 
and Sengupta~\cite{VanDenBroeck:2006ar, VanDenBroeck:2006qu}
found that, when harmonics are included, the SNR hardly changes, but 
is always smaller, relative to a restricted waveform. However, the 
presence of frequencies higher than twice the orbital frequency 
means that it will be possible to observe heavier systems,
increasing the mass reach of ground-based detectors by
a factor of 2 to 3 in advanced LIGO and third generation
detectors~\cite{VanDenBroeck:2006ar, VanDenBroeck:2006qu}. 
The same effect was found in the case
of LISA too, allowing LISA to observe SMBH
masses up to a $\mathrm{few} \times 10^{8}\,M_\odot$~\cite{Arun:2007qv}. 
More than the increased mass reach, the harmonics reduce
the error in the estimation of the chirpmass, symmetric
mass ratio and the time of arrival by more than
an order of magnitude for stellar-mass black hole binaries.
The same is true to a greater extent in the case of
SMBH binaries, allowing as well a far 
greater accuracy in the measurement of the luminosity
distance and sky resolution in LISA's observation of 
these sources~\cite{Arun:2007hu, Trias:2007fp}. 
For instance, Figure~\ref{fig:SkyMap_GAIN}~\cite{Trias:2007fp}
shows the gain in LISA's angular resolution for two massive
black-hole--binary mergers as a consequence of using higher harmonics
for a specific orientation of the binary. 
Improvements of order 10 to 100 
can be seen over large regions of the sky. 
This improved performance of LISA makes it a good probe of 
dark energy~\cite{Arun:2007hu} (see \Secref{sec:cosmography}).

A word of caution is in order with regard to the improvements brought
about by higher harmonics. If the sensitivity of a detector has
an abrupt lower frequency cutoff, or falls off rapidly below a 
certain frequency, then the harmonics
bring about a more dramatic improvement than when the sensitivity
falls off gently. Higher harmonics, nevertheless, always help in
reducing the random errors associated with the measurement of
parameters of a coalescing black-hole binary.

\epubtkImage{SkyMap_gain_77-SkyMap_gain_76.png}{%
  \begin{figure}[htbp]
    \centerline{\includegraphics[width=5.5in]{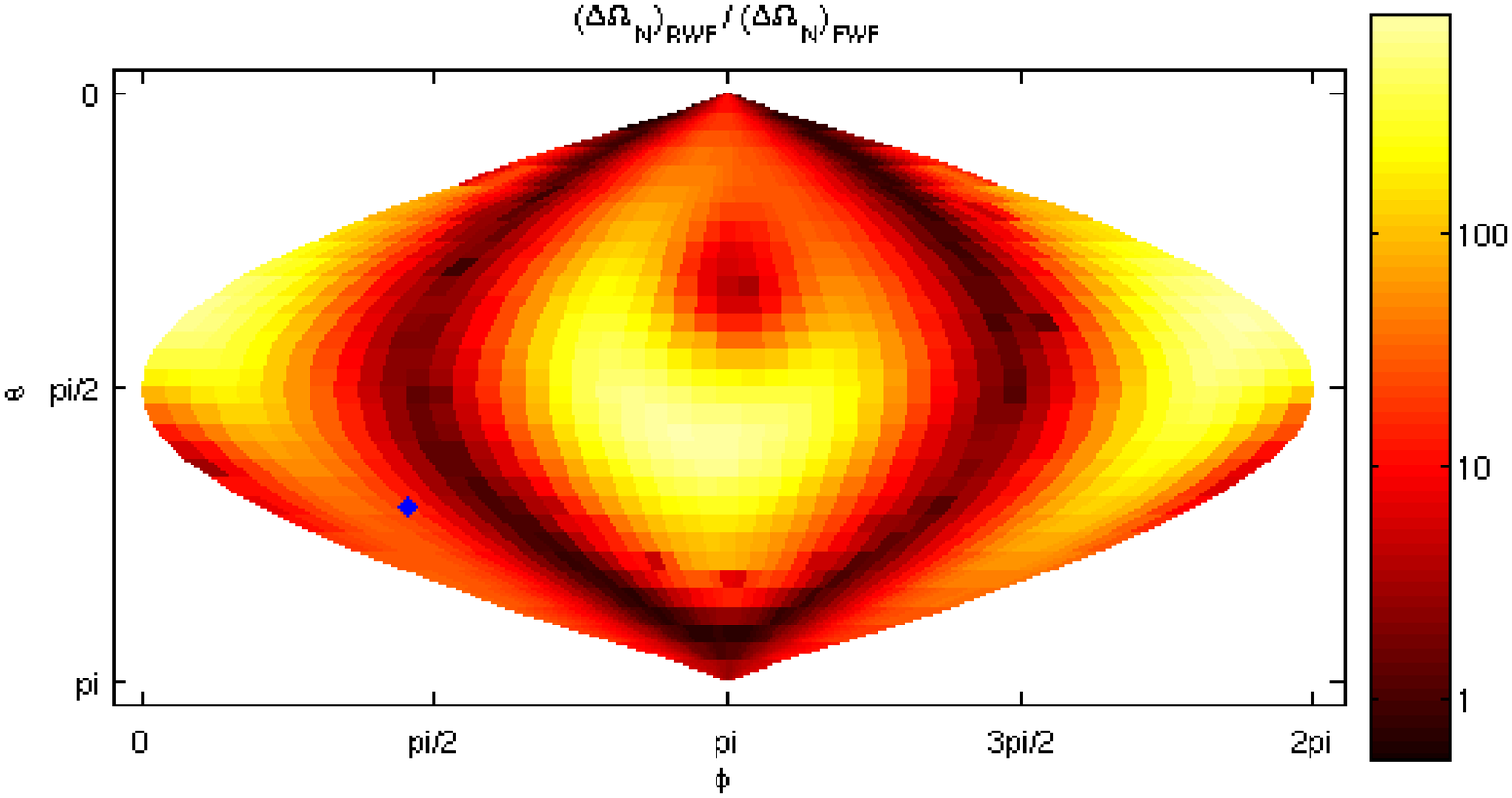}}
    \centerline{\includegraphics[width=5.5in]{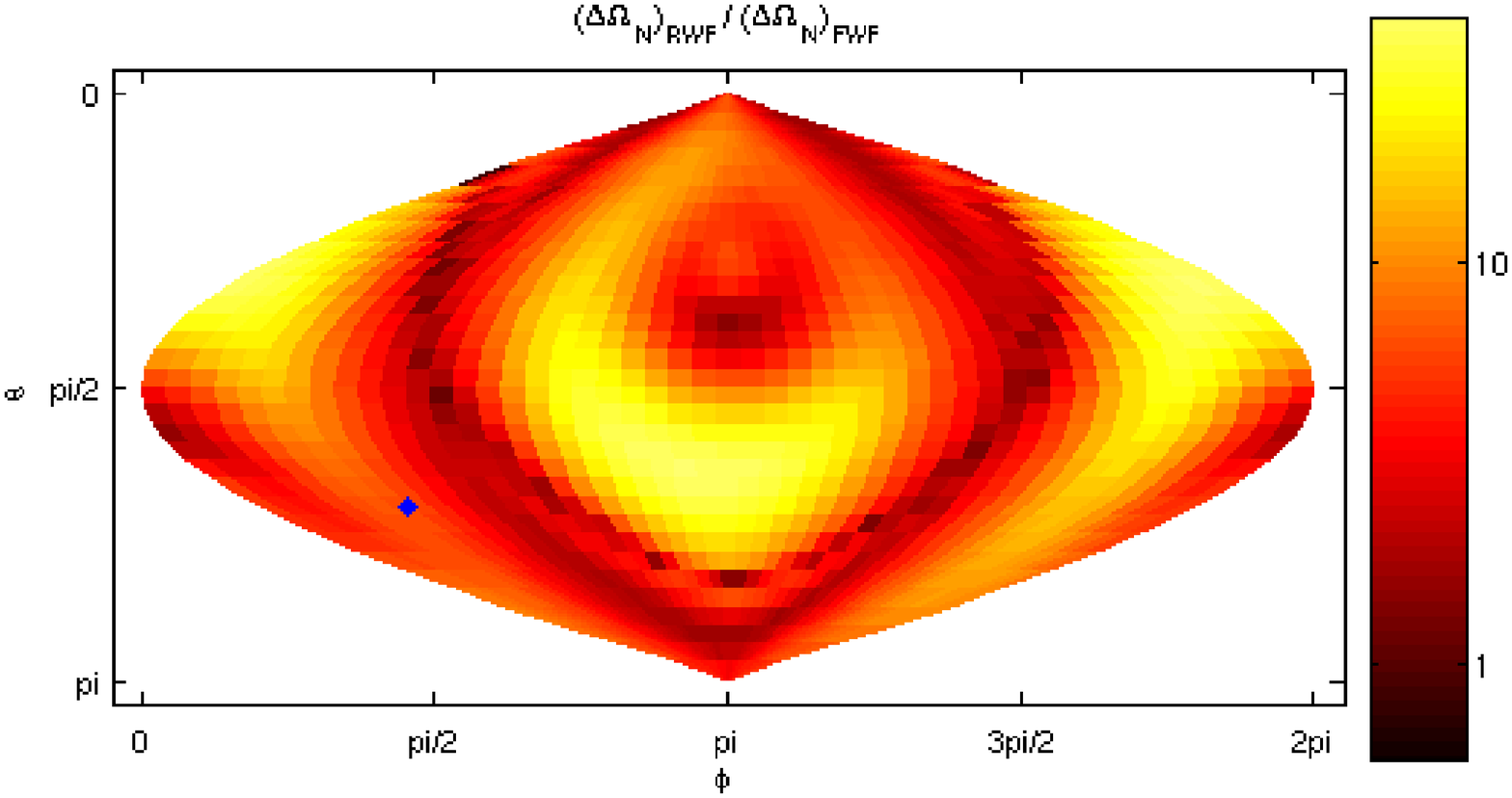}}
    \caption{Sky map of the gain in angular resolution for LISA
    observations of the final year of inspirals using full waveforms
    with harmonics versus restricted post-Newtonian waveforms with
    only the dominant harmonic, corresponding to the equal mass case
    ($m_1 = m_2 = 10^7\,M_{\odot}$, top) and a system with mass ratio
    of 10 ($m_1 = 10^7\,M_{\odot}$, $m_2 = 10^6\,M_{\odot}$,
    bottom). The sources are all at $z=1$, have the same orientation
    ($\cos \theta_L = 0.2$, $\phi_L = 3$) and zero spins
    $\beta=\sigma=0$. Figure reprinted with permission
    from~\cite{Trias:2007fp}. \copyright\
    \href{http://link.aps.org/abstract/PRD/v77/e024030}{The American
    Physical Society}.}
    \label{fig:SkyMap_GAIN}
\end{figure}}

\subsection{Tests of general relativity}

Gravitational wave measurements of black holes automatically  
test general relativity in its strong-field regime. Observations 
of the mergers of comparable-mass black holes will be rich in 
details of their strong-field interactions. If measurements can 
determine the masses and spins of the initial black holes, as well 
as the eccentricity and orientation of their inspiral orbit, then 
one would hope to compare the actual observed waveform with the output 
of a numerical simulation of the same system. If measurements can 
also determine the final mass and spin (say from the ringdown radiation) 
then one can test the Hawking area theorem (the final area must exceed
the sum of the areas of the initial holes) and the Penrose cosmic 
censorship conjecture (the final black hole should have $J/M^2<1$).

Observations of stellar mass black holes inspiraling into SMBHs, the
\emph{extreme mass ratio inspirals} (EMRIs), have an even greater
potential for testing general relativity. The stellar mass black hole
spends thousands of precessing (both of periastron and the orbital
plane) orbits along highly-eccentric trajectories and slowly inspirals
into the larger black hole.  The emitted gravitational radiation
literally carries the signature of the spacetime geometry around the
central object. So fitting the orbit to theoretical templates could
reveal small deviations of this geometry from that of Kerr. For
example, if we know (from fitting the waveform) the mass and spin of
the central black hole, then all its higher multipole moments are
determined. If we can measure some of these and they deviate from
Kerr, then that would indicate that either the central object is not a
black hole or that general relativity needs to be
corrected~\cite{Glampedakis:2005cf, Barack:2006pq}.

\subsubsection{Testing the post-Newtonian approximation}
\label{sec:gravitonmass}

Current tests of general relativity rely on experiments in the 
solar system (using the sun's gravitational field) and observations 
of binary pulsars. In dimensionless units, the gravitational 
potential on the surface of the sun is about one part in a million 
and even in a binary pulsar the potential that each neutron star 
experiences due to its companion is no more than one part in ten 
thousand. These are mildly relativistic fields, with the 
corresponding escape velocity being as large as a thousandth and 
a hundredth that of light, respectively.

Thus, gravitational fields in the solar system or in a binary 
pulsar are still weak by comparison to the largest possible values. 
Indeed, close to the event horizon of a black hole, gravitational 
fields can get as strong as they can ever get, with the 
dimensionless potential being of order unity and the escape 
velocity equal to that of the speed of light. Although general 
relativity has been found to be consistent with experiments in 
the solar system and observations of binary pulsars, phenomena 
close to the event horizons of black holes would be a great 
challenge to the theory. It would be very exciting to test 
Einstein's gravity under such circumstances.

The large SNR that is expected from SMBH binaries makes it possible to test Einstein's 
theory under extreme conditions of gravity~\cite{Arun:2006hn, Arun:2006yw}.
To see how one might test the post-Newtonian structure of Einstein's
theory, let us consider the waveform from a binary in the
frequency domain. Since an inspiral wave's frequency changes
rather slowly (adiabatic evolution) it is possible to apply a
stationary phase approximation to compute the Fourier transform
$H(f)$ of the waveform given in \Eqnref{eq:antenna response}:
\begin{equation}
H(f) = {\cal A}\, f^{-7/6} \exp\left [i \Psi(f) + i \frac{\pi}{4}\right ],
\end{equation}
with the Fourier amplitude ${\cal A}$ and phase $\Psi(f)$ given by
\begin{equation} 
{\cal A} = \frac{\cal C}{D\,\pi^{2/3}} \sqrt{\frac{5\nu}{24}} M^{5/6},\quad
\Psi(f) = 2\pi f t_C + \Phi_C + \frac{3}{128\,\nu}\sum_k \alpha_k\, \left (\pi M f \right )^{(k-5)/3}.
\label{eq:Fourier Phase}
\end{equation}
Here $\nu$ is the symmetric mass ratio defined before (see
Equation~\ref{eqn:chirpmassdef}), ${\cal C}$ is a function of the
various angles, as in Equation (\ref{eq:antenna response}), and $t_C$
and $\Phi_C$ are the fiducial epoch of merger and the phase of the
signal at that epoch, respectively.  The coefficients in the PN
expansion of the Fourier phase are given by
\begin{eqnarray}
\label{eq:psikvsmass1}
\alpha_0 &=&1,\ \ \ \
\alpha_1 =0,\ \ \ \
\alpha_2 =\frac{3715}{756}+\frac{55}{9}\nu ,\ \ \ \
\alpha_3 =-16 \pi,\nonumber\\
\alpha_4 &=&\frac{15293365}{508032}+\frac{27145}{504} \nu+\frac{3085}{72} \nu ^2,\ \ \ \
\alpha_5 =\pi \left(\frac{38645}{756}-\frac{65}{9}\nu\right)
\left[1+\ln\left(6^{3/2}\pi M\,f\right)\right],\nonumber\\
\alpha_6 &=&\frac{11583231236531}{4694215680}-\frac{640}{3}\pi
^2-\frac{6848}{21}\gamma
+\left(-\frac{15737765635}{3048192}+ \frac{2255}{12}\pi ^2\right)\nu
\nonumber\\
& + &\frac{76055}{1728}\nu^2-\frac{127825}{1296}\nu^3
-\frac{6848}{63}\ln\left(64 \pi M\, f\right),\nonumber \\
\alpha_7 & = & \pi\left( \frac{77096675}{254016}+\frac{378515}{1512}
\nu -\frac{74045}{756}\nu ^2\right).
\label{eq:psikvsmass2}
\end{eqnarray}
These are the PN coefficients in Einstein's theory; in an alternative
theory  of gravity they will be different.  In Einstein's
theory the coefficients depend only on the two mass parameters, the total mass $M$ and symmetric mass ratio $\nu$. One of the tests
we will discuss below concerns the consistency of the various 
coefficients. Note, in particular, that in Einstein's gravity the 0.5 PN
term is absent, i.e., the coefficient of the term $v$ is zero. Even
with the very first observations of inspiral events, it should
be possible to test if this is really so.

\Figref{fig:ToG} shows one such test that is possible with SMBH binaries~\cite{Arun:2006hn, Arun:2006yw}. 
The observation of these systems 
in LISA makes it possible to measure the parameters associated 
with different physical effects. For example, the rate at 
which a signal chirps (i.e., the rate at which its
frequency changes) depends on the binary's chirpmass.
Given the chirpmass, the length of the signal depends 
on the system's symmetric mass ratio (the ratio of reduced 
mass to total mass). Another example would be the scattering of 
gravitational waves off the curved spacetime geometry of the binary,
producing the \emph{tail effect} in the emitted signal, 
which is determined principally by the system's 
total mass~\cite{Blanchet:1993ec,Blanchet:1994ez}. 
Similarly, spin-orbit interaction, spin-spin coupling, 
etc. depend on other combinations of the masses.

The binary will be seen with a high 
SNR, which means that we can measure the mass 
parameters associated with many of these physical effects. 
If each parameter is known precisely, we can draw a 
curve corresponding to it in the space of masses. However, 
our observations are inevitably subject to statistical 
(and possibly systematic) errors. Therefore, each parameter 
corresponds to a region in the parameter space (shown in \Figref{fig:ToG} 
for the statistical errors only). If Einstein's theory 
of gravitation is correct,  the regions corresponding to 
the different parameters must all have at least one common 
region, a region that contains the true parameters of the 
binary. This is because Einstein's theory, or an alternative,
has to be used to project the observed data onto the space 
of masses. If the region corresponding to one or more 
of these parameters does not overlap with the common region 
of the rest of the parameters, then Einstein's theory, or 
its alternative, is in trouble.

In Brans--Dicke theory the system is expected to emit dipole radiation
and the PN series would begin an order $v^{-2}$ earlier than in 
Einstein's theory. In the notation introduced above we would have
an $\alpha_{-2}$ term, which has the form~\cite{BBW05a, BBW05b}
\begin{equation}
\alpha_{-2} = - \frac{5 {\cal S}^2}{84\,\omega_{\mathrm{BD}}}.
\end{equation}
Here $\cal S$ is the the difference in the scalar charges of
the two bodies and $\omega_{\mathrm{BD}}$ is the Brans--Dicke
parameter.  Although this term is formally two orders lower than the
lowest-order quadrupole term of Einstein's gravity (i.e., it is
${\cal O}(v^{-2})$ order smaller), numerically its effect will be 
far smaller than the quadrupole term because of the rather large
bound on $\omega_{\mathrm{BD}} \gg 1$. Nevertheless, its importance
lies in the fact that there is now a new parameter on which
the phase depends.  Berti, Buonanno and Will conclude that 
LISA observations of massive black-hole binaries will enable
scientists to set limits on $\omega_{\mathrm{BD}} \sim
10^4\mbox{\,--\,}10^5$.

A massive graviton theory would also alter the phase. The 
dominant effect is at 1~PN order and would change the coefficient 
$\alpha_2$ to
\begin{equation}
\alpha_2 \rightarrow \alpha_2  - \frac{128\nu}{3} 
\frac{\pi^2 D M}{\lambda_g^2 (1+z)},
\end{equation}
where $\nu$ is the symmetric mass ratio. 
This term alters the time of arrival of waves of different frequencies,
causing a dispersion, and a corresponding modulation, in the wave's phase,
depending on the Compton wavelength $\lambda_g$ and the distance $D$
to the binary. 
Hence, by tracking the phase of the inspiral waves, one can bound the graviton's
mass. Will~\cite{bss:will.98} finds that one can bound the mass to 
$1.7 \times 10^{13}$~km using ground-based detectors and
$1.7 \times 10^{17}$~km using space-based detectors, as also
confirmed by more recent and exhaustive calculations~\cite{BBW05a}.
These limits might improve if one takes into account the modulation
of the waveform due to spin-orbit and spin-spin coupling, but the
latter authors~\cite{BBW05a} looked at spinning, but nonprecessing, 
systems only.

\epubtkImage{m1m2_5105-m1m2.png}{%
  \begin{figure}[htbp]
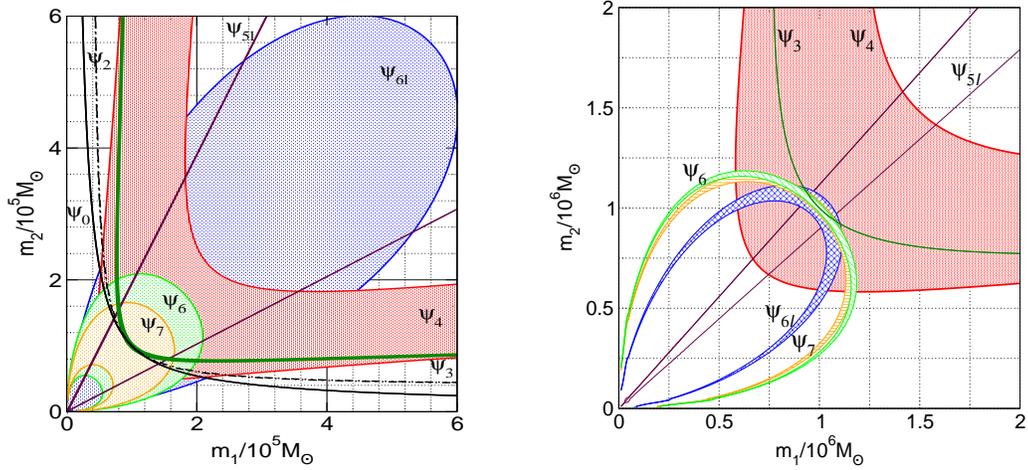

    \centerline{
      \includegraphics[width=2.4in, height=2.4in]{m1m2_5105}\hspace{1cm}
      \includegraphics[width=2.5in,angle=0]{m1m2}
    }
    \caption{By fitting the Fourier transform of an observed signal to
    a post-Newtonian expansion, one can measure the various
    post-Newtonian coefficients $\psi_k(m_1,m_2),\,
    k=0,\,2,\,3,\,4,\,6,\,7$ and coefficients of log-terms
    $\psi_{5l}(m_1,m_2)$ and $\psi_{6l}(m_1,m_2)$. In Einstein's
    theory, all the coefficients depend only on the two masses of the
    component black holes. By treating them as independent parameters
    one affords a test of the post-Newtonian theory. Given a measured
    value of a coefficient, one can draw a curve in the $m_1$--$m_2$
    plane. If Einstein's theory is correct, then the different curves
    must all intersect at one point within the allowed errors.  These
    plots show what might be possible with LISA's observation of the
    merger of a binary consisting of a pair of $10^6\,M_\odot$ black
    holes. In the right-hand plot all known
    post-Newtonian parameters are treated as independent, while in the
    left-hand plot only three parameters $\psi_0,\,
    \psi_2$ and one of the remainingpost-Newtonian parameter are
    treated as independent and the procedure is repeated for each of
    the remaining parameters. The large SNR in LISA for SMBH binaries
    makes it possible to test various post-Newtonian effects, such as
    the tails of gravitational waves, tails of tails, the presence of
    log-terms, etc., associated with these parameters. Left-hand
    figure adapted from~\cite{Arun:2006yw}, right-hand figure
    reprinted with permission from~\cite{Arun:2006hn}. \copyright\
    \href{http://link.aps.org/abstract/PRD/v74/e024006}{The American
    Physical Society}.}
    \label{fig:ToG}
\end{figure}}

\subsubsection{Uniqueness of Kerr geometry}

In Section~\ref{sec:gwsources} we pointed out that LISA should be able to 
see many hundreds of signals emitted by compact objects -- black holes, 
neutron stars, even white dwarfs -- orbiting around and being captured by 
massive black holes in the centers of galaxies. But for LISA to reach its
full potential, we must model the orbits and their emitted radiation 
accurately. Since the wave trains may be many hundreds or thousands of cycles 
long in the LISA band, the challenge of constructing template 
waveforms that remain accurate to within about one radian over the 
whole evolution  is significant. 

The range of mass ratios is also wide. LISA's central black holes 
might have masses between $10^3$ and $10^7\msolar$. Inspiraling 
neutron stars and white dwarfs might have masses between 0.5 and 
$2\msolar$. Inspiraling stellar-population black holes might be 
in the range of $7\mbox{\,--\,}50\msolar$, while intermediate-mass black holes 
formed by the first generation of stars (Population~III stars) 
might have masses around $300\msolar$ or even $1000\msolar$. So the 
mass ratios might be anything in the range $10^{-7}$ to 1. 

The techniques that must be used to compute these signals depend 
on the mass ratio. Ratios near one are treated by post-Newtonian 
methods until the objects are so close that only numerical relativity 
can follow their subsequent evolution. For ratios below 
$10^{-4}$ (a dividing line that is rather very uncertain and that depends on 
the bandwidth being used to observe the system, i.e., on how long the 
approximation must be valid for), systems are treated by 
fully-relativistic perturbation theory, expanding in the mass ratio. 
Intermediate mass ratios have not been studied in much detail yet; 
they can probably be treated by post-Newtonian methods up to a 
certain point, but it is not yet clear whether their final stages can 
be computed accurately by either numerical relativity or perturbation 
theory.

Post-Newtonian methods have been extensively discussed above. The 
basics of perturbation theory underlying this problem are treated 
in two Living Reviews~\cite{Living:Poisson, Living:Sasaki}. Once one 
has sufficiently good waveform templates, there remain the challenge 
of finding weak signals in LISA's noise. This depends on a number 
of factors, including the rates of sources. A recent study by a 
number of specialists~\cite{GairRates} has concluded that the event 
rate is high enough and the detection methods robust enough for us 
to be very optimistic that LISA will detect hundreds of these 
sources. In fact, the opposite problem might materialize: LISA might 
find it has a confusion problem for the detection of these sources, 
as for the galactic binaries. Recent estimates of the magnitude of 
the problem~\cite{BarackCutler2004} suggest that LISA's noise may at 
worst be raised effectively by a factor of two, but in return one 
gets a large number of sources of this kind.

\subsubsection{Quantum gravity}
\label{sec:quantum}

It seems inevitable that general relativity's description of nature will one 
day yield to a quantum-based description, involving uncertainties in 
geometry and probabilities in the outcome of gravitational observations. 
This is one of the most active areas of research in fundamental physics today, 
and there are many speculations about how quantum effects might show up 
in gravitational wave observations.

The simplest idea might be to try to find evidence for ``gravitons'' directly 
in gravitational waves, by analogy with the way that astronomers count individual 
photons from astronomical sources. But this seems doomed to failure. The waves that 
we can observe have very low frequency, so the energy of each graviton is 
extremely small. And the total energy flux of the waves is, as we have seen, 
enormous. So the number of gravitons in a detectable gravitational wave 
is far more than the number of photons in the light from a distant quasar. 

Quantum gravity might involve new gravity-like fields, whose effects might 
be seen indirectly in the inspiral signals of black holes or neutron stars, 
as we have noted above. String theory might lead to the production of 
cosmic strings, which might be observed through their gravitational wave 
emission~\cite{Damour:2004kw}. If our universe is just a 4-dimensional subspace of 
a large-scale 10 or 11-dimensional space, then dynamics in the larger space might 
produce gravitational effects in our space, and in particular
gravitational waves~\cite{RandallServant2007}.

It might be possible to observe the quantum indeterminacy of geometry directly 
using gravitational wave detectors, if Hogan's principle of holographic indeterminacy 
is valid~\cite{Hogan2008}. Hogan speculates that quantum geometry might be manifested 
by an uncertainty in the position of a beam splitter, and that this could be the 
explanation for an unexpectedly large amount of noise at low frequencies in the 
GEO600 detector. In this connection it is interesting to construct, from fundamental 
constants alone, a quantity with the dimensions of amplitude spectral
noise density $(S_h)^{1/2}$. This has units of $\mathrm{s}^{1/2}$, so one
can define the ``Planck noise power'' $S_\text{Pl} = t_\text{Pl} =
(G\hbar/c^5)^{1/2}$. Then the amplitude noise is $S_\text{Pl}^{1/2} =
(G\hbar/c^5)^{1/4} = 2.3\times 10^{-22}\mathrm{\ Hz}^{-1/2}$. This is
comparable to or larger than the instrumental noise in current
interferometric gravitational wave detectors, as shown in
Figure~\ref{fig:noise-curves}. This in itself does not mean that
Planckian noise will show up in gravitational wave detectors, but
Hogan argues that the particular design of GEO600 might indeed make it
subject to this noise more strongly than other large interferometers.

%==================================================================
\newpage

\section{Astrophysics with Gravitational Waves}
\label{sec:gwastronomy}

Gravitational radiation plays an observable role in the dynamics of many 
known astronomical systems.  In some, such as cataclysmic 
variables~\cite{Faulkner1971} and neutron-star--binary systems~\cite{TaylorWeisberg1989}, the role of gravitational radiation has been 
understood for years.  In others, such as young neutron stars~\cite{Andersson1998} and 
low-mass X-ray binaries~\cite{Bildsten:1998ey}, the potential importance of gravitational radiation has 
been understood only recently.  As further observations, particularly at 
X-ray wavelengths, become available, the usefulness of gravitational 
radiation as a tool for modelling astronomical  systems should increase~\cite{Watts2008}.

At this point in the progress of gravitational wave detection, 
the greatest emphasis in calculations of sources is on \emph{prediction}: 
trying to anticipate what might be seen. Not only is this important 
in motivating the construction of detectors, but it also guides details 
of their design and, very importantly, the design of data analysis 
methods. Historically, many predictions 
of emission strengths and the capability of detectors to extract 
information from signals have relied on estimates using the quadrupole 
formula. This was justifiable because, given the uncertainties in 
our astrophysical understanding of potential sources, more accurate 
calculations would be unjustified in most cases. 

But these rough estimates are now being replaced by more 
and more detailed source models where possible.
This applies particularly in two cases. One is binary orbits, where the point-mass 
approximation is good over a large range of observable frequencies, 
so that fully relativistic calculations (using the post-Newtonian methods 
described above) are not only possible, but are necessary for 
the construction of sensitive search templates in the data 
analysis. The second exception is the numerical simulations 
of the merger of black holes and neutron stars, where the 
dynamics is so complex that none of our analytic approximations 
offers us reliable guidance. In fact, these two methods are 
currently being joined to produce uniform models of signal evolution
over as long an observation time as the signal allows. From 
these models we not only improve detection algorithms, but we 
also learn much more about the kinds of information that 
detections will extract from the signals.

Once gravitational waves have been observed, there will of course be a welcome 
shift of emphasis to include \emph{interpretation}. The emphasis 
will be on extracting observable parameters (waveforms, polarizations, 
source location, etc.) from noisy data or data where (in the case of LISA) 
there is source confusion. These issues need considerably more attention 
than they have so far received.

\subsection{Interacting compact binaries} 

The first example of the use of 
gravitational radiation in modelling an observed astronomical system was 
the explanation by Faulkner~\cite{Faulkner1971} of how the activity of 
cataclysmic binary systems is regulated.  Such systems, which include 
many novae, involve accretion by a white dwarf from a companion star. 
Unlike accretion onto neutron stars, where the accreted hydrogen is 
normally processed quickly into heavier elements, on a white dwarf the 
unprocessed material can build up until there is a nuclear chain 
reaction, which results in an outburst of visible radiation from the system.   

Now, in a circular binary system that conserves total mass and angular 
momentum, a transfer of mass from a more massive to a less massive star 
will make the orbit shrink, while a transfer in the opposite direction 
makes the orbit grow.  If accretion onto a white dwarf begins with the 
dwarf as the less massive star, then the stars will draw together, and 
the accretion will get stronger.  This runaway process stops when the stars are 
of equal mass, and then accretion begins to drive them apart again. 
Astronomers observed that in this phase accretion in certain very close 
binaries continued at a more or less steady rate, instead of shutting 
off as the stars separated more and more.  Faulkner pointed out that 
gravitational radiation from the orbital motion would carry away angular 
momentum and drive the stars together.  The two effects together result 
in steady accretion at a rate that can be predicted from the quadrupole 
formula and simple Newtonian orbital dynamics, and which is in good 
accord with observations in a number of systems.

Binaries consisting of two white dwarfs in very tight orbits will be
direct LISA sources: we won't have to infer their radiation indirectly,
but will actually be able to detect it. Some of them will also be close enough 
to tidally interact with one another, leading in some cases to mass 
transfer. Others will be relatively clean systems in which the dominant 
effect will be gravitational radiation reaction.

Observations during the last decade have identified a number of such 
systems with enough confidence to predict that LISA should see their
gravitational waves. These are called verification binaries: if LISA 
does not see them then either the instrument is not working properly 
or general relativity is wrong! For a review of verification binaries, 
see~\cite{Vecchio2006}.

\subsubsection{Resolving the mass-inclination degeneracy}
\label{sec:inclination}

Gravitational-wave--polarization measurements can be very helpful
in resolving the degeneracy that occurs in the measurement of the
mass and inclination of a binary system. As is well known,
astronomical observations of binaries cannot yield the total mass
but only the combination $m \sin \iota$, where $\iota$ is the inclination of
the binary's orbit to the line of sight. However, measurement of
polarization can determine the angle $\iota$ since the polarization state
depends on the binary's inclination with the line of sight.

For instance, consider a circular binary system with total mass $M$ 
at a distance $D$. Suppose its orbital angular momentum vector 
makes an angle $\iota$ with the line of sight (the standard definition
of the inclination of a binary orbit). The two observed polarizations  
are given in the quadrupole approximation by \Eqref{eq:rpn spinless wave}. We can eliminate 
the distance $R$ between the stars that is implicit in the velocity $v=R\omega$ 
(where $\omega$ is the instantaneous angular velocity of the orbit, the derivative of 
the orbital phase function $\varphi(t)$) by using the Newtonian orbital
dynamics equation $\omega^2 = M/R^3$. Then we find
\begin{equation}
h_+ = \frac{2\nu M}{D} [\pi M f(t)]^{2/3} ( 1 + \cos^2 \iota )\, \cos [2\varphi(t)],\quad
h_\times = \frac {4\nu M}{D} [\pi M f(t)]^{2/3} \cos \iota\, \sin [2\varphi(t)],
\label{eqn:binaryradiation}
\end{equation}
where $M$ is the total mass of the binary and, as before, $\nu$ is the 
symmetric mass ratio $m_1m_2/M^2$. The frequency $f = \omega/\pi$ is the gravitational wave 
frequency, twice the orbital frequency. 
Notice that, consistent with \Eqnref{eqn:unequalbinary}, 
the masses of the stars enter these equations only in the combination $\mptr = \nu^{3/5}M$. 

It is clear that the ratio of the two polarization amplitudes determines the angle $\iota$. 
In this connection it is interesting to relate the polarization to the orientation. When the 
binary is viewed from a point in its orbital plane, so that $\iota=\pi/2$, then $h_\times = 0$; 
the radiation has pure $+$~polarization. From the observer's point of view, the motion of the
binary stars projected onto the sky is purely linear; the two stars simply go back and 
forth across the line of sight. This linear projected motion results in linearly polarized 
waves. At the other extreme, consider viewing the system down its orbital rotation axis, where 
$\iota = 0$. The stars execute a circular motion in the sky, and the polarization components 
$h_+$ and $h_\times$ have equal amplitude and are out of phase by $\pi/2$. This is circularly 
polarized gravitational radiation. So, when the radiation is produced in the quadrupole 
approximation, the polarization has a very direct relationship to the motions of the masses 
when projected on the observer's sky plane. If the radiation is produced by higher multipoles
it becomes more complex to make these relations, but it can be
done. For example, see~\cite{SchutzRicci} for the case of current
quadrupole radiation, which is emitted by the r-mode instability
discussed in Section~\ref{sec:rmode} below.

While a single detector is linearly polarized, it can still measure
the two polarizations if the signal has a long enough duration for the
detector to turn (due to the motion of the Earth) and change the
polarization component it measures. Alternatively, a network of three
detectors can determine the polarization and location of the source
even over short observation times.

Such a measurement would lead to a potentially very interesting interplay between gravitational 
and electromagnetic observations, with applications in the observations of 
isolated neutrons stars, binary systems, etc. And would lead to synergies, for 
example, between the LISA and Gaia~\cite{GAIA} missions.

\subsection{Black hole astrophysics}

Black holes are the most relativistic systems possible. Observing
gravitational waves from them, individually or in binaries, helps 
to test some of the predictions of general relativity in 
the strongly nonlinear regime, such as the
tails of gravitational waves, spin-orbit coupling induced precession, 
nonlinear amplitude terms, hereditary effects, 
etc~\cite{bss:mapping.kerr.bh, Blanchet:1994ez, Blanchet:1994ex, BSSAndBFS}.
They are also good test beds to constrain
other theories of gravity.  Gravitational waves -- emitted either during the
inspiral and merger of rotating SMBHs or when a 
stellar-mass compact object falls into a {\small SMBH} --
can be used to map the structure of spacetime and
test uniqueness theorems on rotating black holes~\cite{bss:mapping.kerr.bh}.
{\small LISA} will be able to see the formation of massive black holes at 
cosmological distances by detecting the waves emitted in the 
process~\cite{bss:mapping.kerr.bh}. We give below a brief discussion
of the physics that will follow from
the observation of gravitational waves from black holes.

\subsubsection{Gravitational waves from stellar-mass black holes}

Astronomers now recognize that there is an abundance of black 
holes in the universe.  Observations across the electromagnetic 
spectrum have located black holes in X-ray binary systems in 
our galaxy in the centers of star clusters, and in the centers of galaxies.  

These three classes of black holes have very different masses.  
Stellar black holes typically have masses of around $10\msolar$, 
and are thought to have been formed by the gravitational 
collapse of the center of a large, evolved red giant star, 
perhaps in a supernova explosion. Black holes in clusters have been  
found in the range of $10^4\msolar$, and are called intermediate-mass 
black holes. Black holes in galactic 
centers have masses between $10^6$ and $10^{10}\msolar$, and 
are called SMBHs. The higher masses are found in 
the centers of active galaxies and quasars. The history and method 
of formation of intermediate-mass and supermassive black holes 
are not yet well understood.

All three kinds of black hole can radiate gravitational waves.  
According to \Figref{fig:freq}, stellar black-hole radiation 
will be in the ground-based frequency range, while galactic 
holes are detectable only from space. Intermediate-mass black holes 
may lie at the upper end of the LISA band or between LISA and 
ground-based detectors. The radiation from an 
excited black hole itself is strongly damped, lasting only a 
few cycles at its natural frequency [see \Eqref{eqn:natfreq} 
with $R=2\,M$]:
\[f_{BH}\sim 1000 \fracparen{M}{10\msolar}^{-1} \mathrm{\ Hz}\].

\subsubsection{Stellar-mass black-hole binaries}

Radiation from stellar-mass black holes is expected mainly from 
coalescing binary systems, when one or both of the components 
is a black hole.  Although black holes are formed more rarely 
than neutron stars, the spatial abundance of binary systems 
consisting of neutron stars with black holes, or of two black 
holes, is amplified relative to neutron-star binaries because  
binary systems are much 
more easily broken up when a neutron star forms than when 
a black hole forms. When a neutron star forms, most of the 
progenitor star's mass ($6\msolar$ or more) must be expelled 
from the system rapidly.  This typically unbinds the binary: the companion 
star has the same speed as before but is held to the neutron 
star by only  a  fraction of the original gravitational 
attraction. Observed neutron-star binaries are thought to have 
survived because the neutron star was coincidentally given a kick against its 
orbital velocity when it formed. When a black hole forms, most of the original 
mass may simply go down into the hole, and the binary will 
have a higher survival probability. However, this argument 
may not lead to observable black hole binaries; there is a 
possibility that systems that would form black holes close 
enough to coalesce in a Hubble time do not become binaries, but 
rather the two progenitor stars are so close that they merge before 
forming black holes.

On the other hand, double black-hole binaries may in fact be formed
abundantly by capture processes in globular clusters, which appear to
be efficient factories for black-hole binaries~\cite{Zwart}. Being
more massive than the average star in a globular cluster, black holes
sink towards the center, where three-body interactions can lead to the
formation of binaries. The key point is that these binaries are not
strongly bound to the cluster, so they can easily be expelled by later
encounters.  From that point on they evolve in isolation, and
typically have a lifetime shorter than $10^{10}$~yrs.

The larger mass of stellar black-hole systems makes them 
visible from a greater distance than neutron-star binaries. If the abundance of 
binaries with black holes is comparable to that of neutron-star binaries, black hole events 
will be detected much more frequently than those involving neutron 
stars. They may even be seen by first-generation detectors in the 
S5 science run of the LSC (see Section~\ref{sec:present}), although 
that is still not very probable, even with optimistic estimates of 
the black-hole binary population.  It seems very 
possible, however, that the first observations of binaries by interferometers 
will eventually be of black holes.

More speculatively, black hole binaries may even be part of 
the dark matter of the universe.  Observations of Massive Compact Halo
Objects (MACHOs) -- microlensing of distant stars by compact objects
in the halo of our galaxy -- have indicated that up to half of the
galactic halo could be made up of dark compact objects of
$0.5\msolar$~\cite{MACHO, Sutherland1999}. This is difficult to
understand in terms of stellar evolution, as we understand it today:
neutron stars and black holes should be more massive than this, and
white dwarfs of this mass should be bright enough to have been
identified as the lensing objects. One speculative possibility is that
the objects were formed primordially, when conditions may have allowed
black holes of this mass to form.  If so, there should also be a
population of binaries among them, and occasional coalescences should,
therefore, be expected. In fact, the abundance would be so high that
the coalescence rate might be as large as one every 20~years in each
galaxy, which is higher than the supernova rate. Since binaries are
maximally non-axisymmetric, these systems could be easily detected by
first-generation interferometers out to the distance of the Virgo
Cluster~\cite{Sasaki1997}.

The estimates used here of detectability of black hole systems 
depend mainly on the radiation emitted as the orbit decays, 
during which the point-particle post-Newtonian approximation should 
be adequate.  But the inspiral phase will, of course, be followed 
by a burst of gravitational radiation from the merger of the black holes 
that will depend in detail on the masses and spins of the objects.  
Numerical simulations of such events will be used to interpret this 
signal and to provide templates for the detection of black holes too 
massive for their inspiral signals to be seen.  There is an abundance 
of information in these signals: population studies of the masses 
and spins of black holes, studies of typical kick velocities for 
realistic mergers, tests of general relativity.

\subsubsection{Intermediate-mass black holes}

Intermediate-mass black holes, with masses between $100\msolar$ and $10^4\msolar$,
are expected on general evolutionary grounds, but have proved hard to identify 
because of their weaker effect on surrounding stellar motions. Very recently~\cite{OmegaCenIMBH}
strong evidence has been found for such a black hole in 
the star cluster Omega Centauri. If such 
black holes are reasonably abundant, then they may be LISA sources when they 
capture a stellar-mass black hole or a neutron 
star from the surrounding cluster. For these merger events the mass ratio 
is not as extreme as for EMRIs, and so these are accordingly 
called IMRIs: Intermediate Mass-Ratio Inspirals. 

The problem of modelling the 
signals from these systems has not yet been fully studied. If these signals 
can be detected, they will tell us how important black holes were in the early 
stellar population, and whether these black holes have anything to do with the central 
black holes in the same galaxies.

\subsubsection{Supermassive black holes}
\label{sec:smbhobs}

Gravitational radiation is expected from SMBHs 
in two ways.  In one scenario, two massive black holes spiral 
together in a much more powerful version of the coalescence we 
have just discussed.  The frequency is much lower, in inverse 
proportion to their masses, and the amplitude is higher.  
\Eqref{eq:Fourier Phase} implies that the effective 
signal amplitude (which is what appears in the expression for
the SNR) is almost linear in the masses of the black holes, so 
that a signal from two $10^6\msolar$ black holes will have an 
amplitude $10^5$ times bigger than the signal from two 
$10\msolar$ holes at the same distance.  Even allowing for 
differences in technology, this indicates why space-based 
detectors will be able to study such events with a very 
high SNR, no matter where in the universe they occur.  
Observations of coalescing massive black-hole binaries will 
therefore provide unique insight into the behavior of strong 
gravitational fields in general relativity. 

The event rate for such coalescences is not easy to predict,  
but is likely to be large.  It seems that the central core 
of most galaxies may contain a black hole of at least 
$10^6\msolar$.  This is known to be true for our 
galaxy~\cite{ECKART1996} and for a very large proportion of 
other galaxies that are near enough to be studied in sufficient 
detail~\cite{Richstone}.  SMBHs (up to 
a few times $10^9\msolar$) are believed to power quasars and 
active galaxies, and there is a good correlation between the 
mass of the central black hole and the velocity dispersion of 
stars in the core of the host galaxy~\cite{GebhardtSMBHDispersion}.

If black holes are formed with their galaxies, in a single 
spherical gravitational collapse event, and if nothing happens to 
them after that, then coalescences will never be seen.  But 
this is unlikely for two reasons. First, it is believed that 
galaxies may have formed through the merger of smaller units, 
sub-galaxies of masses upwards of $10^6\msolar$.  If these 
units had their own black holes, then the mergers would have 
resulted in the coalescences of many of the black holes on a 
timescale shorter than the present age of the universe.  
This would give an event rate of several mergers per year 
in the universe, most of which would be observable by LISA, 
if the more massive black hole is not larger than about $10^7\msolar$.  
If the $10^6\msolar$ black holes were formed from smaller black holes 
in a hierarchical merger scenario, then the event rate could be 
hundreds or thousands per year.  The second reason is that 
we see large galaxies merging frequently.  Interacting galaxies 
are common, and if galaxies come together in such a way that 
their central black holes both remain in the central core, then 
dynamical friction with other stars will bring them close 
enough together to allow gravitational radiation to bring 
about a merger on a timescale of less than $10^{10}$~yrs. There
is considerable evidence for black hole binaries in a number of
external galaxies~\cite{Living:Merritt}. There is even a 
recent report of an SMBH having been ejected 
from a galaxy, possibly by the kick following a merger~\cite{Komossa:2008qd}
and of an SMBH binary that will coalesce in
about 10,000~yrs~\cite{Valtonen:2008aj}!

Besides mergers of holes with comparable masses, the capture 
of a small compact object by a massive black hole can also result 
in observable radiation. The tidal disruption of main-sequence 
or giant stars that stray too close to the black hole is thought to provide 
the gas that powers the quasar phenomenon. These disruptions are not 
expected to produce observable radiation. But the clusters will 
also contain a good number of neutron stars and stellar-mass black 
holes.  They are too compact to be disrupted by the black hole, even if 
they fall directly into it.  

Such captures, therefore, emit a gravitational wave signal that 
will be well approximated as that from a point mass near the 
black hole.  This will again be a chirp of radiation, but in 
this case the orbit may be highly eccentric.  The details of 
the waveform encode information about the geometry of spacetime 
near the black hole.  In particular, it may be possible to measure the 
mass and spin of the black hole and thereby to test the uniqueness 
theorem for black holes.  The event rate is not very dependent 
on the details of galaxy formation, and is probably high enough for 
many detections per year from a space-based detector~\cite{AmaroSeoane:2007aw}, provided 
that theoretical calculations give data analysts accurate 
predictions of the motion of these point particles over many hundreds
of thousands of orbits. These Extreme Mass-Ratio Inspiral sources (EMRIs)
are a primary goal of the LISA detector. By observing them, LISA will 
provide information about the stellar population near central black holes. 
When combined with modelling and spectroscopic observations, this will 
facilitate a deep view of the centers of galaxies and their evolution.

\subsection{Neutron star astrophysics}

\subsubsection{Gravitational collapse and the formation of neutron stars}

The event that forms most neutron stars is the gravitational collapse 
that results in a supernova.  It is difficult to predict the 
waveform or amplitude expected from this event.  Although detecting 
this radiation has been a goal of detector development for 
decades, little more is known about what to expect than 
30~years ago.  The burst might be at any frequency between 100~Hz
and 1~kHz, and it might be a regular chirp (from a rotating deformed 
core) or a more chaotic signal (from convective motions in the core).  
Considerable energy is released by a collapse, and on simple energetic 
grounds this source could produce strong radiation: if the emitted energy 
is more than about $0.01\msolar$, then 
second-generation detectors would have no trouble seeing events 
that occur in the Virgo Cluster. This energetic consideration drove 
the early development of bar detectors. 

But numerical simulations tell a different story, and it seems very likely
that radiation amplitudes will be much smaller, as described 
in \Secref{sec:gwsources}. 
Such signals might be detectable by second-generation detectors 
from a supernova in our galaxy, but not from much greater distances. When 
they are finally detected, the 
gravitational waves will be extremely interesting, providing our only 
information about the dynamics inside the collapse, and helping to 
determine the equation of state of hot nuclear matter.

If gravitational collapse forms a neutron star spinning very rapidly, 
then it may be followed by a relatively long period (perhaps 
a year) of emission of nearly monochromatic gravitational 
radiation, as the r-mode instability (\Secref{sec:instabilities}) forces 
the star to spin down to speeds of about 
100\,--\,200~Hz~\cite{Owen1999}. If as few as 10\% of all the neutron 
stars formed since star formation began (at a redshift of 
perhaps four) went through such a spindown, then they may have produced 
a detectable random background of gravitational radiation at frequencies 
down to 20~Hz~\cite{SchneiderStoch}.

\subsubsection{Neutron-star--binary mergers}

When two neutron stars merge, they will almost certainly have too much mass
to remain as a star, and will eventually collapse to a black hole, unless
they can somehow expel a significant amount of mass. The collision heats 
up the nuclear matter to a point where, at least initially, thermal pressure 
becomes significant. Numerical simulations can use theoretical equations of 
state (such as that of Lattimer and Swesty~\cite{Lattimer91}) to predict the merger
radiation, and observations will then test the nuclear physics assumptions that 
go into the equation of state. Simulations show that the choice of equation of state 
makes a big difference to the emitted waveform, as do the masses of the stars: 
there is no mass scaling as there is for black holes~\cite{AEINSMerger2008}.

When a neutron star encounters a black hole in a stellar compact binary 
merger, the star may not be heated very much by the tidal forces, and the 
dynamics may be governed by the \emph{cold} nuclear-matter equation of 
state, about which there is great uncertainty. Again, comparing observed 
with predicted waveforms may provide some insight into this equation of state. 
Simulations suggest that these systems may give rise to many of the observed short, 
hard gamma-ray bursts~\cite{Faber2006, ShibataUryu}. Simultaneous gravitational 
wave and gamma ray detections would settle the issue and open the way 
to more detailed modeling of these systems.

\subsubsection{Neutron-star normal mode oscillations}
\label{sec:nsmodes}

Gravitational wave observations at high frequencies of neutron-star vibrations 
may also constrain the cold-matter equation of state. 
In \Figref{fig:freq} there is a dot for the typical  neutron star.  
The corresponding frequency is the fundamental vibrational frequency 
of such an object.  In fact, neutron stars have a rich spectrum 
of nonradial normal modes, which fall into several 
families: f, g, p, w, and r-modes have all been studied. These 
have been reviewed by Andersson and Comer~\cite{Living:AnderssonComer}. 
If their gravitational wave emissions can be detected, then the 
details of their spectra would be a sensitive probe of their 
structure and of the equation of state of neutron stars, 
in much the same way that helioseismology probes the 
interior of the sun. Even knowing accurately the frequency and decay
time of just the fundamental $\ell=2$ f-mode would be enough to
eliminate most current equations of
state~\cite{AnderssonKokkotas1998}.

This is a challenge to ground-based 
interferometers, which have so far focussed their efforts on frequencies 
below 1~kHz. But Advanced LIGO and the upgraded GEO-HF detector 
(Section~\ref{sec:present}) may have the capability to perform 
narrow-banding and enhance their sensitivity considerably at frequencies up 
to perhaps 2~kHz, which could put the f-modes of neutron stars into range. 

The f-modes of neutron stars, which could 
be excited by glitches or by the nuclear explosions on accreting neutron 
stars that are thought to produce X-ray flares and soft gamma-ray repeater events. 
The rise-time of X-ray emission can be as short as a few milliseconds~\cite{2008Kes75}, which 
might be impulsive enough to excite acoustic vibrations. If the rise time of 
the explosion matches the period of the mode well enough, then a substantial fraction 
of the energy released could go into mechanical vibration, and almost all of this 
fraction would be carried away by gravitational waves, since other mode-damping mechanisms 
inside neutron stars are much less efficient. 

Radio-pulsar glitches seem to release energies of order $10^{35}$~J, and X-ray and gamma ray
events can be much more energetic. Using Equation~(\ref{eq:effective.amplitude}), we can estimate that 
the release of that much energy into gravitational waves at 2~kHz at a distance of 1~kpc would 
create a wave of effective amplitude around $3\times10^{-22}$. (The effective amplitude 
assumes we can do matched filtering, which in this case is not very difficult.) This kind of 
amplitude should be within the reach of Advanced LIGO
(Figure~\ref{fig:noise-curves}) and perhaps GEO-HF, provided they
implement narrowbanding. This will not be easy, either scientifically
or operationally, but the payoff in terms of our understanding of
neutron star physics could be very substantial.

Observations of these modes would 
immediately constrain the cold-matter nuclear equation of state in significant 
ways~\cite{AnderssonKokkotas1998, Living:AnderssonComer}.

In fact, modes of neutron stars may have already been observed in 
X-rays~\cite{WattsStrohmayer2007}. But these are likely to be crustal 
modes, whose restoring force is the shear strength of the crust. While 
the physics of the crust is interesting in itself, such observations 
provide only weak constraints on the interior physics of the neutron star.

\subsubsection{Stellar instabilities}
\label{sec:instabilities}

\paragraph{The CFS instability.}
In 1971 Chandrasekhar~\cite{ChandraCFS} applied the 
quadrupole formula to calculate the corrections to the 
eignefrequencies of the normal mode vibrations of 
rotating stars, and he found to his surprise that some modes were made 
unstable, i.e., that coupling to gravitational radiation could 
destabilize a rotating star.  Subsequent work by Friedman and 
Schutz~\cite{Friedman1978} showed that there was a key signature for 
the mode of a Newtonian star that would be unstable in general relativity.  
This was the pattern speed of the mode, i.e., the angular velocity at 
which the crests of the pattern rotated about the rotation axis 
of the star. If this speed was in the same sense as the rotation of 
the star, but slower than the star, then the mode would be unstable 
in a perfect-fluid star. This instability has come to be known as the 
CFS instability, after the three authors who explained it.  

The basic theory was developed for perfect-fluid stars.  However, 
Lindblom and Detweiler~\cite{LINDBLOM1977} showed that the effect 
of viscosity ran counter to that of radiation reaction, so that the 
instability was strongest in modes with the longest wavelengths, 
i.e., in the quadrupolar modes.  Full numerical calculations on 
Newtonian stellar models with realistic viscosity models showed~\cite{Lindblom:1995zs} that the standard fundamental and acoustic 
modes of rotating neutron stars were not vulnerable to this 
instability.  Subsequent work on fully relativistic models~\cite{Stergioulas:1997ja} has hinted that the instability may be 
stronger than the Newtonian models indicate, but it is still 
at the margins of astrophysical interest. 

\paragraph{The r-mode instability.} \label{sec:rmode}
The situation changed in 1997 when Andersson~\cite{Andersson1998} 
pointed out that there is another class of modes of Newtonian stars 
that should be unstable in the same way, but which had not been 
studied in this context before, the Rossby or r-modes.  
These are momentum-dominated modes, where the gravitational radiation 
comes from the current-quadrupole terms, rather than from the mass 
quadrupole.  Investigations by a number of authors~\cite{Lindblom:1998wf, Andersson1999, Owen1999} have shown that this 
instability could be very strong in hot, rapidly-rotating stars.  
This is particularly relevant to young neutron stars, which may 
well be formed with rapid spin and which will certainly be hot.  
For their first year, stars spinning faster than about 100~Hz could  
spin down to about 100~Hz by losing angular momentum to 
gravitational radiation. The instability might also operate 
in old accreting neutron stars, such as those in LMXB X-ray 
binaries (see the next section). 

However, the instability is, like other 
CFS instabilities, sensitive to viscosity, and there is great 
uncertainty about the amount of viscosity inside 
neutron stars~\cite{LindblomOwen2002, Lackey2006, Living:AnderssonComer}.

\subsubsection{Low-mass X-ray binaries}
\label{sec:LMXB} 

Observations by the 
Rossi satellite (RXTE) have given evidence that the class of X-ray 
sources called Low-Mass X-ray Binaries (LMXB's) contains neutron 
stars with a remarkably narrow range of spins, between perhaps 250~Hz 
and 320~Hz~\cite{vanderKlis1998}.  These are systems in which it 
is believed that neutron stars are spun up from the low angular 
velocities they have after their lifetime as normal pulsars to the 
high spins that millisecond pulsars have.  One would expect, therefore, 
that the spins of neutron stars in such systems would be spread 
over a wide range. The fact that they are not requires an explanation.

The most viable explanation offered so far is the suggestion of 
Bildsten~\cite{Bildsten:1998ey} that gravitational radiation limits 
the rotation rate.  The proposed mechanism is that anisotropic 
accretion onto the star creates a temperature gradient in the 
crust of the neutron star, which in turn creates a gradient in 
the mass of the nucleus that is in local equilibrium, and 
this in turn creates a density gradient that leads, via the 
rotation of the star, to the emission of gravitational radiation.  
This radiation carries away angular momentum, balancing that 
which is accreted, so that the star remains at an approximately 
constant speed.

According to 
the model, the gravitational wave luminosity of the star is 
proportional to the measured flux of X-rays, since the X-ray 
flux is itself proportional to the accreted angular momentum 
that has to be carried away by the gravitational waves.  If 
this model is correct, then the X-ray source Sco X-1 might be 
marginally detectable by advanced interferometers, and other 
similar systems could also be candidates~\cite{Watts2008}.

\subsubsection{Galactic population of neutron stars}

Neutron stars are known to astronomy through the pulsar phenomenon. As radio surveys 
improve, the number of known pulsars is pushing up toward 2000. There is a public 
catalogue on the web~\cite{PsrCat}. But 
the galactic population of neutron stars is orders of magnitude larger, perhaps as 
many as $10^8$. Most are much older than typical pulsars, which seem to stop 
emitting after a few million years. X-ray surveys reveal a number of unidentified 
point sources, which might be hot neutron stars, but older neutron stars are probably 
not even hot enough to show up in such surveys. 

Gravitational wave observations have the potential to discover more neutron stars, 
but in the foreseeable future the numbers will not be large. Spinning neutron stars 
can be found in searches for continuous-wave signals, but there is no a priori reason 
to expect significant deformations that would lead to large gravitational wave amplitudes. 
One mechanism, proposed by Cutler~\cite{CutlerToroidalB}, is that a large 
buried toroidal magnetic field could, by pulling in the waist of a spinning star, 
turn it into a prolate spheroid. This is classically unstable and would tip over and 
spin about a short axis, emitting gravitational waves. Millisecond pulsars could, in 
principle, be spinning down through the emission of gravitational waves in this way. Only 
deep observations by Advanced LIGO could begin to probe this possibility.

In fact, strong emission of gravitational waves is in some sense counterproductive, 
since it causes a neutron star to spin down and move out of the observing band quickly. 
This places important limits on the likely distribution of observable continuous-wave
amplitudes from neutron stars~\cite{KnispelAllen}. This is important input into the 
blind searches for such signals being conducted by the LSC.

Radio observations of pulsars have, of course, revealed a fascinating
population of binary systems containing neutron stars, including the
original Hulse--Taylor pulsar~\cite{HulseTaylor} and the double pulsar
PSR~J0737-3039~\cite{Lyne:doublepulsar}. But radio surveys only cover
a small fraction of our galaxy, so there may be many more interesting
and exotic systems waiting to be discovered, including neutron stars
orbiting black holes. In fact, not all neutron stars are pulsars, so
there are likely to be nearby binary systems containing neutron stars
that are not known as pulsars at all.

LISA has enough sensitivity to detect all such binaries in the galaxy whose 
gravitational wave emission is above 1~mHz, i.e., with orbital periods shorter 
than half an hour. Below that frequency, systems may just blend into the 
confusion noise of the white-dwarf background, unless they are particularly 
close. The Hulse--Taylor system 
is a bit below the LISA band, and even its higher harmonics are likely to be 
masked by the dense confusion noise of white-dwarf galaxies at low frequencies. 
Double pulsars should be detectable by LISA with low SNR (around five in five years) 
above the confusion background at a frequency of 0.2~mHz~\cite{KalogeraLorimer}. In 
all, LISA might detect several tens or even hundreds of double neutron-star systems, 
and potentially even a handful of double black hole binaries.

Neutron stars are the fossils of massive stars, and so a population census of binaries
can help normalize our galaxy's star-formation rate in the past. The mass distribution 
of such systems will also be of interest: do all neutron-star binaries have stars whose
masses are near $1.4\msolar$, or is this only true of systems that become pulsars? 
LISA observations are likely to illuminate many puzzles of stellar evolution.

Finally, it is possible to search for gravitational waves from individual spinning neutron 
stars in binary systems. Although more rare than isolated neutron stars, these systems 
might have a different history and a different distribution of amplitudes. Searches are 
planned by the LSC, but they are difficult to do, since the parameter space is even 
larger than for isolated pulsars.

\subsection{Multimessenger gravitational-wave astronomy}
\label{sec:multimessanger}

Multimessenger gravitational-wave astronomy refers to coordinated observations using
different kinds of radiation and information carriers:
electromagnetic, neutrino, cosmic ray, and gravitational wave. Joint
coordinated observing has much to offer gravitational wave detection,
by allowing it to target known interesting sources or locations,
thereby reducing the parameter space that must be searched and
improving the confidence of a detection. Even more importantly, the
information obtained from gravitational wave observations is typically
complementary to that which one can get from electromagnetic
astronomy, and so there are big science gains to be realized from
coordinated observations.

One can distinguish three broad classes of coordinated observations:
\emph{triggered} gravitational-wave searches, \emph{follow-up}
electromagnetic observations, and \emph{parameter refinement}. 

\begin{itemize}

\item Triggered searches use transient electromagnetic events, such as
  gamma-ray bursts, to narrow down the window of time for a search in
  the gravitational-wave data stream, and possibly also to restrict
  the ranges of various parameters. Since gravitational-wave--signal
  data is recorded, it is no problem to go back to data at the time of
  the triggering event and search it. This helps to lower the
  detection threshold, since gravitational wave events need, in this
  case, to be significant over a time scale of a few minutes rather
  than, say, an entire year.

\item Follow-up searches use a, possibly tentative, gravitational-wave
  detection to mark an area in the sky and a timeframe for an
  electromagnetic search. A very interesting example of this will
  occur with LISA, which will be able to predict the location and time
  of the coalescence of two SMBHs with reasonable
  accuracy at least a week in advance. This will allow sufficient time
  for telescopes in all the electromagnetic wavelengths to prepare to
  observe the event.

\item Parameter refinement refers to the use of electromagnetic
  obeservations of potential gravitational wave sources to improve the
  values of the parameters that must be used in the gravitational wave
  search. This has already been used in LSC searches, for example in
  trying to detect radiation from known radio pulsars: radio
  observations during the gravitational-wave observation period were
  used to track the changing frequency of the pulsar
  \cite{Abbott:2007tda}. 

\end{itemize}

Finding electromagnetic counterparts to gravitational wave
observations is important, of course, for learning about the nature of
the events. But it has a more subtle benefit: it generally improves
significantly the accuracy with which parameters can be estimated from
the gravitational wave observation. The reason is that one of the
biggest sources of parameter uncertainty is the sky location of a
gravitational wave source. Interferometers have broad antenna
patterns, which is helpful in that they can monitor essentially the
entire sky continuously, but which means that directional information
for transient events can come only from time delay information among
different detectors. The simple Rayleigh limit $\lambda_{gw}/D$ for
ground-based interferometers gives angular accuracies on the order of several
degrees, divided by the amplitude SNR (never smaller
than 5 for any reasonable detection). The covariance of angular errors
with uncertainties in other parameters (distance, polarization,
stellar masses, etc) is usually significant. Therefore, if a follow-up
electromagnetic observation can provide a more accurate position, this
can also improve the determination of all the other parameters
measured gravitationally.

Triggered searches are already being performed by the LSC for
gravitational waves associated with gamma-ray
bursts~\cite{Abbott:2008zz}. The nondetection of any gravitational
waves associated with the gamma-ray burst GRB~070201 showed that it
was not created by the merger of neutron stars in the nearby galaxy
M31, despite its positional coincidence on the
sky~\cite{LSC:GRB070201}. In addition, the gravitational wave
detectors are monitoring the triggers provided by both high-energy and
low-energy neutrino detectors in order to get instant warning of a
supernova in our galaxy or of some more exotic event further away. As
we have noted above, X-ray flares from neutron stars may signal
normal-mode radiation from acoustic vibrations.

Triggers may also allow the first detection of gravitational waves
from the normal modes of neutron stars, which as mentioned in
Section~\ref{sec:nsmodes}, would provide our first ``view''
inside these exotic objects. These triggers could be radio-pulsar
glitches, X-ray flares, or even the formation and subsequent ringdown
of a neutron star.

Follow-up observations of neutron-star--binary coalescence events are
likely to be particularly informative. It is possible that these
events are associated with short gamma-ray bursts, in which case most
events are missed because of the narrow beaming of the gamma
rays. Gravitational waves, by contrast, are emitted nearly
isotropically, so that they will pick out essentially all such events
within the range of the detectors, and astronomers can subsequently
search for afterglows and prompt X- and gamma-ray emission. The
ability to study such events from all aspect angles will help model
them reliably. Even if coalescences are not associated with gamma-ray
bursts, it is difficult to imagine that they will not produce visible
afterglows or other transient electromagnetic events that would
presumably not have been recognized before. The same considerations
apply to coalescences of neutron stars and black holes.

Gravitational wave events may also provide our first notice of a
gravitational collapse event, if the event is a strong radiator and is
too far away for neutrino detectors to see it. While supernova
simulations generally suggest that the amplitude of emitted
gravitational waves is small~\cite{Dimmelmeier2002}, numerical
simulations of the aftermath of neutron-star coalescence suggest the
possibility of very powerful gravitational-wave
emission~\cite{AEINSMerger2008}. While this event seems to lead
inevitably to a black hole, because the total mass is too large for a
single neutron star, neutron stars might occasionally be formed in
this way by mergers of white dwarfs, again with strong rotation and the
possibility of the emission of strong gravitational radiation. In this
connection the suggestion of Arons~\cite{Arons2003} that at least some
magnetars are formed in events of some kind that involve strong
magnetic field braking but also strong gravitational wave emission,
and that these events are the source of the ultra-high--energy cosmic
rays whose source, is so far unexplained~\cite{Watson2006}.

LISA offers particularly interesting opportunities for follow-up
observations with electromagnetic waves, beyond the direct monitoring
of the merger events for SMBHs mentioned above.
Because SMBHs often carry accretion disks, the
merger event may be followed by the turning on of accretion after a
delay of, perhaps, a year or so \cite{MilosavljevicPhinney2005}. The
merger may also cause a prompt shock in surrounding gas, due to the
essentially instantaneous loss of several percent of the gravity of the
central mass. These or similar effects may make it possible to
identify the galaxies that host LISA mergers, which in turn will allow
one to associate a redshift with the luminosity distance that the
gravitational wave event provides. This will be important for LISA's
cosmographic capabilities (next section).

LISA will look for close white-dwarf binaries in our galaxy and will probably 
see thousands of them.  White-dwarf binaries never reach the last stable orbit, 
which would occur at roughly 1.5~kHz for these masses. Instead they undergo a tidal 
interaction and can either disrupt at much lower frequency or end up as AM~CVns 
(see, for instance, \cite{Smak:1967,Nelemans:2004pp}).
In the latter case, we have a close white-dwarf binary with orbital
periods of minutes or hours, wherein the smaller of the two 
stars transfers mass to the more massive one. This mass loss 
leads to an \emph{increase} in the orbital period as a result of 
redistribution of the angular momentum.  So far only a handful
of AM CVn systems are known. LISA could potential discover a lot
more of these as their orbital periods are right in the heart of 
LISA's sensitivity band and simultaneous observation of these systems in the
gravitational and electromagnetic window has huge impacts on the science we can
learn about these end products of stellar evolution and their eventual fate.

For each resolved white-dwarf binary LISA can determine the orbital period
and the spatial orientation of the orbit, and it can
give a relatively crude position.  If the orbit is seen to decay
during the observation, LISA can determine the distance
to the binary. If the binary is known from optical or X-ray observations, then 
this can be very valuable additional information about the system, 
again complementary to that which is normally available from the 
electromagnetic observation. Even for systems that have not been 
identified, LISA's census of white-dwarf binaries will provide 
important statistics (on the mass function, distribution of separations, etc) 
that should lead to a better understanding of 
white-dwarf and binary evolution.

In the near term, one of the most practical applications of
multimessenger astronomy is to use electromagnetic observations to
refine the values of key search parameters for the gravitational wave
data analysis. This has been extensively discussed for possible
observations of low-mass X-ray binaries, as described in
Section~\ref{sec:LMXB}. Watts et al.~\cite{Watts2008} surveyed
the known ranges of parameters, such as spin rates and orbital
parameters, and concluded that they need to be narrowed considerably
if a practical search were to be possible, not just because of the
computer power required, but more importantly because of the loss of
significance if too large a parameter space has to be searched.

%==================================================================
\newpage

\section{Cosmology with Gravitational Wave Observations}
\label{sec:gwcosmology}

Gravitational wave observations may inform us about cosmology in at least two 
ways: by studies of individual sources at cosmological distances that give 
information about cosmography (the structure and kinematics of the universe) and 
about early structure formation, and 
by direct observation of a stochastic background of gravitational waves 
of cosmological origin. In turn, a stochastic background could either be 
astrophysical in origin (generated by any of a myriad of astrophysical 
systems that have arisen since cosmological structure formation began, 
as described in \Secref{sec:agwb}), 
or it could come from the Big Bang itself (generated by quantum processes 
associated with inflation or with spontaneous symmetry breaking in the 
extremely early universe, as described in \Secref{sec:cgwb}). 
The observation of a cosmic gravitational 
wave background (CGWB) is probably the most fundamentally important observation 
that gravitational wave detectors can make. But the astrophysical 
gravitational wave background (AGWB) also contains important information 
and may mask the CGWB over much of the accessible spectrum.

The detection of discrete sources at cosmological distances will require 
high sensitivity. Advanced ground-based detectors should be able to see a few  
individual sources (mainly stellar-mass black hole binaries) 
at redshifts approaching 1, with which they may be able to make a good 
determination of the Hubble constant. But LISA's observations of the  
coalescences of massive black hole binaries at all redshifts should make LISA a 
significant tool for cosmography. We examine cosmography 
measurements in \Secref{sec:cosmography}. These high-$z$ 
observations may also contain interesting information about early 
structure formation, such as the relationship between SMBH formation 
and galaxy formation. We have mentioned this already in \Secref{sec:smbhobs}.

Both kinds of detectors will search 
for a stochastic background in their own wave band. As we have seen earlier,
 LISA will almost 
certainly detect an AGWB from binary systems in our galaxy, and both LISA 
and advanced ground-based detectors may see a CGWB, if the more optimistic 
estimates of its strength are correct. But scientists are already sketching 
designs for a mission to follow LISA with much higher sensitivity, dedicated
to observing the CGWB from inflation. Stochastic searches are described in 
\Secref{sec:stoch_gwdets}.

Other detection methods are also being used to probe the spectrum of the background 
radiation at longer wavelengths. Pulsar timing observations (\Secref{sec:stoch_psr}) 
are already being used to 
set limits on the background at periods of a few years, and they will reach much 
greater sensitivity when coherent antenna arrays 
(like the Square Kilometer Array~\cite{SKA_fundamental, SKA}) are 
available. And observations of the temperature fluctuations of the cosmic
microwave background (\Secref{sec:stoch_cmb}) 
have the potential to reveal the gravitational wave content 
of the universe at the redshift of decoupling, which means at 
wavelength scales comparable to the size of the universe~\cite{QUaD_Data, Keating:2008}. 

Before examining the details of detection, we begin by examining the statistics of 
a random gravitational wave background. A good introduction to the 
theory of the CGWB is the set of lectures by Bruce Allen at the 1996 
Les Houches summer school~\cite{Allen1997}. The first paper of the LSC on searches for a stochastic background~\cite{LSC_Stoch1} also contains a brief introduction.

\subsection{Detecting a stochastic gravitational wave background}
\label{sec:gwbackground}

\subsubsection{Describing a random gravitational wave field}
\label{sec:describebackground}

By definition, a stochastic background of gravitational waves is 
a superposition of waves arriving at random times and from 
random directions, overlapping so much that individual waves are not 
identifiable. We assume that there are so many sources (either astrophysical sources
or the quantum fluctuations that create the CGWB) that individual ones
are not distinguishable. Such a gravitational wave field 
will appear in detectors as a time-series noise, 
which by the central limit theorem should have a Gaussian-normal distribution 
function if there are enough overlapping sources. 
This kind of background will compete 
with instrumental noise. It will be detectable by a single detector, if it is 
stronger than instrumental noise, but a weaker background could still be detected 
by using a pair of detectors and looking for a correlated component of their 
``noise'' output, on the assumption that their instrumental noise is not correlated.

As a random phenomenon, the gravitational wave fields at two 
different locations are uncorrelated, because gravitational waves arrive from all 
directions and at all frequencies. It might, therefore, be thought that two 
detectors' responses would be correlated only if they were located at the same 
position. But if one considers one component of the wave field with a single 
frequency, then it is clear that there will be strong 
correlations between points if they are separated along the wave's propagation 
direction by much less than a 
wavelength. We shall see that these frequency-dependent correlations 
allow one to 
detect a background by cross-correlating the output of two separated detectors, 
albeit with less sensitivity than if they were co-located. We shall consider 
cross-correlation as a detection method in \Secref{sec:stoch_gwdets}.

Random gravitational waves are conventionally described 
in terms of their energy density spectrum $\rhogw(f)$, rather than 
their mean amplitude. It is convenient 
to normalize this energy density to the critical 
density $\rho_c$ required to close the universe, which is given in terms 
of the Hubble constant $H_0$ as 
\[\rho_c=3H_0^2/8\pi G.\]  
We then define 
\begin{equation}\label{eqn:cosdef}
\Omega\textsub{gw}:= \oderiv{\rhogw/\rho_c}{\ln f}.
\end{equation}
This can be interpreted as the fraction of the closure energy density that is in 
random gravitational waves between the frequency $f$ and $e\times f$.  
If the source of radiation is \emph{scale-free} (which means that 
there is no preferred length or time scale in the process), then 
it will produce a power-law spectrum, i.e., one in which 
$\Omega\textsub{gw}(f)$ depends on a power of $f$. Inflation, 
as we describe below, predicts
a \emph{flat} energy spectrum, one in which $\Omega\textsub{gw}$ 
is essentially independent of frequency~\cite{Allen1997}.

The energy in the cosmological background is, of course, related to the 
spectral density of the noise that the background would produce in 
a gravitational wave detector. Since we describe the gravitational wave 
noise in terms of amplitude rather than energy, there are scaling factors 
involving the frequency between the two. An isotropic gravitational 
wave background incident on an interferometric detector will induce 
a strain spectral noise density equal to~\cite{Thorne1987, Allen1997} 
\begin{equation}\label{eqn:omega_strain}
S\textsub{gw}(f) = \frac{3H_0^2}{10\pi^2} f^{-3}\; \Omegagw(f).
\end{equation}
Note that the explicit dependence on frequency is $f^{-3}$: two factors 
come from the relation of energy and squared-strain, and one factor from 
the fact that $\Omegagw$ is an energy distribution per unit logarithmic 
frequency. Note also that there are no explicit factors of $c$ or $G$ 
needed in this formula if one wants to work in nongeometrized units.

If we scale $H_0$ by $h_{100}=H_0/(100\;\text{km}\;\text{s}^{-1}\; \text{Mpc}^{-1})$, 
and we note that $100\;\text{km}\;\text{s}^{-1}\; \text{Mpc}^{-1}=
3.24\times10^{-18}\; \text{s}^{-1}$, then this equation implies that the strain noise is 
\begin{equation}\label{eqn:shstoch}
S\textsub{gw}^{1/2} = 5.6\E{-22}\;\text{Hz}^{-1/2}\; \Omegagw^{1/2}\fracparen{f}{100\;\text{Hz}}^{-3/2}h_{100}.
\end{equation}

\subsubsection{Observations with gravitational wave detectors}
\label{sec:stoch_gwdets}

To be observed by a single gravitational wave detector, the gravitational 
wave noise must be larger than the instrumental noise. This is a bolometric method of detection 
of the background, and it requires great confidence in the understanding 
of the detector, in order to believe that the observed noise is external.  
This is how the cosmic microwave background was originally 
discovered in a radio telescope by Penzias and Wilson.

If there are two detectors, then one may be able to get better sensitivity 
by cross-correlating their output, as mentioned in
Section~\ref{sec:crosscorrelation} above. This works best when the two
detectors are close enough together to respond to the same random wave
field. Even when they are separated, however, they are correlated well
at lower frequencies. 

From Equation~\ref{eqn:shstoch} and the discussion in
Section~\ref{sec:crosscorrelation} it is straightforward to deduce
that two co-located detectors, each with spectral noise density $S_h$
and fully uncorrelated instrumental noise, observing over a bandwidth
$f$ at frequency $f$ for a time $T$, can detect a stochastic
background with energy density 
\begin{equation}\label{eqn:cosbackcorr}
  \Omegagw^{1/2}h_{100} = \fracparen{S_h^{1/2}}{3.1\times
  10^{-18}\;\text{Hz}^{-1/2}}\fracparen{f}{10\;\text{Hz}}^{5/4}\fracparen{T}{3\;\text{yrs}}^{-1/4}.
\end{equation}

The two LIGO detectors (separated by about 
10~ms in light-travel time) are reasonably well placed for performing
such correlations, particularly when upgrades push their lower 
frequency limit to 20~Hz or less. Two co-located first-generation LIGO 
instruments operating at 100~Hz could, in a one-year correlation, reach 
a sensitivity of $\Omegagw \sim 1.7\times 10^{-8}$. But the separation 
of the actual detectors takes its toll at this frequency, so that 
they can actually only reach $\Omegagw \sim 10^{-6}$.  Advanced LIGO 
may improved this by two or three orders of magnitude, going well 
below the nucleosynthesis bound. 
The third-generation instrument ET, with instrumental noise as 
shown in Figure~\ref{fig:noise-curves}, 
can go even deeper. Two co-located ETs, observing at 10~Hz for three years, 
could reach $\Omegagw \sim 10^{-12}$. At this frequency the detectors could be 
as far apart as 5000~km without a substantial loss in correlation sensitivity. 
The numbers given here are reflected in the curves in Figure~\ref{fig:noise-curves}.

Correlation searches are also possible between resonant detectors or 
between one resonant and one interferometric detector~\cite{Astone:1994cz}. 
This has been implemented with bar detectors~\cite{Astone1999} and between 
LIGO and the ALLEGRO bar detector~\cite{Whelan2003}. 

LISA does not gain by a simple correlation between any two of its 
independent interferometers, since they share a common arm, which contributes
common noise that competes with that of the background. A gravitational 
wave background of $\Omegagw\sim 10^{-10}$ would compete with LISA's 
expected instrumental noise. However, using 
all three interferometers together can improve things for LISA at low
frequencies, assuming that the LISA instrumental noise is
well behaved~\cite{Hogan:2001}. This might enable LISA to go below
$10^{-11}$.

\subsubsection{Observations with pulsar timing}
\label{sec:stoch_psr}

Other less-direct methods are also being used to search for primordial
gravitational waves. As we saw in \Secref{sec:pulsartiming}, pulsar
timing can, in principle, detect gravitational-wave--induced
fluctuations in the arrival times of pulses. Millisecond pulsars are
such stable clocks when averaged over years of observations that they
are being used to search for gravitational waves with periods longer
than one year. A single pulsar can set limits on a stochastic
background by removing the slow spindown and looking for random timing
residuals. Although one would never have enough confidence in the
stability of a single pulsar to claim a detection, this sets upper
limits in the important frequency range below that accessible to
man-made instruments. The best such limits are on pulsar PSR~B1855+09,
with an upper limit (at 90\% confidence) of $\Omegagw <
4.8\times10^{-9}$ at $f=4.4\mathrm{\ nHz}$~\cite{Millisecond}.

Arrays of pulsars offer the possibility of cross-correlating their 
fluctuations, which makes it possible to distinguish between intrinsic variability
and gravitational-wave--induced variability. Pulsars are physically separated 
by much more than a wavelength of the gravitational waves even with periods 
of 10~yrs, so that the correlated fluctuations come from the wave amplitudes at 
Earth. It will soon be possible to monitor many pulsars simultaneously 
with multibeam instruments, as mentioned in \Secref{sec:pulsartiming}. 
This method could push the limits on $h_c\equiv (fS_{\mathrm{gw}})^{1/2}$ 
[cf.\ \Eqref{eqn:omega_strain}] down to $10^{-16}$ at 10~nHz~\cite{Kramer:2003xs}, 
which translates into a limit on $\Omegagw$ of around $10^{-12}$.

\subsubsection{Observations using the cosmic microwave background}
\label{sec:stoch_cmb}

Observations of the cosmic microwave background (CMB) may in fact make the first
detections of stochastic (or any other!) gravitational waves. The temperature
fluctuations first detected by the Cosmic Background Explorer 
(COBE)~\cite{SMOOT1992} and more recently measured with great 
precision by the Wilkinson Microwave Anisotropy Probe 
(WMAP)~\cite{BENNETT2003} are produced by both density perturbations
and long-wavelength gravitational waves in the early universe (see the 
next Section~\ref{sec:cgwb}). Inflation 
suggests that the gravitational wave component may be almost as large as 
the density component, but it can only be separated from the density 
perturbations by looking at the polarization of the 
cosmic microwave background~\cite{KeatingPGWB_CMB}. WMAP made the first  
measurements of polarization~\cite{WMAP_3}, but it did not have the 
sensitivity to see the weak imprint of gravitational waves, which appears 
in the \emph{B-component} of the polarization, the part that is 
divergence-free on the whole sky. The best limits on the B-component
so far (early 2008) have been made by the QUaD\epubtkFootnote{QUaD stands for 
QUEST (Q and U Extragalactic Survey Telescope) at DASI (Degree Angular 
Scale Interferometer.} detector~\cite{QUaD_Data},
a cryogenic detector that operated for three seasons in Antarctica.  These have
not yet shown any evidence for gravitational waves. Results are expected soon
from the Background Imaging of Cosmic Extragalactic Polarization
(BICEP) detector, also in Antarctica~\cite{Keating:2008}. The next satellite
to study the microwave background will be Planck, due for launch by the 
European Space Agency in 2009~\cite{PLANCK}.

The gravitational waves detectable in the CMB have wavelengths a good 
fraction of the horizon size at the time of decoupling, and today they
have been redshifted to much longer wavelengths. They are, therefore, 
much lower frequency than the radiation that would be observed directly 
by LISA or ground-based detectors, or even by pulsar timing.

\subsection{Origin of a random background of gravitational waves}

\subsubsection{Gravitational waves from the Big Bang}
\label{sec:cgwb}

Gravitational waves have traveled almost unimpeded through the 
universe since they were generated.  The cosmic microwave 
background~\cite{BENNETT2003} is a picture of the universe at a 
time $3\times10^5$~yrs after the Big Bang, and studies of 
nucleosynthesis~\cite{Steigman:2007xt} (how the primordial hydrogen, 
helium, deuterium, and lithium were created) reveal conditions in 
the universe a few minutes after the Big Bang. Gravitational 
waves, on the other hand, were produced at times earlier 
than $10^{-24}$~s after the Big Bang.  Observing this background 
would undoubtedly be one of the most important measurements that 
gravitational wave astronomy could make. It would provide a 
test of inflation, and it would have the potential to give 
information about the fundamental interactions of physics at 
energies far higher than we can reach with accelerators.

The most well-defined predictions about the energy in the cosmological 
gravitational wave background come from inflationary models.
Inflation is an attractive scenario for the early universe because,  
among other things, it provides a natural mechanism for producing the initial 
density perturbations that evolved into 
galaxies and galaxy clusters as the universe expands. These perturbations 
start out as quantum fluctuations in the (hypothetical) scalar inflaton 
field that is responsible for the inflationary expansion of the universe. 
The fluctuations are parametrically amplified by the 
expansion~\cite{Grishchuk:1974ny, Mukhanov1992, Allen1997} and lead
to fluctuations in the density of normal matter after inflation ends. 

Several strands of evidence -- among them the statistical distribution
of density perturbations seen in the cosmic microwave background (most
recently by WMAP~\cite{SPERGEL2006}), the present distribution of
galaxies~\cite{Lahav2004}, and numerical simulations of structure
formation in the early universe~\cite{VIRGO} -- are fully consistent
with the now-standard model of a universe dominated by dark energy and
whose matter density is dominated by some kind of cold (i.e., massive)
dark matter particles~\cite{Sumner2002} with density perturbations
consistent with those that inflation could have produced.

The scalar inflaton fluctuations are accompanied by tensor quantum
fluctuations in the gravitational field that similarly get amplified
by inflation and form a random background~\cite{ALLEN1988,
  Allen1997}. Different models of inflation make different predictions
about the relative strength of the scalar and tensor components.

Although inflation is in excellent agreement with observation, other
mechanisms in the early universe may have led to the additional
production of gravitational waves. Defects that arise from symmetry
breaking as the presumed early unified interactions separate from one
another can lead to \emph{cosmic strings}~\cite{Shellard2000}, which
can produce both a continuous observable gravitational wave
background~\cite{CALDWELL1996} and characteristic isolated bursts of
gravitational waves~\cite{Damour:2000wa, Damour:2001bk,
  Damour:2004kw}. String theory~\cite{BRUSTEIN1995, Buonanno1997} and
brane theory~\cite{HOGAN1998, Maartens2004} may also provide
mechanisms for generating observable radiation.

The various models usually predict significantly different spectra for
background radiation. Standard inflationary models predict 
that the spectrum of $\Omega\textsub{gw}$ should  be nearly flat, 
independent of frequency, but variants exist that allow a spectrum that rises 
with frequency (positive spectral index) or falls.  Symmetry-breaking and brane model 
cosmologies can make very different predictions, even  leading to narrow spectral features.
It is, therefore, important to measure the spectrum at as many frequencies as 
possible. Limits on power at one frequency (such as at the very low-frequency end in the 
cosmic microwave background) do not necessarily predict the power at other 
frequencies (such as at ground-based frequencies, a factor $10^{20}$ times higher).

It is even possible that there will be a feature in the spectrum in 
the observing band of ground-based or space-based detectors.  
In standard cosmologies, the radiation observable by LISA (1~mHz) 
had a wavelength comparable to the (then) horizon size 
at around the time when the temperature of the universe was equal to 
the electroweak symmetry-breaking energy. If electroweak symmetry breaking 
led to a first-order phase transition, where density fluctuations occurred on 
the length scale of the typical symmetry domain size, then it is likely 
that these density fluctuations produced gravitational waves with wavelengths 
of the size of the horizon, which would be in the LISA band
today~\cite{Megevand2004}. Detection of this radiation would have deep
implications for fundamental physics.

The other expected phase transition is the GUTs (Grand Unified Theory)
transition, whose energy might have been $10^{13}$ times higher. Any
gravitational radiation from this transition today would then be at a
frequency $10^{13}$ times that from the electroweak transition, i.e.,
at centimeter wavelengths. This is one motivation for building
microwave-based table-top detectors aimed at high
frequencies~\cite{Cruise2006}. For this radiation to be observable by
standard interferometers, the GUTs transition would have to have an
energy $10^7$ times smaller than expected, i.e., around $10^9$~GeV. We
shall have to wait for observations at these frequencies to tell us if
it is there!

In addressing the possibility of new physics, observation of gravitational waves in 
the cosmic microwave background would play a unique role. These waves 
originated long after nucleosynthesis, at energies where physics is 
presumably well understood. They would, therefore, normalize 
the amount of power in the initial tensor perturbations.  
Then observations at higher frequencies can use this normalization 
to measure the excess energy due to any exotic effects 
due to string theory, phase transitions, or other unknown 
physics~\cite{Grishchuk:1996ek, Buonanno1997}.

Pulsar timing arrays (see Section~\ref{sec:stoch_psr}) will also be used to 
search for a CGWB at frequencies of a few nanoHertz. As for the microwave background, 
the physics of the universe when gravitational waves at these frequencies 
originated is well understood, so they could be used to normalize the 
spectrum. If the power at pulsar frequencies and that in the microwave 
background are not consistent, then this could indicate something about 
the conditions in the universe \emph{before} inflation began.

The predicted spectrum from inflation, strings, and symmetry breakings
is highly nonthermal. Any thermal radiation produced in the Big Bang
(for example, if, hypothetically, there was some kind of equipartition between 
gravitational degrees of freedom and other fields in the initial 
data at the singularity, whatever that might mean!) would have 
been redshifted away to unobservability by the subsequent inflationary
expansion. If inflation did not in fact occur, then this radiation today  
would  have a temperature only a little below that of the cosmological 
microwave background.  So far no instrument has been proposed that would 
be sensitive to this radiation, but its detection would presumably be inconsistent
with inflation.

\subsubsection{Astrophysical sources of a stochastic background}
\label{sec:agwb}

After galaxy formation, it is possible that many systems arose that
have been radiating gravitational waves in the bands observable by
pulsar timing, LISA, and ground-based detectors. There are likely to
be strong extra-galactic backgrounds in the LISA band from compact
binary systems, which would limit searches for a
CGWB~\cite{Schneider:2000sg} by LISA, even if the sensitivity were
better. At lower frequencies, even down to pulsar timing frequencies,
black hole binaries may make the strongest background, while at
frequencies above the LISA band (i.e., above 0.1~Hz) the universe
should be relatively free of serious backgrounds~\cite{Ferrari,
  UngarelliVecchio2001a}.

In the LISA band our galaxy is a strong source of
backgrounds~\cite{Hils:1990vc}. This presents a serious confusion
noise in searching for other sources at frequencies below 1~mHz. It
should be possible to distinguish this from a CGWB by its intrinsic
anisotropy~\cite{UngarelliVecchio2001b}.

\subsection{Cosmography: gravitational wave measurements of
  cosmological parameters}
\label{sec:cosmography}

Since inspiral signals are standard candles~\cite{SCHUTZ1986}, as 
described in Section~\ref{sec:gwphysics}, observations of
massive black hole coalescences at cosmological distances 
by space-based  detectors can facilitate an accurate determination of the
distance to the source. Our earlier expressions for the chirp waveform 
can be generalized to the cosmological case (a source at redshift $z$) 
by multiplying all masses by $1+z$ and by replacing the physical distance $D$ 
by the cosmological luminosity distance 
$D_L$~\cite{KrolakSchutz1987}. If the wave amplitude, 
frequency, and chirp rate of the binary can be measured, then its 
luminosity distance can be inferred. It is not, however, possible to infer the 
redshift $z$ from the observed signal: the scale-invariance of black hole 
solutions means that a signal with a redshift of two and a chirp mass ${\cal M}$
looks identical to a signal with no redshift and a chirp mass of ${\cal M}/3$. 
To use these distance measures for cosmography, one has to obtain redshifts 
of the host galaxies.

Before considering how this might be done, 
we should ask about the accuracy with which the distance can be measured. 
The relative error in the distance is dominated by the relative error in 
the measurement of the intrinsic amplitude of the gravitational wave, 
because the masses will normally be much more accurately measured (by 
fitting the evolving phase of the signal) than the 
amplitude. Several factors contribute to the amplitude uncertainty:

\begin{itemize}

\item {\bf Signal-to-noise ratio.} The intrinsic measurement uncertainty in the
  amplitude of the detector's response is simply the inverse of the
  SNR. Since LISA can have an SNR of several thousand when it observes
  an SMBH coalescence at high redshift, LISA has
  great potential for cosmography.

\item {\bf Position error.} From the detector response one must infer the
  intrinsic amplitude of the wave, which means projecting it on the
  antenna pattern of the detectors. This requires a knowledge of the
  source position, and this will be potentially a bigger source of
  uncertainty because the sensitivity of LISA depends on the location
  of the source in its antenna pattern. Recent work~\cite{Lang:2007ge,
    Babak:2008bu} has shown that LISA may be able to achieve position
  accuracies between one and ten arcminutes. At, say, three arcminutes error,
  the amplitude uncertainty will be of order 0.1\%. This error can be
  reduced to the SNR-limited error if the source can be
  identified. Although the coalescence of two SMBHs
  itself may not have an immediate effect on the visible light from a
  galaxy, the host galaxy might be identifiable either because it
  shows great irregularity (mergers of black holes follow from mergers
  of galaxies) or because some years after the merger an X-ray source
  turns on (accretion will be disrupted by the tidal forces of the
  orbiting black holes, but will start again after they
  merge)~\cite{MilosavljevicPhinney2005}. Other effects that might
  lead to an identification include evidence that stars have been
  expelled from the core of a galaxy, fossil radio jets going in more
  than two directions from a common center, and evidence for accretion
  having stopped in the recent past.

\item {\bf Microlensing.} If the source is at a redshift larger than one, as
  we can expect for LISA, then random microlensing can produce a
  magnification or demagnification on the order of a few
  percent~\cite{HH,BBHSirens}. The measured intrinsic amplitude then
  does not match the amplitude that the signal would have in an ideal
  smooth cosmology.

\end{itemize}

The relatively small error boxes within which the LISA coalescences can 
be localized are promising for identifications, especially if the 
X-ray indicators mentioned above pick out the host in the error box. 
These factors and their impact on cosmography measurements have been 
examined in detail by Holz and Hughes~\cite{HH}, who coined the term
``standard siren'' for the chirp sources whose distance can be 
determined by gravitational wave measurements. The potential for 
cosmographic measurements by advanced ground-based detectors 
have been considered in a further paper by the same authors and 
collaborators~\cite{BBHSirens}. Nearby coalescences and IMRIs should provide
an accurate determination of the Hubble Constant~\cite{Living:Jackson,
  MacLeodHogan2008}. Perhaps the most interesting measurement 
will be to characterize the evolution of the dark energy, which 
is usually characterized by inserting a parameter $w$ 
in the equation of state of dark energy, $p=w\rho$. If $w=-1$, then 
the dark energy is equivalent to a cosmological constant~\cite{Living:Carroll} and the 
energy density will be the same at all epochs. If $w>-1$, the dark 
energy is an evolving field whose energy density diminishes in time.
According to~\cite{BBHSirens}, gravitational wave measurements 
have the potential to measure $w$ to an accuracy better than 10\% 
(for advanced ground-based detectors) and around 4\% (for LISA).
The accuracy with which parameters can be measured improves greatly
when one includes in the computation of the covariance matrix
the harmonics of the binary inspiral signal that is normally
neglected~\cite{VanDenBroeck:2006ar}. Arun et al.~\cite{Arun:2007hu} 
have shown that the source 
location in the sky can be greatly improved when the signal harmonics 
(up to fifth harmonic) are included, which further helps in measuring 
the parameter $w$ even better.

%\subsection{Catalogue of binary coalescences and cosmological models}

%\subsection{Lensing of gravitational waves}

%\subsection{Supernova background and the epoch of galaxy formation}

%\subsection{Primordial spectrum of density fluctuations}
%Via observations of MACHO black hole binaries

%==================================================================
\newpage

\section{Conclusions and Future Directions}
\label{sec:conclusions}

The development of gravitational wave detectors to their present
capability has required patience, ingenuity, and dedication by an
entire generation of experimental physicists. No less dedication and
vision have been required by scientific funding organizations of a
half-dozen nations and two major space agencies. The initial data runs
of the LIGO and VIRGO detectors at their first sensitivity goals
(bursts with amplitudes of $10^{-21}$) have not so far yielded any
detections, but this is certainly not surprising. The operation of
these detectors at this sensitivity level has demonstrated that the
technology is understood, and the analysis of the data has provided
important early experience and the opportunity to organize the efforts
into the LSC and VIRGO collaborations. As the detectors are upgraded
during the period 2008\,--\,2014, the first detection could occur at
any time; if the advanced detectors do not make early detections, then
there will inevitably be serious questions about general
relativity. The field of gravitational wave detection has never before
been at the point where it could test the fundamental theory.

Once the first detection is made, there will be increasing emphasis on
the fundamental physics and astrophysics that will follow from further
detections. As we have discussed in this review, one can look forward
soon thereafter to a detailed comparison of black hole mergers with
theory, to exploring the relationship between compact-object mergers
and gamma-ray bursts, to using this association to make a precise and
calibration-free measurement of the Hubble constant, and to population
studies of neutron stars and black holes. In this early phase of
gravitational wave astronomy there are very exciting (but less
certain) potential observations: an unexpectedly strong cosmological
background, which would revolutionize early-universe physics; the
detection of mass asymmetry or normal-mode oscillations of rotating
neutron stars, either of which would for the first time probe the
interior physics of these complex objects and would help unravel the
mystery of the pulsar phenomenon; the first studies of the interior
core dynamics of a supernova, if one happens to occur nearby; the
detection of populations of compact dark objects, like cosmic strings
or small black holes; the discovery of exceptionally-massive black
holes, around $100\msolar$; or the association of gravitational wave
events with transient phenomena other than gamma-ray bursts, such as
transient radio bursts.

When LISA is launched, the physics and astrophysics consequences
become even richer. LISA will study black hole mergers during the
early phases of galaxy formation, exploring the mysterious link
between the two. It will map in detail black hole spacetimes and
verify the black-hole uniqueness and area theorems of general
relativity. It is likely to map the history of the expansion of the
universe through measuring the distances to massive black hole
mergers, and from that look for evidence that the dark energy has been
evolving with time. It will discover every short-period binary system
in our galaxy, calibrating white-dwarf masses, mapping their mass
distribution, determining the population of neutron stars in
binaries. As with ground-based detectors, LISA might make other
discoveries that are harder to predict, such as a cosmological
background, cosmic strings, intermediate-mass black holes, even
$g$-mode oscillations of the sun. LISA has enough sensitivity to be
able to make discoveries even of sources for which there are no signal
models to aid data analysis. And if LISA does not see its verification
binary sources, that will be fatal for general relativity.

Gravitational wave detections may also come from other technologies,
such as pulsar timing searches or observations of the cosmic microwave
background. The spectrum of gravitational waves is enormous, and
present technologies can explore only a tiny fraction of it. Beyond
the LISA timeframe, say after 2020, new technologies may come into
the field and make possible detectors that extend the ground-based
detection band to lower frequencies (such as the Einstein Telescope project),
observing in space in the 0.1~Hz band, going up to megaHertz
frequencies.

The present review has attempted to give a good overview of the
science that can be done with gravitational waves, but it is certainly
not complete. Future revisions are planned to add more on LISA, more
on data analysis issues, and considerably more on detectors that might
go beyond Advanced LIGO and VIRGO. This is a field that is developing
rapidly. For example, the launch of LISA is 10~years away (at the time
of writing, 2008), but already the scientific literature contains many
hundreds of refereed papers on LISA science and technology, and every
second year there is a major international symposium on the
subject. This is probably unprecedented among space
missions. \emph{Living Reviews in Relativity} is planning to release a
suite of articles in the near future on LISA, which will cover
cosmology, tests of general relativity, galactic astrophysics, black
hole astrophysics, and observations of low-frequency gravitational
wave sources with LISA. Until the next revision, readers interested in
keeping up with the field should also consult the proceedings of the
regular conferences on gravitational waves: the Amaldi meetings, GWDAW
(Gravitational Wave Data Analysis Workshops), GWADW (Gravitational
Wave Advanced Detectors Workshops), and the LISA Symposium.

%==================================================================
\newpage

\section{Acknowledgements}

We would like to thank all authors who granted permission to reproduce
their figures. We greatly acknowledge our GEO colleagues, from whom we
have learned so much over the years. Finally, this article would not
have seen the light of day without the insistence of Professor Bala
Iyer; we are thankful to him for his patience and encouragement.

%==================================================================
\newpage

\bibliography{refs}

\end{document}